\documentclass[12pt]{report}

\usepackage{ODUthesisFall2023}
\usepackage{indentfirst}
\usepackage{cite}
\usepackage{latexsym}
\usepackage{amsfonts,amssymb,amsmath}
\usepackage[table]{xcolor}
\usepackage{array,booktabs,ragged2e}
\newcolumntype{P}[1]{>{\centering\arraybackslash}p{#1}}
\newcolumntype{M}[1]{>{\centering\arraybackslash}m{#1}}
\newcolumntype{R}[1]{>{\raggedleft\arraybackslash}m{#1}}
\PassOptionsToPackage{hyphens}{url}\usepackage[pdfauthor={John A. Snoap},
			pdftitle={Deep-Learning-Based Classification of Digitally Modulated Signals},
			pdfsubject={Deep-Learning-Based Classification of Digitally Modulated Signals},
			pdfkeywords={Modulation Classification, Cyclic Cumulants, Neural Networks, Deep-Learning}]{hyperref}
\usepackage[labelsep=period, labelfont=bf]{caption} 
\hypersetup
{
    colorlinks=true, 
    linktoc=all,     
    linkcolor=black, 
    allcolors=black, 
    bookmarksopen=true,
    bookmarksnumbered=true,
    pdfstartview=FitH,
    pdfremotestartview=FitH,
}
\usepackage{bookmark}

\usepackage [english]{babel}
\usepackage [autostyle, english = american]{csquotes}
\MakeOuterQuote{"}

\usepackage{subfigure}
\usepackage[subfigure, titles]{tocloft}
\cftsetindents{figure}{0em}{2.55em} 
\cftsetindents{table}{0em}{2.55em} 
\usepackage{graphicx}
\graphicspath{{./figures/}}
\DeclareGraphicsExtensions{.pdf,.eps}
\usepackage{epstopdf}

\usepackage{algpseudocode}

\begin{document}

\title{Deep-Learning-Based Classification\\of Digitally Modulated Signals}

\author{John A. Snoap}
\principaladviserOne{Dimitrie C. Popescu}
\principaladviserTwo{Chad M. Spooner}
\member{W. Steven Gray}
\member{Jiang Li}

\degrees{B.S. April 2014, Oklahoma Christian University \\ M.S. December 2016, Old Dominion University}

\dept{Electrical \& Computer Engineering}          

\submitdate{December 2023}

\phdtrue



\vita{
{John Andrew Snoap was born on January 7, 1992, in Orlando, Florida.  Because of his father's position in the military, John lived in many different states but finished growing up in Chesapeake, Virginia.  After graduating from Hickory High School in Chesapeake, Virginia in 2010, he went on to earn his Bachelor of Science degree in Electrical Engineering from Oklahoma Christian University in April 2014.  While earning his bachelor's degree, John worked two summer internships at Newport News Shipbuilding (NNS) in Newport News, Virginia.  After earning his bachelor's degree, John began working for NNS, where he is currently employed full-time.  In the Fall of 2014, he enrolled at Old Dominion University as a part-time student, earning his Master of Science degree in Electrical and Computer Engineering in December 2016.  In the Spring of 2017, he continued at Old Dominion University as a part-time student, while still employed full-time at NNS, earning his Doctor of Philosophy in Electrical and Computer Engineering in December 2023.}
\\
{\indent John's notable publications to date include:
\begin{itemize}
\item J. A. Snoap, D. C. Popescu, and C. M. Spooner, ``Novel Nonlinear Neural-Network Layers for High Performance and Generalization in Modulation-Recognition Applications,'' in \emph{Proceedings 2023 IEEE Military Communications Conference (MILCOM)}, Boston, MA, November 2023, pp. 353–358.
\item J. A. Snoap, D. C. Popescu, J. A. Latshaw, and C. M. Spooner, ``Deep-Learning-Based Classification of Digitally Modulated Signals Using Capsule Networks and Cyclic Cumulants,'' \emph{Sensors}, vol. 23, no. 12, 2023. [Online]. Available: https://www.mdpi.com/1424-8220/23/12/5735
\item J. A. Snoap, J. A. Latshaw, D. C. Popescu, and C. M. Spooner, ``Robust Classification of Digitally Modulated Signals Using Capsule Networks and Cyclic Cumulant Features,'' in \emph{Proceedings 2022 IEEE Military Communications Conference (MILCOM)}, Rockville, MD, December 2022, pp. 298–303.
\end{itemize}}
}

\abstract
{
{This dissertation presents several novel deep-learning (DL)-based approaches for classifying digitally modulated signals, one method of which involves the use of capsule networks (CAPs) together with cyclic cumulant (CC) features of the signals.  These were blindly estimated using cyclostationary signal processing (CSP) and were then input into the CAP for training and classification.  The classification performance and the generalization abilities of the proposed approach were tested using two distinct datasets that contained the same types of digitally modulated signals but had distinct generation parameters.  The results showed that the classification of digitally modulated signals using CAPs and CCs proposed in this dissertation outperformed alternative approaches for classifying digitally modulated signals that included conventional classifiers that employed CSP-based techniques, as well as alternative DL-based classifiers that used various conventional neural networks (NNs) with in-phase/quadrature (I/Q) data used for training and classification.}
\\
{\indent Another method of digital modulation classification presented in this dissertation showcases two novel DL-based classifiers that each use a CAP with custom-designed feature extraction layers.  The classifiers take I/Q data as input, and the feature extraction layers are inspired by CSP techniques, which extract the CC features employed by conventional CSP-based approaches to blind modulation classification and signal identification.  Specifically, the feature extraction layers implement a proxy of the mathematical functions used in the calculation of the CC features and include a squaring layer, a raise-to-the-power-of-three layer, and a fast-Fourier-transform (FFT) layer, along with additional normalization and warping layers to ensure that the relative signal powers are retained and to prevent the trainable NN layers from diverging in the training process.  The performance of the proposed CAPs are tested using the same previous two distinct datasets, and numerical results obtained reveal that the proposed CAPs with novel feature extraction layers achieve high classification accuracy while also outperforming conventional DL-based approaches for signal classification in generalization abilities.  Suggestions for further research are also provided.}
}

\beforepreface

\prefacesection{Acknowledgments}
I would like to acknowledge and express my gratitude to Dr. Dimitrie C. Popescu for being both my constructive advisor and, foremost, encouraging me to pursue the enthralling subject of cyclostationary signal processing as my main topic of research.  I am indebted to him for his guidance and criticism during the writing of the many papers that led to this dissertation.  Through his encouragement, I sought to implement my own cyclostationary signal processing software to ensure this research was both valid and meaningful.  It was due to this search that I discovered the Cyclostationary Signal Processing Blog, authored by Dr.~Chad~M.~Spooner, whose general mentality agrees with Napoleon Bonaparte, ``If you want a thing done well, do it yourself.''

I would like to acknowledge and express my gratitude to Dr. Chad M. Spooner for being my most critical advisor and always encouraging me to continue improving my cyclostationary signal processing software so we could completely recreate necessary prior works and subsequently further improve upon them.  Even with modern electronic mail filters, it is difficult to count the number of emails we have back and forth with his guidance on what did not appear accurate, how to proceed with resolving an issue, or encouragement to keep moving on towards a better solution; all of which are primary reasons I was able to accomplish the technical efforts required for this research.

I would like to thank one of Dr. Dimitrie C. Popescu's graduate students, James A. Latshaw, for introducing me to new techniques for optimizing neural networks and for his support using Old Dominion University's High-Performance Computing facilities along the journey, it was very satisfying to help each other out.

I would like to thank Dr. W. Steven Gray and Dr. Jiang Li for serving on my dissertation committee and for how they have always pointed me in an encouraging direction.  I am grateful that they accepted.

I would like to thank Newport News Shipbuilding (NNS) for reimbursing me for the cost of tuition and I would also like to thank my managers and colleagues at NNS for their generous flexibility which allowed me to complete this research.

My entire family, especially my parents for the initial years of this research and my beloved wife for the later years of this research, have always been there for me.  It is not possible to perfectly express how thankful I am for the truly unlimited and continual support.  I am looking forward to raising our children in the nurture and admonition of the Lord (Ephesians~6:4; Matthew~12:48-50).
\afterpreface

\chapter{Introduction}\label{sec:Intro}
Blind classification of digitally modulated signals is a problem that occurs in both military and commercial applications such as signal intelligence, electronic warfare, or spectrum monitoring \cite{Dobre_Sarnoff2005}, and implementing conventional approaches to modulation classification in modern software-defined and cognitive-radio receivers can prove challenging~\cite{Dobre_InstrumentMag2015}.  These approaches use signal processing techniques and are grouped into three distinct classes:
\begin{itemize}
\item Likelihood-based methods \cite{Hameed_etal_TW2009, Xu_etal_TSMC2011}, in which the likelihood function of the received signal is calculated under multiple hypotheses that correspond to the various signals that are expected to be received, and the classification decision is made based on the maximum of this function.  However, likelihood-based approaches are sensitive to variations in signal parameters, which are expected to be estimated, and estimation errors can lead to significant performance degradation \cite{Panagiotou_etal_MILCOM2000}.
\item Feature-based methods that extract information such as amplitude, phase, frequency, kurtosis, stationary moments, stationary cumulants~\cite{HOS_10, Shakra_NRSC_2015, Tang_WRI_2009, Swami_Sadler_TCOM2000}, and subsequently use a decision-tree or polynomial classifier approach for automatic modulation recognition, where the estimated features are often compared to theoretical values derived from stationary stochastic process models.  However, the stochastic processes of these feature-based approaches are not ergodic for digital modulation schemes, resulting in feature estimates that can poorly match their theoretical values, even when the features are estimated correctly.
\item Feature-based methods that use cyclostationary signal processing (CSP) techniques~\cite{Gardner_CSP_pt1_1994, Spooner_CSP_pt2_1994, Dandawate_TSP1994} in which cyclic cumulant (CC) features \cite{Spooner_Asilomar1995, Spooner_Asilomar2001} are extracted from the received signal and classification is accomplished by comparing the values of these features with prescribed values corresponding to the signals that are expected to be received~\cite{Spooner_Asilomar2000}.  As noted in \cite{Dobre_etal_WPC2010}, the performance of CC-based approaches to modulation classification is affected by the presence of multipath fading channels, and robust CC-based classifiers for multipath channels are discussed in \cite{Wu_Saquib_TWC2008, Yan_etal_ComLet2017, Yan_etal_ComLet2018}.
\end{itemize}

As an alternative to the conventional methods mentioned above, in recent years deep-learning (DL)-based techniques using neural networks (NNs) have been explored for classifying digitally modulated signals. Neural networks require extensive training using specific signal data to learn how to distinguish the different types of digitally modulated signals that are expected to be received.  In this direction, there are works such as~\cite{Tim2018, Tim2017}, which train the NNs to classify signals using the unprocessed I/Q signal data, as well as the approaches in \cite{Sun2018, Zhang2020,Kulin2018, Rajendran2018, Bu2020}, in which the NNs are trained using the amplitude/phase or frequency domain representations of the digitally modulated signals for classification.

Since NNs require extensive training on a dataset to become proficient, overfitting and lack of generalization can affect the robustness of the trained NNs such that their performance degrades when presented with inputs that have different probability distributions than those in the training dataset.  This aspect is referred to as the dataset shift problem \cite{Djolonga_etal_CVPR2021} and has been studied for DL-based classification of digitally modulated signals in \cite{Latshaw_COMM2022, Snoap_CCNC_2022}, where it is noted that NNs trained using the I/Q signal data have difficulty maintaining classification performance when presented with digitally modulated signals whose underlying parameters have different probability distributions than those of signals in the training dataset.  There is an infinite set of possibilities for signal generation parameters, so the dataset shift problem cannot simply be overcome by creating a large enough dataset of digitally modulated signals that accounts for all possibilities. 

Motivated by the above-mentioned aspects, this dissertation presents a new approach to DL-based classification of digitally modulated signals that involves the use of capsule NNs~\cite{Sabour_etal_NIPC2017}, which have been used in alternative approaches for DL-based classification of digitally modulated signals \cite{sang2018application, li2020automatic} and outperform other types of NNs \cite{Latshaw_COMM2022}.  The proposed approach uses the capsule networks together with CC features obtained by CSP \cite{Spooner_Asilomar2001}, which are known for their robustness to co-channel signals  \cite{Spooner_Asilomar1995} and to variations in noise models~\cite{Hazza_etal_EurasipWCN2011}.  There is a related approach in \cite{SCF_NN_2021}, which uses a convolutional NN and the spectral correlation function (SCF) \cite{Gardner_CSP_pt1_1994, Spooner_CSP_pt2_1994} in the context of spectrum sensing to classify wireless signals based on standards such as GSM, UMTS, and LTE.  To assess the robustness and generalization abilities of the proposed DL-based digital modulation classifier two distinct datasets are used, which are publicly available from~\cite{CSPblog_DataSets}, that include signals with similar digital modulation schemes but which have been generated using distinct parameters, such that data from the testing dataset is not used in the training dataset.

\newpage

\section{Dissertation Contributions}\label{sec:DissContributions}
This dissertation is composed of proposed solutions to a critical problem that arises when using deep-learning neural networks to classify digitally modulated signals.

A critical issue with deep-learning neural networks is their inability to generalize (or properly function) on new data that conforms to different distributions than the training dataset.  This issue is introduced for digital modulation classification in Chapter~\ref{ch:DLCoDMSUIQSD} where examples are presented showing the failure of I/Q-trained NNs to perform well on datasets they were not trained on, specifically when the underlying distribution of signal generation parameters are slightly different between the training and testing datasets.  While a new capsule network (CAP) is developed that improves over other NN types on classifying I/Q data, it too fails to generalize to out-of-distribution data.  The results of this chapter are published in the following two conference papers:  \cite{Snoap_CCNC_2022, Latshaw_COMM2022}.

Chapters~\ref{ch:CSP_BackInfo} and~\ref{ch:CSPModClass} introduce cyclostationary signal processing and how it can be used, without NNs, to extract cyclic cumulants (CCs) from I/Q data and perform reliable modulation classification.  In Chapter~\ref{ch:CCsNNsModClass} a solution is proposed to use extracted CCs to train capsule networks such that the trained CAPs can classify new CCs, even when the signal generation parameters are different from those used to train the NN.  The results of this study are published in conference paper~\cite{Snoap_MILCOM_2022}.  Also, in journal article~\cite{Snoap_Sensors_2023}, more details are provided concerning the CC-trained neural network in~\cite{Snoap_MILCOM_2022} and its results are compared with a CSP baseline classification model which does not use neural networks for classification, but rather calculates distances to theoretical CC features for classification.

The generalization problem discussed in Chapter~\ref{ch:DLCoDMSUIQSD} along with the proposed solution of using extracted CCs to solve the generalization problem in Chapters~\ref{ch:CSPModClass} through~\ref{ch:CCsNNsModClass} are completed and published.  The proposed combination of CCs with CAPs provides better results than the other methods it is compared to, especially when classifying out-of-distribution data (test signals formed from different signal generation distributions).

The use of low-quality datasets for NN modulation classification appears to be prevalent in recent literature.  Many recent works~\cite{Tim2018, Bu2020, Sun2018, Kulin2018, Zhang2020, Rajendran2018} have used publicly available datasets which are not appropriate for evaluating the out-of-distribution data shift problem, yet their work is presented as though it is superior to methods that properly apply CSP.  So, in Chapter~\ref{ch:HQDatasets} of this dissertation, explanations on how to use two high-quality datasets~\cite{DataSetforMLChallenge, ShiftedDataSetforMLChallenge} are provided to encourage the research community to use high-quality datasets when training and testing new NN structures to perform modulation classification, the details of which have also been furnished in IEEE DataPort~\cite{IEEE_DataPort}.

Finally, proposed solutions for how to enable an NN to generalize well with I/Q data rather than CCs which are extracted from I/Q data are presented in Chapter~\ref{ch:NovelLayers}.  The first proposed solution is published in conference paper~\cite{Snoap_MILCOM_2023} and the final proposed solution has been submitted to the IEEE journal Transactions on Pattern Analysis and Machine Intelligence (TPAMI).  The final proposed novel neural network layers enable the CAP to exceed the performance of the CSP baseline model and are close to matching the performance of the CC-trained CAP model, showing that these novel neural network layers enable a CAP to achieve high classification performance and good generalization on I/Q data, without the need to estimate cyclic cumulants.

\newpage

\chapter{Deep-Learning Classification of Digitally Modulated Signals Using I/Q Signal Data}\label{ch:DLCoDMSUIQSD}
DL-based approaches to digital modulation classification implement neural networks (NNs) and rely on their extensive training to make the distinction among different classes of digitally modulated signals. Usually, the available dataset is partitioned into subsets that are used for training/validation and testing, respectively, and it is not known in general how well NNs trained on a specific dataset respond to similar signals that are generated differently than those in the dataset used for training and validation. However, the problem of the dataset shift, also referred to as out-of-distribution generalization \cite{Djolonga_etal_CVPR2021}, is an important problem in machine-learning-based approaches, which implies that the training and testing datasets are distinct, and that data from the testing environment is not used for training the classifier.  This problem motivates the work presented in this dissertation, which studies the performance of DL-based classification of digitally modulated signals when the training and testing data are taken from distinct datasets.

\section{Neural Network Models for Digital Modulation Classification}\label{sec:NNModels}
DL NNs consist of multiple interconnected layers of neuron units and include an input layer, which takes as input the available data for processing, several hidden layers that provide various levels of abstractions of the input data, and an output layer, which determines the final classification of the input data \cite{Goodfellow-et-al-2016}. The hidden layers of an NN include:
\begin{itemize}
\item Convolutional layers, which may be followed by batch normalization to increase regularization and avoid overfitting.
\item Fully connected layers, which may be preceded by dropout layers.
\item Nonlinear layers, with the two common types of nonlinearities employed being the rectified linear unit (ReLU) and the scaled exponential linear unit (SELU).
\item Pooling layers that use average or maximum pooling to provide invariance to local translation.
\item A final {SoftMax} layer that establishes the conditional probabilities for input data classification by determining the output unit activation function for multi-class classification problems.
\end{itemize}

\begin{table}
\centering
\caption[I/Q-trained Residual Neural Network Layout.]{\textbf{I/Q-trained Residual Neural Network Layout.}}
{
\begin{tabular}{ c c }
\toprule
\textbf{Layer} & \textbf{Output Dimensions} \\
\midrule
Input & $2 \times 1024$ \\
Residual Stack & $32 \times 512$ \\
Residual Stack & $32 \times 256$ \\
Residual Stack & $32 \times 128$ \\
Residual Stack & $32 \times 64$ \\
Residual Stack & $32 \times 32$ \\
Residual Stack & $32 \times 16$ \\
Drop($50\%$)/FC/SELU & $128$ \\
Drop($50\%$)/FC/SELU & $128$ \\
Drop($50\%$)/FC/SoftMax & \# Classes \\
\bottomrule
\end{tabular}
}\label{table:RNnetwork}
\end{table}

\begin{table}
\centering
\caption[Residual Stack Layers.]{\textbf{Residual Stack Layers.}}
{
\begin{tabular}{ c c }
\toprule
\textbf{Layer} & \textbf{Output Dimensions} \\
\midrule
Input & $X \times Y$ \\
$1 \times 1$ Conv & $32 \times Y$ \\
Batch Normalization & $32 \times Y$ \\
ReLU & $32 \times Y$ \\
Residual Unit & $32 \times Y$ \\
Residual Unit & $32 \times Y$ \\
Maximum Pooling & $32 \times Y/2$ \\
\bottomrule
\end{tabular}
}\label{table:ResStack}
\end{table}

\begin{table}
\centering
\caption[Residual Unit Layers.]{\textbf{Residual Unit Layers.}}
{
\begin{tabular}{ c c }
\toprule
\textbf{Layer} & \textbf{Output Dimensions} \\
\midrule
Input & $X \times Y$ \\
Conv & $32 \times Y$ \\
Batch Normalization & $32 \times Y$ \\
ReLU\_1 & $32 \times Y$ \\
Conv & $32 \times Y$ \\
Batch Normalization & $32 \times Y$ \\
ReLU\_2 & $32 \times Y$ \\
Addition(Input, ReLU\_2) & $32 \times Y$ \\
\bottomrule
\end{tabular}
}\label{table:ResUnit}
\end{table}

\begin{table}
\centering
\caption[I/Q-trained Convolutional Neural Network Layout.]{\textbf{I/Q-trained Convolutional Neural Network Layout.}}
{
\begin{tabular}{ c c }
\toprule
\textbf{Layer} & \textbf{Output Dimensions} \\
\midrule
Input					& $2 \times 1024$ \\
Conv					& $16 \times 1024$ \\
Batch Normalization		& $16 \times 1024$ \\
ReLU					& $16 \times 1024$ \\
Maximum Pooling			& $16 \times 512$ \\
Conv					& $24 \times 512$ \\
Batch Normalization		& $24 \times 512$ \\
ReLU					& $24 \times 512$ \\
Maximum Pooling			& $24 \times 256$ \\
Conv					& $32 \times 256$ \\
Batch Normalization		& $32 \times 256$ \\
ReLU					& $32 \times 256$ \\
Maximum Pooling			& $32 \times 128$ \\
Conv					& $48 \times 128$ \\
Batch Normalization		& $48 \times 128$ \\
ReLU					& $48 \times 128$ \\
Maximum Pooling			& $48 \times 64$ \\
Conv					& $64 \times 64$ \\
Batch Normalization		& $64 \times 64$ \\
ReLU					& $64 \times 64$ \\
Maximum Pooling			& $64 \times 32$ \\
Conv					& $96 \times 32$ \\
Batch Normalization		& $96 \times 32$ \\
ReLU					& $96 \times 32$ \\
Average Pooling			& $96$ \\
Drop($0\%$)/FC/SoftMax	& \# Classes \\
\bottomrule
\end{tabular}
}\label{table:CNNnetwork1}
\end{table}

Commonly used types of NNs in DL include convolutional neural networks (CNNs) and residual networks (RNs), with the latter including bypass connections between layers that enable features to operate at multiple scales and depths throughout the NN \cite{Goodfellow-et-al-2016}.

The study on DL-based classification of digitally modulated signals presented in this dissertation aims to assess the generalization ability of trained NNs and to demonstrate the need for more robust testing of NNs trained to recognize digital modulation schemes.  In this direction, both an RN with the layout specified in Tables~\ref{table:RNnetwork}--\ref{table:ResUnit} and a CNN with the layout specified in Table~\ref{table:CNNnetwork1} are considered.  For both types of NNs (RN and CNN) the convolutional layers use a filter size of $23$, as increasing the filter size increases the computational cost with no performance gain while decreasing it lowers performance. 

The NN structures outlined in Tables~\ref{table:RNnetwork}--\ref{table:CNNnetwork1} are similar to those in~\cite{Tim2018}, and the novelty of this chapter consists in the use of two distinct datasets for training and testing, respectively, which will allow evaluation of the NNs' ability to generalize and continue to distinguish the classes of digitally modulated signals for which they have been trained when tested on signals that are generated differently than the signals in the training dataset.

\section{Datasets for NN Training and Testing the Classification of Digitally Modulated Signals}\label{sec:Datasets1}
The DL NNs used for classification of digitally modulated signals with the structure outlined in Section~\ref{sec:NNModels} are trained and tested using digitally modulated signals in two distinct datasets that are publicly available for general use, which are referred to as \texttt{RADIOML 2018.01A}~\cite{DeepSigDataSet} and \texttt{CSPB.ML.2018}~\cite{DataSetforMLChallenge}, respectively.  These datasets contain collections of the I/Q data corresponding to signals generated using common digital modulation schemes that include BPSK, QPSK, 8PSK, 16QAM, 64QAM, and 256QAM, with different signal-to-noise ratios (SNRs).  Brief details on the signal generation methods are also provided in the descriptions of the two datasets, and the signals include overlapping excess bandwidths and the use of square-root raised-cosine (SRRC) pulse shaping with similar roll-off parameters.

According to \cite{Tim2018}, \texttt{RADIOML 2018.01A} includes both simulated signals and signals captured over-the-air, with $24$ different modulation types and $26$ distinct SNR values for each modulation type, ranging from $-20$~dB to $30$~dB in $2$~dB increments.  For each modulation type and SNR value combination, there are $4,096$ signals included in the dataset, with $1,024$ I/Q samples for each signal, and the total number of signals in \texttt{RADIOML 2018.01A} is $2,555,904$.

\texttt{CSPB.ML.2018} \cite{DataSetforMLChallenge} contains only simulated signals corresponding to $8$~different digital modulation types with SNRs varying between $0$~dB and $13$~dB.  The dataset includes a total of $112,000$ signals, with $32,768$ I/Q samples for each signal.

One key difference between \texttt{RADIOML 2018.01A} and \texttt{CSPB.ML.2018} is that in the former the total SNR (i.e., the ratio of the signal power to the total noise power within the sampling bandwidth) is used, while in the latter the in-band SNR (i.e., the ratio of the signal power to the power of the noise falling within the signal's actual bandwidth) is used.  Because the listed SNRs for the signals in \texttt{CSPB.ML.2018} correspond to in-band SNR values, the total SNR corresponding to the sampling bandwidth for signals in \texttt{CSPB.ML.2018} has also been estimated. This turned out to be about $8$~dB below the in-band SNRs and is useful for a side-by-side comparison with the signals in \texttt{RADIOML 2018.01A}.

Although the two datasets considered contain common modulation types, the signals in \texttt{RADIOML 2018.01A} have been generated independently and by different means than the signals in \texttt{CSPB.ML.2018}, which implies that they are well suited for this study to assess the out-of-distribution generalization ability of a trained NN to classify digitally modulated signals.

\section{NN Training and Numerical Results}\label{sec:NumericalResults}
Numerical results have been obtained by training both the RN and CNN twice, first using \texttt{RADIOML 2018.01A} and then again separately using \texttt{CSPB.ML.2018}.  To make the signals in \texttt{CSPB.ML.2018} compatible with a DL NN trained on \texttt{RADIOML 2018.01A}, each signal in \texttt{CSPB.ML.2018} is split into $32$ separate signals with the same class labels to both increase the total number of signals and reduce the samples per signal to $1,024$, which yielded a total of $3,584,000$ digitally modulated signals, each with $1,024$ samples, similar to the ones in \texttt{RADIOML 2018.01A}.

For each training, the datasets have been divided into subsets that have been used as follows:  $75\%$ of the signals in a given dataset have been used for training, $12.5\%$ for validation during training, and the remaining $12.5\%$ for testing after training has been completed.  Prior to training, the signals were normalized to unit power, and the stochastic gradient descent with momentum (SGDM) algorithm~\cite{SGDM} was implemented for training with a mini-batch size of $256$ training signals.  Twelve training epochs were used, as further training beyond 12~epochs did not appear to result in significant performance improvement.

The NNs have been implemented in MATLAB and trained on a high-performance computing cluster with 21 graphical processing unit (GPU) nodes available, consisting of Nvidia K40, K80, P100, and V100 GPU(s) with 128 GB of memory per node.  NN training is computationally intensive and takes about 34 hours to complete training for one NN on a single Nvidia K80 GPU.

\begin{figure}
\begin{center}
\subfigure[Training and Testing Using \texttt{RADIOML 2018.01A}.]
{\includegraphics[width=0.49\linewidth]{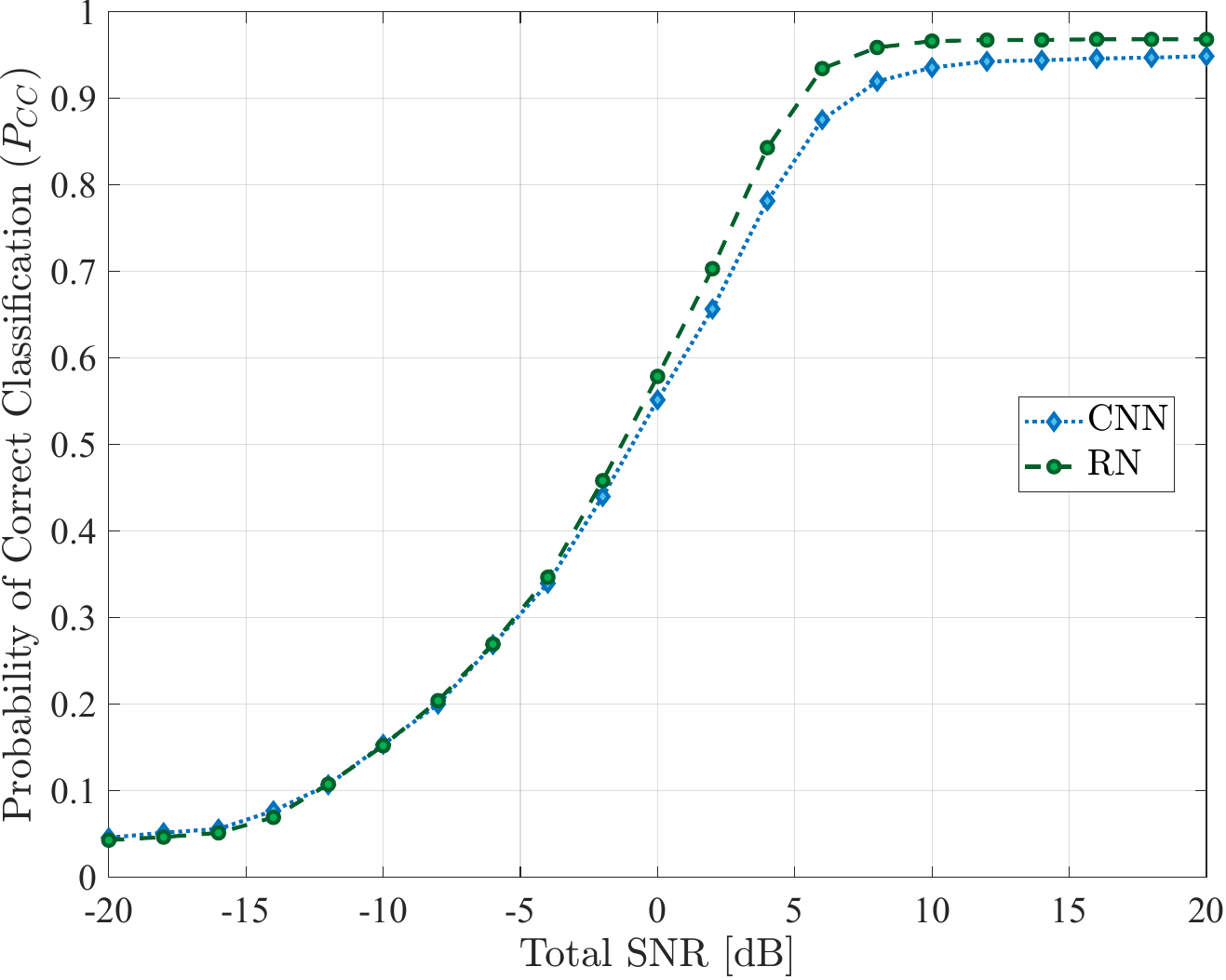}}
\subfigure[Training and Testing Using \texttt{CSPB.ML.2018}.]
{\includegraphics[width=0.49\linewidth]{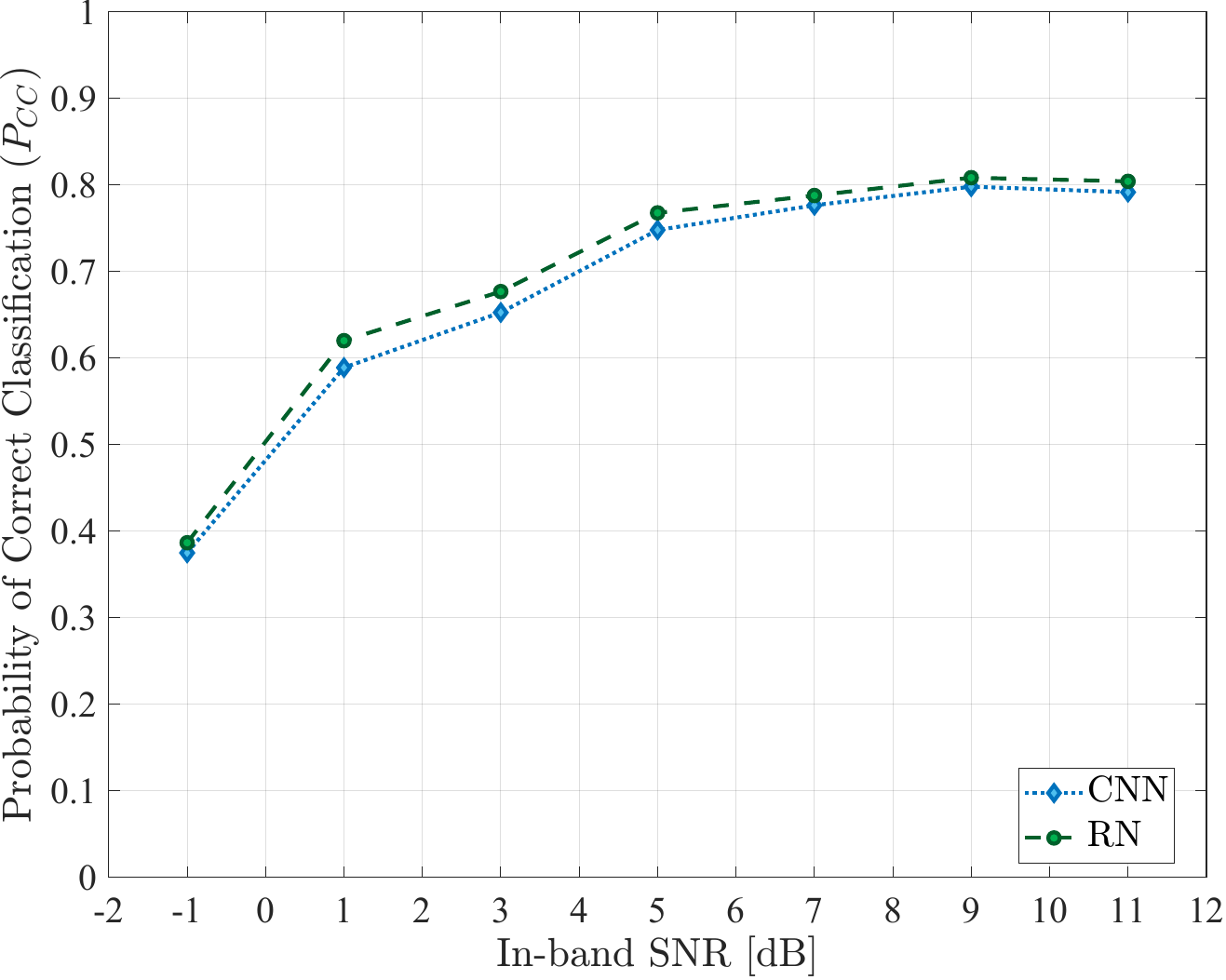}}
\caption[Classification Performance of Neural Networks Trained on \texttt{RADIOML 2018.01A} or \texttt{CSPB.ML.2018} I/Q Data When Training and Testing Signals Come From the Same Dataset.]{\textbf{Classification Performance of Neural Networks Trained on \texttt{RADIOML 2018.01A} or \texttt{CSPB.ML.2018} I/Q Data When Training and Testing Signals Come From the Same Dataset.}}\label{fig:Train=Test}
\end{center}
\end{figure}

\begin{figure}
\centering
\includegraphics[width=\linewidth]{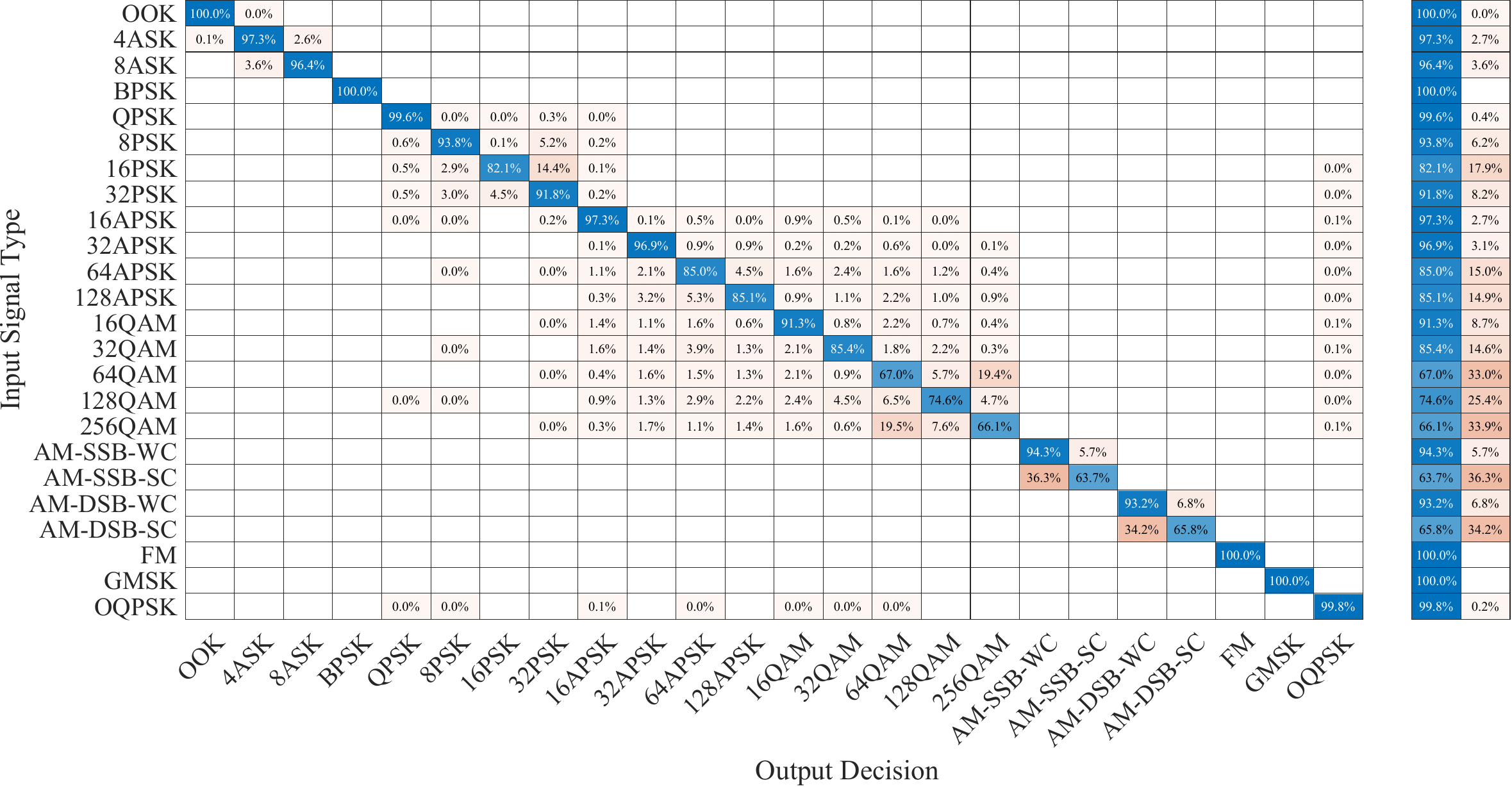}
\caption[Confusion Matrix Results of the CNN Trained and Tested on \texttt{RADIOML 2018.01A} I/Q Data.]{\textbf{Confusion Matrix Results of the CNN Trained and Tested on \texttt{RADIOML 2018.01A} I/Q Data.}}\label{fig:DS_CM_CNN}
\end{figure}

\begin{figure}
\centering
\includegraphics[width=\linewidth]{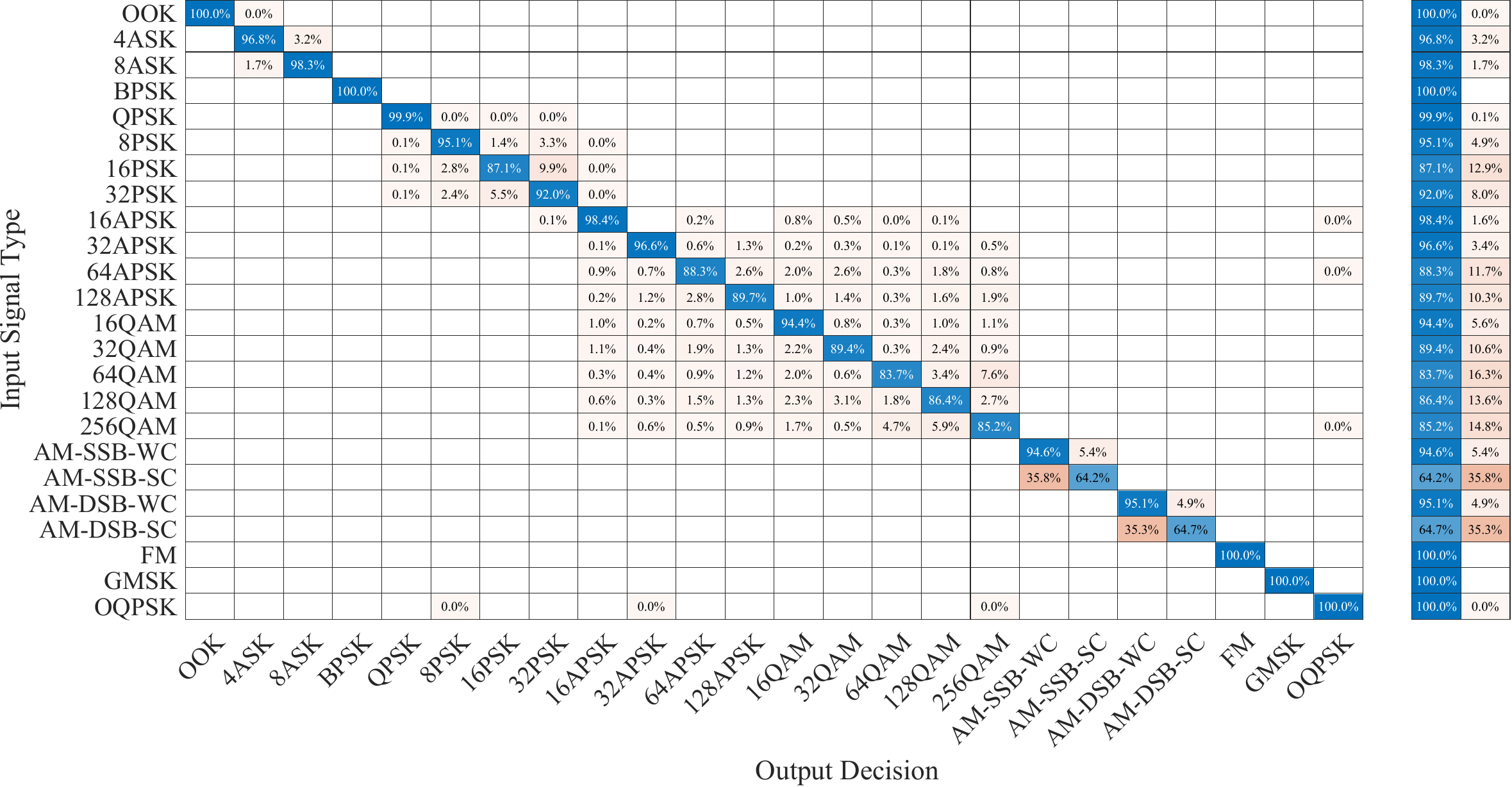}
\caption[Confusion Matrix Results of the RN Trained and Tested on \texttt{RADIOML 2018.01A} I/Q Data.]{\textbf{Confusion Matrix Results of the RN Trained and Tested on \texttt{RADIOML 2018.01A} I/Q Data.}}\label{fig:DS_CM_RN}
\end{figure}

The first experiment performed involves evaluating the performance of the trained RN and CNN using test signals coming from the same dataset as the signals used for training. Results from this experiment are shown in Figure~\ref{fig:Train=Test}, from which it can be seen that, as expected for a signal processing algorithm, the classification performance improves with increasing SNR. It can also be seen that the use of \texttt{RADIOML 2018.01A} for NN training and testing corresponds to the same scenario considered in \cite{Tim2018}, and the corresponding plot shown in Figure~\ref{fig:Train=Test}(a) is similar to the probability of classification plots presented in \cite{Tim2018}.  This is further corroborated by confusion matrix results, which are similar to those in \cite{Tim2018} and are shown in Figure~\ref{fig:DS_CM_CNN} and Figure~\ref{fig:DS_CM_RN}.

When \texttt{CSPB.ML.2018} is used for NN training and testing, the corresponding probability of classification plot shown in Figure~\ref{fig:Train=Test}(b) levels at a slightly lower value than when \texttt{RADIOML 2018.01A} is used. This is because the two datasets are different, with more random variables in \texttt{CSPB.ML.2018} including a larger range of SRRC roll-off values, CFOs, randomized symbol rates, and signal power levels.  Also, the architectures of the RN and CNN employed were very similar to those considered in \cite{Tim2018} and have been optimized for \texttt{RADIOML 2018.01A}; further improvement on \texttt{CSPB.ML.2018} may be possible by changing the NN architecture.

\begin{figure}
\begin{center}
\subfigure[Training Using \texttt{RADIOML 2018.01A}, Testing Using \texttt{CSPB.ML.2018}.]
{\includegraphics[width=0.49\linewidth]{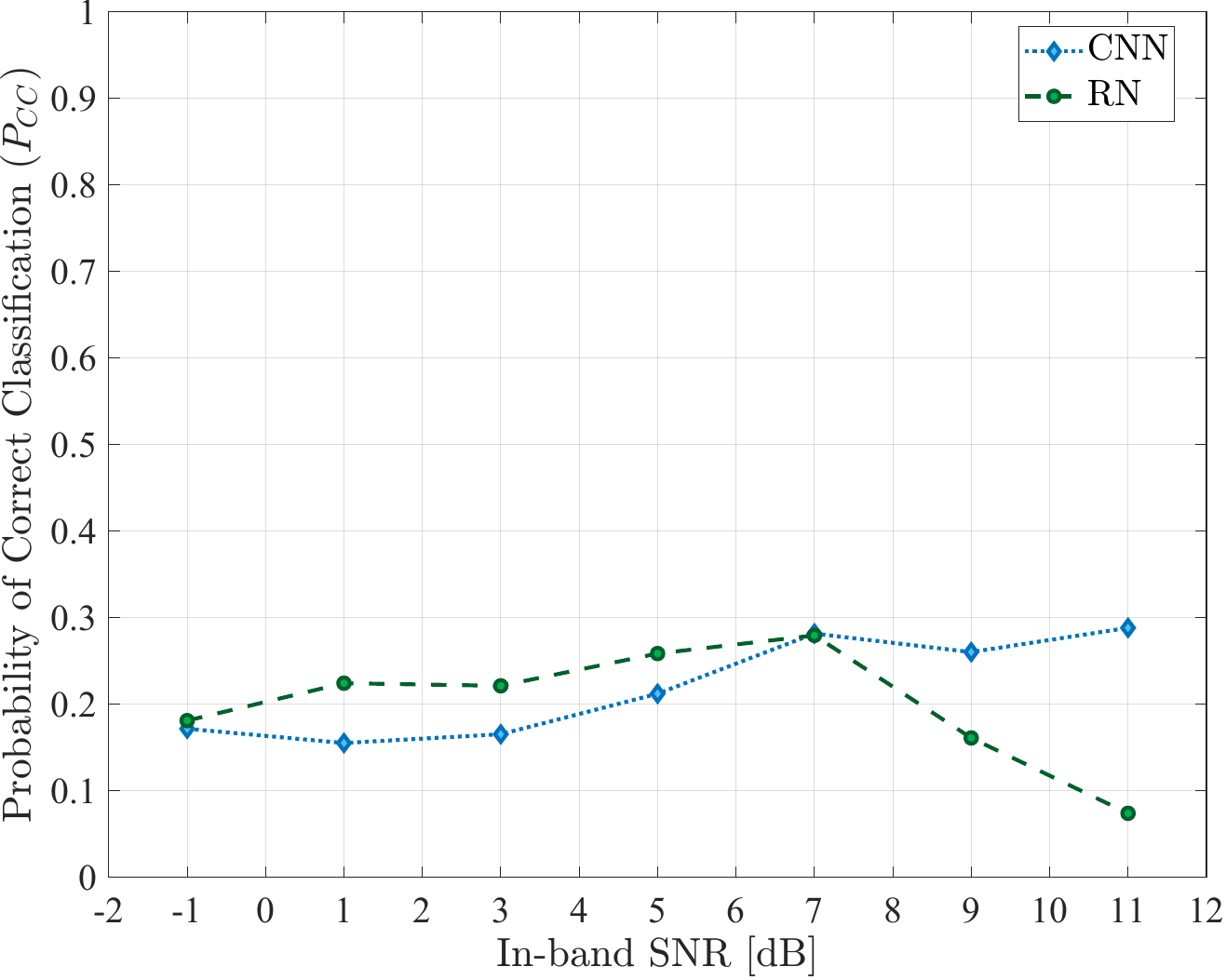}}
\subfigure[Training Using \texttt{CSPB.ML.2018}, Testing Using \texttt{RADIOML 2018.01A}.]
{\includegraphics[width=0.49\linewidth]{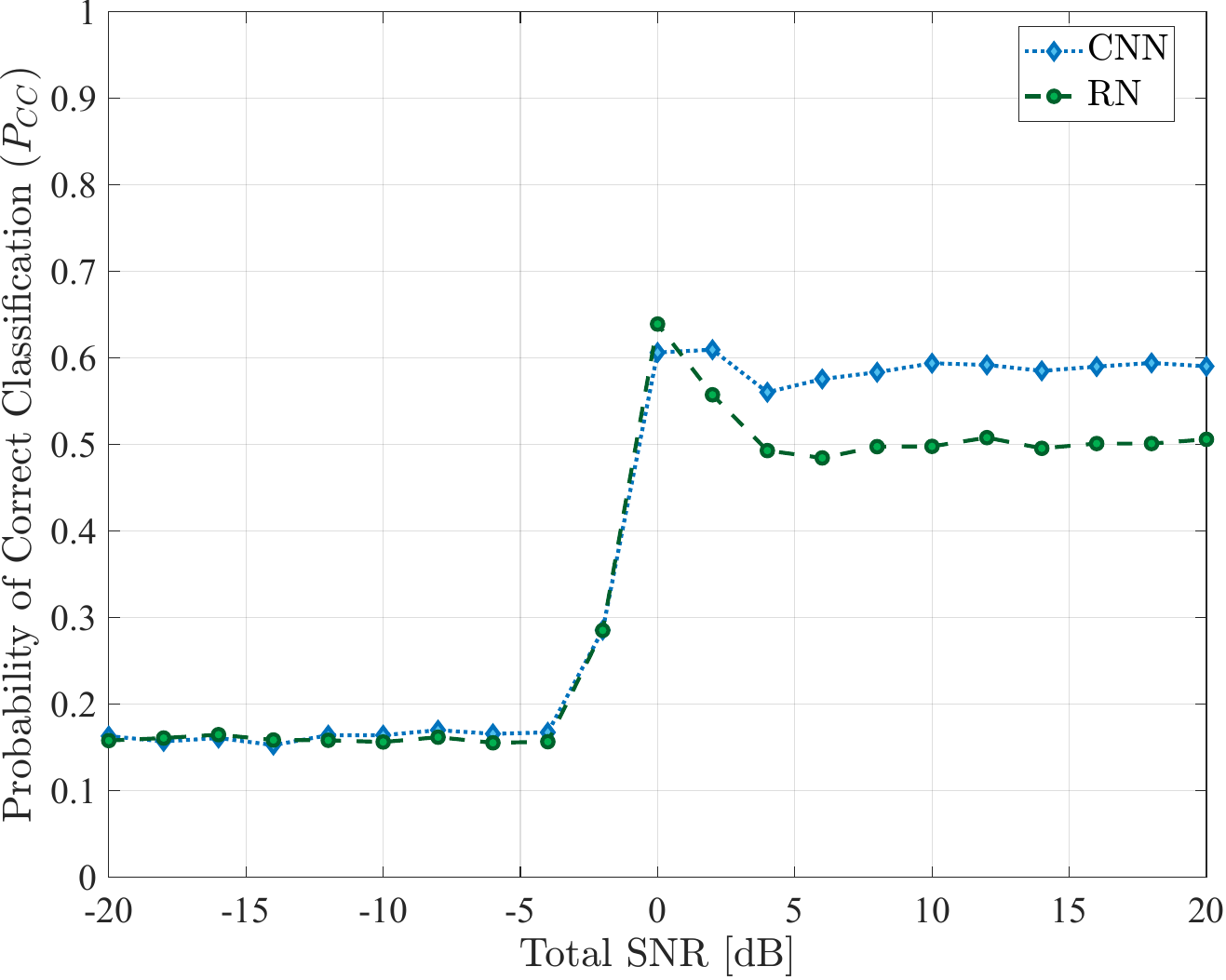}}
\caption[Classification Performance of Neural Networks Trained on \texttt{RADIOML 2018.01A} or \texttt{CSPB.ML.2018} I/Q Data When Training and Testing Signals Come From Different Datasets.]{\textbf{Classification Performance of Neural Networks Trained on \texttt{RADIOML 2018.01A} or \texttt{CSPB.ML.2018} I/Q Data When Training and Testing Signals Come From Different Datasets.}}\label{fig:Train!=Test}
\end{center}
\end{figure}

In the second experiment, the performance of trained RN and CNN is evaluated using test signals coming from a different dataset than the one used for training.  Results from this experiment are shown in Figure~\ref{fig:Train!=Test}, from which it is apparent that the generalization performance of the trained NNs is poor, and that they fail to recognize digitally modulated signals belonging to classes they have been trained on when these signals come from a dataset that is different than the one the NNs have been trained on. The use of \texttt{RADIOML 2018.01A} for training the NNs does not seem to imply any generalization abilities as the corresponding probability of correct classification shown in Figure~\ref{fig:Train!=Test}(a) is essentially flat around $0.2$ for all SNR values.

When using \texttt{CSPB.ML.2018} for training the NNs, the corresponding probability of correct classification shown in Figure~\ref{fig:Train!=Test}(b) displays a jump around $0$~dB SNR from a low value of $0.2$ leveling at values around $0.5$ and $0.6$ for the RN and CNN, respectively.  Thus, in this case, while the overall classification performance is below that obtained in the first experiment or reported in \cite{Tim2018}, the trained NNs seem to have some limited generalization ability.


%

\section{Capsule Networks for Digital Modulation Classification}\label{sec:CAPsModels}
Capsule networks aim to focus on learning desirable characteristics of the input pattern or signal, which correspond to a specific input class, and they have been used in attempts to emulate human vision. When an eye receives a visual stimulus, the eye does not focus on all available inputs, instead points of fixation are established and these points are used to identify or reconstruct a mental image of the object of focus \cite{Sabour_etal_NIPC2017}.  This specificity is achieved by means of capsules, which are multiple parallel and independent nodes that learn class-specific characteristics and represent points of fixation in the form of a capsule vector.  In the context of classifying digitally modulated signals, the capsules are expected to discover characteristics that are intrinsic to a modulation type \cite{li2020automatic, sang2018application}.

\begin{figure}
\centering
\includegraphics[width=0.75\linewidth]{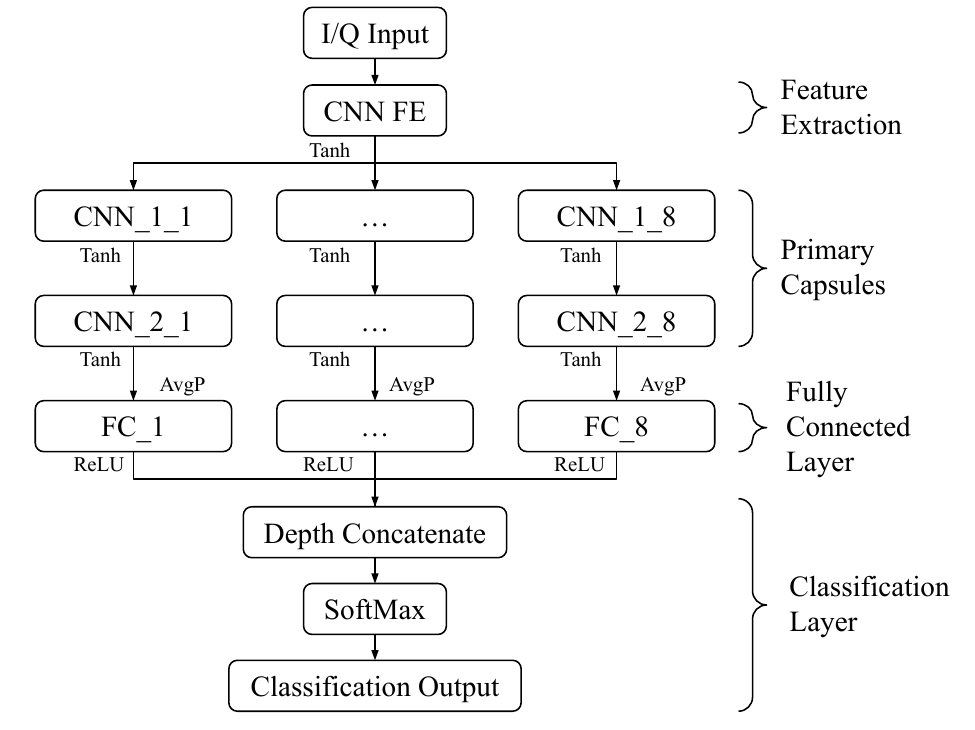}
\caption[Topology of the I/Q-trained Capsule Network for Classification of Digitally Modulated Signals With Eight Branches.]{\textbf{Topology of the I/Q-trained Capsule Network for Classification of Digitally Modulated Signals With Eight Branches.}}\label{fig:topology}
\end{figure}

In general, a capsule network (CAP) is a shallow convolutional neural network (CNN) that consists of a feature-extracting layer followed by parallel CNN layers referred to as primary capsules.  Each of the parallel primary capsule layers has a ``capsule vector'' as the output, which is referred to as a digit capsule, and has dimension $1 \times N$ neurons.  This is different than in the case of CNN approaches, which rely on a single output neuron per class.  The value of $N$ is a design parameter and corresponds to the points of fixation that the capsule network may learn, which are class-specific attributes discovered by the network during training. Ideally, the magnitude of the capsule vector corresponds to the probability that the input matches the output corresponding to this digit capsule and its orientation carries information related to the input properties.  The neurons in the digit capsules have connections to neurons in the primary capsule layers and can be determined iteratively using the dynamic routing by agreement algorithm in \cite{Sabour_etal_NIPC2017}, which has also been applied for modulation classification of digitally modulated signals \cite{li2020automatic, sang2018application}.

In this dissertation, the capsule network with topology illustrated in Figure~\ref{fig:topology} is considered, and its use is studied in the context of classification of digitally modulated signals.  Furthermore, instead of using dynamic routing by agreement algorithm to update connections between a given capsule vector and higher layer neurons, all higher layer neurons are fully connected to each neuron in a $1\times N$ neuron vector.  These neurons will discover desirable attributes in the previous primary capsules and will activate on these characteristics.  The considered topology allows for easy implementation from a design perspective and for efficient training as matrix operations are efficiently processed by graphic processing units (GPUs), whereas iterative learning through dynamic routing by agreement may result in increased computational complexity.

The capsule network shown in Figure~\ref{fig:topology} takes as input sampled versions of the normalized in-phase (I) and quadrature (Q) components of a digitally modulated signal, which must be classified as corresponding to one of the following eight digital modulation schemes: BPSK, QPSK, 8PSK, $\pi/4$-DQPSK, MSK, 16QAM, 64QAM, and 256QAM.  The various components of the capsule network considered in Figure~\ref{fig:topology} include:
\begin{itemize}
\item \textbf{Feature Extraction Layer:} This is the first layer of the network that performs a general feature mapping of the input signal, and its parameters are inspired by the CNNs used for classification of digitally modulated signals in \cite{Tim2018, Zhou_etal_EURASIP_SP2019, Snoap_CCNC_2022}, to include a convolutional layer followed by a batch normalization layer and an activation function.
\item \textbf{Primary Capsules:} This layer consists of a number of primary capsules that is equal to the number of digital modulation classes considered, which are operating in parallel using as inputs the output  from the feature extraction layer.  Each primary capsule in this layer includes two convolutional layers with customized filter and stride, and an activation function, and is followed by a fully connected layer.
\item \textbf{Fully Connected Layer:} This layer consists of a $1 \times N$  neuron vector with the weights connecting to the previous layer.  Each neuron in the last layer of the primary capsules will be fully connected to each neuron in this layer.  These connections are expected to, ideally, discover characteristics specific to each capsule's class.  To make the output of the network compatible with a {SoftMax} classification layer, each neuron within this layer is fully connected to a single output neuron, and the output neurons for all primary capsules will be combined depth-wise to produce an \linebreak[4] 8-dimensional vector \textbf{n}, which is passed to the classification layer.  The value of each respective element of \textbf{n} will be representative of the likelihood that its corresponding modulation type is present in the I/Q input data.
\item \textbf{Classification Layer:} This vector \textbf{n} is passed to the {SoftMax} layer, which will map each element $n_i$, $i=1,\ldots,8$, in \textbf{n} to a value $\sigma_i (\textbf{n})$ that is between $[0,1]$, with each element representing the probability of occurrence, such that the sum of elements in \textbf{n} adds up to $1$ \cite{luce_2008}:
\begin{align}
\sigma_i (\textbf{n}) =  \frac{e^{n_{i}}}{\displaystyle \sum_{j=1}^{8} e^{n_{j}}}. \label{eq:sm1}
\end{align}
This provides a convenient way to determine which modulation type is most likely to correspond to the signal with the I/Q data at the input of the capsule network.
\end{itemize}

More specific details on the capsule network parameters, such as filter sizes, strides, output dimensions, etc., are given in Table~\ref{table:CNNnetwork2}.

\begin{table}
\centering
\caption[I/Q-trained Capsule Network Layout.]{\textbf{I/Q-trained Capsule Network Layout.}}
{
\begin{tabular}{ c c c c}
\toprule
\textbf{Layer} & \textbf{Filter} & \textbf{Stride} & \textbf{Size/Weights} \\
\midrule
Input & & & 2x32,768\\
Conv & [1,22] & [1,9] & 22x2x64 \\
Batch Normalization		    & $ $ \\
Tanh					        & $ $ \\
Conv-1-(i)	&[1 23]& [1,7] & 23x64x48	 \\
Batch Normalization-1-(i)		& \\
Tanh-1-(i)					    & \\
Conv-2-(i)	& [1 22] & [1,8]	 & 22x48x64 \\
Batch Normalization-2-(i)		& \\
Tanh-2-(i)					    &  \\
Average Pool (i) & [1,8] & [1,1]  &  \\
FC-(i)& &		                & 32 \\
Batch Normalization-3-(i)		& \\
ReLu-1-(i)					    &  \\
Point FC-(i)		   & &       	& 1\\
Depth Concatenation(i=1:8)     & & & 8  \\
SoftMax	                    &   \\
\bottomrule
\end{tabular}
}\label{table:CNNnetwork2}
\end{table}

\newpage

\section{Datasets for Capsule Network Training and Testing}\label{sec:Datasets2}
A capsule network with the structure outlined in Section~\ref{sec:CAPsModels} is trained and tested using digitally modulated signals in two distinct datasets that are publicly available for general use~\cite{CSPblog_DataSets}.  These are referred to as \texttt{CSPB.ML.2018}~\cite{DataSetforMLChallenge} and \texttt{CSPB.ML.2022}~\cite{ShiftedDataSetforMLChallenge}, respectively, and each of them contains collections of the I/Q data corresponding to a total of $112,000$ computer-generated digitally modulated signals that include BPSK, QPSK, 8PSK, $\pi/4$-DQPSK, 16QAM, 64QAM, 256QAM, and MSK modulation schemes.  Signals employ square-root raised-cosine (SRRC) pulse shaping with a roll-off factor in the interval $[0.1, 1.0]$ and a total of $32,768$ samples for each signal are included in the datasets.

The listed signal-to-noise ratios (SNRs) for the signals in both \texttt{CSPB.ML.2018} and \texttt{CSPB.ML.2022} correspond to in-band SNR values, and a band-of-interest (BOI) detector \cite{BOIdetector} was used to validate the labeled SNRs, CFOs, and SRRC roll-off values for the signals in both datasets.

\subsection{\texttt{CSPB.ML.2018}}
\texttt{CSPB.ML.2018} is available for download from \cite{CSPblog_DataSets}, and additional characteristics for signals in this dataset include:
\begin{itemize}
\item Symbol rates vary between $1$ and $23$ samples/symbol.
\item The in-band SNR varies between $0$ and $12$~dB.
\end{itemize}

\subsection{\texttt{CSPB.ML.2022}}
\texttt{CSPB.ML.2022} is available for download from \cite{CSPblog_DataSets}, and for this dataset, the additional signal characteristics include:
\begin{itemize}
\item Symbol rates vary between $1$ and $29$ samples/symbol.
\item The in-band SNR varies between $1$ and $18$~dB. 
\end{itemize}
The symbol rates and CFOs differ between the two datasets, and this difference will be used to study the generalization ability of the trained capsule networks.  The CFOs are distributed over disjoint intervals, whereas the symbol rates overlap substantially.

\newpage

\subsection{Data Augmentation}
It is expected that the trained digital modulation classifier will perform well at classifying high in-band SNR signals and perform less desirably as the SNR decreases.  This is visible in similar digital modulation classification approaches \cite{Snoap_CCNC_2022, Tim2017, Tim2018, Zhou_etal_EURASIP_SP2019}.  However, for the two datasets used, there are few samples for the lower SNR values, and the small sample size for the lower SNR values is not meaningful in evaluating the capsule network performance for the lower SNR values.  To overcome this aspect, a process of data augmentation was used.  The process consists of adding random noise to higher SNR signals to reduce the overall in-band SNR and implies after data augmentation the following in-band ranges:
\begin{itemize}
\item From $-2$~dB to $12$~dB for \texttt{CSPB.ML.2018}.
\item From $3.5$~dB to $14$~dB for \texttt{CSPB.ML.2022}.
\end{itemize}

\section{Capsule Network Training and Numerical Results}\label{sec:benchmarkTesting}
To illustrate the performance of the chosen capsule network for classifying digitally modulated signals it has been trained and tested with signals from the two datasets available.  Each dataset was divided into three subsets that were used for training, validation, and testing, such that for both \texttt{CSPB.ML.2018} and \texttt{CSPB.ML.2022}, 70\% of the data is used for training, 5\% of the data is used for validation and 25\% is used for testing.  All data is normalized to unit power before training begins, and the stochastic gradient descent with momentum (SGDM) algorithm~\cite{SGDM} was used for training with a mini-batch size of 250.  The capsule networks have been implemented in MATLAB and trained on a high-performance computing cluster with $18$~NVidia V100 graphical processing unit (GPU) nodes available, with each node having $128$~GB of memory.  Training is computationally intensive however, if the resources are leveraged correctly and the entire dataset is loaded into RAM, training can be completed in several hours.

The classification performance of the capsule network is compared to that of the alternative deep-learning approaches for classifying digitally modulated signals in \cite{Snoap_CCNC_2022}, which also use the I/Q signal data but employ a convolutional neural network (CNN) and a residual network (RN) for signal classification.  The CNN and RN used in \cite{Snoap_CCNC_2022} are similar to the neural network structures considered in~\cite{Tim2018} and yield similar results to those in \cite{Tim2018} when tested with signals that come from the same dataset as the one used for training.  However, as discussed in \cite{Snoap_CCNC_2022}, these NNs do not display meaningful generalization ability, and both the CNN and the RN fail to identify most digital modulation schemes that they have been trained to recognize when tested with signals from a dataset that was generated independently from the training one.

\begin{figure}
\begin{center}
\subfigure[Probability of Correct Classification Versus SNR.]
{\includegraphics[width=0.49\linewidth]{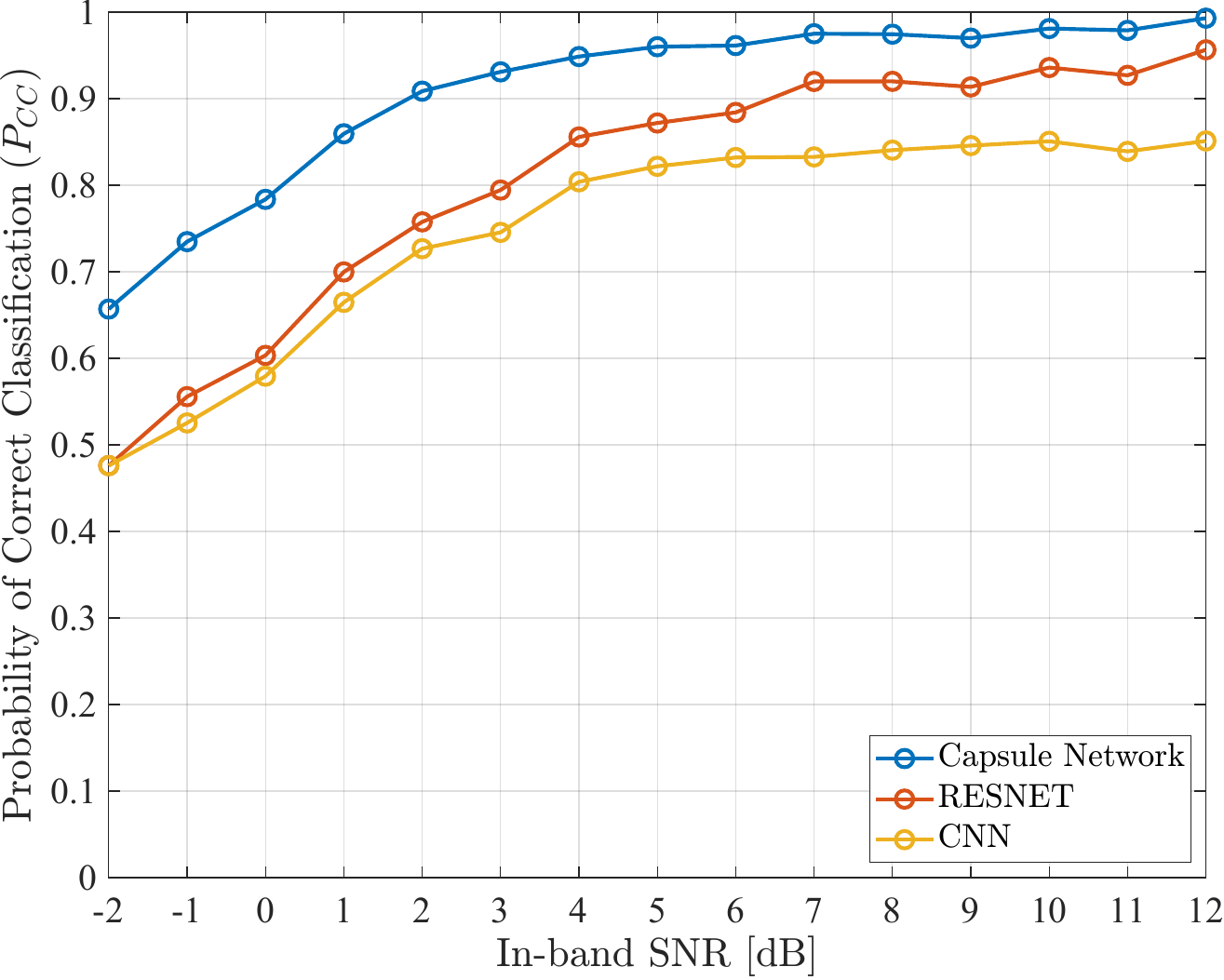}}
\subfigure[Capsule Network Confusion Matrix.]
{\includegraphics[width=0.49\linewidth]{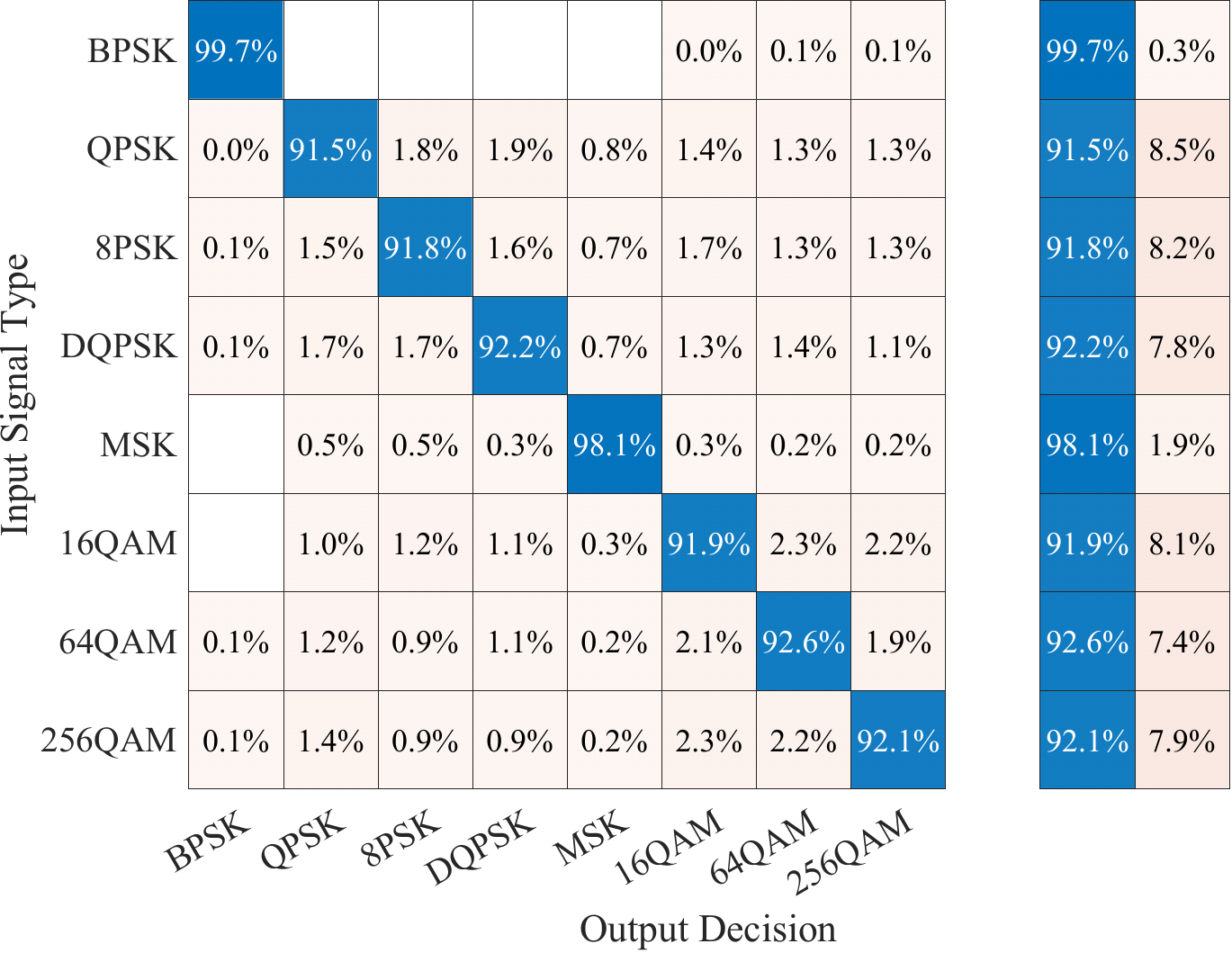}}
\caption[Probability of Correct Classification Versus SNR and Confusion Matrix When Both Training and Test Data Come From \texttt{CSPB.ML.2018} I/Q Data.]{\textbf{Probability of Correct Classification Versus SNR and Confusion Matrix When Both Training and Test Data Come From \texttt{CSPB.ML.2018} I/Q Data.}}\label{fig:snr}
\end{center}
\end{figure}

In the first experiment performed, the capsule network outlined in Figure~\ref{fig:topology} is trained using \texttt{CSPB.ML.2018}.  Results from this experiment showing the probability of correct classification and the confusion matrix are given in Figure~\ref{fig:snr}(a) and Figure~\ref{fig:snr}(b), respectively, from which can be seen an overall performance of $93.7\%$ correct classification with probabilities of correct classification of individual modulation schemes ranging from $91.9\%$ for 16QAM to $99.7\%$ for BPSK.

\begin{figure}
\begin{center}
\subfigure[Probability of Correct Classification Versus SNR.]
{\includegraphics[width=0.49\linewidth]{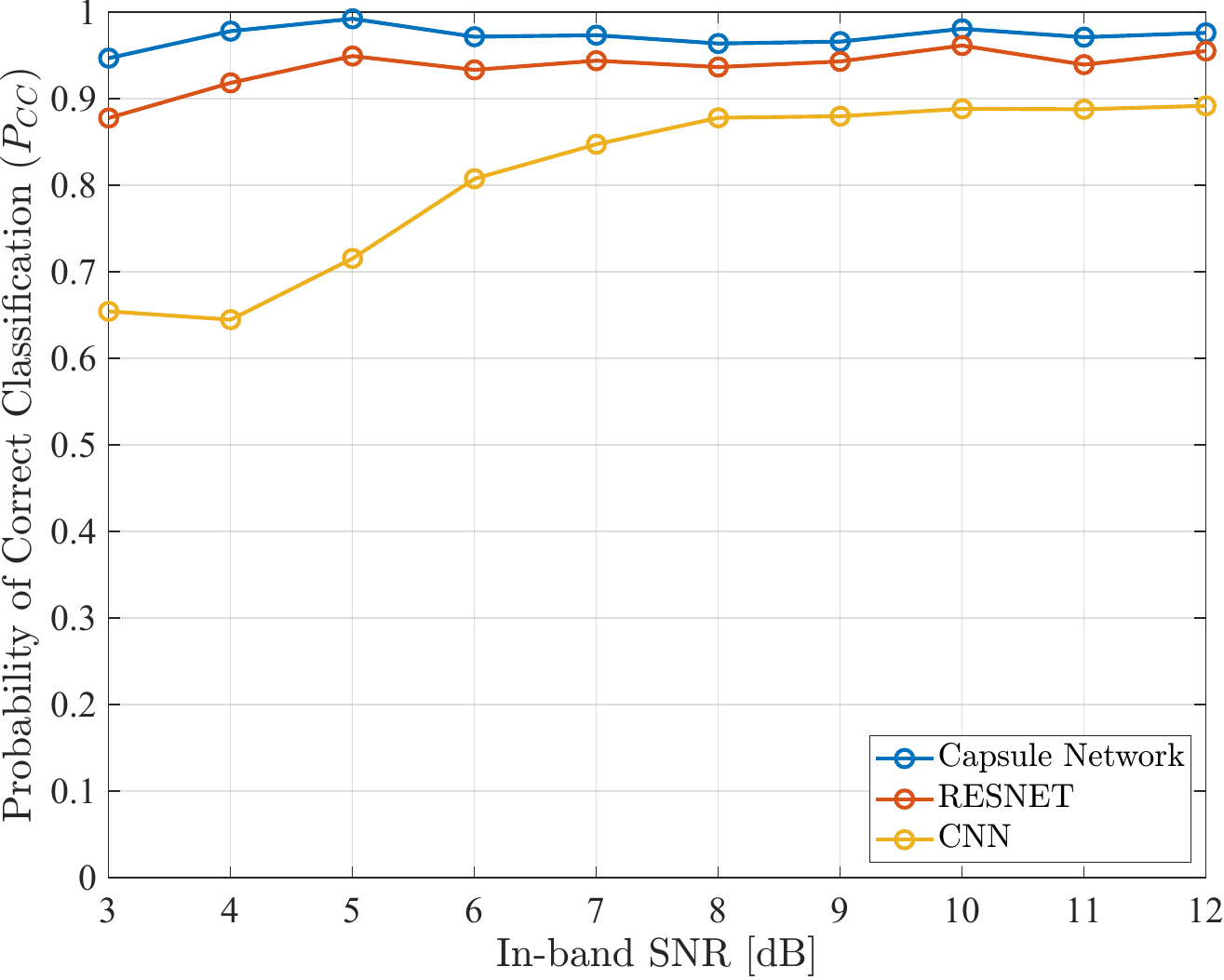}}
\subfigure[Capsule Network Confusion Matrix.]
{\includegraphics[width=0.49\linewidth]{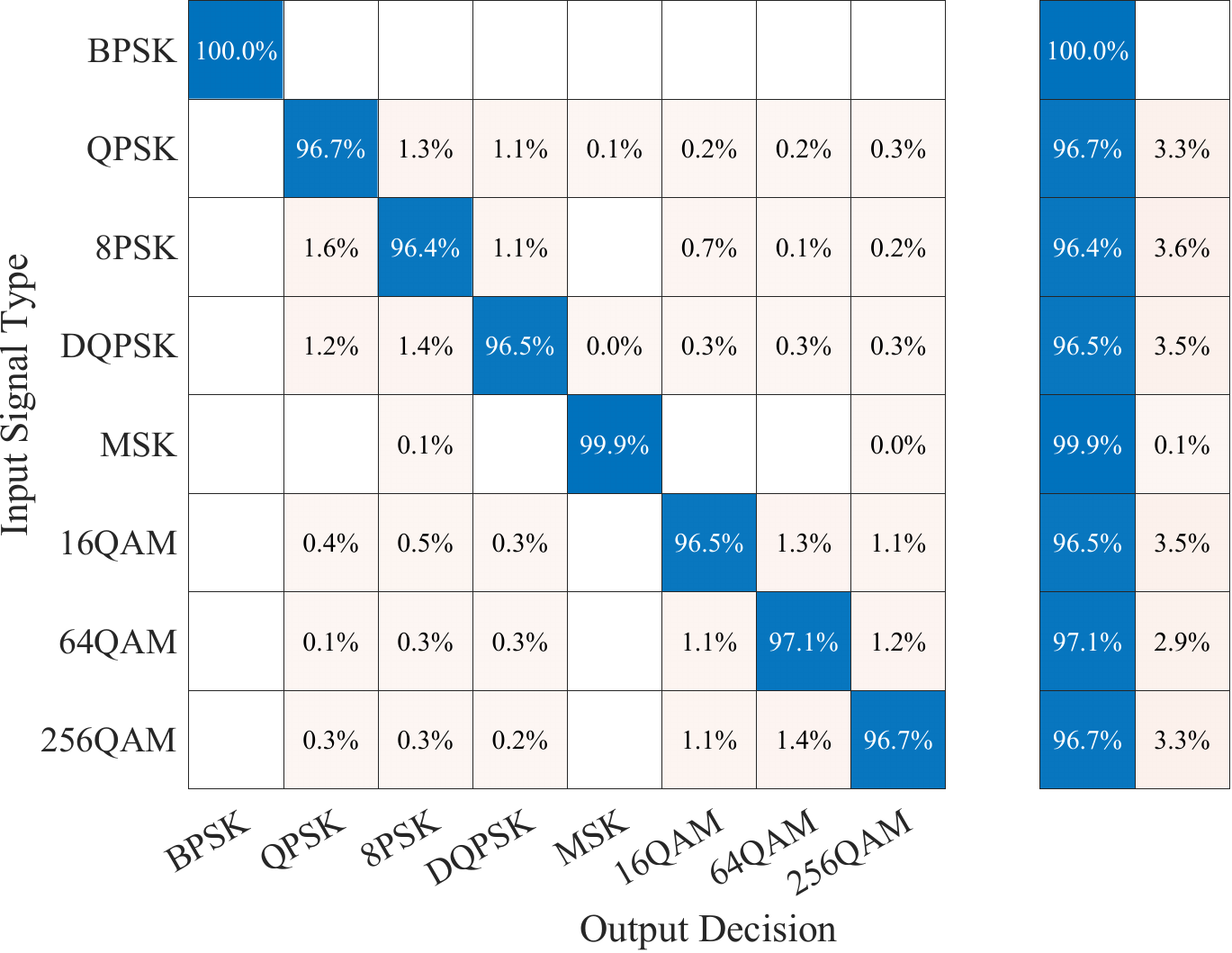}}
\caption[Probability of Correct Classification Versus SNR and Confusion Matrix When Both Training and Test Data Come From \texttt{CSPB.ML.2022} I/Q Data.]{\textbf{Probability of Correct Classification Versus SNR and Confusion Matrix When Both Training and Test Data Come From \texttt{CSPB.ML.2022} I/Q Data.}}\label{fig:snr_gc}
\end{center}
\end{figure}

In a following experiment, the capsule network is re-trained using \texttt{CSPB.ML.2022}, and similar results are obtained as can be seen from Figure~\ref{fig:snr_gc}(a) and Figure~\ref{fig:snr_gc}(b), showing the probability of correct classification and the confusion matrix, respectively, that are achieved when \texttt{CSPB.ML.2022} is used.  In this case an overall performance of $97.5\%$ correct classification with probabilities of correct classification of individual modulation schemes ranging from $96.4\%$ for 8PSK to $100.0\%$ for BPSK.  The slight improvement in classification performance displayed in this case is due to the fact that the SNR range for signals in \texttt{CSPB.ML.2022} is more favorable than the SNR range for signals in \texttt{CSPB.ML.2018}.

\begin{figure}
\begin{center}
\subfigure[Probability of Correct Classification Versus SNR.]
{\includegraphics[width=0.49\linewidth]{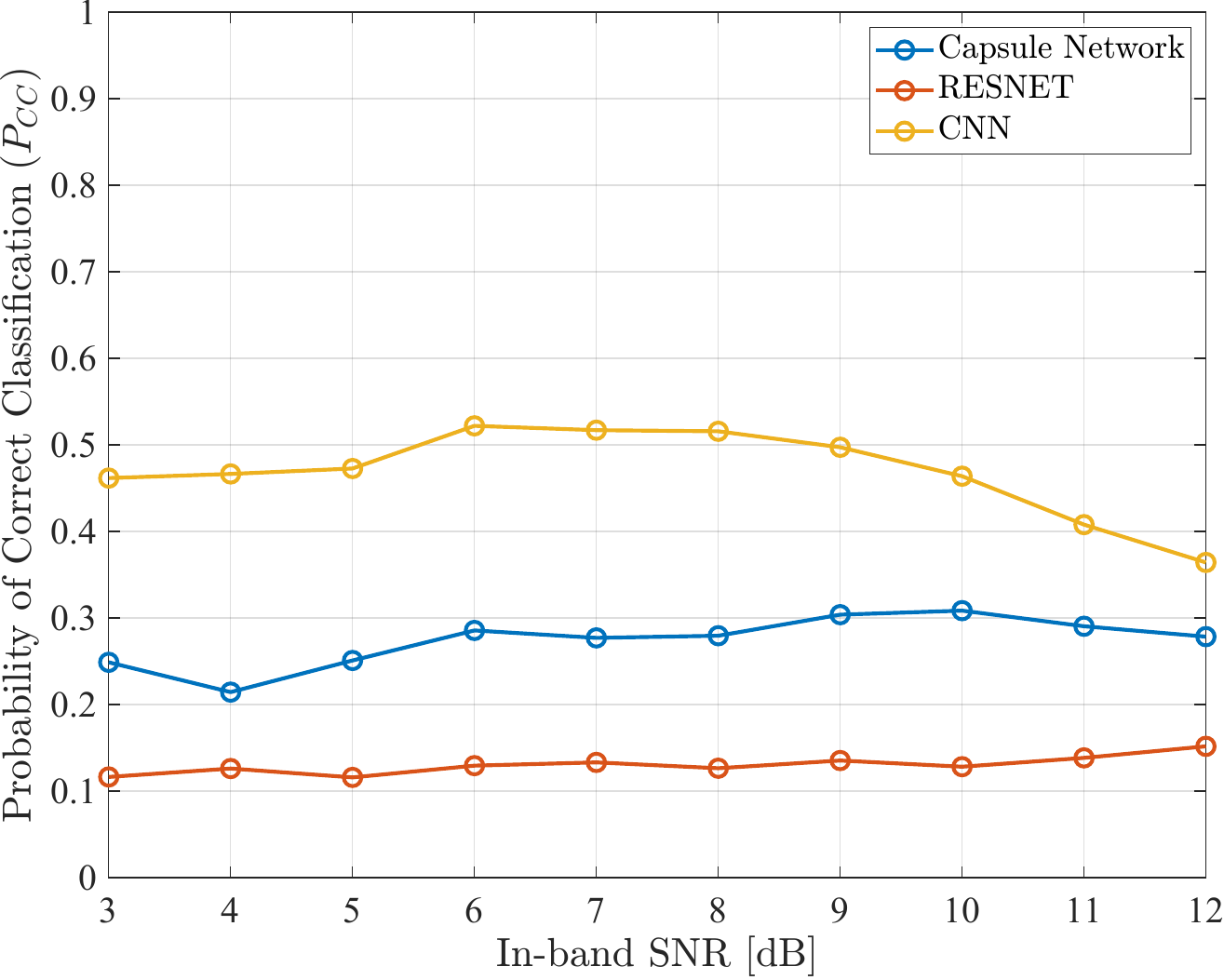}}
\subfigure[Capsule Network Confusion Matrix.]
{\includegraphics[width=0.49\linewidth]{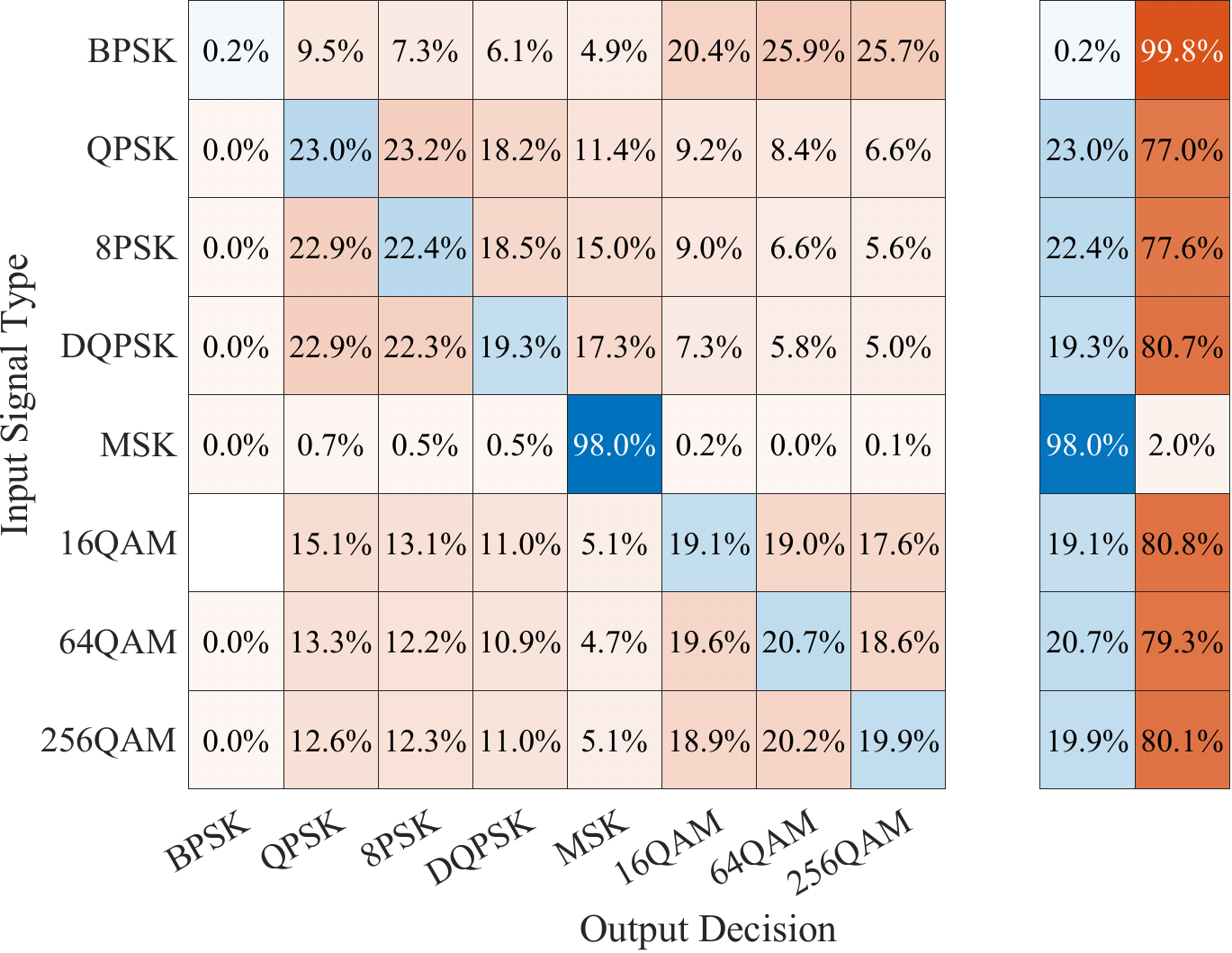}}
\caption[Probability of Correct Classification Versus SNR and Confusion Matrix When Training Is Done Using \texttt{CSPB.ML.2018} I/Q Data and Testing Uses \texttt{CSPB.ML.2022} I/Q Data.]{\textbf{Probability of Correct Classification Versus SNR and Confusion Matrix When Training Is Done Using \texttt{CSPB.ML.2018} I/Q Data and Testing Uses \texttt{CSPB.ML.2022} I/Q Data.}}\label{fig:snr_gentest}
\end{center}
\end{figure}

Next, the generalization ability of the capsule network was studied by using signals in \texttt{CSPB.ML.2018} for training the network followed by testing with signals in \texttt{CSPB.ML.2022}.  Results from this experiment are shown in Figure~\ref{fig:snr_gentest}(a) and Figure~\ref{fig:snr_gentest}(b).  When the capsule network trained using signals in \texttt{CSPB.ML.2018} was tested with signals in \texttt{CSPB.ML.2022}, the classification performance degraded significantly to an overall probability of correct classification of only $27.9\%$ as can be observed from Figure~\ref{fig:snr_gentest}, indicating that the capsule network trained using \texttt{CSPB.ML.2018} does not appear to generalize training to a different dataset.

The lack of ability to generalize training is also observed if the capsule network is re-trained using signals from \texttt{CSPB.ML.2022} and then tested using signals in \texttt{CSPB.ML.2018}, in which case a similar overall probability of correct classification of $26.2\%$ is obtained.  The plots showing the variation of probability of correct classification vs. SNR and the confusion matrix for this experiment are omitted, as these are similar to the ones shown in Figure~\ref{fig:snr_gentest}(a) and Figure~\ref{fig:snr_gentest}(b), respectively. Thus, while the capsule network is able to adapt well to the dataset change and re-learn to correctly classify modulation types in a new dataset with high accuracy, it is not able to generalize its training to maintain good classification performance when presented with signals in a different dataset.  Nevertheless, the capsule network does appear to learn some baseline signal features that are common to both \texttt{CSPB.ML.2018} and \texttt{CSPB.ML.2022}, as the overall probability of correct classification in both cases is more than double that of a random guess\footnote{With $8$ digital modulation schemes to be classified there is a $12.5\%$ chance of a random guess being correct.}, but the classification performance is sensitive to the different CFOs or symbol rates in the two datasets.

\begin{figure}
\begin{center}
\subfigure[Probability of Correct Classification Versus SNR.]
{\includegraphics[width=0.49\linewidth]{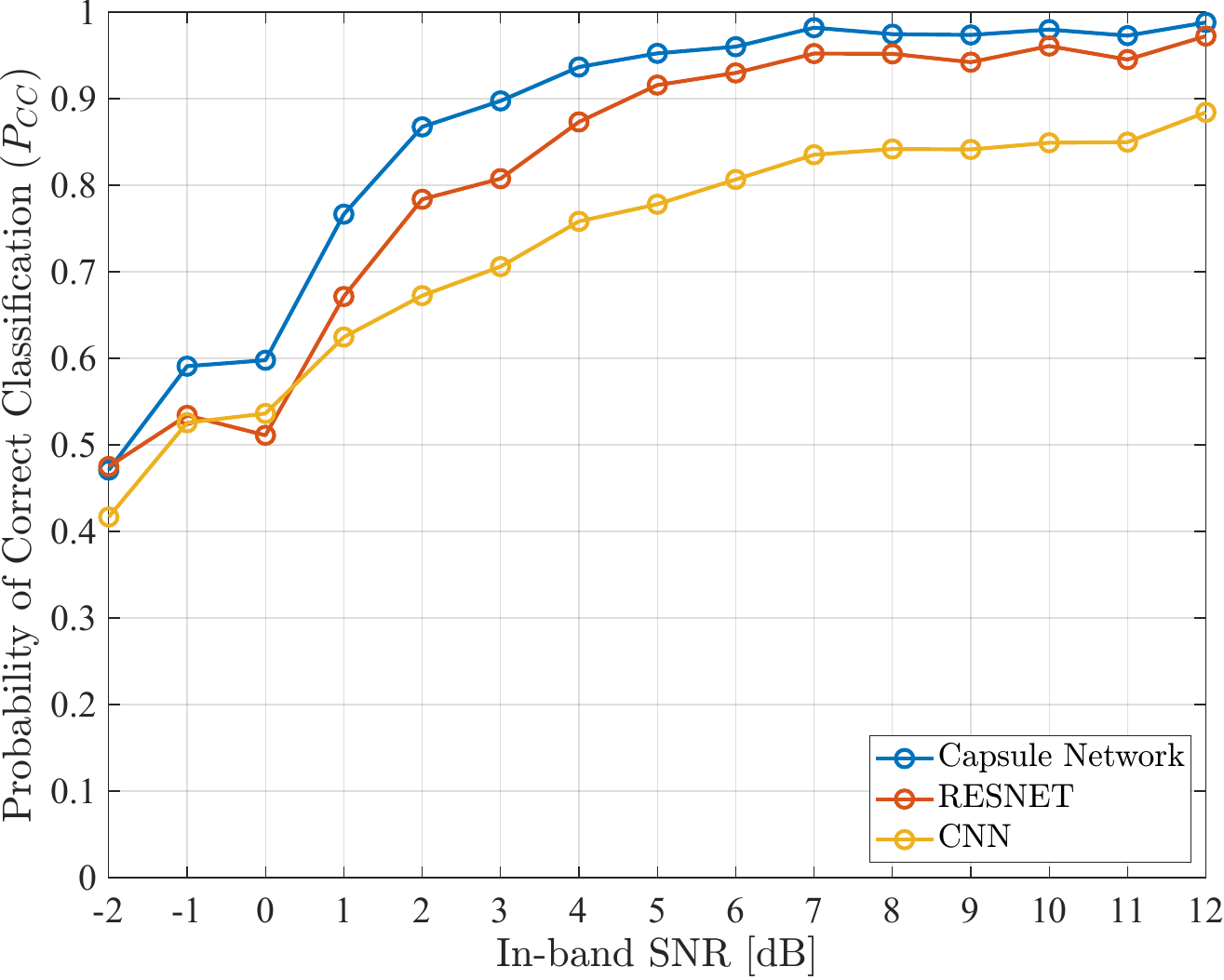}}
\subfigure[Capsule Network Confusion Matrix.]
{\includegraphics[width=0.49\linewidth]{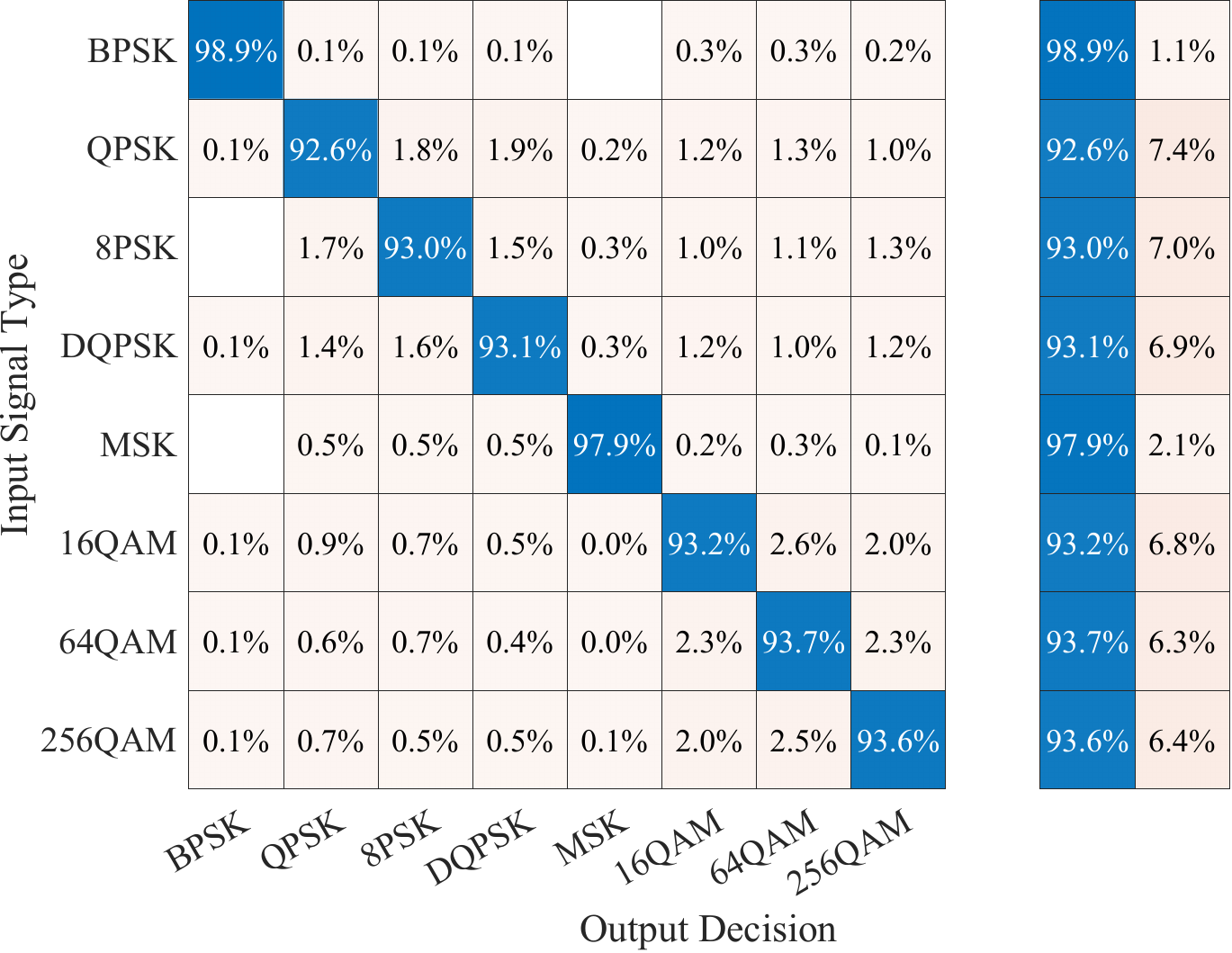}}
\caption[Probability of Correct Classification Versus SNR and Confusion Matrix When Training and Test Data Come From a Mix of \texttt{CSPB.ML.2018} I/Q Data and \texttt{CSPB.ML.2022} I/Q Data.]{\textbf{Probability of Correct Classification Versus SNR and Confusion Matrix When Training and Test Data Come From a Mix of \texttt{CSPB.ML.2018} I/Q Data and \texttt{CSPB.ML.2022} I/Q Data.}}\label{fig:snr_gc_c}
\end{center}
\end{figure}

In a final experiment, the signals in \texttt{CSPB.ML.2018} and \texttt{CSPB.ML.2022} were combined and the capsule network was trained with this new, mixed dataset.  While a total of $224,000$ signals are available in the two datasets, due to storage and memory constraints imposed by the high-performance cluster hardware and operation, only $160,000$ signals were included in the combined dataset, randomly taking $80,000$ signals from \texttt{CSPB.ML.2018} and  $80,0000$ from \texttt{CSPB.ML.2022} to make up the mixed dataset containing $160,000$ digitally modulated signals.  Following a similar approach as for previous experiments, the $160,000$ signals in the combined dataset were divided into three categories, with $70\%$ of signals used for training, $5\%$ for validation, and the remaining $25\%$ of the signals used for testing.  The results of this experiment are shown in Figure~\ref{fig:snr_gc_c}(a) and Figure~\ref{fig:snr_gc_c}(b), with an overall performance of $94.5\%$ correct classification and with probabilities of correct classification for individual modulation schemes ranging from $92.6\%$ for QPSK to $98.9\%$ for BPSK.

This experiment explored the use of capsule networks for classification of digitally modulated signals using the unprocessed I/Q components of the modulated signal.  The overall classification performance implied by capsule networks is on par or exceeds that obtained in related work where CNNs or residual networks (RNs) are used \cite{Tim2018, Snoap_CCNC_2022}, indicating that capsule networks are a meaningful alternative for machine-learning approaches to digitally modulated signal classification. Similar to CNNs and RNs, when trained with the unprocessed I/Q signal data, capsule networks are able to learn characteristics of the signals in the dataset used for training, but they are not able to generalize their learning to new datasets, which contain similar types of digitally modulated signals but with differences in some of their characteristics such as the CFO or symbol period.

\chapter{Cyclostationary Signal Processing Background Information}\label{ch:CSP_BackInfo}
Prior to introducing cyclostationary signal processing (CSP), an introduction to cyclostationarity and stationarity helps clarify how CSP is useful in practice.  Section~\ref{sec:StochProcess} will discuss stochastic processes with emphasis as they pertain to communication signals or mathematical time-series, followed by a description of moments in Section~\ref{sec:Moments}.  Definitions of stationarity and cyclostationarity are then provided in Sections~\ref{sec:Stationarity} and~\ref{sec:Cyclostationarity}, respectively, with examples of how the two stochastic process models differ when estimating signal parameters of communication signals.  Section~\ref{sec:StochProcess} through~\ref{sec:Cyclostationarity} is material largely drawn from and adapted from~\cite{S_vs_CS}.

\section{Stochastic Processes}\label{sec:StochProcess}
In the context of communication signals, a stochastic (or random) process is a collection of random variables indexed by time, $t$.  Let $X \left( t \right)$ denote a random process and $x \left( t \right)$ denote a signal.  For each instant in time, $t_j$, $X \left( t_j \right)$ is a random variable with a cumulative distribution function (CDF), $F_{X_j} \left( x \right)$, and a probability density function (PDF), $f_{X_j} \left( x \right)$.  The CDF and PDF at each time instant, $t_j$, may or may not be identical since the two random variables may or may not be independent, correlated, uncorrelated, etc.  Furthermore, every collection of random variables $\{X(t_j)\}_{j=1}^n$ has an $n^{\mbox{\footnotesize th}}$-order joint PDF and CDF when $X \left( t \right)$ is real-valued.  When two random variables are under consideration $\left\lbrace X \left( t_1 \right), X \left( t_2 \right) \right\rbrace$, the $2^{\mbox{\footnotesize nd}}$-order joint PDF is denoted by $f_{X_1, X_2} \left( x_1, x_2 \right)$.

With random variables, a particular occurrence of the variable, such as the outcome of a particular card draw or die throw, is called an \emph{instance} or \emph{sample} of the variable.  For a random process, a particular occurrence of the process is a function and is called a~\emph{sample~path}, a~\emph{sample~function}, a~\emph{time-series}, or a~\emph{signal}.

\pagebreak

\section{Moments}\label{sec:Moments}
The equation for computing an $n^{\mbox{\footnotesize th}}$-order moment of a random variable, $X \left( t_j \right)$, is defined as follows~\cite{Papoulis_Pillai_2002}:
\begin{align}
M_{X}^{n}(t_j) = E[X^n(t_j)] = \int_{-\infty}^\infty x^n f_{X_j}(x) \, dx.
\end{align}

When we have knowledge of a random process, $X \left( t \right)$, then we know the PDFs for $X \left( t_j \right)$ for all $t_j$: the collection of functions $f_{X_j} \left( x \right)$.  This knowledge allows us to compute the first moment, or the mean value, for each $X \left( t_j \right)$,
\begin{align}
M_X(t_j) = E[X(t_j)] = \int_{-\infty}^\infty x f_{X_j}(x) \, dx,
\end{align}
where any particular $M_X(t_j)$ may or may not equal some other $M_X(t_k)$.

Likewise, we can look at the correlation between two elements of the random process,
\begin{align}
R_X(t_1, t_2) = E[X(t_1) X(t_2)],
\end{align}
which may turn out to be independent of $t_1$ and $t_2$, dependent on them in some general way, or dependent only on their difference:  $t_2 - t_1$.

The covariance is the correlation of the two variables after their means are removed~\cite{Papoulis_Pillai_2002},
\begin{align}
K_X(t_1, t_2) = E[(X(t_1)-M_X(t_1))(X(t_2)-M_X(t_2))].
\end{align}

There are many practical scenarios where we do not know the PDFs of the process itself; however, we are still able to find the mean, auto-correlation, and covariance functions.  For example, a random process may be characterized by a few specific random variables, such as a sine-wave amplitude ($A$) or phase ($\Phi$).  Then, $X \left( t \right)$ can be considered as a function of the random variables $\{ A, \Phi \}$ and we know that the expected value of the process is the expected value of the function of the random variables.  So, all we need are the joint PDFs of the random variables $\{ A, \Phi \}$.  This will be clarified by way of example.

\pagebreak

\subsection{The Random-Phase Sine-Wave Example}\label{RPSWE}
Consider the following random process defined by
\begin{align}
X(t) = A \cos(2 \pi f_c t + \Phi),
\end{align}
where $A$ and $f_c$ are constants and the only random variable is the sine-wave phase $\Phi$, which is uniform on the interval $(0, 2\pi)$.  This random process models the situation where we receive a sine wave, and we know the amplitude and frequency, but since we do not know the phase, we model it with an equal probability of being any valid value.

The mean value (and first moment) of this random process is
\begin{align}
M_X(t) = E[X(t)] = E \left[ A \cos(2 \pi f_c t + \Phi) \right]. \label{eq:MeanUniformPhi}
\end{align}
Evaluating the expectation in~(\ref{eq:MeanUniformPhi}) requires the joint distribution of all involved random variables.  Since the only random variable is $\Phi$, only that distribution is needed, so with $t = t_1$, the expectation is
\begin{align}
M_X(t_1)	&= \int_{-\infty}^\infty A\cos(2\pi f_c t_1 + \phi) f_{\Phi}(\phi)\, d\phi, \nonumber \\
			&= \frac{A}{2\pi} \int_0^{2\pi} \cos(2\pi f_c t_1 + \phi)\, d\phi, \nonumber \\
			&= \frac{A}{2\pi} \int_{2\pi f_c t_1 + 0}^{2\pi f_c t_1 + 2\pi} \cos(u)\, du, \nonumber \\
			&= \frac{A}{2\pi} \left. \sin(u) \right|_{u=2\pi f_c t_1}^{2\pi f_c t_1 + 2\pi}, \nonumber \\
			&= \frac{A}{2\pi} \sin(2\pi f_c t_1+2\pi) - \frac{A}{2\pi} \sin(2\pi f_c t_1), \nonumber \\
			&= 0,
\end{align}
and the first moment of the random process is identically zero and independent of time; the mean is zero for every time instant because the average is over the ensemble of sample functions, where each sample function is delayed by some amount, $\phi/(2\pi f_c)$, and all delays modulo the period, $1/f_c$, are equally represented in that ensemble.  $X(t)$ is therefore a zero-mean random process.

The autocorrelation function (and second moment) for the random-phase sine-wave is the expected value of the product of two values of the process:
\begin{align}
R_X(t_1, t_2)	&= E[X(t_1)X(t_2)], \nonumber \\
				&= E[A\cos(2\pi f_c t_1 + \Phi) A\cos(2 \pi f_c t_2 + \Phi)], \nonumber \\
				&= \frac{A^2}{2} E[\cos(2\pi f_c t_1 - 2\pi f_c t_2) + \cos(2\pi f_c t_1 + 2\pi f_c t_2 + 2\phi)], \nonumber \\
				&= \frac{A^2}{2} \cos(2\pi f_c (t_1 - t_2)) + \frac{A^2}{2} E[\cos(2\pi f_c (t_1 + t_2) + 2\phi)], \nonumber \\				
				&= \frac{A^2}{2} \cos(2\pi f_c (t_1 - t_2)) + \frac{A^2}{2} \int_{-\infty}^\infty \cos(2\pi f_c (t_1 + t_2) + 2\phi) f_{\Phi}(\phi)\, d\phi, \nonumber \\
				&= \frac{A^2}{2} \cos(2\pi f_c (t_1 - t_2)) + \frac{A^2}{4\pi} \int_{0}^{2\pi} \cos(2\pi f_c (t_1 + t_2) + 2\phi) \, d\phi, \nonumber \\
				&= \frac{A^2}{2} \cos(2\pi f_c (t_1 - t_2)) + \frac{A^2}{8\pi} \int_{2\pi f_c (t_1 + t_2)}^{2\pi f_c (t_1 + t_2) + 4\pi} \cos(u) \, du, \nonumber \\
				&= \frac{A^2}{2} \cos(2\pi f_c (t_1 - t_2)) + \frac{A^2}{8\pi} \left. \sin(u) \right|_{u=2\pi f_c (t_1 + t_2)}^{2\pi f_c (t_1 + t_2) + 4\pi}, \nonumber \\
				&= \frac{A^2}{2} \cos(2 \pi f_c(t_1-t_2)).
\end{align}
The autocorrelation is, therefore, a function only of the difference between the two considered times $t_1 - t_2 = \tau$, and is also independent of time.  Next, the first and second moments will be computed using only a single sample path with infinite-time averages.  The single sample-path mean corresponds to the infinite-time average,
\begin{align}
m_X = \lim_{T\rightarrow\infty} \frac{1}{T} \int_{-T/2}^{T/2} x(t) \, dt = \lim_{T\rightarrow\infty} \frac{1}{T} \int_{-T/2}^{T/2} A\cos(2 \pi f_c t + \phi) \, dt, \label{eq:infTimeAvgMean}
\end{align}
where $\phi$ is a number rather than a random variable because it is an instance of the random variable $\Phi$.  This time-average mean is easily shown to be zero for any $\phi$:
\begin{align}
m_X = \lim_{T\rightarrow\infty} \frac{A \cos(\phi) \sin(\pi f_c T)}{\pi f_c T} = 0.
\end{align}
The time-average autocorrelation is defined by the following infinite-time average,
\begin{align}
R_x(\tau) = \lim_{T\rightarrow\infty} \frac{1}{T} \int_{-T/2}^{T/2} x(t) x(t-\tau) \, dt,
\end{align}
where $t_1 = t$ and $t_2 = t-\tau$ such that $\tau = t_1 - t_2$.  This infinite-time average autocorrelation evaluates as follows:
\begin{align}
R_x(\tau)	&= \lim_{T\rightarrow\infty}\frac{1}{T} \int_{-T/2}^{T/2} A\cos(2\pi f_c t + \phi)A\cos(2\pi f_c (t-\tau) + \phi) \, dt, \nonumber \\
				&= \frac{A^2}{2} \lim_{T\rightarrow\infty} \frac{1}{T} \int_{-T/2}^{T/2} \cos(2 \pi f_c \tau) + \cos(2\pi 2f_c t - 2\pi f_c \tau + 2\phi) \, dt, \nonumber \\
				&= \frac{A^2}{2} \cos(2\pi f_c \tau).
\end{align}

In this example, the ensemble mean matches the time-average mean and the ensemble correlation matches the time-average correlation.  However, this is not always the case as will be shown in the next example.

\subsection{The Random-Amplitude Sine-Wave Example}\label{RASWE}
Consider another random process given by
\begin{align}
X(t) = A \cos(2\pi f_c t + \phi),
\end{align}
where $f_c$ and $\phi$ are constants and $A$ is a uniform random variable on the interval $[0, 2 B]$ with $B>0$.  The ensemble average evaluates to a function of $t$:
\begin{align}
M_X(t)	&= E[X(t)] = E[A\cos(2\pi f_c t + \phi)], \nonumber \\
		&= \cos(2\pi f_c t + \phi) E[A], \nonumber \\
		&= \cos(2 \pi f_c t + \phi) \int_{-\infty}^\infty a f_A(a) \, da, \nonumber \\
		&= \cos(2 \pi f_c t + \phi) \int_{0}^{2B} a\left(\frac{1}{2B} \right) \, da, \nonumber \\
		&= \frac{\cos(2\pi f_c t + \phi)}{2B} \left. \frac{a^2}{2} \right|_{a=0}^{2B}, \nonumber \\
		&= \frac{\cos(2\pi f_c t + \phi)}{4B} (4B^2), \nonumber \\
		&= B \cos(2\pi f_c t + \phi).
\end{align}
However, the infinite-time average of a sample path where $A$ takes on a concrete value $a \in [0, 2B]$ is
\begin{align}
m_X = \lim_{T\rightarrow\infty} \frac{1}{T} \int_{-T/2}^{T/2} x(t) \, dt = \lim_{T\rightarrow\infty} \frac{1}{T} \int_{-T/2}^{T/2} a \cos(2 \pi f_c t + \phi) \, dt,
\end{align}
which evaluates to zero, similar to~(\ref{eq:infTimeAvgMean}).  So, the ensemble mean and time-average mean are not equal.  Comparing the probabilistic (ensemble-average) autocorrelation and the temporal (time-average) autocorrelation yields a similar result.

The probabilistic autocorrelation is
\begin{align}
R_X(t_1, t_2)	&= E[X(t_1)X(t_2)], \nonumber \\
				&= E[A^2] \cos(2\pi f_c t_1 + \phi)\cos(2\pi f_c t_2 + \phi), \nonumber \\
				&= \left(\int_{0}^{2B} \left(\frac{1}{2B} \right) a^2 \, da \right) \cos(2\pi f_c t_1 + \phi) \cos(2\pi f_c t_2 + \phi), \nonumber \\
				&= \left. \frac{a^3}{6B}\right|_{a=0}^{2B} \cos(2\pi f_c t_1 + \phi) \cos(2\pi f_c t_2 + \phi), \nonumber \\
				&= \frac{4B^2}{3} \cos(2\pi f_c t_1 + \phi) \cos(2\pi f_c t_2 + \phi).
\end{align}
Letting  $t_1 = t$ and $t_2 = t-\tau$,
\begin{align}
R_X(t, t-\tau) = \frac{4B^2}{3} \cos(2\pi f_c t + \phi) \cos(2\pi f_c(t-\tau) + \phi),
\end{align}
which is a function of both $t$ and $\tau$.  The temporal correlation is, however,
\begin{align}
R_x(\tau)	&= \left\langle x(t) x(t-\tau) \right\rangle = \lim_{T\rightarrow\infty} \frac{1}{T} \int_{-T/2}^{T/2} x(t) x(t-\tau) \, dt, \nonumber \\
				&= \lim_{T\rightarrow\infty} \frac{1}{T} \int_{-T/2}^{T/2} a^2 \cos(2\pi f_c t + \phi)\cos(2\pi f_c (t - \tau) + \phi) \, dt, \nonumber \\
				&= \lim_{T\rightarrow\infty} \frac{a^2 \sin(2\pi f_c T) \cos(2\pi f_c \tau - 2\phi)}{4\pi f_c T} + \frac{a^2}{2} \cos(2\pi f_c \tau), \\
R_x(\tau)	&= \frac{a^2}{2} \cos(2\pi f_c \tau),
\end{align}
which is not, and cannot be, a function of time $t$.  Therefore, for the random-amplitude sine-wave random process, the probabilistic and temporal autocorrelation functions also do not match.  

\newpage

\section{Stationarity}\label{sec:Stationarity}
When the probabilistic functions of a random process are independent of time, the time-invariance of the process is called \emph{stationarity}.  Even though a signal may correspond to a stationary random process, the signal itself is not a constant as a function of time, its underlying ensemble is what possesses the time-invariance in its statistical functions.  Strict-sense stationarity (SSS) requires all the moments of the random process to be time-invariant, which can be difficult to verify since the process must be simple enough that all the moments can be checked to ensure they are not a function of time $t$.  Wide-sense stationarity (WSS), sometimes referred to as weak-sense stationarity, only requires the first two moments of the random process to be time-invariant.  So for a WSS process, the mean must be a constant,
\begin{align}
M_X(t_1) = E[X(t_1)] = M,
\end{align}
and the autocorrelation must be a function only of the time difference between the two involved times $t_1$ and $t_2$,
\begin{align}
R_X(t_1, t_2) = E[X(t_1)X^*(t_2)] = R_X(t_1-t_2).
\end{align}

For the two presented sine-wave examples, only the one with the random-phase variable is WSS.  The random-amplitude ensemble produces a time-varying first moment (mean) and time-varying second moment (autocorrelation) and is therefore neither SSS nor WSS.

\subsection{Ergodicity}\label{sec:Ergodicity}
Ergodicity is a property of a random process that ensures the time-averages of the sample paths correspond to the ensemble averages. This is the property that must be invoked to pursue real-world utility of ensemble sample paths.

Ergodic random processes must be stationary because the time averages of sample paths are not functions of time.  By construction, the time-averages are an integration over time. Therefore, if those averages have any chance to match the ensemble averages, those ensemble averages also cannot be functions of time, and so the process must be stationary.

The example ensemble with the random-phase variable is stationary and ergodic because the mean and the autocorrelation functions in both ensemble- and temporal-averaging agree.

\newpage

\section{Cyclostationarity}\label{sec:Cyclostationarity}
A wide-sense cyclostationary random process is one for which the mean and autocorrelation are periodic (or almost-periodic) functions of time, rather than being time-invariant.  A second-order cyclostationary signal, or cyclostationary time-series, is simply a signal for which the infinite-time average defined by either the non-conjugate cyclic autocorrelation function,
\begin{align}
R_x^\alpha(\tau) = \lim_{T\rightarrow\infty} \frac{1}{T} \int_{-T/2}^{T/2} x(t+\tau/2) x^*(t-\tau/2) e^{-i2\pi \alpha t} \, dt, \label{eq:2ndOrderNonConjCycloSignal}
\end{align}
or defined by the conjugate cyclic autocorrelation function,
\begin{align}
\displaystyle R_{x^*}^\alpha(\tau) = \lim_{T\rightarrow\infty} \frac{1}{T} \int_{-T/2}^{T/2} x(t+\tau/2) x(t-\tau/2) e^{-i2 \pi \alpha t} \, dt, \label{eq:2ndOrderConjCycloSignal}
\end{align}
is not identically zero for all non-trivial cycle frequencies $\alpha$ and delays $\tau$. The sole trivial cycle frequency is $\alpha=0$ for (\ref{eq:2ndOrderNonConjCycloSignal}).  There are no trivial cycle frequencies for the conjugate case in (\ref{eq:2ndOrderConjCycloSignal}).

It is also possible that there are no non-trivial cycle frequencies for which either the non-conjugate cyclic autocorrelation function (\ref{eq:2ndOrderNonConjCycloSignal}) or the conjugate cyclic autocorrelation function (\ref{eq:2ndOrderConjCycloSignal}) is non-zero, and yet the signal is still a cyclostationary signal.  This cyclostationarity can occur if some other, higher-order, moment function possesses a non-trivial cycle frequency.  A real-world example is a square-root raised-cosine QPSK signal with a roll-off of zero: it is a wide-sense stationary signal but possesses many higher-order non-trivial cyclic cumulants.

\subsection{Cycloergodicity}\label{sec:Cycloergodicity}
Similar to ergodicity for a stationary random process, when using stochastic machinery to define a cyclostationary random process, cycloergodicity must be invoked or checked to ensure the time-averages of the sample paths correspond to the ensemble averages.  For cycloergodicity, a new tool is needed involving time averages to compare with the time-varying ensemble-average autocorrelation.  This is the fraction-of-time (FOT) expectation, also called the multiple-sine-wave extractor (or similar), $E^{\{\alpha\}} [\cdot]$.  This operator returns all finite-strength additive sine-wave components in its argument.  The FOT expectation is shown to be an expected-value operation, provided with a cumulative distribution function and associated probability density function in \cite{StatisticalSpectralAnalysis}.

A few examples illustrate the FOT expectation.  The FOT mean value of a single sine wave is just that sine wave,
\begin{align}
E^{\{\alpha\}} [Ae^{i 2 \pi f_0 t + i \phi_0}] = Ae^{i 2 \pi f_0 t + i \phi_0}.
\end{align}
Given a periodic function $y(t)$, such as a single sine wave, a square wave, a radar signal, etc., with period $T_0$.  Then $y(t)$ can be represented in a Fourier series:
\begin{align}
y(t) = \sum_{k=-\infty}^\infty c_k e^{i 2 \pi (k/T_0) t}.
\end{align}
The FOT sine-wave extractor is linear, such that
\begin{align}
E^{\{\alpha\}}[y(t)] = y(t).
\end{align}
Furthermore, suppose some periodic signal $y(t)$ is embedded in noise and interference that does not possess any periodic component,
\begin{align}
z(t) = y(t) + w(t),
\end{align}
then
\begin{align}
E^{\{\alpha\}}[z(t)] = y(t).
\end{align}

To ensure the cycloergodic property of a random process, the ensemble-average mean value must equate to the FOT mean value:
\begin{align}
M_X(t) = E[X(t)] \stackrel{?}{=} E^{\{\alpha\}}[x(t)]. \label{eq:CycloErgodicity}
\end{align}
In (\ref{eq:CycloErgodicity}), the ensemble average on the left and the temporal (non-stochastic) average on the right can be functions of time $t$.  Also, the ensemble-average second moment must equate to the FOT second moment,
\begin{align}
R_X(t+t_1, t+t_2) = E[X(t+t_1)X^*(t+t_2)] \stackrel{?}{=} E^{\{\alpha\}}[x(t+t_1)x^*(t+t_2)].
\end{align}
If these averages match, and the process is cyclostationary, then the process is a cycloergodic process.

A commonly used mathematical model of communication signals is a pulse-amplitude modulated (PAM) signal given by
\begin{align}
s(t) = \sum_{k=-\infty}^\infty a_k p(t - kT_0), \label{eq:PAM}
\end{align}
where $p(t)$ is the pulse function (for example, rectangular or square-root raised-cosine), $k$ is the symbol index, and $\{a_k\}$ is the symbol sequence, typically drawn from a finite alphabet such as $\{+1, -1\}$, where the bits are independent and identically distributed (IID).  For the binary alphabet $a_k \in \{+1, -1\}$,  $s(t)$ is the complex envelope of an RF BPSK signal.  When the alphabet is $\{+1, -1, +i, -i\}$, the signal is QPSK.  In (\ref{eq:PAM}), averaging over the ensemble produces periodically time-varying results (cyclostationarity), very similar to the random-amplitude sine-wave example given earlier.  As such, digital communication signals do not have stationarity or ergodicity, but rather have cyclostationarity and can be shown to have cycloergodicity.  For these reasons, a cyclostationary stochastic process model will produce more reliable and accurate results when processing individual digital communication signals, or sample paths, than a stationary stochastic process model.

Furthermore, cycloergodicity only needs to be verified if one attempts to use the stochastic process framework to develop algorithms/estimators with the intent of performing those algorithms on sample paths.  If a stochastic process model is never employed, then cycloergodicity need not be proved; instead, one only needs to postulate that functions~(\ref{eq:2ndOrderNonConjCycloSignal}) and~(\ref{eq:2ndOrderConjCycloSignal}) are non-zero for the data.

\chapter{Cyclostationary Signal Processing for Digital Modulation Classification}\label{ch:CSPModClass}
CSP provides a set of analytical tools for estimating distinct features that are present in various modulation schemes and that can be used for blind signal classification in various reception scenarios, which include stationary noise and/or co-channel interference.  These tools enable the estimation of higher-order CCs \cite{Spooner_Asilomar2001, Spooner_Asilomar1995} and of the SCF \cite{Gardner_CSP_pt1_1994, Spooner_CSP_pt2_1994} from the received signal, such that the estimates of the CC or the SCF can then be compared to a set of theoretical values of the CCs or the SCF, for classifying the corresponding digital modulation scheme embedded in the received signal.  In addition to being affected by propagation, which includes noise and/or co-channel interference, the received signal features are also affected by other aspects with underlying random variables whose probability distributions vary widely in practice and influence the decisions of blind classifiers.  These may include (but are not limited to):  the symbol interval or the corresponding data rate, the carrier frequency offset (CFO), the excess bandwidth of the signal implied by variations of the pulse-shaping function parameters, and the received signal power level, which directly impacts the in-band signal-to-noise ratio (SNR).  While some of these parameters that affect signal characteristics may have limited practical ranges, such as the roll-off parameter $\beta$ in the case of square-root raised-cosine (SRRC) pulse-shaping, which is typically in the $[0.2, 0.5]$ range, others such as the symbol interval or the CFO possess an infinite number of valid practical choices.

The emphasis here is that CCs, although conceptually and mathematically obscure and complex, are intimately related to the set of $n^{\mbox{\footnotesize th}}$-order probability density functions (PDFs) governing the behavior of communication signals \cite{Gardner_CSP_pt1_1994, Spooner_CSP_pt2_1994}.  As outlined in Section~\ref{sec:Proposed_Model_CC_FE}, CCs are the Fourier series components of the power-series-expansion coefficients of the logarithm of the characteristic function, which itself is simply the Fourier transform of a PDF.  CCs are not the sort of features that are typically associated with machine-learning and data mining, where voluminous datasets are searched with fast computers for correlations between mathematical transformations of the data and the signal-class label.  Being strongly related to all joint PDFs associated with the signals' samples, the use of CCs as features has much more in common with decision-theoretic approaches than with modern feature-based approaches, for which there may not be any provided mathematical rationale \cite{Spooner_Asilomar2000, Gardner_Spooner_TCom1993}.

\section{CC Feature Extraction}\label{sec:Proposed_Model_CC_FE}
Consider the generic digitally modulated signal written as
\begin{align}\label{eq:generic_signal}
x \left( t \right) = {a s(t) e^{i (2 \pi f_0 t + \phi)} + w \left( t \right),}
\end{align}
where $a$ denotes the signal amplitude, $s \left( \cdot \right)$ is the complex envelope of the signal, $f_0$ is the CFO, and $w \left( t \right)$ is additive white Gaussian noise (AWGN).  The CC features of this signal are extracted using a CSP-based approach \cite{Spooner_Asilomar1995, Spooner_Asilomar2001, Spooner_Asilomar2000} that starts with the $n^{\mbox{\footnotesize th}}$-order temporal moment function defined by
\begin{align}\label{eq:tmf}
R_x \left( t, \boldsymbol{{\tau}}; n, m \right) = \widehat{E}^{ \left \lbrace \gamma \right\rbrace } { \left \lbrace \prod_{j=1}^{n}{x^{ \left( \ast \right) j } \left( t + \tau_j \right) } \right \rbrace },
\end{align}
where $m$ of the factors are conjugated, $\left( \ast \right)$ represents an optional conjugation, and $\widehat{E}^{ \left\lbrace \gamma \right \rbrace }$ is the multiple sine-wave extraction operator (the multiple sine-wave extraction operator is the direct analog of the stochastic process expected value operator in the fraction-of-time probability framework) \cite{Gardner_CSP_pt1_1994},
\begin{align}\label{eq:Ebeta}
\widehat{E}^{ \left \lbrace \gamma \right \rbrace }{\left\lbrace g \left( t \right) \right \rbrace} = \sum_{\gamma}{g_{\gamma} e^{i 2 \pi \gamma t}},
\end{align}
for arbitrary $g(t)$, with
\begin{align}
g_{\gamma} \triangleq \lim_{T \to \infty}{\dfrac{1}{T} \int_{\frac{-T}{2}}^{\frac{T}{2}}{g \left( u \right) e^{- i 2 \pi \gamma u}}\,du} \equiv \langle g \left( u \right) e^{- i 2 \pi \gamma u} \rangle,
\end{align}
and the summation in (\ref{eq:Ebeta}) is taken over all $\gamma$ for which $g_\gamma \ne 0$.  The corresponding $n^{\mbox{\footnotesize th}}$-order temporal cumulant function (TCF) is given by
\begin{align}\label{eq:tcf}
C_x \left( t, \boldsymbol{{\tau}}; n, m \right) = \sum_{P_n}{\left[ h \left( p \right) \prod_{j=1}^{p}{R_{x_{\nu_j}} \left( t, \boldsymbol{{\tau}}_{\nu_j}; n_j, m_j \right) } \right], }
\end{align}
where the sum is over all distinct partitions $ \left\lbrace \nu_j \right\rbrace _{j=1}^{p} $ of the index set $ \left\lbrace 1,2,...,n \right\rbrace $ and \linebreak[4]
$h \left( p \right) = \left( -1 \right) ^{p-1} \left( p-1 \right) !$.  The $n^{\mbox{\footnotesize th}}$-order moment functions are polyperiodic functions of time, which implies that the $n^{\mbox{\footnotesize th}}$-order cumulant functions are as well, and thus, each of them can be represented in terms of a generalized Fourier series.

The coefficients of the cumulant Fourier series represent the CC of the signal (\ref{eq:generic_signal}) and are given by \cite{Spooner_Asilomar1995}
\begin{align}\label{eq:CCexpr}
C_x^{\alpha} \left( \boldsymbol{{\tau}}; n, m \right) = \langle C_x \left( t, \boldsymbol{{\tau}}; n, m \right) e^{- i 2 \pi \alpha t} \rangle,
\end{align}
where $\alpha$ is an $n^{\mbox{\footnotesize th}}$-order cycle frequency (CF) of the signal.

The CFs $\alpha$ for which the CC (\ref{eq:CCexpr}) is not zero for typical digitally modulated signals include the harmonics of the symbol rate $1/T_0$, multiples of the CFO $f_0$, and combinations of these two sets, such that $\alpha$ can be written as
\begin{align}\label{eq:basic_cf_psk_qam}
\alpha = (n-2m)f_0 \pm k/T_0.
\end{align}
For second-order CFs ($n=2$), the non-conjugate CFs corresponding to $m=1$ depend only on the symbol rate $1/T_0$, while the conjugate CFs corresponding to $m=0$ depend on both the symbol rate $1/T_0$ and the CFO~$f_0$.

To obtain accurate estimates of the CC features for signal classification, knowledge of signal parameters such as the symbol rate and CFO is necessary, as these parameters define the CFs needed for CC computation.  Estimates of these signal parameters are obtained using CSP techniques such as the strip spectral correlation analyzer (SSCA)~\cite{Brown_SSCA_1993} or the time- and frequency-smoothing methods in \cite{Gardner_CSP_pt1_1994, Spooner_CSP_pt2_1994} and can be further refined using additional parameters such as the excess bandwidth and the in-band SNR, which can be estimated using energy-based band-of-interest (BOI) detectors that do not require CSP \cite{BOIdetector}.

To extract the necessary signal parameters for CC estimates, the following procedure was used:
\begin{enumerate}
\item Use the BOI detector \cite{BOIdetector} to evaluate the signal bandwidth and obtain a low-resolution estimate of the center frequency.
\item Frequency shift the BOI to the baseband using the low-resolution CFO estimate.
\item \label{step:MaxBW} Downsample/upsample the data as necessary such that the signal bandwidth is maximized, but keep the fractional bandwidth of the result strictly less than $1$.
\item \label{step:SSCA} Apply the SSCA to the data provided by Step~\ref{step:MaxBW} to detect the second-order CFs.
\item \label{step:NCsymbolRate} Use the non-conjugate second-order CFs (if these are present) to obtain a high-resolution estimate of the symbol rate $1/T_0$.
\item \label{step:CsymbolRate} If no non-conjugate CFs are present, the symbol rate may be estimated from any conjugate CFs present, which can also be used to provide a high-resolution estimate of the CFO.
\item \label{step:CFpatternBOI} Determine the basic pattern of the second-order CFs present in the BOI. 
\item \label{step:noC-CFs} If conjugate CFs are not present from Step~\ref{step:SSCA}, then the data from Step~\ref{step:MaxBW} are raised to the fourth power and Fourier transformed, and further CSP is applied to determine the CF pattern and to estimate the symbol rate if not provided by Step~\ref{step:NCsymbolRate} and to obtain a high-resolution estimate of the CFO.
\end{enumerate}

\begin{figure}
\centering
\includegraphics[width=1.0\linewidth]{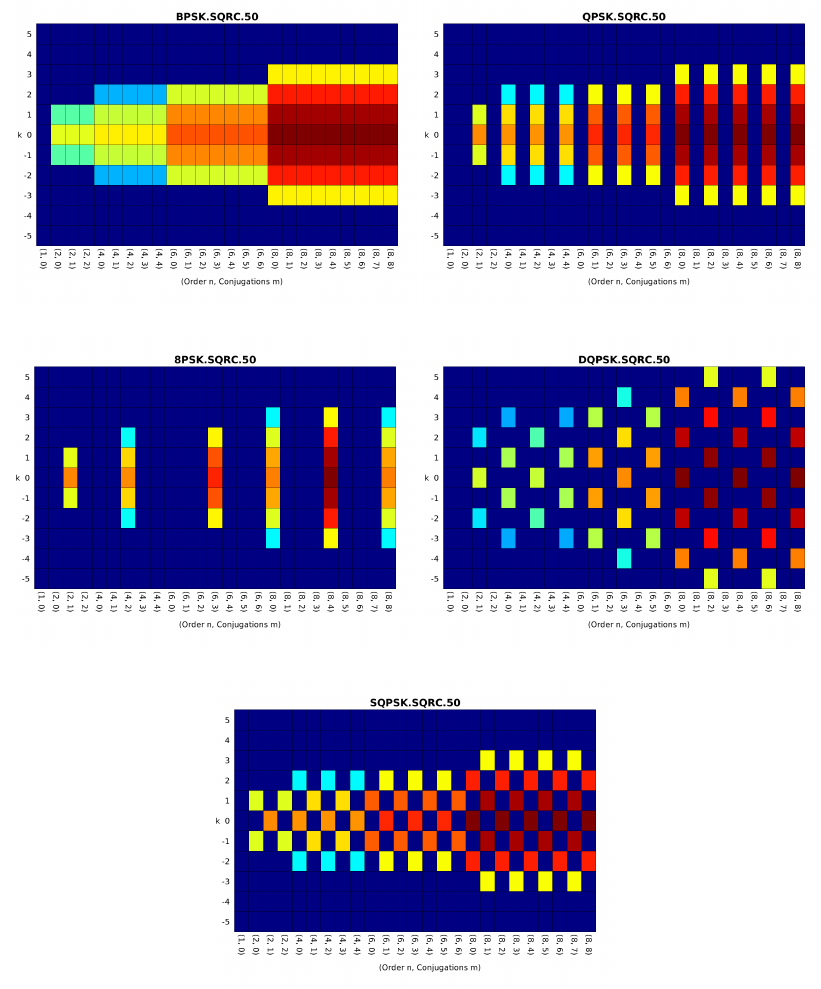}
\caption[Five Common CF Patterns for Digital QAM and PSK Modulation Types.]{\textbf{Five Common CF Patterns for Digital QAM and PSK Modulation Types.}  The CF pattern is more general than the constellation, with the shaping pulse bandwidth controlling the number of significant CC values in the harmonic number ($k$) dimension, and the probabilistic structure of the symbol random variable controls the pattern across the $(n,m)$ dimension~\cite{Spooner_CSP_pt2_1994}.
}\label{fig:CFpatterns}
\end{figure}

Steps~\ref{step:CFpatternBOI}--\ref{step:noC-CFs} are a key element of the procedure, which aims at identifying the basic second-order CF pattern of the signal implied by (\ref{eq:basic_cf_psk_qam}), which for typical digital modulation schemes is one of the following:  BPSK-like, QPSK-like, $\pi/4$-DQPSK-like, 8PSK-like, and staggered QPSK (SQPSK)-like.  Furthermore, all digital QAM signals with balanced (symmetric) constellations and more than two constellation points map to the QPSK-like pattern.  Amplitude-shift-keyed signals, BPSK, and OOK map to the BPSK-like pattern.  For an illustration, some of these CF patterns are shown in Figure~\ref{fig:CFpatterns}.

Once the CF pattern is identified, one can also determine the actual number of CFs needed to fully characterize the modulation type through its set of associated CC values.

To reduce computations for the CCs (\ref{eq:CCexpr}), the following parameters were used:
\begin{itemize}
\item The delay vector $\boldsymbol{{\tau}} = 0$;
\item The orders of CC features were limited to the set $n = \left \lbrace 2, 4, 6 \right \rbrace$ ({The number of conjugation choices was constrained by the order $n$ to $n+1$.}); 
\item For each $\left( n, m \right)$ pair, the CFs where CCs are non-zero are related to the CFO ($f_0$) and symbol rate ($1/T_0$) by eq.~(\ref{eq:basic_cf_psk_qam}), where the set of non-negative integers $k$ is restrained to a maximum value of five.
\end{itemize}

These settings imply a total of $11$ potential CFs for each of the $15$ $\left( n, m \right)$ pairs or a maximum of $165$ CC estimates for each digitally modulated signal to be classified.  The actual number will depend on the blindly estimated second-order CF pattern.

\pagebreak

\section{The Cyclic Cumulant Estimate}\label{sec:CC_estimate}
After the band-of-interest detection, blind key parameter estimation, and CF pattern determination are performed, the cyclic cumulants can be estimated.  This was performed by combining appropriate estimates of cyclic temporal moment functions (CTMFs), which are the Fourier series coefficients of the temporal moment functions (\ref{eq:tmf}).  The CTMF for cycle frequency $\gamma$ is given by
\begin{align}
R_x^\gamma(\boldsymbol{\tau};n,m) = \lim_{T\rightarrow\infty} \frac{1}{T} \int_{-T/2}^{T/2} R_x(t, \boldsymbol{\tau};n,m) e^{-i2\pi \gamma t} \, dt.
\label{eq:ctmf}
\end{align}
It can be shown that the cyclic cumulant (\ref{eq:CCexpr}) is given by 
\begin{align}
\label{eq:CCexpr_2}
C_x^\alpha (\boldsymbol{\tau};n,m) = \sum_{P_n} \left[ h(p) \sum_{\boldsymbol{\gamma}\boldsymbol{1}^\dagger = \alpha} \prod_{j=1}^p R_{x_{\nu_j}}^{\gamma_j} (\boldsymbol{\tau}_{\nu_j}; n_j, m_j) \right],
\end{align}
where $\boldsymbol{\gamma} = [\gamma_1\ \gamma_2 \ldots \gamma_p]$, and the sum over $\boldsymbol{\gamma}$ requires the inclusion of all such distinct CTMF cycle frequency vectors whose components sum to the target cyclic cumulant cycle frequency~$\alpha$~\cite{Gardner_CSP_pt1_1994,Spooner_CSP_pt2_1994}.

An estimate of the CC (\ref{eq:CCexpr_2}) (equivalently (\ref{eq:CCexpr})) is given by simply replacing the infinite-time averages in the definition of the Fourier coefficients (\ref{eq:ctmf}) with finite-time averages and by replacing the TMF in (\ref{eq:ctmf}) with the corresponding homogeneous delay product,
\begin{align}
\label{eq:ctmf_est}
\hat{R}_x^{\gamma} (\boldsymbol{\tau};n,m) = \frac{1}{T} \int_0^{T} \left[ \prod_{j=1}^n x^{(*)_j}(t+\tau_j) \right] e^{-i 2 \pi \gamma t} \, dt, 
\end{align}
where the shift in the integration interval results in a phase shift of the CTMF relative to (\ref{eq:ctmf}) and is easily accommodated.  The estimate of the cyclic cumulant is then given by the properly weighted sum of products of CTMF estimates:
\begin{align}
\label{eq:ctcf_est}
\hat{C}_x^\alpha (\boldsymbol{\tau};n,m) = \sum_{P_n} \left[ h(p) \sum_{\boldsymbol{\gamma}\boldsymbol{1}^\dagger = \alpha} \prod_{j=1}^p \hat{R}_{x_{\nu_j}}^{\gamma_j} (\boldsymbol{\tau}_{\nu_j}; n_j, m_j) \right].
\end{align}
Because we have estimated the CF pattern and the key signal parameters, we can find all required lower-order cycle frequencies implied by the sum over the lower-order cycle frequency vector~$\boldsymbol{\gamma}$.

\section{Baseline Classification Model}\label{sec:Baseline_Model}
The baseline classification model is a conventional CC-based classifier as outlined in~\cite{Spooner_Asilomar2001, Spooner_Asilomar2000}, where CC values are estimated for each signal to be classified and a modulation classification decision is made based on the closest proximity of the estimated CCs to a modulation's theoretical CC values to assign a signal modulation label based as discussed in~\cite{Spooner_Asilomar2001, Spooner_Asilomar2000}.  Eight digital modulation schemes of interest are considered:  BPSK, QPSK, 8PSK, $\pi/4$-DQPSK, MSK, 16QAM, 64QAM, and 256QAM, and the performance of the CSP baseline classification model is assessed on the two datasets available from \cite{CSPblog_DataSets}.  Details on these datasets are discussed in Section~\ref{sec:Dataset}, and all of the signal processing techniques used in the CSP baseline classification model are fully blind and all parameters needed (such as the symbol interval or rate, CFO, or signal bandwidth) are estimated from the I/Q data using signal processing as outlined in Section~\ref{sec:Proposed_Model_CC_FE}.  This allows a fair comparison with the performance of the proposed DL-based classifier, which uses capsule networks and is described in the following section, because it is also not provided any prior information when making an inference.

\chapter{Cyclic Cumulants and Capsule Networks for Digital Modulation Classification}\label{ch:CCsNNsModClass}
Capsule networks (CAPs) are a particular set of NNs that have been introduced in the context of emulating human vision \cite{Sabour_etal_NIPC2017} because of their proven ability to focus on learning desirable characteristics of the input pattern or signal, which correspond to a specific input class.  In the case of the human visual system, when the eye is excited by a visual stimulus, it does not focus on all available inputs but rather establishes points of fixation instead, which can be thought of as characteristics present in the input data that are useful for classification.  CAPs are a special class of shallow CNNs in which the learned desirable characteristics of the training dataset are captured by means of capsules consisting of multiple parallel and independent units that can learn class-specific characteristics of the training data.  CAPs differ from CNNs, which rely on a single output neuron per class, as well as from multi-branch NNs, in which the multiple branches processed are recombined into fully connected layers.

\begin{figure}
\centering
\includegraphics[width=0.75\linewidth]{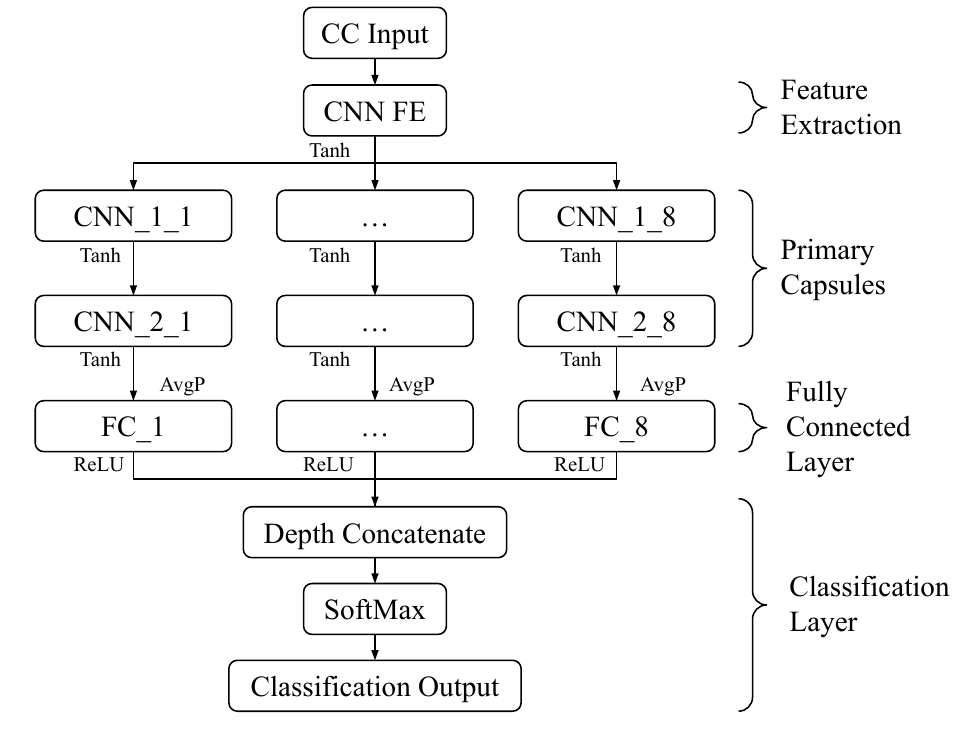}
\caption[Topology of the CC-trained Capsule Network for Classification of Digitally Modulated Signals With Eight Branches.]{\textbf{Topology of the CC-trained Capsule Network for Classification of Digitally Modulated Signals With Eight Branches.}}\label{fig:CAP_Topology1}
\end{figure}

In recent years, CAPs have been successfully used in DL-based modulation recognition systems \cite{li2020automatic, sang2018application} and have been shown to display better classification performance than CNNs and RNs in the I/Q-based classification of digitally modulated signals~\cite{Latshaw_COMM2022}.  The apparent superiority of CAPs over other types of NNs has prompted the proposed approach for digital modulation classification, in which the CC features of digitally modulated signals were used as the inputs to the CAP to train it to classify the same eight digital modulation schemes of interest mentioned in Section~\ref{sec:Baseline_Model} (BPSK, QPSK, 8PSK, $\pi/4$-DQPSK, MSK, 16QAM, 64QAM, and 256QAM).  Consequently, the CAP used in this approach consists of eight capsules, as illustrated in Figure~\ref{fig:CAP_Topology1}, taking as inputs the $11 \times 15 = 165$ cyclic cumulant (CC) values of the received signal, which matches the dimension of the input layers for the defined capsules.

Because in general, the higher-order CCs have larger magnitudes than the lower-order CCs and because the CCs also scale with the signal power, the CC estimates $\widehat{C}_x^{\alpha} \left( \boldsymbol{{\tau}}; n, m \right)$ were further processed as follows prior to use with the proposed CAP for training and classification:
\begin{itemize}
\item {\em Warping}:  This involves using the order $n$ of the CC estimates to obtain ``warped'' versions $\widehat{C}_x^{\alpha} \left( \boldsymbol{{\tau}}; n, m \right)^{ \left( 2/n \right) }$.  CSP-based blind modulation classification also employs warped CC estimates~\cite{Spooner_Asilomar2000}.
\item {\em Scaling}:  The warped CC estimates were subsequently scaled to a signal power of unity, using a blind estimate of the signal power.  This provided consistent values for the capsule network to train on and prevented varying signal powers from causing erroneous classification results due to neuron saturation---a common issue with input data that do not go through some normalization process.
\end{itemize}

After these pre-processing steps, the warped and scaled CC estimates can be used to train the CAP with the structure shown in Figure~\ref{fig:CAP_Topology1} and parameters outlined in Table~\ref{table:CC_CAPnetwork}, and subsequently, the trained CAP can be employed to blindly classify digitally modulated signals.  The various components of the proposed CAP include:
\begin{itemize}
\item \textbf{Feature Extraction Layer:}  This first layer of the network performs a general feature mapping of the input signal, and its parameters are similar to those used in other DL-based approaches to classification of digitally modulated signals \cite{Tim2018, Zhou_etal_EURASIP_SP2019, Snoap_CCNC_2022}.  This layer includes a convolutional layer followed by a batch normalization layer and an activation function.
\item \textbf{Primary Capsules:}  This layer consists of eight primary capsules, which is equal to the number of digital modulation classes of interest.  These capsules operate in parallel using as the input the output from the feature extraction layer, and each primary capsule includes two convolutional layers with a customized filter, stride, and activation function, followed by a fully connected layer.
\item \textbf{Fully Connected Layer:}  This layer consists of a $1 \times 8$ neuron vector with the weights connecting to the previous layer.  Each neuron in the last layer of the primary capsules will be fully connected to each neuron in this layer. These neurons are expected to discover characteristics specific to the capsules' class. To make the output of the network compatible with a {SoftMax} classification layer, each neuron within this layer is fully connected to a single output neuron, and the output neurons for all primary capsules are combined depthwise to produce an eight-dimensional vector \textbf{n}, which is passed to the classification layer.  The value of each respective element of \textbf{n} will be representative of the likelihood that its corresponding modulation type is present in the received digitally modulated signal.
\item \textbf{Classification Layer:}  In this layer, vector \textbf{n} is passed to the {SoftMax} layer, which will map each element $n_i$, $i=1,\ldots,8$, in \textbf{n} to a value,
\begin{align}
\sigma_i (\textbf{n}) = \frac{e^{n_{i}}}{\displaystyle \sum_{j=1}^{8} e^{n_{j}}} \label{eq:sm2},
\end{align}
where $\sigma_i (\textbf{n}) \in [0,1]$, with each element representing the probability of occurrence, such that the sum of elements in \textbf{n} adds up to $1$ \cite{luce_2008}.  This provides a convenient way to determine which modulation type is most likely to correspond to the signal at the input of the CAP.
\end{itemize}

\begin{table}
\centering
\caption[CC-trained Capsule Network Layout.]{\textbf{CC-trained Capsule Network Layout.}}
{
\begin{tabular}{ c c c c }
\toprule
\textbf{Layer} & \textbf{(\# Filters)[Filter Size]} & \textbf{Stride} & \textbf{Activations} \\
\midrule
Input					&									&					& $11 \times 15 \times 1$ \\
Conv					& ($56$)[$6 \times 4 \times 1$]	& [$1 \times 2$]	& $11 \times 8 \times 56$ \\
Batch Norm				& \\
Tanh					& \\
Conv-1-(i)				& ($56$)[$4 \times 4 \times 56$]	& [$1 \times 2$]	& $11 \times 4 \times 56$ \\
Batch Norm-1-(i)		& \\
Tanh-1-(i)				& \\
Conv-2-(i)				& ($72$)[$4 \times 6 \times 56$]	& [$1 \times 2$]	& $11 \times 2 \times 72$ \\
Batch Norm-2-(i)		& \\
Tanh-2-(i)				& \\
FC-(i)					& &	& $7$ \\
Batch Norm-3-(i)		& \\
ReLu-1-(i)				& \\
Point FC-(i)			& & & $1$ \\
Depth Concat(i=1:8)		& & & $8$ \\
SoftMax					& \\
\bottomrule
\end{tabular}
}\label{table:CC_CAPnetwork}
\end{table}

{
Similar to the CAP used in \cite{Latshaw_COMM2022}, the CAP described above was inspired by~\cite{Sabour_etal_NIPC2017}, and its structure and topology were established using a custom genetic algorithm that determined the CAP parameters shown in Table~\ref{table:CC_CAPnetwork} (convolutional layer filter size, filter stride, and the number of layers).  The algorithm would randomly choose a value (over a defined interval) for each of the above-listed parameters, randomly turning off a layer or adding a new one.  This ``pseudo-randomly'' generated network would then be trained and evaluated against the testing dataset, repeating the experiment multiple times, with the best-performing networks being noted.  The hyper-parameters of the best-performing network(s) were noted, and their likelihood of selection for subsequent experiments was slightly increased.  Over the course of many such experiments, networks having a specific layer structure and layer hyper-parameters began to emerge, which helped to inform the chosen topology.
}

\section{CAP Training and Performance Evaluation}\label{sec:Dataset}
To train and assess the modulation classification performance of the proposed CAP (including its out-of-distribution generalization ability), two datasets were used that both contain the eight modulation types of interest (BPSK, QPSK, 8PSK, $\pi/4$-DQPSK, MSK, 16QAM, 64QAM, and 256QAM) and are publicly available from \cite{IEEE_DataPort}.



\subsection{The Training/Testing Datasets}\label{sec:Datasets}

The two datasets are referred to as \texttt{CSPB.ML.2018} and \texttt{CSPB.ML.2022} \cite{IEEE_DataPort}, and details about their signal-generation parameters are given in Table~\ref{table:SigGenParms} in Chapter~\ref{ch:HQDatasets} of this dissertation.  To summarize their characteristics, each of the datasets contains collections of the I/Q data corresponding to a total of 112,000 synthetic digitally modulated signals that include equal numbers of the eight digital modulation schemes of interest.  With the exception of MSK-modulated signals, all other signals employ square-root raised-cosine (SRRC) pulse-shaping with roll-off factor~$\beta$, and 32,768 samples for each instance of each signal are provided.  The listed SNRs for the signals in both datasets correspond to in-band SNR values and a BOI detector \cite{BOIdetector} was used to validate the labeled SNRs, CFOs, and SRRC roll-off values for the signals in both datasets.

Reviewing the signal-generation parameters of these two datasets outlined in Table~\ref{table:SigGenParms} in Chapter~\ref{ch:HQDatasets} of this dissertation confirms that they are suited for testing the generalization abilities of the proposed capsule network as the signals in the two datasets were generated with distinct non-overlapping ranges for the CFO.  Specifically, the maximum CFO in \texttt{CSPB.ML.2018} is $0.001$, while the minimum CFO in \texttt{CSPB.ML.2022} is an order of magnitude larger at $0.01$.  As the other signal-generation parameters are similar for the two datasets, the differences in the CFO will enable the observation of the generalization abilities of the trained CAPs, as will be discussed in detail in Section~\ref{sec:Performance}:
\begin{itemize}
\item For the CAP that uses the I/Q signal data for training and testing, the CFO shift in the testing dataset relative to the training dataset resulted in significant degradation of the classification performance of the CAP and indicated that it was unable to generalize its training to new datasets that contain similar types of signals, but with differences in some of their digital modulation characteristics.  This aspect was also reported in~\cite{Latshaw_COMM2022}, and similar results have been reported for CNNs and RNs in~\cite{Snoap_CCNC_2022}.
\item As will be seen in Section~\ref{sec:Performance}, the CAP that uses the CC features for training and testing the CFO shift in the testing dataset relative to the training dataset resulted in similar classification performance and indicated that the CAP trained using CC features was resilient to variations of the CFO from the training dataset.
\end{itemize}

\subsection{CAP Training}\label{sec:Training}
The proposed CAP was implemented in MATLAB and trained on a high-performance computing cluster with 18~NVidia V100 graphical processing unit (GPU) nodes available, with each node having 128 GB of memory.  While the DL network training process is computationally intensive, if the available computing resources are leveraged appropriately such that the entire training dataset is loaded into the available memory, training can be completed in several minutes for the CC-trained networks ({provided the CC estimates are readily available}), as compared to several hours in the case of a CAP that uses the I/Q signal data \cite{Latshaw_COMM2022}.

The CC-trained CAPs obtained the best results with an adaptive moment estimation (Adam) optimizer \cite{ADAM_ML} using ten epochs while shuffling the training data before each training epoch, a mini-batch size of $250$, an initial learning rate of $0.001$, a piecewise learning schedule involving a multiplicative learning rate drop factor of $0.9$ every five epochs, an $L_2$ regularization factor of $0.0001$, a gradient decay factor of $0.9$, a squared gradient decay factor of $0.999$, an epsilon denominator offset of $10^{-8}$, and a final batch normalization using the entire training data population statistics.

Two distinct training/testing instances were performed as follows:
\begin{itemize}
\item In the first training instance, dataset \texttt{CSPB.ML.2018} was used, splitting the available signals into $70\%$ for training, $5\%$ for validation, and $25\%$ for testing.  The corresponding objective and loss functions for the trained CAP are shown in Figure~\ref{fig:obj_loss_2018_1}, and the probability of the correct classification for the test results was obtained using the $25\%$ test portion of the signals in \texttt{CSPB.ML.2018}.

\begin{figure}
\begin{center}
\subfigure[Objective Function.]
{\includegraphics[width=\linewidth]{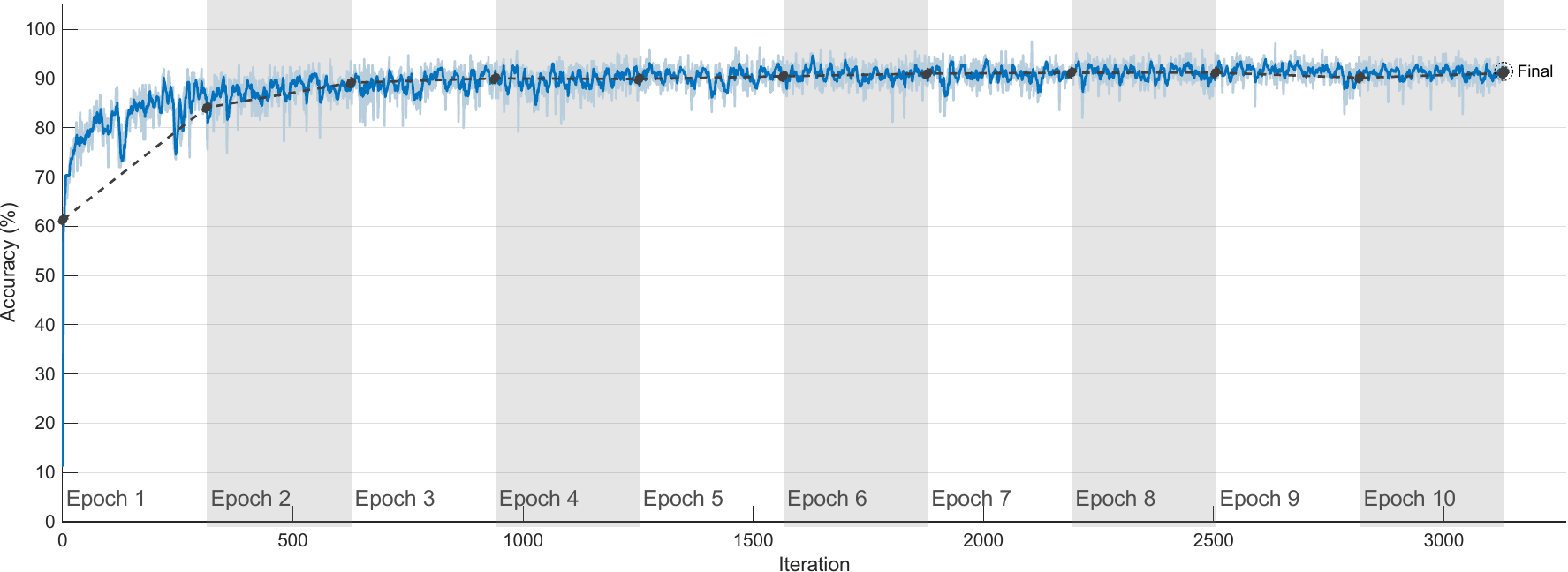}}
\subfigure[Loss Function.]
{\includegraphics[width=\linewidth]{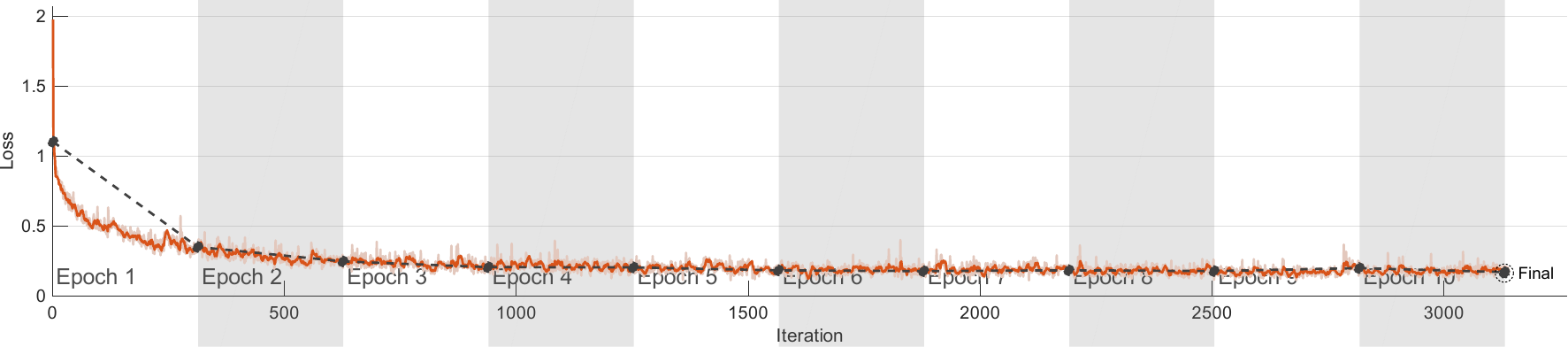}}
\caption[Objective and Loss Functions for the CC-trained CAP Using the \texttt{CSPB.ML.2018} Dataset.]{\textbf{Objective and Loss Functions for the CC-trained CAP Using the \texttt{CSPB.ML.2018} Dataset.}}\label{fig:obj_loss_2018_1}
\end{center}
\end{figure}

\begin{figure}
\begin{center}
\subfigure[Objective Function.]
{\includegraphics[width=\linewidth]{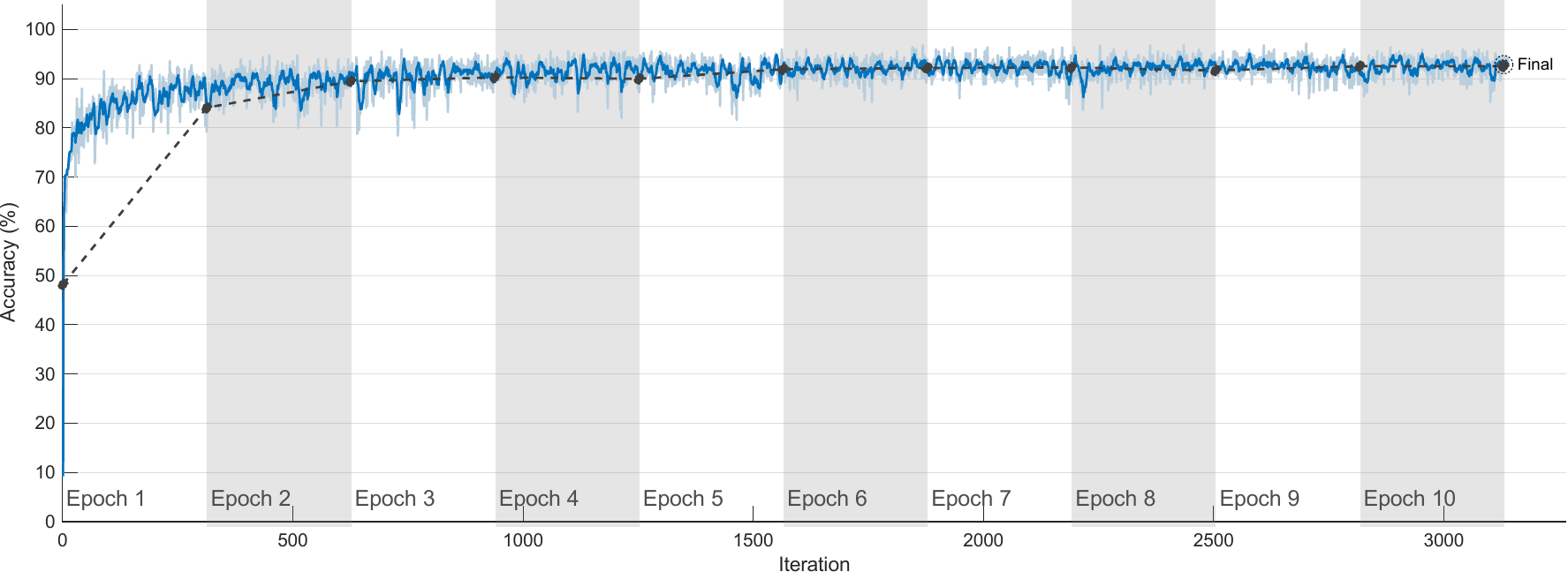}}
\subfigure[Loss Function.]
{\includegraphics[width=\linewidth]{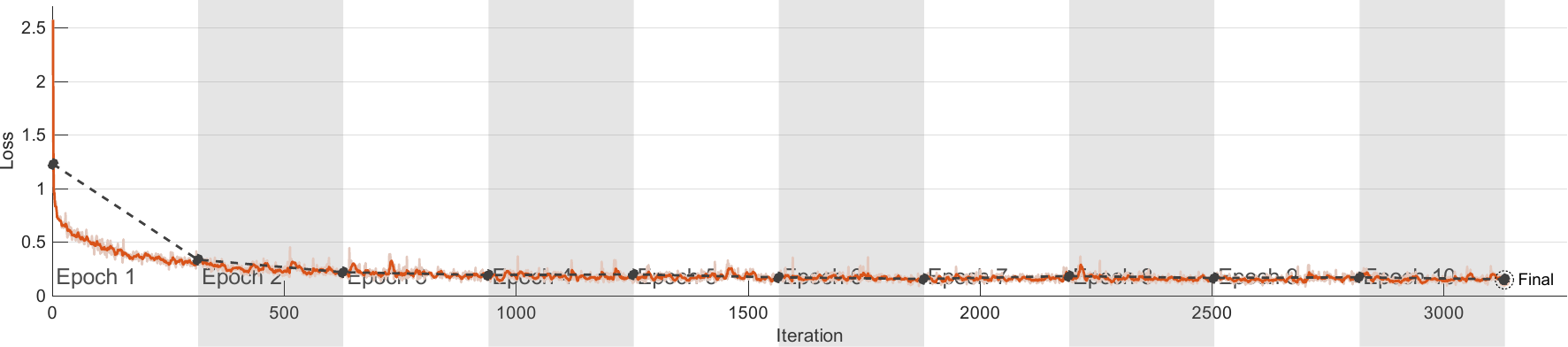}}
\caption[Objective and Loss Functions for the CC-trained CAP Using the \texttt{CSPB.ML.2022} Dataset.]{\textbf{Objective and Loss Functions for the CC-trained CAP Using the \texttt{CSPB.ML.2022} Dataset.}}\label{fig:obj_loss_2022_1}
\end{center}
\end{figure}

\hspace{0.5cm} The CAP trained on \texttt{CSPB.ML.2018} was then tested on dataset \texttt{CSPB.ML.2022} to assess the generalization abilities of the trained CAP in classifying all signals available in \texttt{CSPB.ML.2022}.

\item In the second training instance, the CAP was reset and trained anew using the signals in dataset \texttt{CSPB.ML.2022}, with a similar split of $70\%$ of signals used for training, $5\%$ for validation, and $25\%$ for testing.  The corresponding objective and loss functions for the trained CAP were similar to the ones in Figure~\ref{fig:obj_loss_2018_1}.  The probability of correct classification for the test results was obtained using the $25\%$ test portion of the signals in \texttt{CSPB.ML.2022}.

\hspace{0.5cm} The CAP trained on \texttt{CSPB.ML.2022} was then tested on dataset \texttt{CSPB.ML.2018} to assess the generalization abilities of the re-trained CAP when classifying all signals available in \texttt{CSPB.ML.2018}.
\end{itemize}

\newpage

\subsection{Assessing Generalization Abilities}\label{sec:Generalization}
The CFOs for both datasets were generated randomly with a uniform distribution and the CFO distribution interval for \texttt{CSPB.ML.2018} was non-intersecting with the CFO interval for \texttt{CSPB.ML.2022}.  This was performed to assess the ability of a trained CAP to generalize:
\begin {itemize}
\item If the CAP was trained on a large portion of \texttt{CSPB.ML.2018} and its performance when classifying a remaining subset of \texttt{CSPB.ML.2018} was high, but its performance when classifying \texttt{CSPB.ML.2022} was low, then the CAP's ability to generalize was low, and its performance was vulnerable to shifts in the signal parameter distributions.
\item By contrast, if the classification performance of the CAP on both the remaining subset of \texttt{CSPB.ML.2018} and on all of \texttt{CSPB.ML.2022} was high, then its generalization ability was high, and the CAP was resilient to shifts in signal parameter distributions.
\end{itemize}



\section{Numerical Results and Performance Analysis}\label{sec:Performance}
\begin{figure}
\begin{center}
\subfigure[Initial and Generalization Test Results for CAPs Trained on \texttt{CSPB.ML.2018}.]
{\includegraphics[width=0.49\linewidth]{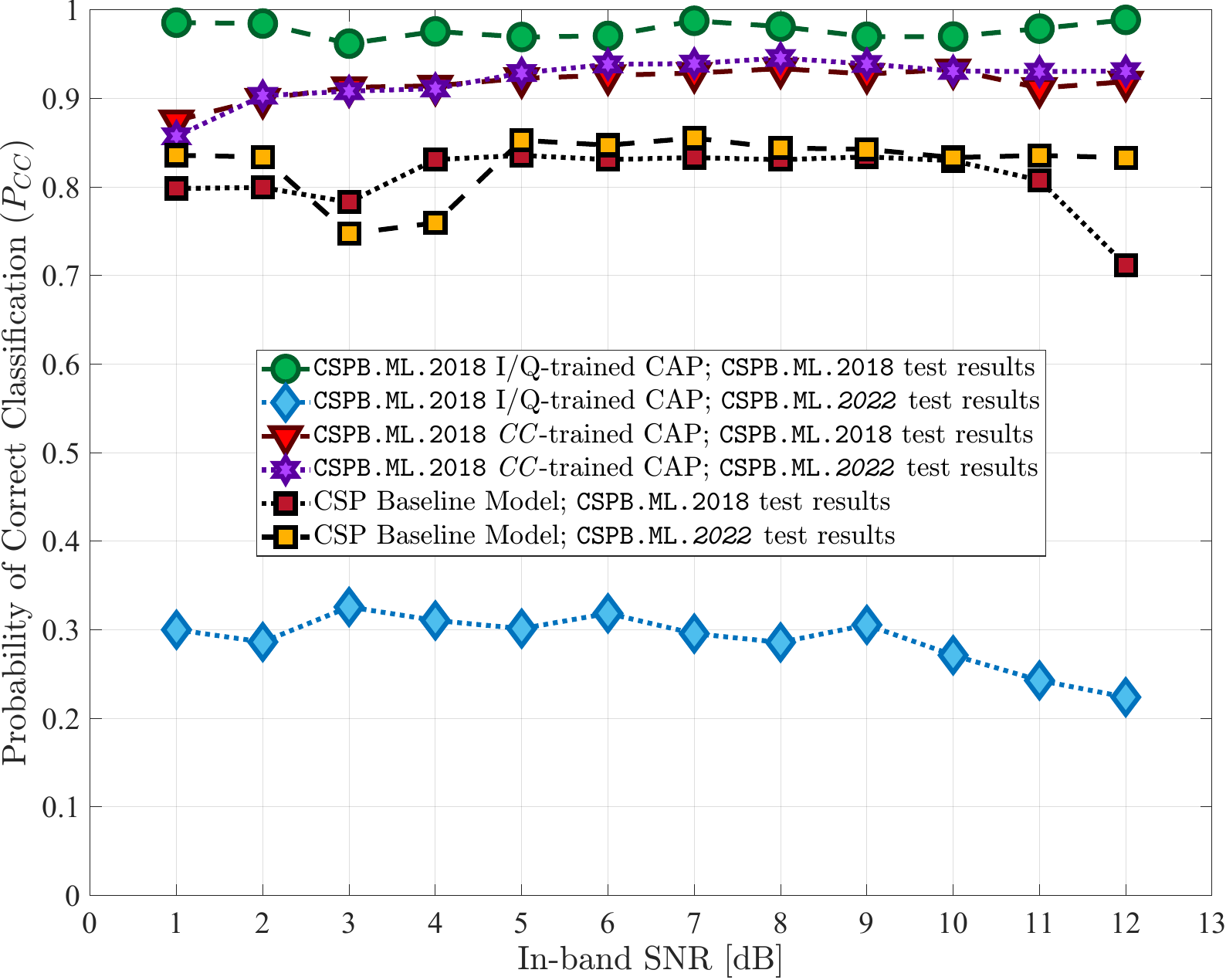}}
\subfigure[Initial and Generalization Test Results for CAPs Trained on \texttt{CSPB.ML.2022}.]
{\includegraphics[width=0.49\linewidth]{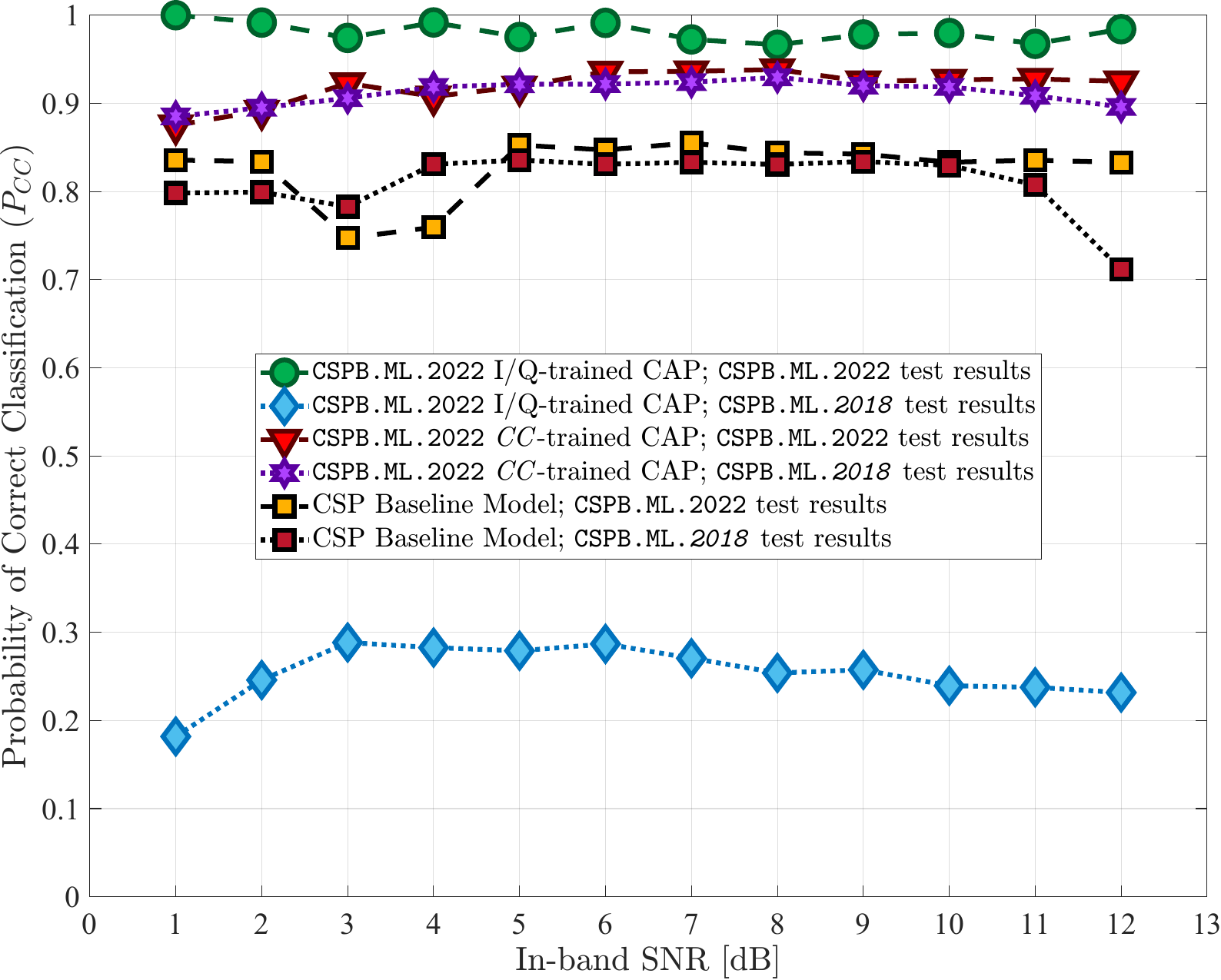}}
\caption[Initial and Generalization Test Results for CAPs Trained on \texttt{CSPB.ML.2018} or \texttt{CSPB.ML.2022}.]{\textbf{Initial and Generalization Test Results for CAPs Trained on \texttt{CSPB.ML.2018} or \texttt{CSPB.ML.2022}.}  The CC-trained CAP exhibited high classification performance under both testing scenarios.}\label{fig:SNR_CTrained_All}
\end{center}
\end{figure}

Simulations were run to test the classification performance of the proposed CAP and compared it to the performance of the CSP baseline classification model and to the performance of a CAP that used I/Q data as the input \cite{Latshaw_COMM2022}.  The results are summarized in Table~\ref{table:OverallPerformance1} and Figures~\ref{fig:SNR_CTrained_All}(a) and \ref{fig:SNR_CTrained_All}(b) and are discussed in the following sections.

\begin{table}
\centering
\caption[Overall CC-trained CAP Classification Performance ($P_{CC}$).]{\textbf{Overall CC-trained CAP Classification Performance ($P_{CC}$).}}
{
\setlength{\tabcolsep}{6.5mm}
\begin{tabular}{ c c c }
\toprule
\textbf{Classification Model} & \begin{tabular}{@{}c@{}} \textbf{Results for Dataset} \\ \texttt{CSPB.ML.2018} \end{tabular} & \begin{tabular}{@{}c@{}} \textbf{Results for Dataset} \\ \texttt{CSPB.ML.2022} \end{tabular} \\
\midrule
CSP Baseline Model & $82.0\%$ & $82.0\%$ \\
\hline
\begin{tabular}{@{}c@{}} \texttt{CSPB.ML.2018} \\ I/Q-trained CAP\end{tabular} & $97.5\%$ & $23.7\%$ \\
\begin{tabular}{@{}c@{}} \texttt{CSPB.ML.2022} \\ I/Q-trained CAP\end{tabular} & $25.7\%$ & $97.7\%$ \\
\hline
\begin{tabular}{@{}c@{}} \texttt{CSPB.ML.2018} \\ CC-trained CAP\end{tabular} & $92.3\%$ & $93.1\%$ \\
\begin{tabular}{@{}c@{}} \texttt{CSPB.ML.2022} \\ CC-trained CAP\end{tabular} & $91.6\%$ & $92.5\%$ \\
\bottomrule
\end{tabular}
}\label{table:OverallPerformance1}
\end{table}

\newpage

\subsection{Baseline Model Performance}\label{sec:Performance_Baseline}
The CSP baseline classification model requires no training as it was based solely on comparing the estimated CCs of the digitally modulated signal with theoretical CC values corresponding to different modulation schemes.  This conventional CSP approach to the classification of digitally modulated signals resulted in good performance for both datasets, \texttt{CSPB.ML.2018} and \texttt{CSPB.ML.2022}.  By utilizing all of the 32,768 samples for each signal in a dataset, the CSP baseline model yielded quite accurate estimates of the CC features and resulted in an overall modulation classification accuracy of $82.0\%$ for both \texttt{CSPB.ML.2018} and \texttt{CSPB.ML.2022}.

The CSP baseline model performance curves shown in Figures~\ref{fig:SNR_CTrained_All}(a) and \ref{fig:SNR_CTrained_All}(b) do not display a steady increase in $P_{CC}$ with increasing SNR, which highlights the difficulty associated with performing modulation classification on these two datasets (\texttt{CSPB.ML.2018} and \texttt{CSPB.ML.2022}) using conventional CSP-based approaches.  Nevertheless, the results of the CSP baseline model set a clear standard for performance and generalization because it was based on CSP and the minimum distance to theoretical CCs, with no training required, such that the classification results can be obtained with equally high performance, independent of signal parameter distribution functions.

\subsection{CC-trained CAP Performance}\label{sec:Performance_DL_CC_CAP}
For the proposed CAP that was trained using CC features extracted from signals in dataset \texttt{CSPB.ML.2018}, the variation of $P_{CC}$ versus the SNR is shown in Figure~\ref{fig:SNR_CTrained_All}(a).  The overall $P_{CC}$ achieved by this network was $92.3\%$, which was about a $10\%$ improvement over the CSP baseline model.  When the CAP trained using CC features from signals in the \texttt{CSPB.ML.2018} dataset was used to classify signals in the \texttt{CSPB.ML.2022} dataset, the overall $P_{CC}$ value continued to remain high at $93.1\%$, which was about $11\%$ larger than that of the CSP baseline classification model and implied excellent generalization abilities.

When the CAP was trained using CC features from the signals in the \texttt{CSPB.ML.2022} dataset, the variation of $P_{CC}$ versus SNR is shown in Figure~\ref{fig:SNR_CTrained_All}(b), and its classification performance was similar to the previous case.  The overall $P_{CC}$ achieved by this network was $92.5\%$, which was again about a $10\%$ improvement over the CSP baseline model.  Furthermore, the CAP was able to generalize and maintained an overall $P_{CC}$ value of $91.6\%$ when tested with signals in the \texttt{CSPB.ML.2018} dataset.

These results showed that the proposed CAP can be successfully trained using CC features to perform modulation classification with an overall $P_{CC}$ that exceeded that of conventional CSP-based classification approaches, such as the one used in the CSP baseline model.  Moreover, the proposed CC-trained CAP was able to generalize training and continued to perform better than the CSP baseline classification model even when the signal-generation parameters differed or were out-of-distribution from the signals of the training dataset.  The generalization performance of the proposed approach was due to the fact that the CC features for distinct signals,
\begin{align}
x(t) = a s(t) e^{i (2\pi f_1 t + \phi_1)} + w_x(t), \\
y(t) = a s(t) e^{i (2 \pi f_2 t + \phi_2)} + w_y(t),
\end{align}
where $w_x(t)$ and $w_y(t)$ are independent AWGN processes and $f_1$, $f_2$, are randomly distributed (but not necessarily with the same distribution) and are identical to within the measurement error, and thus, the disjoint probability density functions for the CFO in the two datasets had little effect on the classification performance.  That is, the CFs differed, but the CC values were invariant to this difference.

\newpage

\subsection{I/Q-Trained CAP Performance}\label{sec:Performance_Comparisons}
Also included in Figures~\ref{fig:SNR_CTrained_All}(a) and \ref{fig:SNR_CTrained_All}(b) are plots for the $P_{CC}$ values versus the SNR for the CAPs trained using the I/Q data discussed in \cite{Latshaw_COMM2022}.  The performance results were similar to those reported in \cite{Latshaw_COMM2022} and, despite outperforming both the CSP baseline model and the CC-trained CAP and achieving very good classification performance with signals that have similar generation characteristics as those in the training dataset, the CAP trained using the I/Q data failed to generalize and had poor performance when tested on signals coming from the alternative datasets with characteristics that had not been used in training.  Thus, employing CAPs that use the I/Q signal data for training and modulation classification is not feasible for practical settings since it would only work reliably under signal conditions that fall exactly within its training dataset.

\subsection{Confusion Matrix Results}\label{sec:Confusion_Matrices}
To gain further insight into the classification performance of the proposed CAPs that use CCs for training and testing, the corresponding confusion matrices were also looked at, comparing them with those corresponding to the CSP baseline classification model, as well as to those of the CAPs that used the I/Q data \cite{Latshaw_COMM2022}.  The results for the \texttt{CSPB.ML.2018} dataset are illustrated in Figures~\ref{fig:CM_CSPB_MSSA_C}--\ref{fig:CM_GCTest_GCTrained_IQ_CAP}(b), from which the following observations are made:
\begin{itemize}
\item The confusion matrix for the CSP baseline classification model in Figure~\ref{fig:CM_CSPB_MSSA_C}(a) showed that, for 5 out of the 8 digital modulation schemes of interest (BPSK, QPSK, 8PSK, DPSK, and MSK), the classification exceeded 95\% accuracy, while for the remaining 3 schemes, which were all QAM-based, the classification accuracy was at 72.5\% for 16QAM, 55.9\% for 256QAM, and 41.7\% for 64QAM.
{
The ``unknown'' classification label appears in the confusion matrix of the CSP baseline model because this was not trained, but rather made its classification decision based on the proximity of the estimated CCs to a modulation's theoretical CC values as outlined in Section~\ref{sec:Baseline_Model} and discussed in~\cite{Spooner_Asilomar2001}.  Thus, when the CSP baseline classification model was not able to match a signal with a known pattern, it declared it ``unknown'' instead of confusing it with a different type of signal as DL-based classifiers do.
}
\item For the CC-trained CAPs, Figure~\ref{fig:CM_GCTest_GCTrained_CC_CAP}(b) shows the confusion matrix corresponding to the generalization experiment, in which the capsule network was trained on the \texttt{CSPB.ML.2022} dataset followed by testing using all signals in the \texttt{CSPB.ML.2018} dataset.  The CAP showed almost perfect accuracy (exceeding 99\%) for the BPSK, QPSK, 8PSK, DPSK, and MSK modulation schemes, with significant improvement over the CSP baseline model for the remaining QAM modulation schemes, for which the classification accuracy increased to 97.5\% for 16QAM, 74\% for 256QAM, and 62.3\% for 64QAM, which implied about 20\% or more improvement over the CSP baseline model classification performance.
\item In contrast, the I/Q-trained CAP confusion matrix shown in Figure~\ref{fig:CM_GCTest_GCTrained_IQ_CAP}(b) corresponding to the generalization experiment (the CAP was trained on the \texttt{CSPB.ML.2022} dataset followed by testing on all signals in the \texttt{CSPB.ML.2018} dataset) showed very poor classification accuracy, despite having excellent accuracy when classifying the 25\% test portion of the signals in the \texttt{CSPB.ML.2018} dataset \cite{Latshaw_COMM2022}.
\end{itemize}

Similar classification accuracies were observed for the signals in the \texttt{CSPB.ML.2022} dataset and are also shown in Figures~\ref{fig:CM_CSPB_MSSA_C}--\ref{fig:CM_GCTest_GCTrained_IQ_CAP}.

\subsection{Computational Aspects}\label{sec:Computation}
From a computational perspective, both the CSP baseline model and  the proposed CAP-based classifier required the estimation of the CC features, for which the computational burden was variable depending on the CF pattern determined during processing (the computation was data-adaptive, unlike simpler signal-processing operations such as the FFT).  When the processing is blind, as it is in this work, second- and higher-order processing is applied to find high-accuracy estimates of the rate and carrier offset. Once these key parameter values are known and the CF pattern is determined, the CC computation can commence.

The computational cost of obtaining a CC feature was determined by the following costs for the major computational steps:

\begin{figure}
\begin{center}
\subfigure[Confusion Matrix of the CSP Baseline Model on All \texttt{CSPB.ML.2018} Signals.]
{\includegraphics[width=0.49\linewidth]{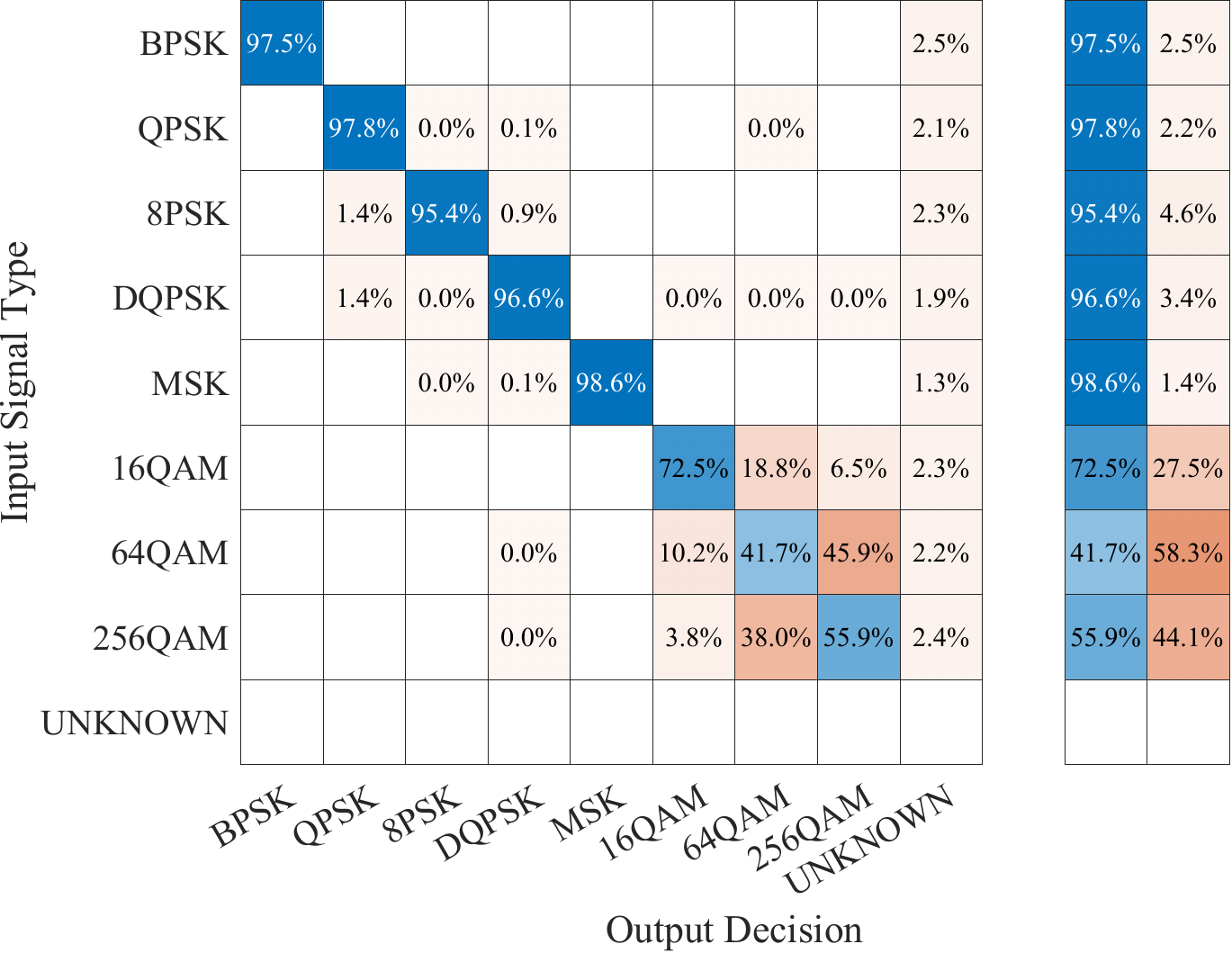}}
\subfigure[Confusion Matrix of the CSP Baseline Model on All \texttt{CSPB.ML.2022} Signals.]
{\includegraphics[width=0.49\linewidth]{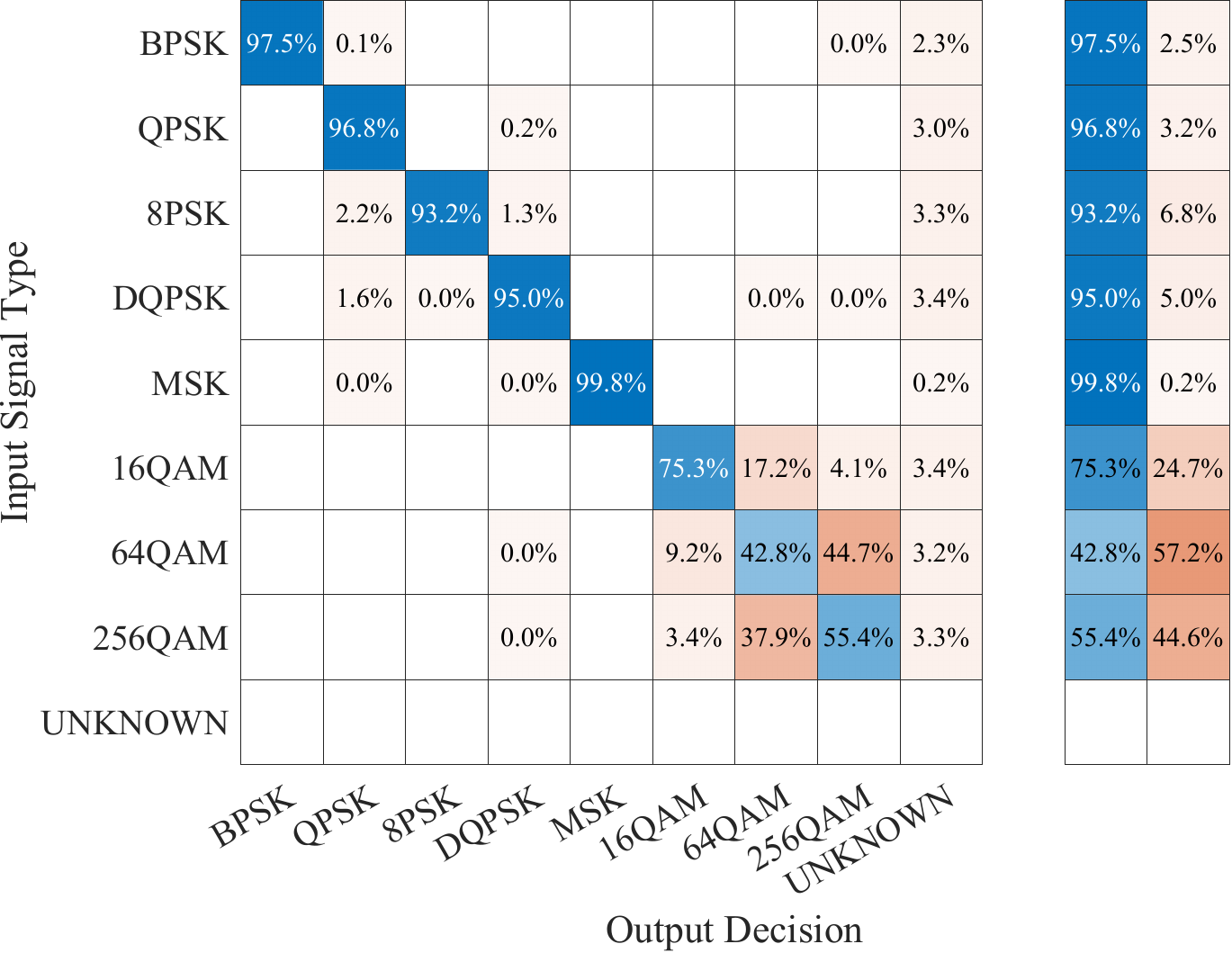}}
\caption[Confusion Matrices of the CSP Baseline Model on All \texttt{CSPB.ML.2018} and \texttt{CSPB.ML.2022} Signals.]{\textbf{Confusion Matrices of the CSP Baseline Model on All \texttt{CSPB.ML.2018} and \texttt{CSPB.ML.2022} Signals.}}\label{fig:CM_CSPB_MSSA_C}
\end{center}
\end{figure}

\begin{figure}
\begin{center}
\subfigure[Confusion Matrix of the \texttt{CSPB.ML.2018} CC-trained CAP Classifying \texttt{CSPB.ML.2018} Test Signals.]
{\includegraphics[width=0.49\linewidth]{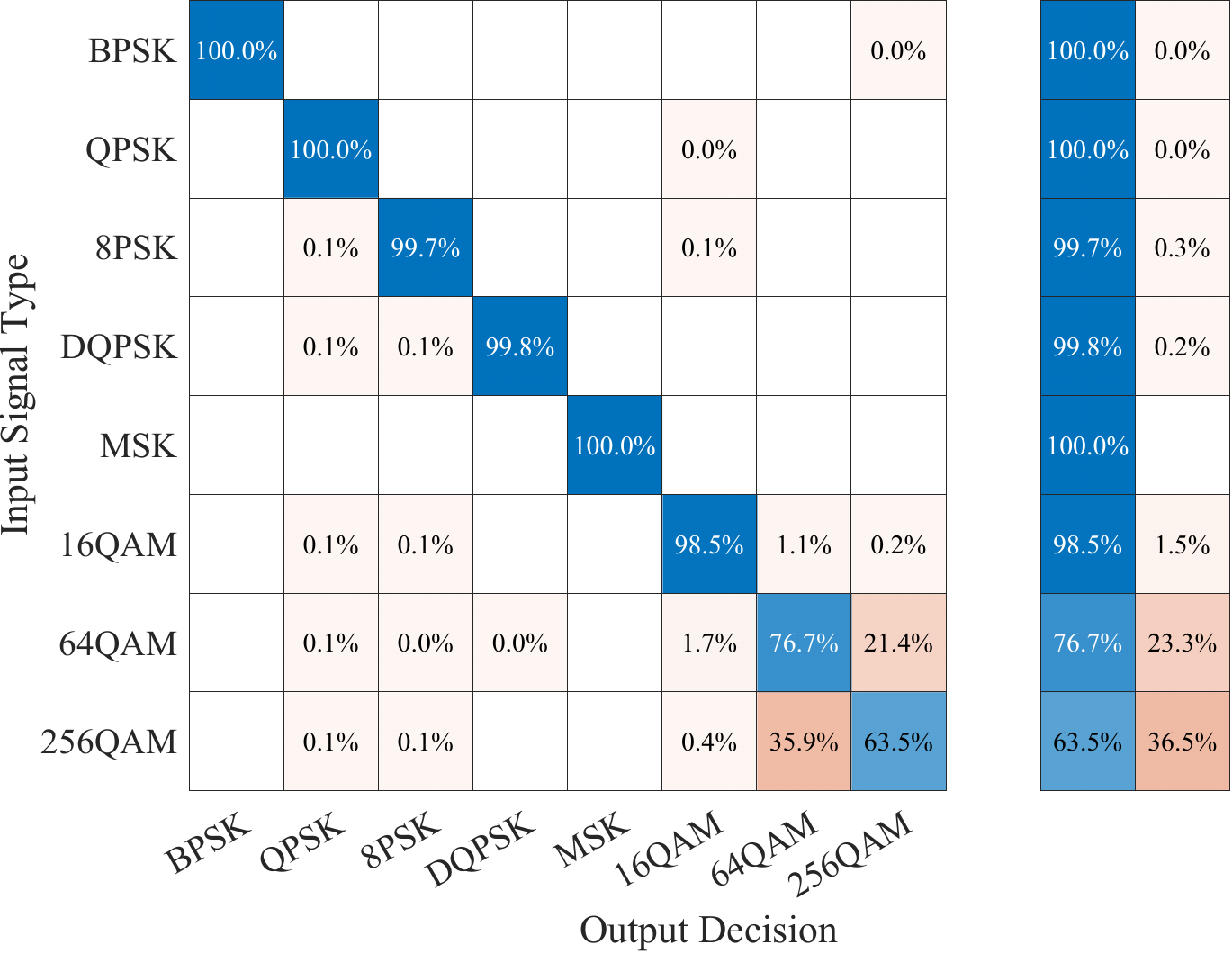}}
\subfigure[Confusion Matrix of the \texttt{CSPB.ML.2018} CC-trained CAP Classifying All \texttt{CSPB.ML.2022} Signals.]
{\includegraphics[width=0.49\linewidth]{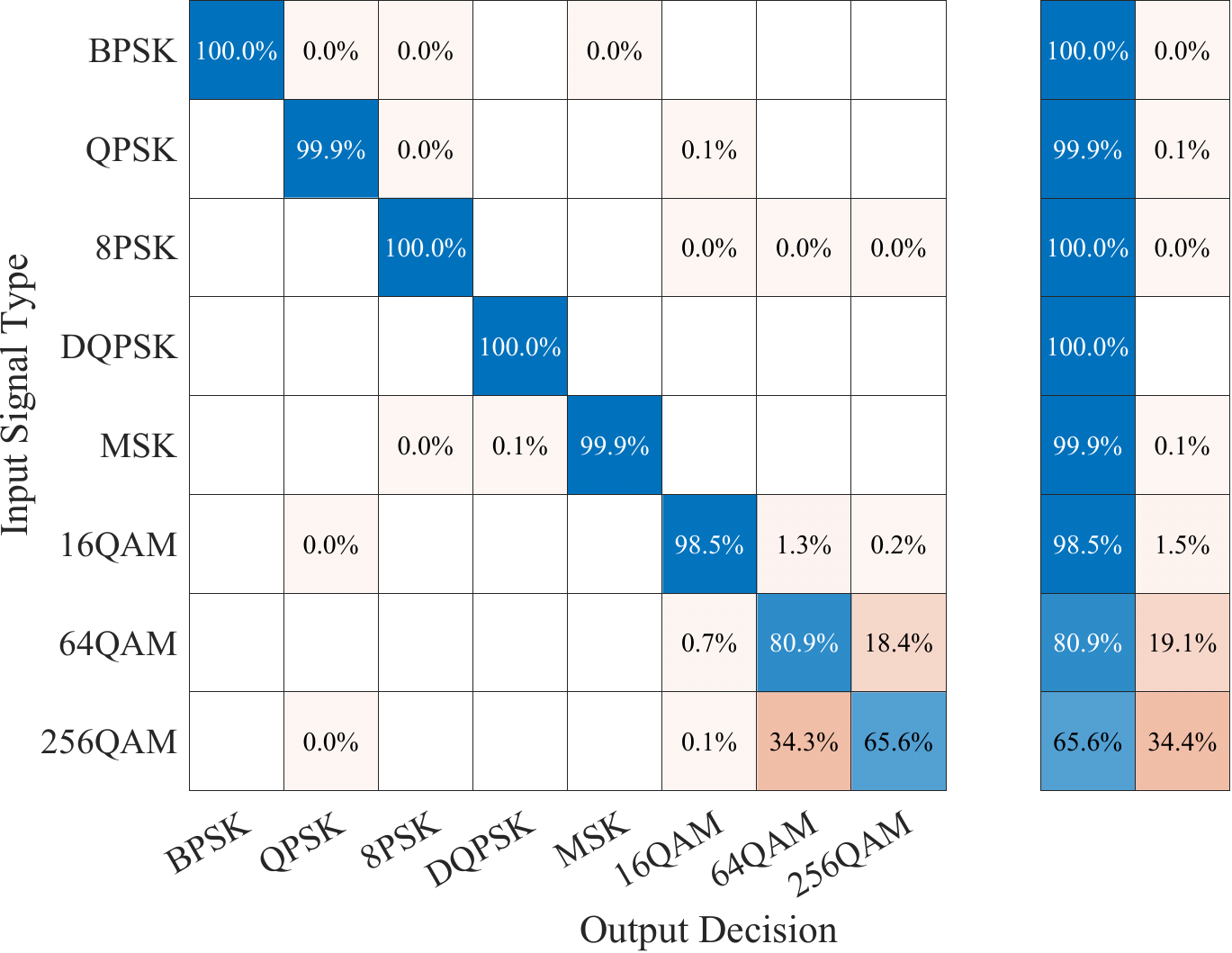}}
\caption[Confusion Matrices of the \texttt{CSPB.ML.2018} CC-trained CAP Classifying \texttt{CSPB.ML.2018} Test Signals and All \texttt{CSPB.ML.2022} Signals.]{\textbf{Confusion Matrices of the \texttt{CSPB.ML.2018} CC-trained CAP Classifying \texttt{CSPB.ML.2018} Test Signals and All \texttt{CSPB.ML.2022} Signals.}}\label{fig:CM_CTest_CTrained_CC_CAP}
\end{center}
\end{figure}

\begin{figure}
\begin{center}
\subfigure[Confusion Matrix of the \texttt{CSPB.ML.2022} CC-trained CAP Classifying \texttt{CSPB.ML.2022} Test Signals.]
{\includegraphics[width=0.49\linewidth]{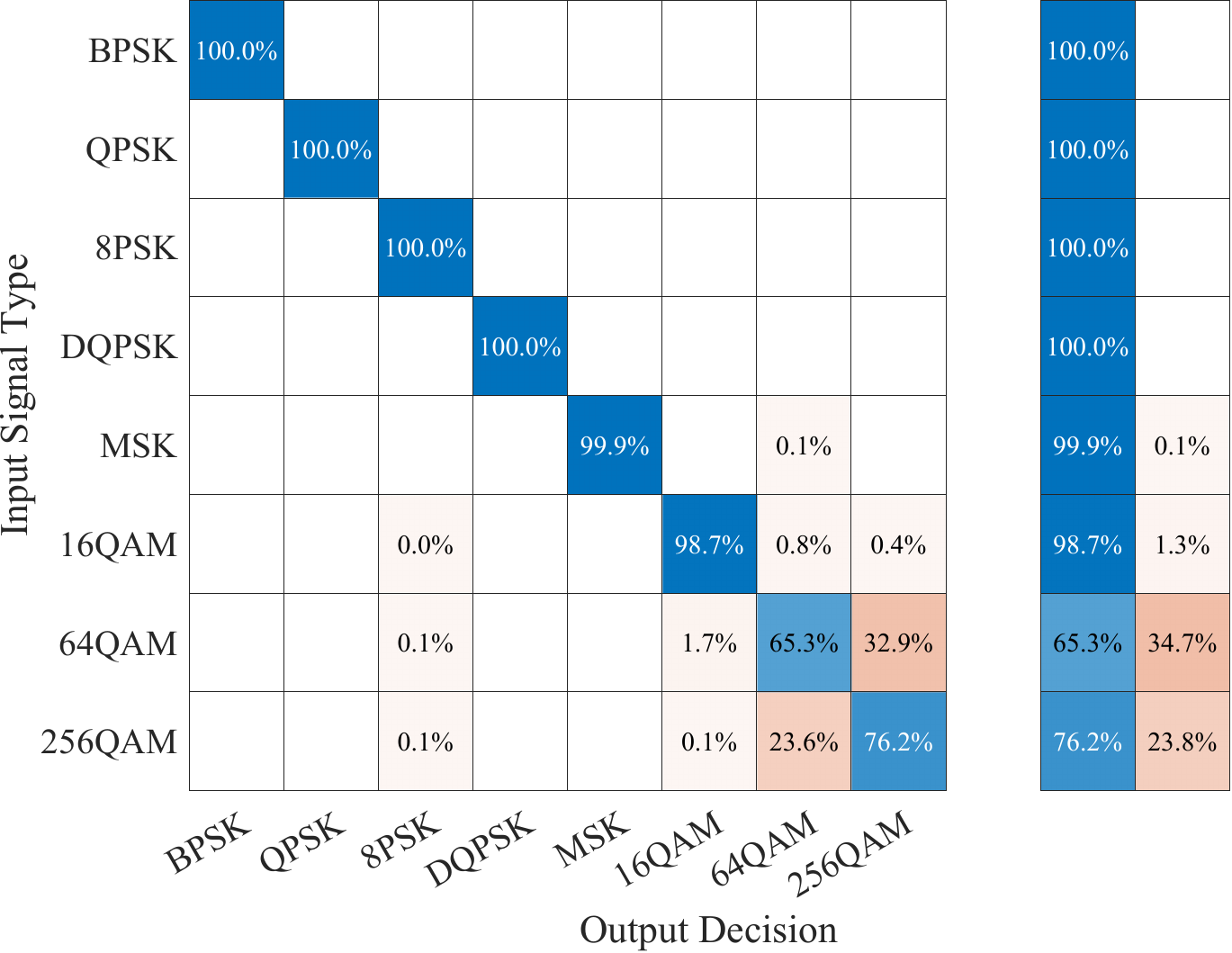}}
\subfigure[Confusion Matrix of the \texttt{CSPB.ML.2022} CC-trained CAP Classifying All \texttt{CSPB.ML.2018} Signals.]
{\includegraphics[width=0.49\linewidth]{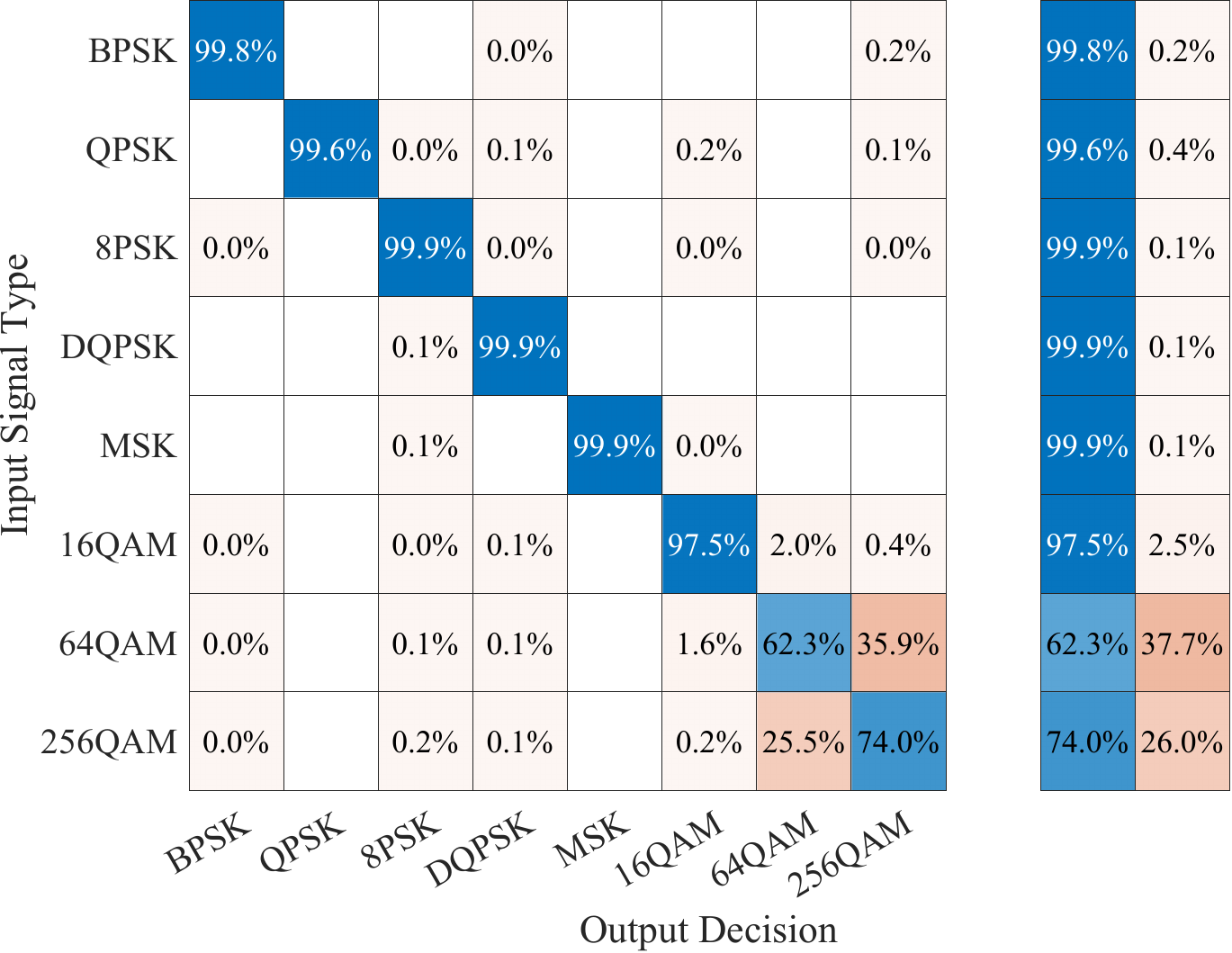}}
\caption[Confusion Matrices of the \texttt{CSPB.ML.2022} CC-trained CAP Classifying \texttt{CSPB.ML.2022} Test Signals and All \texttt{CSPB.ML.2018} Signals.]{\textbf{Confusion Matrices of the \texttt{CSPB.ML.2022} CC-trained CAP Classifying \texttt{CSPB.ML.2022} Test Signals and All \texttt{CSPB.ML.2018} Signals.}}\label{fig:CM_GCTest_GCTrained_CC_CAP}
\end{center}
\end{figure}

\begin{figure}
\begin{center}
\subfigure[Confusion Matrix of the \texttt{CSPB.ML.2018} I/Q-trained CAP Classifying \texttt{CSPB.ML.2018} Test Signals.]
{\includegraphics[width=0.49\linewidth]{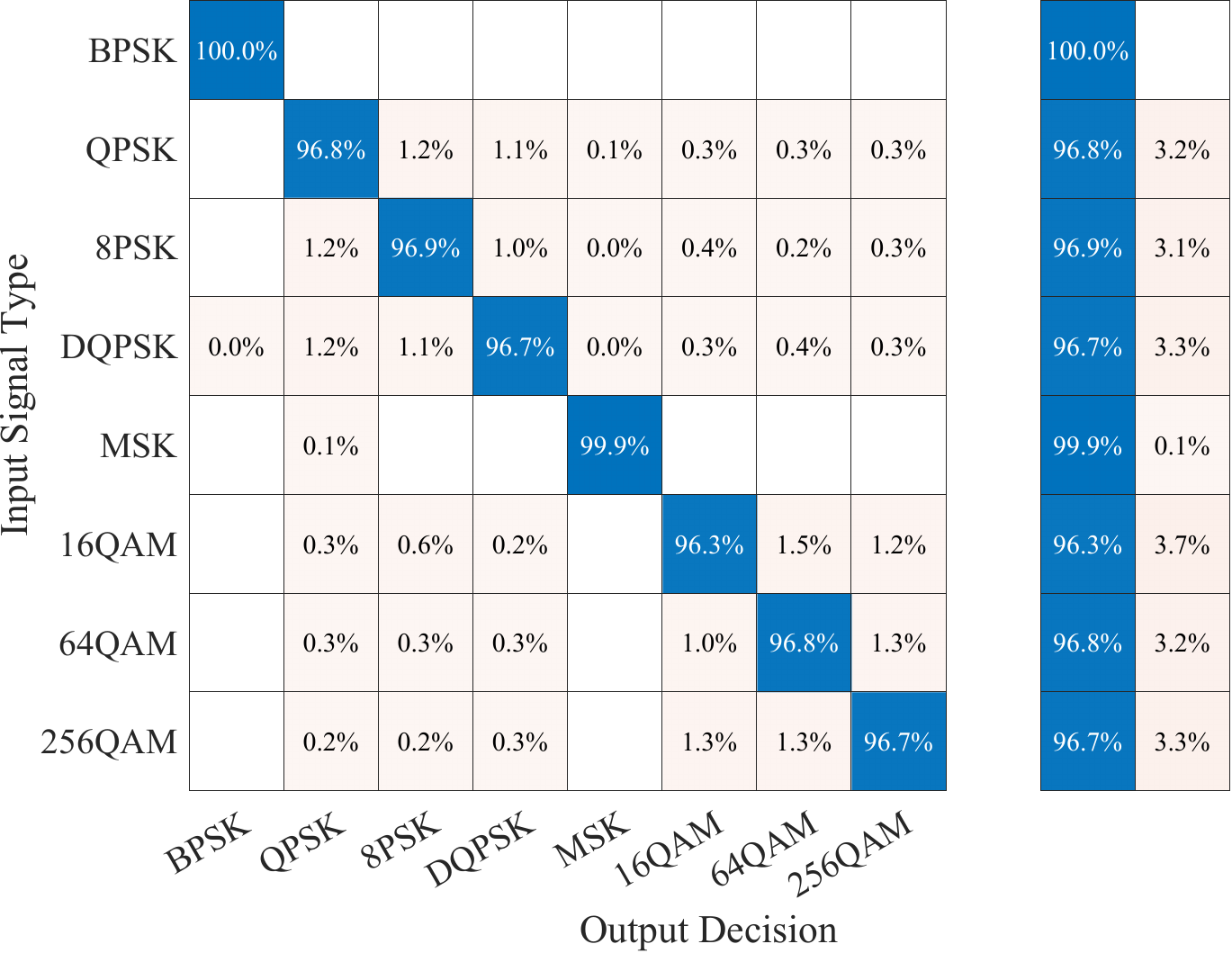}}
\subfigure[Confusion Matrix of the \texttt{CSPB.ML.2018} I/Q-trained CAP Classifying All \texttt{CSPB.ML.2022} Signals.]
{\includegraphics[width=0.49\linewidth]{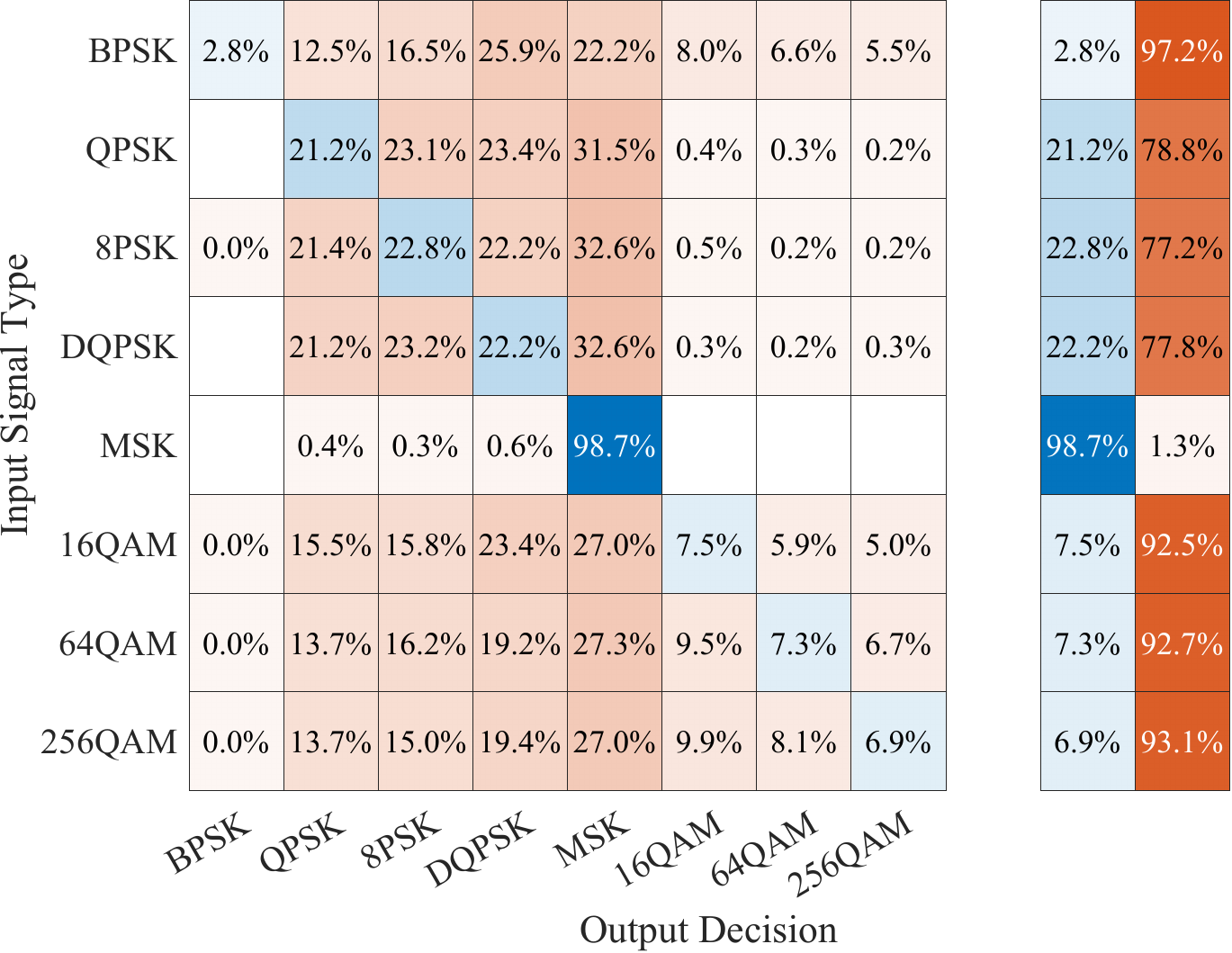}}
\caption[Confusion Matrices of the \texttt{CSPB.ML.2018} I/Q-trained CAP Classifying \texttt{CSPB.ML.2018} Test Signals and All \texttt{CSPB.ML.2022} Signals.]{\textbf{Confusion Matrices of the \texttt{CSPB.ML.2018} I/Q-trained CAP Classifying \texttt{CSPB.ML.2018} Test Signals and All \texttt{CSPB.ML.2022} Signals.}}\label{fig:CM_CTest_CTrained_IQ_CAP}
\end{center}
\end{figure}

\begin{figure}
\begin{center}
\subfigure[Confusion Matrix of the \texttt{CSPB.ML.2022} I/Q-trained CAP Classifying \texttt{CSPB.ML.2022} Test Signals.]
{\includegraphics[width=0.49\linewidth]{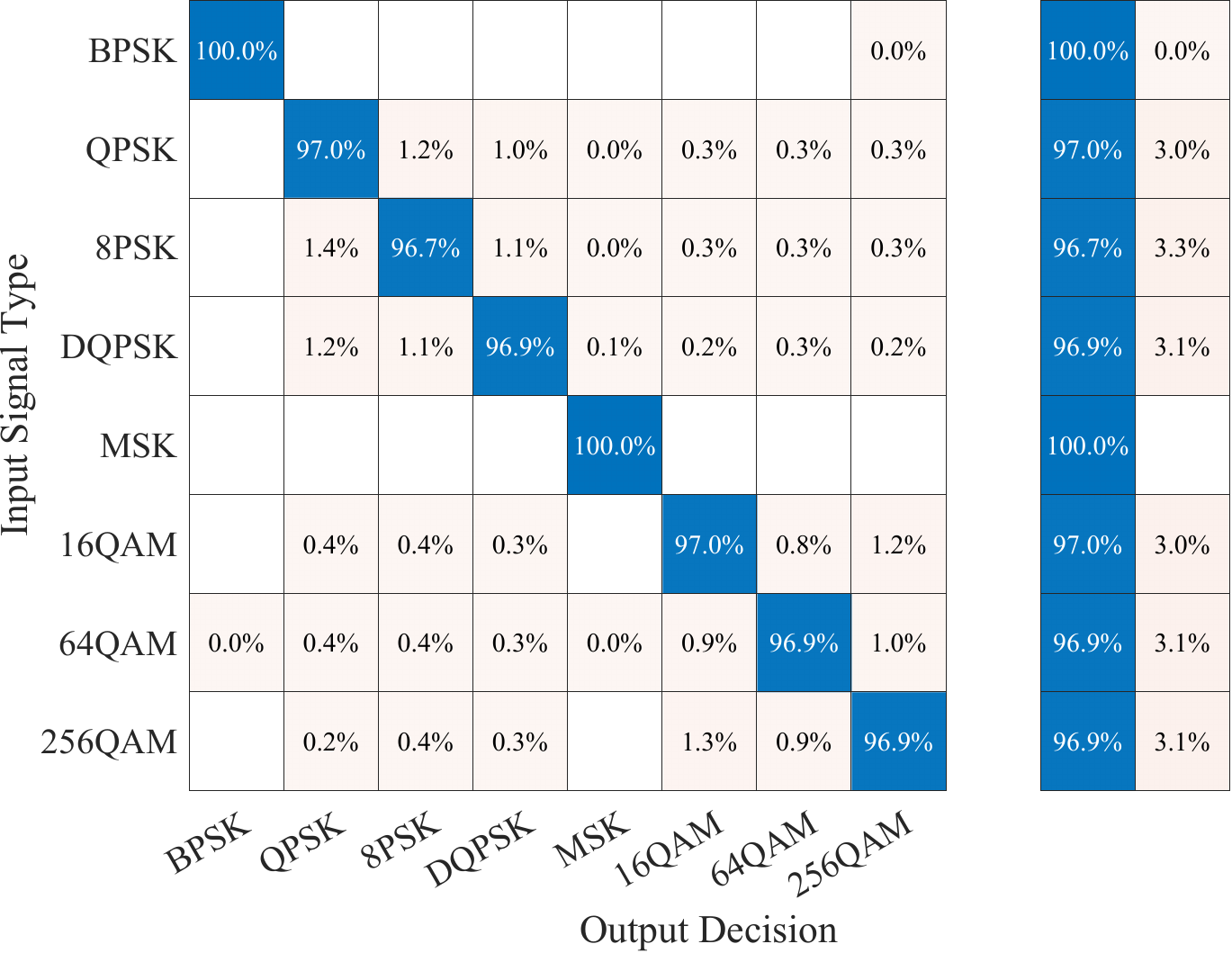}}
\subfigure[Confusion Matrix of the \texttt{CSPB.ML.2022} I/Q-trained CAP Classifying All \texttt{CSPB.ML.2018} Signals.]
{\includegraphics[width=0.49\linewidth]{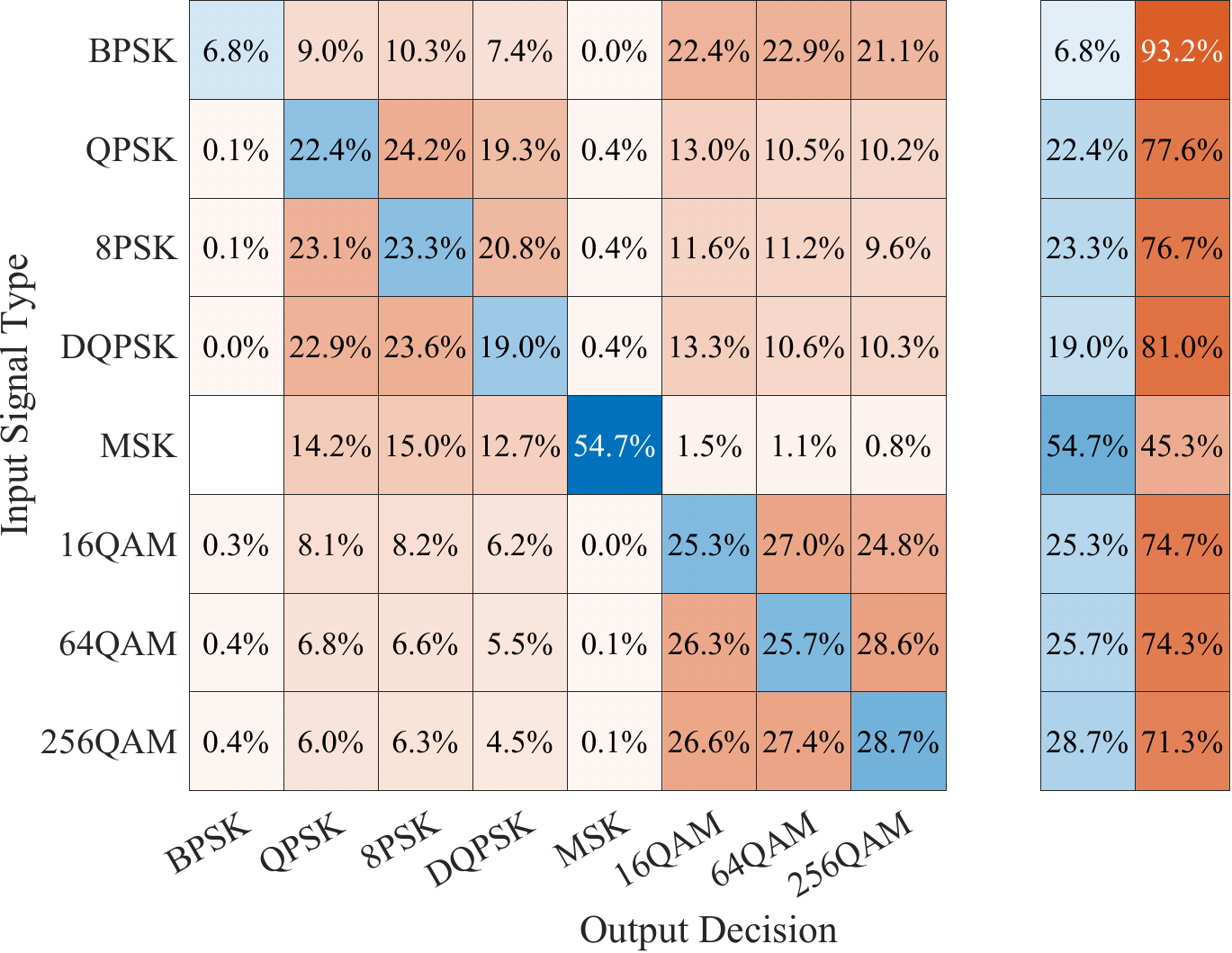}}
\caption[Confusion Matrices of the \texttt{CSPB.ML.2022} I/Q-trained CAP Classifying \texttt{CSPB.ML.2022} Test Signals and All \texttt{CSPB.ML.2018} Signals.]{\textbf{Confusion Matrices of the \texttt{CSPB.ML.2022} I/Q-trained CAP Classifying \texttt{CSPB.ML.2022} Test Signals and All \texttt{CSPB.ML.2018} Signals.}}\label{fig:CM_GCTest_GCTrained_IQ_CAP}
\end{center}
\end{figure}

\begin{enumerate}
\setlength{\itemsep}{0.0in}
\item Blind exhaustive spectral correlation and coherence analysis for $N$ complex values and $N^\prime$ strips in the strip spectral correlation analyzer (SSCA) algorithm \cite{Brown_SSCA_1993}: $NN^\prime (\log_2(N^\prime) + \log_2(N) + 4)$.
\item The cost of estimating the quadruple carrier from the FFT of $x^4(t)$, and therefore the carrier offset, if the CF pattern was not determined to be BPSK-like or SQPSK-like after SSCA analysis: $3N + N\log_2(N)$.
\item The cost of the cyclic moments in (\ref{eq:ctcf_est}) was determined by the cost of creating the needed delay products (such as $x(t)x(t)x^*(t)x^*(t)$) and the DFTs for each combination of lag product and needed cycle frequency.  The number of required lag-product vectors is $P$, which was maximum for BPSK-like and minimum for 8PSK-like, where $P = 3$.  Assuming $K$ CFs across all orders $n$ and lag products, the cost of this step was $PN + KN\log_2(N)$.
\item The cost of combining the CTMFs after their computation was negligible compared to the previously sketched costs.
\item Total cost (blind processing): $NN^\prime(\log_2(N^\prime) + \log_2(N) + 4) + (P+3)N + N(K+1)\log_2(N) $.
\end{enumerate}

For example, when operating on an Intel Xeon E3-1535 laptop using C-language implementations of all operations, the total elapsed time when obtaining a BPSK CC feature for a maximum cumulant order of six, $N$ = 32,768, and $N^\prime = 64$ was 0.18 s and for the same parameters, but for an 8PSK signal, the elapsed time was 0.10 s.

Once the CC features were available, the subsequent processing to classify the signals was minimal in the context of modern processors capable of performing billions of floating-point operations per second:
\begin{itemize}
\item In the case of the CSP baseline model, the subsequent processing involved a comparison of the estimated CC features to theoretical features to identify the closest, in terms of a distance metric, theoretical CF pattern, as outlined in Section~\ref{sec:Baseline_Model}.
\item In the case of the CAP classifier, the classifier was presented with the CC feature at the input, and the classification decision corresponded to the CAP output.  The one-time computational cost for utilizing the CAP should also be included in the cost in this case.
\end{itemize}

Furthermore, when comparing the proposed CAP classifier with the CAP-based classifier in \cite{Latshaw_COMM2022}, which uses the I/Q data of the signal to be classified, the total number of trainable parameters of the latter classifier was 2,079,304 and was significantly larger than that of the proposed classifier, which only had 1,269,072 learnable parameters.  This difference impacted the one-time training cost of the classifier, but once trained, the two CAPs should be able to reach rapid classification decisions.

Concluding the discussion on the computational aspects, the parallel branches of the CAP are well suited for FPGA implementations, since FPGAs are designed for parallel computations.  While this does not reduce the number of operations performed, it does significantly reduce the latency of calculations, which is also an important consideration in practical implementations.

\section{Discussion}\label{sec:Disc}
The results of this work suggested that a high degree of generalization cannot be obtained using DL-based approaches applied to the modulation recognition problem if the input to the DL neural network is constrained to be sampled time domain I/Q data.  On the other hand, if the inputs to the DL neural networks were carefully selected features estimable from I/Q data, such as cyclic cumulants, the observed degree of generalization was very high, and the performance was also high.  The fundamental research question is: Why do I/Q-based DL neural networks not generalize well?  A second urgent question is: Why do I/Q-trained neural networks not learn features like cyclic cumulants?

A speculation here for the reason I/Q-trained neural networks do not learn simultaneously high-performing and high-generalization features is due to their structure and hyperparameters.  Currently, most RF domain machine-learning systems (known to the author) have adopted the structure (layers and the order of layers) and the hyperparameters that have proven capable of performing high-quality image and natural language recognition, such as AlexNet.  However, modulation recognition using sampled data is a problem that differs from image recognition in that the sought-after label is not associated with an additive component of the input; there is no BPSK part of the I/Q sequence; the whole sequence has a BPSK nature.

What separates a BPSK signal from a QPSK signal is the underlying probability density function for the transmitted symbol.  That symbol random variable is binary for BPSK and quaternary for QPSK.  This one density difference then leads to divergent $n^{\mbox{\footnotesize th}}$-order probability density functions for the signal's samples.  If a neural network could be trained to learn several of these density functions from labeled data, high performance and high generalization might be obtained.  However, estimating higher-order probability density functions likely requires explicit nonlinear layers rather than multiple linear convolutional layers.  Therefore, possibly no amount of training or adjustment of the hyperparameters will lead to high generalization for a DL neural network with I/Q input; structural changes are needed.

This chapter presented a novel deep-learning-based classifier for digitally modulated signals that uses capsule networks and blindly estimated cyclic cumulant features as the input.  The proposed classifier outperformed conventional (non-machine-learning) classifiers employing CSP and had very good generalization abilities, unlike conventional CNNs using I/Q sample inputs, which can achieve excellent performance, but have not been made to generalize.  This dissertation so far has shown that the combination of conventional NNs and sampled-data inputs did not lead to both good classification performance and good generalization.  The use of principled features as inputs, such as CCs, did in fact lead to simultaneous good performance and good generalization.  Chapter~\ref{ch:NovelLayers} in this dissertation presents a significant step towards attaining that simultaneity without having to perform all signal-processing feature-extraction steps outside of the network.

\chapter{The Importance of High-Quality Datasets}\label{ch:HQDatasets}
When training an NN to perform a task, whether learning classifications or regressions, the quality of the dataset being used for training the NN directly impacts the quality of the final trained NN.  As presented during the CSP background information, the longer the data record of a sample path, the easier it is to extract cyclic information related to PDFs about the underlying digital communication scheme.  It currently seems to be common practice when hosting I/Q datasets for modulation classification to only provide a small number of samples per signal, such as $1,024$ or less, \cite{DeepSigDataSet, DeepSigDataSetAnalysis}, which prevents signal processing techniques from being able to verify the claims of the truth parameters listed in the hosted datasets.

The datasets \texttt{CSPB.ML.2018}, \texttt{CSPB.ML.2022}, and \texttt{CSPB.ML.2023}~\cite{CSPblog_DataSets} have been provided together in IEEE DataPort~\cite{IEEE_DataPort}, and this chapter of the dissertation highlights the proven accuracy of the truth labels in these datasets such that future works can have confidence accessing them through IEEE DataPort.  The remainder of this chapter is organized as follows:  the generating parameters for the signal datasets~\cite{IEEE_DataPort} are presented in Section~\ref{sec:Generate}, followed by a tutorial on dataset utilization in Section~\ref{sec:DataSetsUsage}.

\section{Generating Parameters for the Three Datasets}\label{sec:Generate}
\subsection{Datasets \texttt{CSPB.ML.2018} and \texttt{CSPB.ML.2022}}\label{sec:2018_and_2022_parameters}
The signals in the \texttt{CSPB.ML.2018} and \texttt{CSPB.ML.2022} datasets are generated following a generic model of a digitally modulated signal,
\begin{align}
x \left( t \right) = {a s(t) e^{i (2 \pi f_0 t + \phi)} + w \left( t \right)},
\end{align}
where $a$ denotes the signal amplitude, $s \left( t \right)$ is the complex envelope of the signal, $f_0$ is the CFO, and $w \left( t \right)$ is an additive white Gaussian noise (AWGN) that corrupts the signal.  The average power of $s \left( t \right)$ is unity, so that the signal power is $a^2$.  The AWGN noise process has zero mean and power spectral density $\sigma^2$, which along with the signal amplitude can be set to determine a specific value for the SNR.

Both datasets contain collections of digitally modulated signals that include eight modulation schemes:  BPSK, QPSK, 8PSK, $\pi/4$-DQPSK, MSK, 16QAM, 64QAM, and 256QAM.  Each dataset contains $14,000$ signals for each modulation type for a total of $112,000$ signals per dataset, and for each instance of each signal, $32,768$ samples are provided.

The signals employ square-root raised-cosine (SRRC) pulse shaping with roll-off factor~$\beta$, and their generation includes random parameters whose probability distributions are designed to simulate effects that occur in practical scenarios such as variations due to the propagation channel or other effects.  These are listed in Table~\ref{table:SigGenParms} and include the symbol period $T_0$; the CFO $f_0$; the excess bandwidth implied by the SRRC roll-off parameter~$\beta$; the received signal power level; and the AWGN spectral density.

The received signal power level along with the AWGN spectral density directly impact the SNR, and it is emphasized that the in-band SNR is listed in Table~\ref{table:SigGenParms}.  This is the ratio of signal-power to noise-power falling within the \emph{signal bandwidth}, and out-of-signal-band noise is excluded from the ratio.  The in-band SNR conveys more consistent information across each data record than the total SNR, which is the ratio of signal-power to noise-power falling within the \emph{receiver total bandwidth} and includes out-of-signal-band noise whenever the signal is sampled in excess of its Nyquist rate.  This is because, for the same signal, a receiver with a wider/narrower total bandwidth will decrease/increase the total SNR, whereas the in-band SNR will remain unaffected by a change in the receiver's total bandwidth since out-of-signal-band noise is excluded.  For example, a data record with an in-band SNR of 10~dB, indicative of a very strong wireless signal, could have a total SNR of 5~dB, or 0~dB, or less; depending on how wide the receiver bandwidth is and how much out-of-signal-band noise is included in the ratio.  To obtain signals with the SNR ranges listed in Table~\ref{table:SigGenParms}, the AWGN spectral density was kept at 0~dB for each generated signal while the various SNR values were created by varying the signal power through its corresponding amplitude value $a$.

\begin{table}[!htbp]
\centering
\caption[Dataset Signal Generation Parameters:  \texttt{CSPB.ML.2018} and \texttt{CSPB.ML.2022}.]{\textbf{Dataset Signal Generation Parameters:  \texttt{CSPB.ML.2018} and \texttt{CSPB.ML.2022}.}}
{
\setlength{\tabcolsep}{8mm}
\begin{tabular}{ l c c }
\toprule
\begin{tabular}{ c } \textbf{Parameter} \end{tabular} & \begin{tabular}{@{}c@{}} \textbf{Dataset} \\ \textbf{\texttt{CSPB.ML.2018}} \end{tabular} & \begin{tabular}{@{}c@{}} \textbf{Dataset} \\ \textbf{\texttt{CSPB.ML.2022}} \end{tabular} \\
\midrule
Sampling Frequency, $f_s$ & 1 Hz & 1 Hz \\
\hline
\begin{tabular}{@{}l@{}} Carrier Frequency \\ Offset (CFO), $f_0$ \end{tabular} & \begin{tabular}{@{}c@{}} Uniformly \\ Distributed \\ $\in (-0.001, 0.001)$ Hz \end{tabular} & \begin{tabular}{@{}c@{}} Uniformly \\ Distributed \\ $\in (0.01, 0.02)$ Hz \end{tabular} \\
\hline
Symbol Period, $T_0$ & $\in [1, 23]$ s & $\in [1, 29]$ s \\
\hline
\begin{tabular}{@{}l@{}} SRRC Pulse-Shaping \\ Roll-Off Factor, $\beta$ \end{tabular} & $\in [0.1, 1]$ & $\in [0.1, 1]$ \\
\hline
In-Band SNR & $\in [0, 12]$ dB & $\in [1, 18]$ dB \\
\hline
\begin{tabular}{@{}l@{}} In-Band SNR PDF \\ Center of Mass \end{tabular} & $9$ dB & $12$ dB \\
\hline
Message Bits & IID & IID \\
\bottomrule
\end{tabular}
}\label{table:SigGenParms}
\end{table}

The CFO, $T_0$, $\beta$, signal power level, and the IID message (modulating) bits were all varied according to their listed distributions as the signals were generated.  The values of the CFO, $T_0$, $\beta$, and signal power level were recorded along with the signal they correspond to, such that they can be verified via blind signal processing or used for NN regression training.  While some of these parameters may have limited practical ranges (such as the SRRC roll-off~$\beta$ and the implied excess bandwidth), others (such as $T_0$ or $f_0$) possess an infinite number of valid practical choices.

\pagebreak

\subsection{Dataset \texttt{CSPB.ML.2023}}\label{sec:2023}
The newest dataset, \texttt{CSPB.ML.2023} has much in common with the other two datasets and also one major difference. The dataset comprises $120,000$ signal examples, the first $60,000$ of which are single signals in noise (as in the \texttt{CSPB.ML.\{2018,2022\}} datasets), and the last $60,000$ of which are two-signal combinations of the signals in the first $60,000$.  The parameters governing the generation of the first $60,000$ examples (the single-signal examples) are shown in Table \ref{table:SigGenParams_2023}.  The eight involved signal types are BPSK, QPSK, 8PSK, 16QAM, 64QAM, SQPSK, MSK, and GMSK.  The BPSK, QPSK, 8PSK, and QAM signals possess a single square-root raised-cosine pulse function with a roll-off of 0.35, which is a major difference relative to the other two datasets, where the roll-off is a variable parameter.

\begin{figure}
\centering
\includegraphics[width=\linewidth]{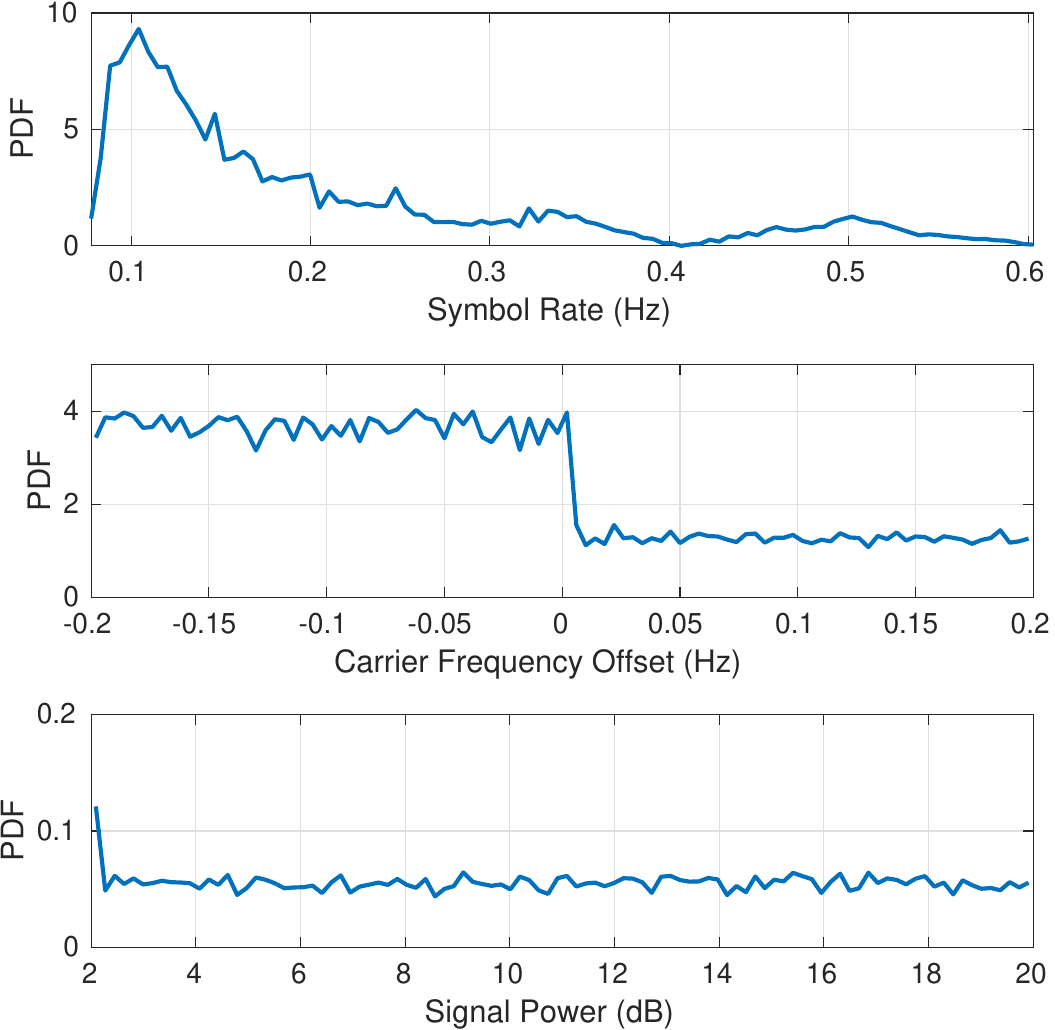}
\caption[Estimated Probability Density Functions for Key Signal-Generation Variables in the Co-Channel-Signal \texttt{CSPB.ML.2023} Dataset.]{\textbf{Estimated Probability Density Functions for Key Signal-Generation Variables in the Co-Channel-Signal \texttt{CSPB.ML.2023} Dataset.}}\label{fig:CSPB.ML.2023_pdfs}
\end{figure}

The distributions for the three key parameters of symbol rate ($1/T_0$), carrier frequency offset ($f_0$), and SNR (in decibels) are shown in Figure~\ref{fig:CSPB.ML.2023_pdfs}.

Each of the two-signal examples in the dataset is constructed by simply adding together two adjacent (in terms of index into the dataset) single-signal files.  No scaling was performed, so that the noise floor for the two-signal examples is twice that of the single-signal cases.

\begin{table}[!htbp]
\centering
\caption[Dataset Signal Generation Parameters:  \texttt{CSPB.ML.2023}.]{\textbf{Dataset Signal Generation Parameters:  \texttt{CSPB.ML.2023}.}}
{
\setlength{\tabcolsep}{8mm}
\begin{tabular}{ l c }
\toprule
\begin{tabular}{ c } \textbf{Parameter} \end{tabular} & \begin{tabular}{@{}c@{}} \textbf{Dataset} \\ \textbf{\texttt{CSPB.ML.2023}} \end{tabular} \\
\midrule
Sampling Frequency, $f_s$ & 1 Hz \\
\hline
\begin{tabular}{@{}l@{}} Carrier Frequency \\ Offset (CFO), $f_0$ \end{tabular} & $\in [-0.2, 0.2]$ Hz \\
\hline
Symbol Period, $T_0$ & $\in [1.7, 13.3]$ s \\
\hline
\begin{tabular}{@{}l@{}} SRRC Pulse-Shaping \\ Roll-Off Factor, $\beta$ \end{tabular} & $0.35$ \\
\hline
In-Band SNR & $\in [2, 20]$ dB \\
\hline
Message Bits & IID \\
\hline
\begin{tabular}{@{}l@{}} Single Signals AWG \\ Noise Spectral Density, $N_0$ \end{tabular} & $0$ dB \\
\hline
\begin{tabular}{@{}l@{}} Two Signals AWG \\ Noise Spectral Density, $N_0$ \end{tabular} & $3$ dB \\
\bottomrule
\end{tabular}
}\label{table:SigGenParams_2023}
\end{table}

\pagebreak

\section{Utilizing the Datasets}\label{sec:DataSetsUsage}
\subsection{\texttt{CSPB.ML.2018} and \texttt{CSPB.ML.2022}}\label{sec:2018_and_2022_utilizing}
The \texttt{CSPB.ML.2018} and \texttt{CSPB.ML.2022} datasets from~\cite{CSPblog_DataSets} have been packaged together, along with example code showing how to utilize the datasets, and uploaded to IEEE DataPort~\cite{IEEE_DataPort}.  The example code provided consists of MATLAB scripts and requires the Deep Learning Toolbox.

The system requirements include at least 128~GB of free disk space for file storage and 128~GB of memory (RAM).  A high-performance Graphics Processing Unit (GPU) is a recommended optional component to speed up computation.  The examples have been developed on a high-performance computing cluster with 18 NVidia V100 GPU nodes available where each node has 128 GB of memory.

\subsection{Example Script Files}
The MATLAB script files included with the two datasets give examples of how the datasets can be effectively loaded in system memory (RAM), processed, used to train NNs, and show valid test results.

The scripts should be run in the following order:
\begin{enumerate}
\item \label{step:Create.m} CSPB\_ML\_Data\_Create\_MAT\_files.m
\item \label{step:Prepare.m} CSPB\_ML\_Prepare\_Data\_for\_NN.m
\item \label{step:Train.m} CSPB\_ML\_Train\_NN.m
\item \label{step:Results.m} CSPB\_ML\_Plot\_Results.m
\end{enumerate}

\noindent
\underline{\bf Script~\ref{step:Create.m}}: takes the original data files in each dataset and converts them into binary MATLAB files (these files have a~\texttt{.mat} extension) that store workspace variables for efficient loading into the RAM.  Script~\ref{step:Create.m} allows the user to specify how many files to split each dataset into, giving the user control over how much RAM is needed for any signal processing that might be performed by subsequent Script~\ref{step:Prepare.m}.

\noindent
\underline{\bf Script~\ref{step:Prepare.m}}: provides an example of how the data is processed and prepared for use with NN training.  This step is performed to facilitate rapid NN development by alleviating time-consuming re-processing of signal data immediately prior to NN training.  Once the processed data is saved, the signal processing does not have to be performed again unless the user wants to change the pre-processing steps.

The example provided in Script~\ref{step:Prepare.m} performs a normalizing of each signal corrupted by AWGN to unit power, separating the normalized data into in-phase and quadrature (I/Q) components for use with the NN, and then performs simple blind signal processing to coarsely estimate signal generation parameters and calculate absolute errors.  These results are then saved so the normalized I/Q data can be readily available for NN training and so that an example of calculating mean absolute error (MAE) can be shown.

\noindent
\underline{\bf Script~\ref{step:Train.m}}: provides an example of how an NN can be trained to perform modulation classification using readily available processed data and saves the trained NNs and results.  

\noindent
\underline{\bf Script~\ref{step:Results.m}}: provides an example of results visualization, which includes calculating and displaying MAEs for the estimated signal parameters, calculating the overall probability of correct classification ($P_{CC}$) from each test, plotting $P_{CC}$ vs. SNR, and displaying confusion matrices.  The $P_{CC}$ vs. SNR plots generated by Script~\ref{step:Results.m} will appear similar to the ones in Figures~1--4 in~\cite{Snoap_MILCOM_2022} and the confusion matrices will appear similar to Figures~3,~5, and~7 in~\cite{Latshaw_COMM2022}.

\subsection{Blind Estimation of Signal Parameters}\label{sec:Verify}
Upon executing Script~\ref{step:Prepare.m}, which includes only simple blind signal processing performed on each signal to determine its parameters; coarse estimates of the CFO, symbol rate, occupied bandwidth, excess bandwidth, and the in-band SNR will be available for each of the $112,000$ signals in each dataset.  These coarse estimates have been further refined using more reliable blind signal processing as discussed in~\cite{Snoap_MILCOM_2022}, and the resulting estimated values for each parameter are compared to their labeled truth values to determine an absolute error for each signal parameter in the two datasets, with an additional breakdown of the absolute errors for CFO estimates based on the modulation type.  The MAEs for each parameter obtained following~\cite{Snoap_MILCOM_2022} are listed in Table~\ref{table:MAEs_1} in a logarithmic scale and the accuracy of the resulting estimates is implied by the small order of magnitudes of their corresponding MAEs.  Table~\ref{table:MAEs_1} can be used as a baseline comparison for results obtained through NN regression training or alternative signal processing approaches.

\subsection{Neural Network Training and Performance Testing}
The example in Script~\ref{step:Train.m} trains an NN from scratch twice:
\begin{itemize}
\item First, the \texttt{CSPB.ML.2018} dataset is used for NN training, validation, and initial testing, and the \texttt{CSPB.ML.2022} dataset is used to assess the generalization performance of the trained NN with out-of-distribution data.
\item Then, the roles of the two datasets are reversed, and the \texttt{CSPB.ML.2022} dataset is used for NN training, validation, and initial testing, while the \texttt{CSPB.ML.2018} dataset is used to assess the generalization performance of the trained NN with out-of-distribution data.
\end{itemize}

Script~\ref{step:Train.m} is the only example that uses 128 GB of RAM, as the entirety of each dataset is loaded into system memory for NN training and testing. For each training session, the training dataset is split up into $70\%$ for training, $5\%$ for validation, and $25\%$ for testing, where the training and validation portions are utilized during NN training.  After training, the $25\%$ testing portion is classified by the trained NN as an initial test result.  The remaining dataset is then loaded into system memory and the trained NN classifies all $112,000$ of the signals in the remaining dataset to test the generalization abilities of the trained NN with the out-of-distribution data.

The NN example provided in Script~\ref{step:Train.m} recreates the capsule network (CAP) from~\cite{Latshaw_COMM2022}, which was documented to have excellent performance during the initial test but performs poorly on the out-of-distribution data shift test.

\begin{table}
\centering
\caption[Mean Absolute Errors for Estimated Signal Generation Parameters.]{\textbf{Mean Absolute Errors for Estimated Signal Generation Parameters.}}
{
\setlength{\tabcolsep}{6mm}
\begin{tabular}{ l c c }
\toprule
\begin{tabular}{ c } \textbf{Estimated Parameter} \\ (10 Outliers Removed) \end{tabular} & \begin{tabular}{@{}c@{}} \textbf{$\log_{10}(\mbox{MAE})$ for} \\ \textbf{\texttt{CSPB.ML.2018}} \end{tabular} & \begin{tabular}{@{}c@{}} \textbf{$\log_{10}(\mbox{MAE})$ for} \\ \textbf{\texttt{CSPB.ML.2022}} \end{tabular} \\
\midrule
CFO (All) & $-3.88$ & $-4.01$ \\
CFO (BPSK) & $-6.54$ & $-6.56$ \\
CFO (QPSK) & $-6.66$ & $-6.74$ \\
CFO (8PSK) & $-3.02$ & $-3.12$ \\
CFO ($\pi/4$-DQPSK) & $-6.72$ & $-6.80$ \\
CFO (MSK) & $-6.57$ & $-6.58$ \\
CFO (16QAM) & $-6.33$ & $-6.52$ \\
CFO (64QAM) & $-5.92$ & $-6.44$ \\
CFO (256QAM) & $-5.64$ & $-6.34$ \\
\hline
Symbol Rate & $-3.08$ & $-3.37$ \\
\hline
Occupied Bandwidth & $-1.58$ & $-1.74$ \\
\hline
Excess Bandwidth & $-0.85$ & $-0.88$ \\
\hline
In-Band SNR (dB) & $-0.31$ & $-0.45$ \\
\bottomrule
\end{tabular}
}\label{table:MAEs_1}
\end{table}

\pagebreak

\subsection{\texttt{CSPB.ML.2023}}
\label{sec:2023_utilizing}
All provided sample scripts can be used as a starting point for the $60,000$ single-signal examples in \texttt{CSPB.ML.2023}, as the sample scripts are provided specifically for the other two datasets.  For the two-signal examples in the second half of \texttt{CSPB.ML.2023}, new scripts will be needed if it is desired to extract parameter estimates blindly from the data.  This is because there are twice as many parameters and patterns to estimate.  The first half of the \texttt{CSPB.ML.2023} dataset can be used as a further evaluation of the generalization properties of networks trained on \texttt{CSPB.ML.\{2018,2022\}}, although the signal sets differ slightly.

A significant complication for machine-learning is that each input is now associated with two output labels (e.g., BPSK and SQPSK).  This can be sidestepped by simply creating all possible pairs of the eight signal types (all of which occur in the dataset), such as BPSK\_BPSK, BPSK\_SQPSK, QPSK\_16QAM, etc. The two-signal eight-class problem is then transformed into a single-signal ${8 \choose 2} = 28$-class problem.

\begin{figure}
\centering
\includegraphics[width=\linewidth]{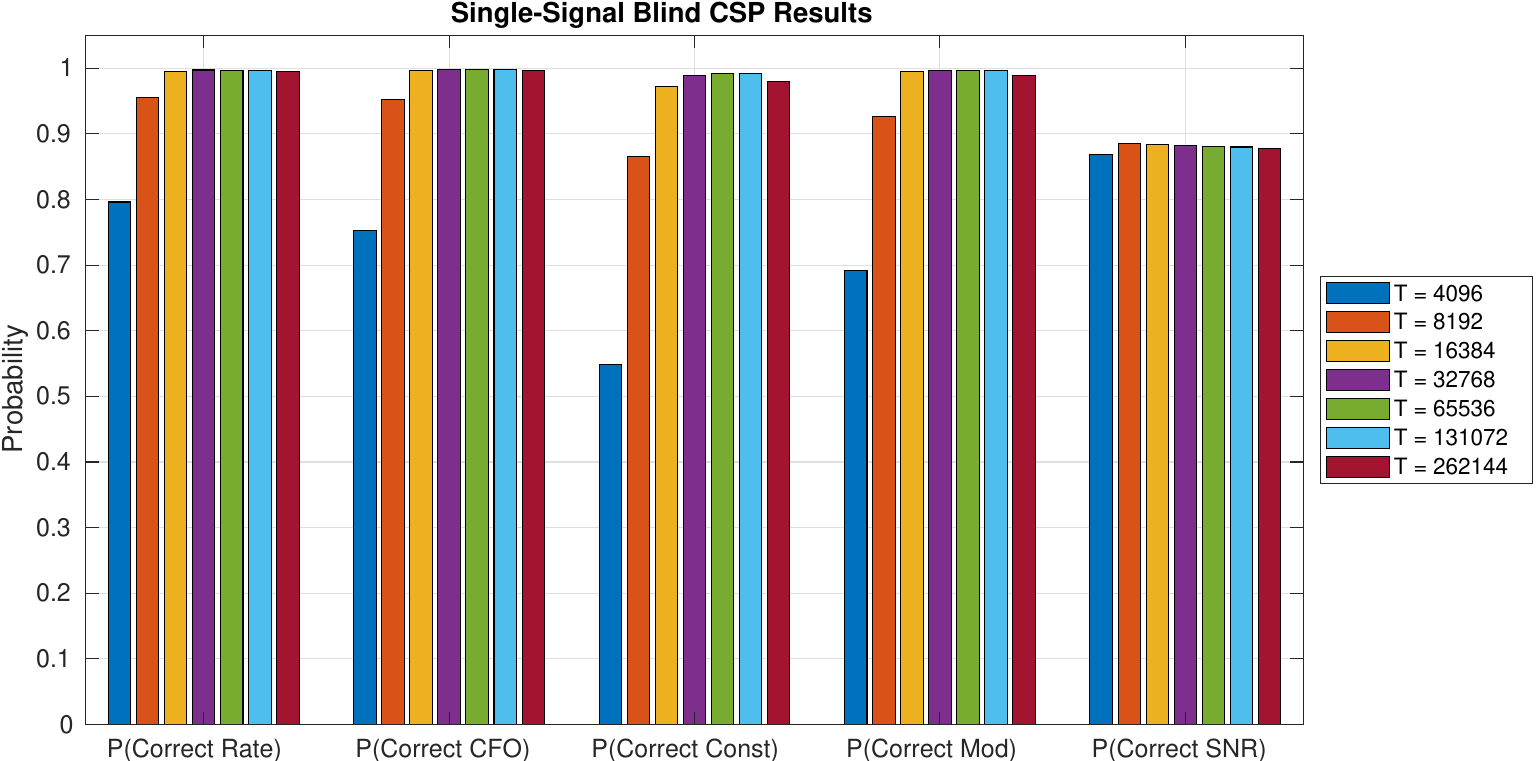}
\caption[Single-Signal \texttt{CSPB.ML.2023} Results Using the CSP Baseline Model.]{\textbf{Single-Signal \texttt{CSPB.ML.2023} Results Using the CSP Baseline Model.}}\label{fig:psk_mixtures_one_signal}
\end{figure}

\begin{figure}
\centering
\includegraphics[width=\linewidth]{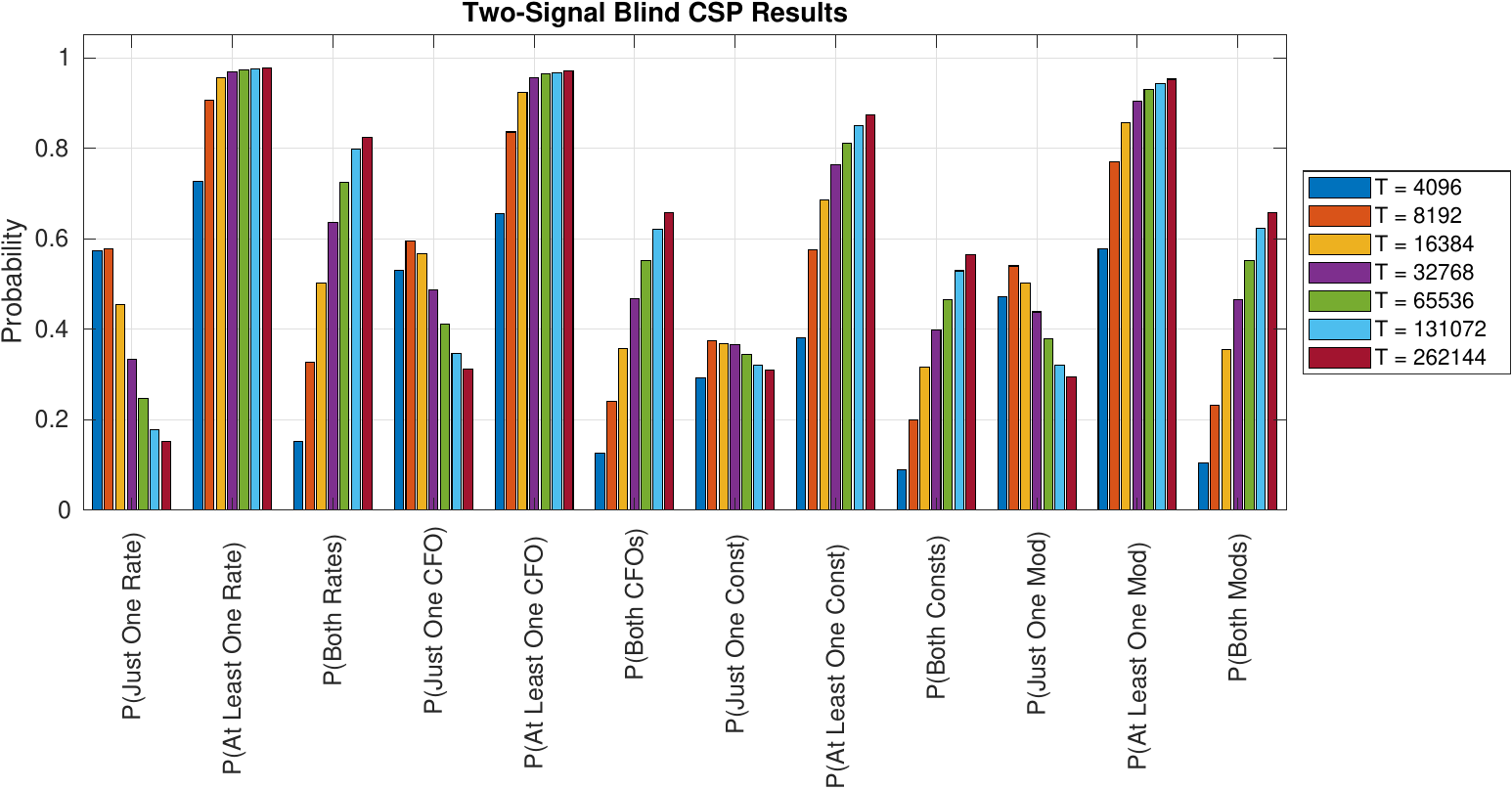}
\caption[Two-Signal \texttt{CSPB.ML.2023} Results Using the CSP Baseline Model.]{\textbf{Two-Signal \texttt{CSPB.ML.2023} Results Using the CSP Baseline Model.}}\label{fig:psk_mixtures_two_signals}
\end{figure}

While the provided sample scripts do not include training and evaluating recognition neural networks for \texttt{CSPB.ML.2023}, the CSP baseline classification model has been applied to the dataset~\cite{CochannelChallenge}.  CSP, and more generally signal-processing, algorithms do not require the combinatorial explosion of output labels because the number of signals present in the processed data can be relatively simply determined using the first stages of a suitable algorithm, such as that in \cite{Spooner_Asilomar2000}.  Taking this, as done before, as the baseline (conventional) modulation-recognition approach, every tenth example in the \texttt{CSPB.ML.2023} dataset has been processed.  It is important to note that the applied algorithm is the same for all of the examples--the algorithm is not afforded a different set of configuration parameters for the single-signal examples relative to the two-signal examples.  A summary of results is shown for the single-signal case in Figure~\ref{fig:psk_mixtures_one_signal} and for the two-signal case in Figure~\ref{fig:psk_mixtures_two_signals}.  Here, $T$ represents the processed block length in samples.  The performance of the CSP approach increases with increasing block length, and there is room for improvement in the two-signal situation, possibly with the aid of machine-learning.

\chapter{Novel Neural Network Layers for \\In-Phase and Quadrature Data}\label{ch:NovelLayers}
The results of this dissertation so far suggest that a high degree of generalization cannot be obtained using DL-based approaches applied to the modulation recognition problem if the input to the DL neural network is constrained to be sampled time-domain I/Q data.  On the other hand, if the inputs to the DL neural networks are carefully selected features estimable from I/Q data, such as cyclic cumulants, the observed degree of generalization is very high, and performance is also high.  

A common theme of the DL-based approaches proposed for signal classification involves the use of NNs with structures developed in the context of image processing, specifically convolutional neural networks (CNNs) \cite{Tian_Access2020} and CAPs \cite{Sabour_etal_NIPC2017}, both of which were explicitly designed based on the growing understanding of how the human eye-brain vision system works and how it is able to recognize images.  Despite their excellent performance with image classification, these ``image-processing'' NNs appear to be poorly suited to work with the minimally processed I/Q data of digitally modulated signals, being unable to generalize their training to new I/Q data when the distribution of the signal parameters of the new data differs slightly from the training data \cite{Latshaw_COMM2022, Snoap_CCNC_2022}.  However, pre-processing the I/Q data to obtain CCs and using the CC features in conjunction with ``image processing'' CAPs having conventional NN layers results in robust signal classification and generalization performance~\cite{Snoap_Sensors_2023, Snoap_MILCOM_2022}.

Motivated by the brittleness of DL-based classifiers that feed minimally processed I/Q components of digitally modulated signals into conventional image processing NN layers and leveraging the knowledge that DL-based classifiers using CC features of digitally modulated signals are robust and display excellent classification performance, a new approach to DL-based classification of digitally modulated signals is pursued in the remainder of this dissertation.  Proposed novel custom-designed NN layers perform specific nonlinear mathematical functions on the input I/Q data and are designed to enable downstream conventional image processing NN capsules to distinguish signal features that are akin to the CC estimates.  The proposed approach obviates the computational expense of pre-processing the I/Q components of digitally modulated signals to estimate the CC features \cite{Snoap_Sensors_2023, Snoap_MILCOM_2022}, allowing DL-based classifiers to use less extensively processed I/Q signal components as inputs and still obtain comparable classification performance and generalization abilities.  To the best of the author's knowledge, this is the first study of its kind, using DL-based classifiers with feature-extraction layers that are different from those used in conventional image processing NNs, as have been used in previous studies of DL-based classification of digitally modulated signals.  The robustness and generalization abilities of the proposed DL-based digital modulation classifiers are assessed on the same two distinct datasets (\texttt{CSPB.ML.2018} and \texttt{CSPB.ML.2022}) that have been used and described throughout this dissertation and are publicly available \cite{IEEE_DataPort}.

The remainder of this chapter is organized as follows:  related background information on CSP and the cycle frequency (CF) features relevant to digital modulation classification is provided in Section~\ref{sec:Background}, followed by the presentation of an initial set of novel NN function layers and proposed NN structure in Section~\ref{sec:Details}.  The datasets and the pre-processing performed on the datasets used for NN training and testing are discussed in Section~\ref{sec:Datasets3}.  In Section~\ref{sec:NumericalResults1}, numerical performance results for this initial set of novel NN function layers are presented.  Next, the presentation of a second and more complete set of novel NN function layers and proposed NN structure is given in Section~\ref{sec:Details2}, followed by the details of dataset pre-processing in Section~\ref{sec:Datasets4}, and numerical performance results in Section~\ref{sec:NumericalResults2}.

\section{Extracting Periodic Features for Classifying Digitally Modulated Signals}\label{sec:Background}
Salient features of digitally modulated signals that are useful in their classification include second- and higher-order CCs~\cite{Spooner_Asilomar2000} as well as the spectral correlation function (SCF)~\cite{Gardner_CSP_pt1_1994}.  They can be estimated using various CSP techniques~\cite{Spooner_Asilomar2001, Spooner_Asilomar1995, Spooner_CSP_pt2_1994} and their estimates can be compared to a set of theoretical CCs or SCFs for classifying the digital modulation scheme embedded in a noisy signal.  This approach provides a CSP baseline classification model that has been used as a comparison reference for DL-based signal classifiers~\cite{Snoap_Sensors_2023}.

To accurately estimate CC features for robust DL-based classification of digitally modulated signals, specific signal parameters must be either known or accurately estimated from the I/Q data prior to estimating the CC features~\cite{Snoap_Sensors_2023, Snoap_MILCOM_2022}.  These parameters include the symbol interval, $T_0$, and symbol rate $1/T_0$, the carrier-frequency offset (CFO), $f_0$, the excess bandwidth of the signal\footnote{For square-root raised-cosine (SRRC) pulse shaping this is implied by the roll-off parameter, $\beta$.}, and the signal power level\footnote{This directly impacts the in-band signal-to-noise (SNR) ratio.}.  These signal parameters define the cycle frequencies (CFs) needed for CC computation, with a key processing step involving blind estimation of the coarse second-order {\em CF pattern} for the signal as discussed in Section~\ref{sec:Proposed_Model_CC_FE} of this dissertation.  CFs $\alpha$ for which a CC is non-zero for  typical PSK/QAM/SQPSK communication signals include harmonics of the symbol rate~$1/T_0$, multiples of the carrier frequency~$f_0$, and combinations of these two sets, such that $\alpha$ can be written~as
\begin{align}
\label{eq:basic_cf_psk_qam2}
\alpha = (n-2m)f_0 \pm k/T_0,
\end{align}
where $n$ is the order, $m$ is the number of conjugations, and $k$ is the set of non-negative integers typically constrained to a maximum value of $k=5$.

Although the set~(\ref{eq:basic_cf_psk_qam2}) of possible CFs for various digitally modulated signals is determined by the symbol rate and the CFO, most signals exhibit only a subset of the possible CF set.  Moreover, as discussed in \cite{Snoap_Sensors_2023}, there are only five basic CF patterns, and if one can identify the CF pattern corresponding to a digitally modulated signal, one can also determine the actual number of CFs needed to fully characterize the modulation type through its set of associated CC values.

The  mathematical functions that are meaningful for identifying which of the five basic CF patterns are present in a signal of interest are the Fourier transform of the squared signal and of the quadrupled signal \cite{Gardner_CSP_pt1_1994}.  Furthermore, the squared signal, the quadrupled signal, and the signal raised to a power of six (and eight) are mathematical operations that are meaningful for estimating even-order CCs  \cite{Spooner_Asilomar2001, Spooner_Asilomar1995}.

In this context, the goal of this chapter is twofold:
\begin{itemize}
\item Establish custom ``feature-extraction'' NN layers implementing the generic mathematical equations that bring out the salient features of digitally modulated signals not apparent in the I/Q signal data.  The outputs of these custom layers are then selectively applied to a series of parallel CAPs whose subsequent layers and hyper-parameters are trained to recognize and classify signals corresponding to eight digital modulation schemes commonly used in practical systems:  BPSK, QPSK, 8PSK, $\pi/4$-DQPSK, MSK, 16QAM, 64QAM, and 256QAM.
\item Study the classification and generalization performance of DL-based classifiers using the custom ``feature-extraction'' layers in conjunction with the trained CAPs, providing a comparison with a CSP baseline classification model that uses conventional CSP techniques as well as with alternative DL-based signal classifiers that use ``image processing'' CAPs with conventional NN layers.
\end{itemize}

For the initial set of novel NN function layers and NN structure discussed in Section~\ref{sec:Details}, the input is I/Q data that has been normalized to unit total power (UTP).  The final novel NN structure and function layers in Section~\ref{sec:Details2} will instead require I/Q data that has been normalized to unit-signal-power (USP).

\section{First CAP with Custom Feature Extraction NN Layers for Modulation Classification}\label{sec:Details}
\begin{figure}
\centering
\includegraphics[width=\linewidth]{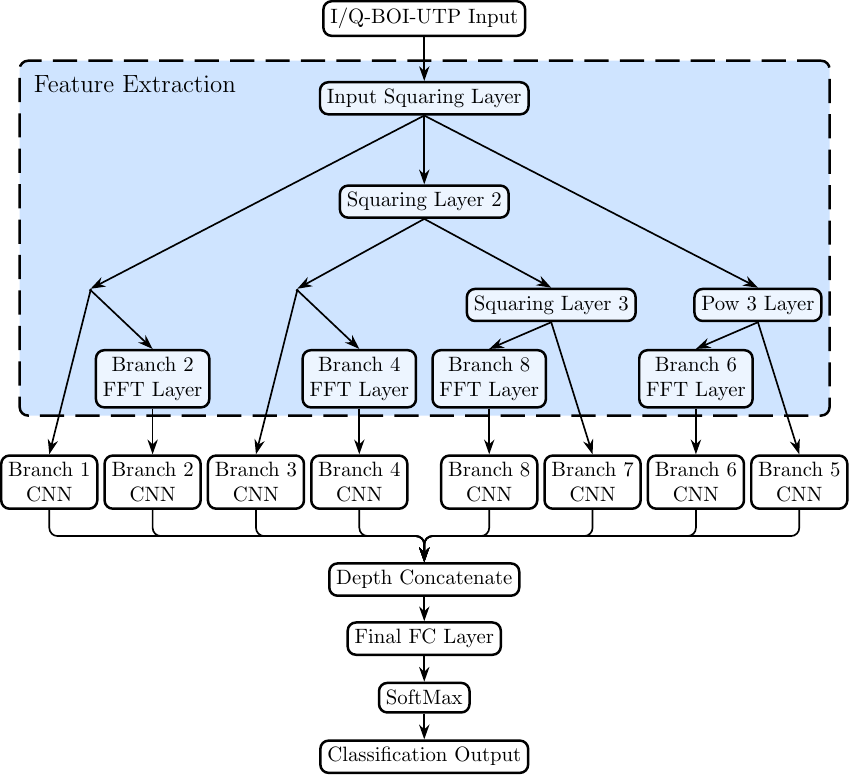}
\caption[I/Q-BOI-UTP CAP Topology With Novel Custom NN Layers Performing Feature Extraction for Each Branch.]{\textbf{I/Q-BOI-UTP CAP Topology With Novel Custom NN Layers Performing Feature Extraction for Each Branch.}}\label{fig:CAP_Topology2}
\end{figure}

To consistently extract the second, fourth, sixth, and eighth-order signal features in both the time-domain and frequency-domain using the I/Q signal data normalized to UTP, we only need to implement three custom NN layers for the needed nonlinear computations:  a squaring layer, a raise-to-the-power-of-three (or Pow3) layer, and a fast Fourier transform (FFT) layer.  These custom layers can then be connected together as shown in Figure~\ref{fig:CAP_Topology2}, such that their outputs branch out until all eight feature types have been extracted.

Noting that the I/Q signal data is input as pairs of real numbers $(I,Q)$, the squaring and power-of-three layers can be implemented as follows:

--~Squaring Layer,
\begin{align}
I_{Output} &= \left( I \times I \right) - \left( Q \times Q \right), \\
Q_{Output} &= 2 \times I \times Q.
\end{align}

--~Pow3 Layer,
\begin{align}
I_{Output} &= \left( I \times I \times I \right) - \left( 3 \times I \times Q \times Q \right), \\
Q_{Output} &= \left( 3 \times I \times I \times Q \right) - \left( Q \times Q \times Q \right).
\end{align}

For the FFT layer, only the magnitude is needed at the output, with the zero frequency bin in the center, as this information is most likely to reveal CF patterns present in the signal.  Therefore, after taking the FFT, its absolute value is obtained to ensure the output magnitude is a real number, suitable for use by downstream trainable NN layers.

\begin{table}
\centering
\caption[CNN Branch Layout.]{\textbf{CNN Branch Layout.}}
{
\begin{tabular}{ c c c }
\toprule
\textbf{Layer} & \textbf{(\# Filters)[Filter Size]} & \textbf{Activations} \\
\midrule
Input			& 							& $32,768 \times Y$ \\
ConvMaxPool		& ($16$)[$23 \times Y$]	& $16,384 \times 16$ \\
ConvMaxPool		& ($24$)[$23 \times 16$]	& $8,192 \times 24$ \\
ConvMaxPool		& ($32$)[$23 \times 24$]	& $4,096 \times 32$ \\
ConvMaxPool		& ($48$)[$23 \times 32$]	& $2,048 \times 48$ \\
ConvMaxPool		& ($64$)[$23 \times 48$]	& $1,024 \times 64$ \\
ConvAvgPool		& ($96$)[$23 \times 64$]	& $96$ \\
FC				& 							& \# Classes \\
\bottomrule
\end{tabular}
}\label{table:CNN_Branch}
\end{table}

\begin{table}
\centering
\caption[ConvMaxPool Layers.]{\textbf{ConvMaxPool Layers.}}
{
\begin{tabular}{ c c c c }
\toprule
\textbf{Layer} 		& \textbf{(\# Filters)[Filter Size]} & \textbf{Stride} & \textbf{Activations} \\
\midrule
Input		& ($A$)[$B \times C$]	& 					& $X \times Y$ \\
Conv		& ($A$)[$B \times C$]	& [$1 \times 1$] 	& $X \times (Y \cdot A/C)$ \\
Batch Norm	& 						& 					& $X \times (Y \cdot A/C)$ \\
ReLU		& 						& 					& $X \times (Y \cdot A/C)$ \\
Max Pool	& ($1$)[$1 \times 2$]	& [$1 \times 2$]	& $(X/2) \times (Y \cdot A/C)$ \\
\bottomrule
\end{tabular}
}\label{table:ConvMaxPoolLayer}
\end{table}

\begin{table}
\centering
\caption[ConvAvgPool Layers.]{\textbf{ConvAvgPool Layers.}}
{
\begin{tabular}{ c c c c }
\toprule
\textbf{Layer} 		& \textbf{(\# Filters)[Filter Size]} & \textbf{Stride} & \textbf{Activations} \\
\midrule
Input		& ($A$)[$B \times C$]	& 					& $X \times Y$ \\
Conv		& ($A$)[$B \times C$]	& [$1 \times 1$] 	& $X \times (Y \cdot A/C)$ \\
Batch Norm	& 						& 					& $X \times (Y \cdot A/C)$ \\
ReLU		& 						& 					& $X \times (Y \cdot A/C)$ \\
Avg Pool	& ($1$)[$1 \times X$]	& [$1 \times 1$]	& $1 \times (Y \cdot A/C)$ \\
\bottomrule
\end{tabular}
}\label{table:ConvAvgPoolLayer}
\end{table}

The CAP includes eight convolutional neural network (CNN) branches that implement the primary capsules and will be trained to classify eight common digital modulation schemes:  BPSK, QPSK, 8PSK, $\pi/4$-DQPSK, MSK, 16QAM, 64QAM, and 256QAM.  The primary capsules contain the majority of the network's learnable hyperparameters, with each branch having as its input one of the eight desired features.  At the output of each branch, there is a fully connected layer with eight outputs, which is used to reduce the branch output size to the number of modulation classes in the training dataset; this is done to ensure that each branch has the ability to distinguish between all modulation types in the training dataset.  This ability is important because, while not every branch will be able to fully distinguish between all modulation types, each branch will be able to identify which classes it can distinguish and this determination can be made during NN training.  For example, due to the nature of the CF patterns, it is expected that the branch with second-order frequency-domain inputs will be able to distinguish between BPSK and MSK, but all other modulation types will not be distinguishable on this branch.  Likewise, the branch with fourth-order frequency-domain inputs should be able to distinguish between 8PSK-like, $\pi/4$-DQPSK-like, and SQPSK-like patterns, but QAM-like and BPSK-like CF patterns may not be distinguishable to this branch.  Each branch, having opportunity to distinguish between all eight modulation types, then has its outputs concatenated together with the other branches followed by a final fully connected layer.  This last fully connected layer learns the appropriate weights to apply to the outputs of each CNN branch so that each branch's ability is combined together to obtain the minimum error on the training dataset.

The overall structure of the proposed CAP with custom feature extracting NN layers is shown in Figure~\ref{fig:CAP_Topology2}, with the characteristics of the subsequent CNN branches for classifying digitally modulated signals outlined in Tables~\ref{table:CNN_Branch}~--~\ref{table:ConvAvgPoolLayer}.  The number of filters $A$ and filter size $\left[ B \times C \right]$ for each convolutional layer are defined in Table~\ref{table:CNN_Branch} as are the number of output activations (e.g., $X \times Y$) for each layer.

\section{Datasets Used for NN Training and Testing the First CAP}\label{sec:Datasets3}
To train and test the performance of the proposed CAP in Figure~\ref{fig:CAP_Topology2} (including its generalization ability), two publicly available datasets are used that both contain signals corresponding to the eight digital modulation schemes of interest (BPSK, QPSK, 8PSK, $\pi/4$-DQPSK, MSK, 16QAM, 64QAM, and 256QAM).  The two datasets are available from \cite{IEEE_DataPort} as \texttt{CSPB.ML.2018} and \texttt{CSPB.ML.2022}, and their signal generation parameters are listed in Table~\ref{table:SigGenParms} in Chapter~\ref{ch:HQDatasets} of this dissertation.

The CFO ranges corresponding to signals in the two datasets are non-intersecting, which allows evaluation of the generalization ability of the trained CAP.  Specifically, if the CAP trained on a large portion of the \texttt{CSPB.ML.2018} dataset displays high classification accuracy on the remaining signals in the dataset, and its performance when classifying signals in the \texttt{CSPB.ML.2022} dataset is at similarly high levels, then the CAP is robust and has a high ability to generalize\footnote{Likewise, if its classification accuracy on signals in \texttt{CSPB.ML.2018} is high, while on signals in \texttt{CSPB.ML.2022} is low, then its generalization ability is also low.}.

Prior to training the proposed CAP on these datasets, a blind band-of-interest (BOI) detector~\cite{BOIdetector} is used to locate the bandwidth for signals in the datasets, filter out-of-band noise, and center the I/Q data at zero frequency based on the CFO estimate provided by the BOI detector.  Finally, the I/Q data is normalized to unit total power (UTP) so that it does not prevent the activation functions of the NN from converging.  To show that using a blind BOI detector to shift the signal to an estimated zero frequency does not necessarily enable an I/Q-trained CAP such as the one in~\cite{Latshaw_COMM2022} to generalize between the \texttt{CSPB.ML.2018} and \texttt{CSPB.ML.2022} datasets, this data is also used to retrain the image-processing CAP in~\cite{Latshaw_COMM2022} as an alternative to provide a point of comparison with the proposed CAP.

\section{First CAP Training and Numerical Results}\label{sec:NumericalResults1}
The proposed CAP and the image-processing CAP in~\cite{Latshaw_COMM2022} have been implemented in MATLAB and trained on a high-performance computing cluster, such that each CAP was trained and tested two separate times as follows:

\begin{figure}
\begin{center}
\subfigure[Performance of the \texttt{CSPB.ML.2018} I/Q-BOI-UTP-trained CAPs When Classifying \texttt{CSPB.ML.2018} Test Signals.]
{\includegraphics[width=0.49\linewidth]{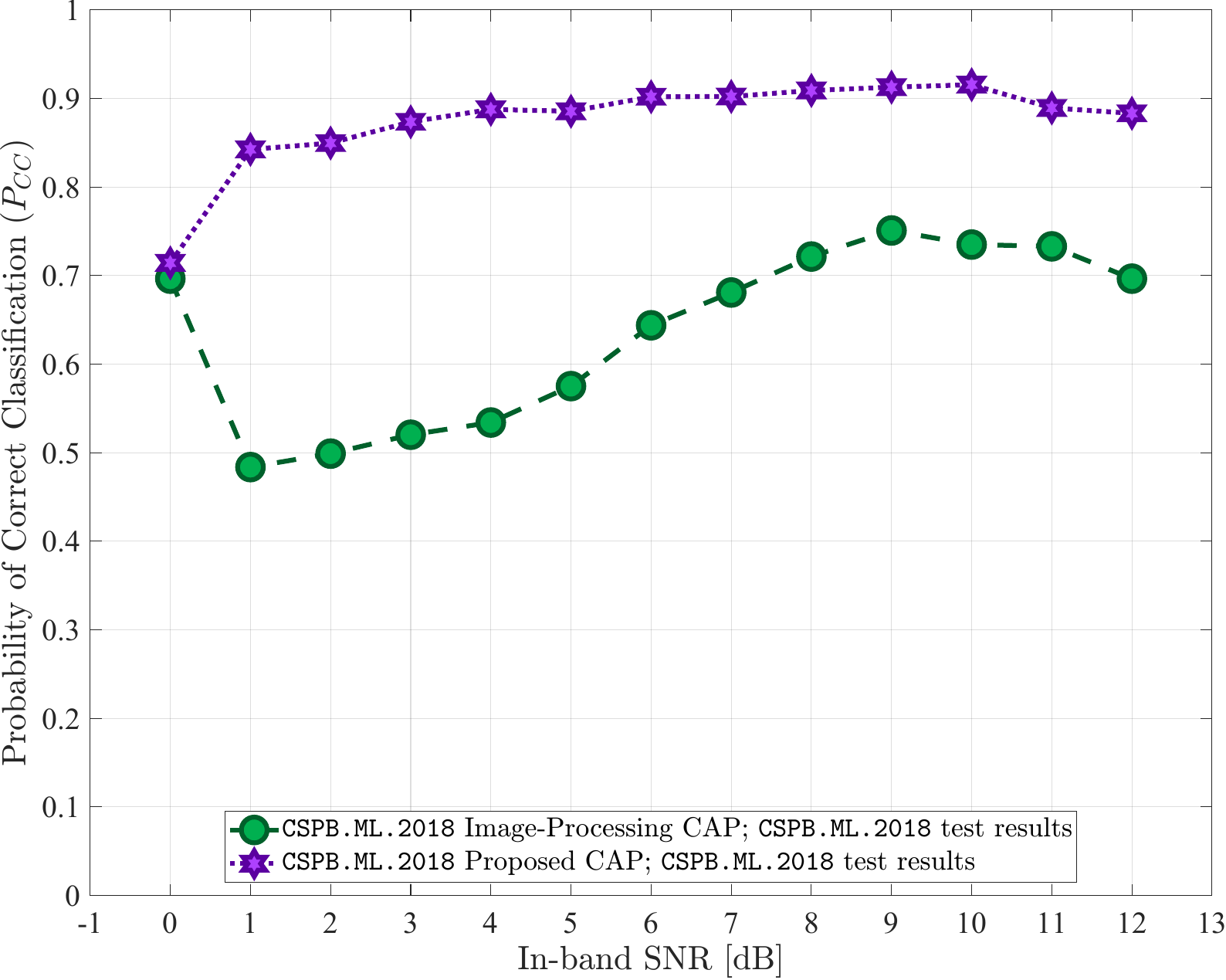}}
\subfigure[Generalization Performance of the \texttt{CSPB.ML.2018} I/Q-BOI-UTP-trained CAPs Classifying All \texttt{CSPB.ML.2022} Signals.]
{\includegraphics[width=0.49\linewidth]{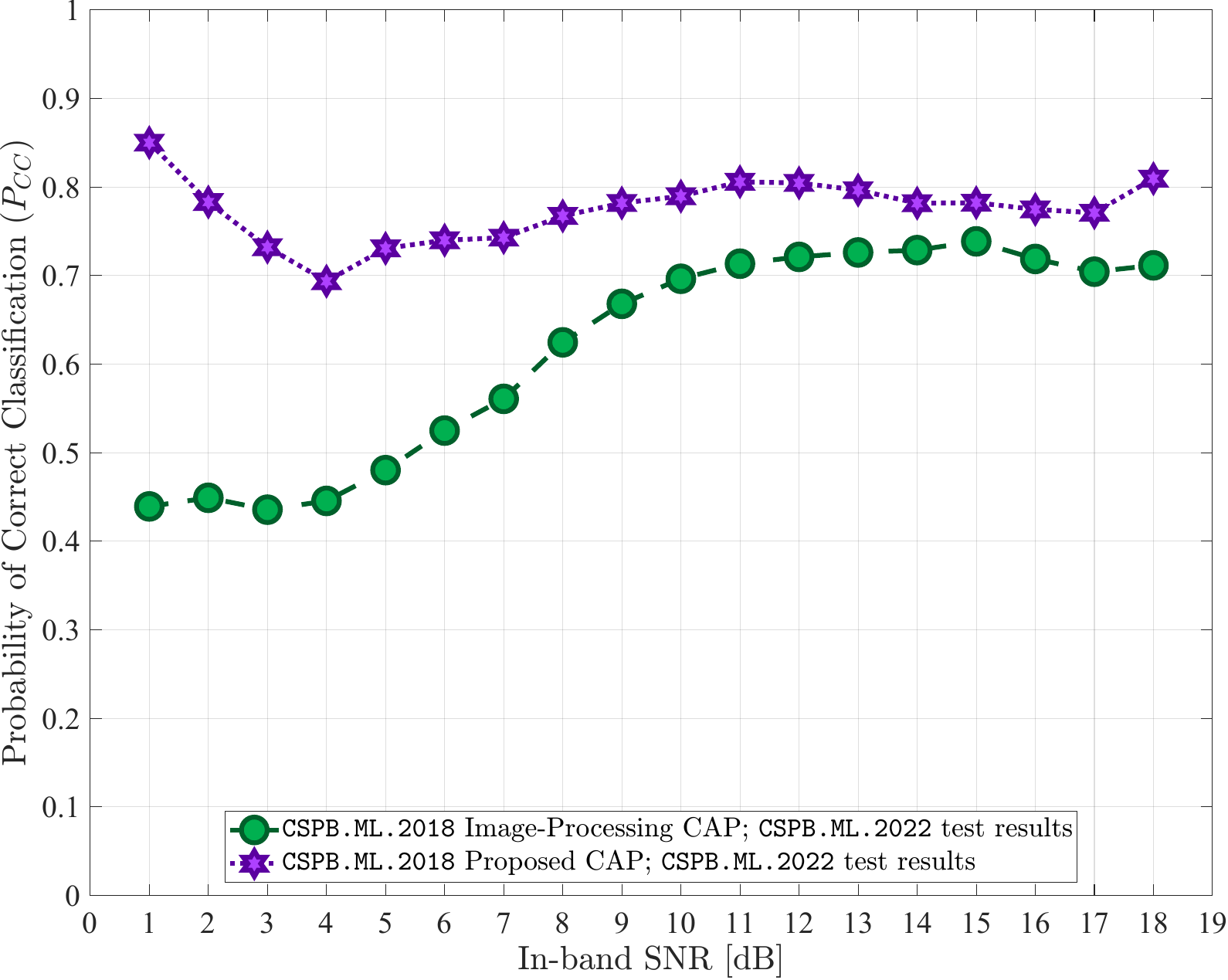}}
\caption[Initial and Generalization Test Results for \texttt{CSPB.ML.2018} I/Q-BOI-UTP-trained CAPs.]{\textbf{Initial and Generalization Test Results for \texttt{CSPB.ML.2018} I/Q-BOI-UTP-trained CAPs.}}\label{fig:SNR_CTest_CTrained_All_1}
\end{center}
\end{figure}

\begin{figure}
\begin{center}
\subfigure[Performance of the \texttt{CSPB.ML.2022} I/Q-BOI-UTP-trained CAPs When Classifying \texttt{CSPB.ML.2022} Test Signals.]
{\includegraphics[width=0.49\linewidth]{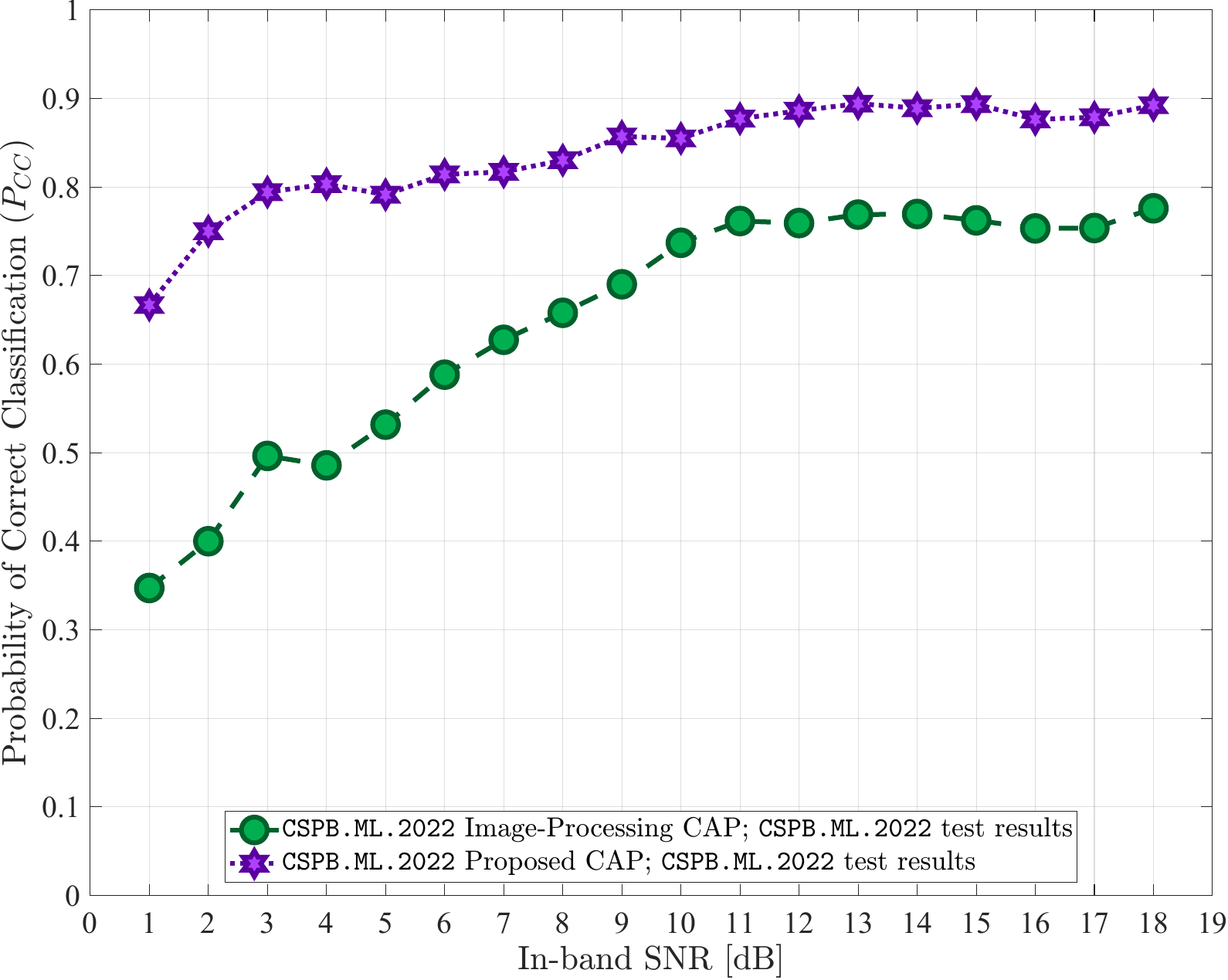}}
\subfigure[Generalization Performance of the \texttt{CSPB.ML.2022} I/Q-BOI-UTP-trained CAPs Classifying All \texttt{CSPB.ML.2018} Signals.]
{\includegraphics[width=0.49\linewidth]{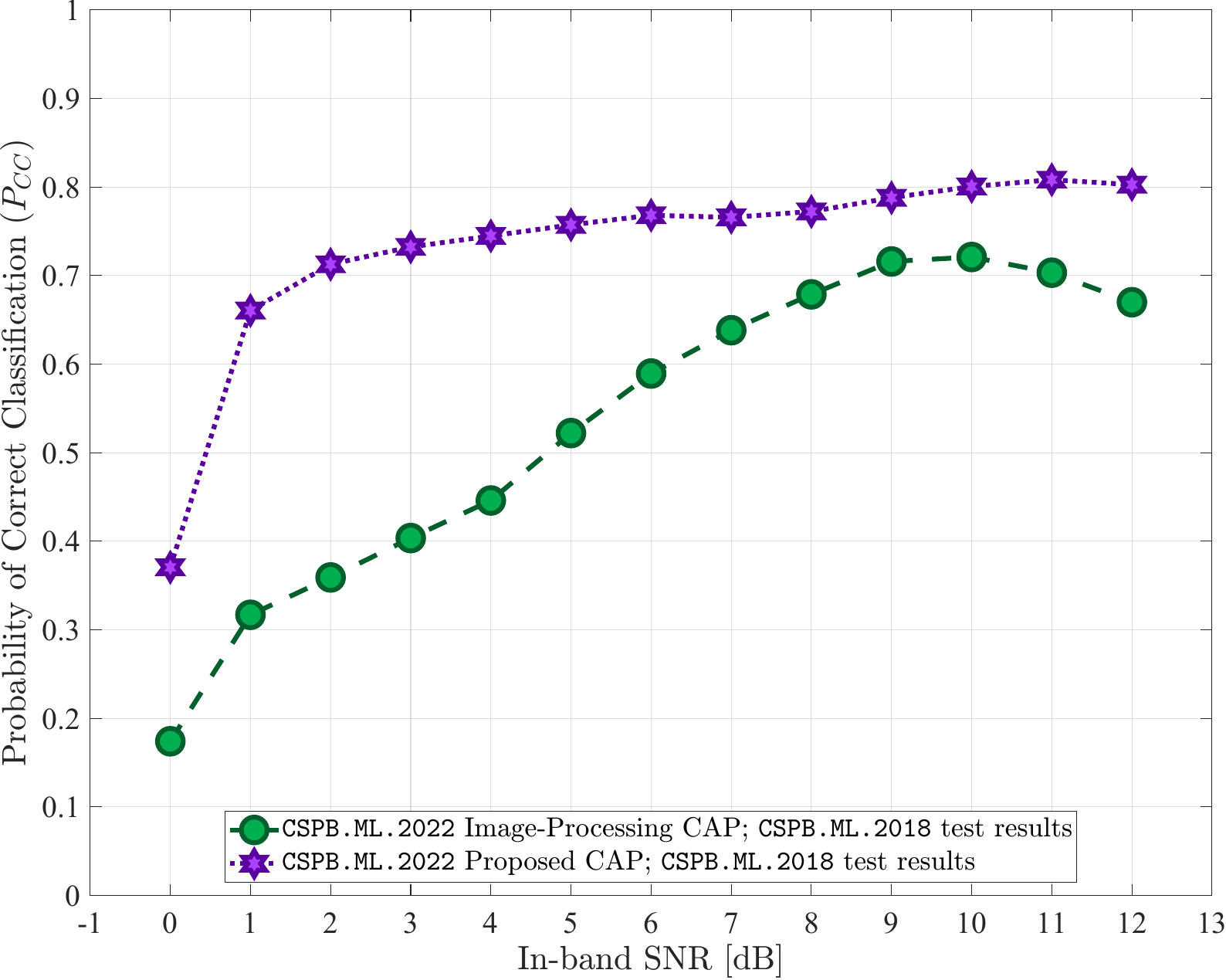}}
\caption[Initial and Generalization Test Results for \texttt{CSPB.ML.2022} I/Q-BOI-UTP-trained CAPs.]{\textbf{Initial and Generalization Test Results for \texttt{CSPB.ML.2022} I/Q-BOI-UTP-trained CAPs.}}\label{fig:SNR_GCTest_GCTrained_All_1}
\end{center}
\end{figure}

\begin{itemize}
\item In the first training instance, dataset \texttt{CSPB.ML.2018} was used, splitting the available signals into $70\%$ for training, $5\%$ for validation, and $25\%$ for testing.  The probability of correct classification for the test results obtained using the $25\%$ test portion of signals in \texttt{CSPB.ML.2018} is shown in Figure~\ref{fig:SNR_CTest_CTrained_All_1}(a).

\hspace{0.25cm} The CAPs trained on \texttt{CSPB.ML.2018} are then tested on dataset \texttt{CSPB.ML.2022} to assess the generalization abilities of a trained CAP when classifying all signals available in \texttt{CSPB.ML.2022}.  The probability of correct classification for this test is shown in Figure~\ref{fig:SNR_CTest_CTrained_All_1}(b).
\item In the second training instance, the NN was reset and trained anew using signals in dataset \texttt{CSPB.ML.2022}, with a similar split of $70\%$ signals used for training, $5\%$ for validation, and $25\%$ for testing.  The probability of correct classification for the test results obtained using the $25\%$ test portion of signals in \texttt{CSPB.ML.2022} is shown in Figure~\ref{fig:SNR_GCTest_GCTrained_All_1}(a).

\hspace{0.25cm} The CAPs trained on \texttt{CSPB.ML.2022} are then tested on dataset \texttt{CSPB.ML.2018} to assess the generalization abilities of the trained NN when classifying all signals available in \texttt{CSPB.ML.2018}.  The probability of correct classification for this test is shown in Figure~\ref{fig:SNR_GCTest_GCTrained_All_1}(b).
\end{itemize}

In summary, from the plots shown in Figures~\ref{fig:SNR_CTest_CTrained_All_1}~--~\ref{fig:SNR_GCTest_GCTrained_All_1}, the proposed CAP with custom feature extraction layers outperforms the image-processing CAP from~\cite{Latshaw_COMM2022} in all training and testing scenarios, displaying high classification performance and generalization abilities.

The overall probability of correct classification ($P_{CC}$) for all experiments performed is shown in Table~\ref{table:OverallPerformance2} and provides further confirmation that the proposed CAP containing the novel custom layers is able to perform feature extraction tailored to the modulated signals of interest more effectively than the image-processing CAP in~\cite{Latshaw_COMM2022}, which employs only the conventional NN layers used in the context of image processing.  The lowest classification performance seen by the proposed CAP came from training on \texttt{CSPB.ML.2022} and testing on \texttt{CSPB.ML.2018} with an overall correct classification probability $P_{CC} = 77.1\%$, while the highest classification performance observed for the image-processing CAP in~\cite{Latshaw_COMM2022} came from training on \texttt{CSPB.ML.2022} and testing on \texttt{CSPB.ML.2022}, resulting in a corresponding overall $P_{CC} = 72.1\%$. Thus, even the lowest performance of the proposed CAP exceeds by $5\%$  the best performance of the image-processing CAP in~\cite{Latshaw_COMM2022}.

\begin{figure}
\begin{center}
\subfigure[Confusion Matrix of the \texttt{CSPB.ML.2018} I/Q-BOI-UTP-trained Proposed CAP Classifying \texttt{CSPB.ML.2018} Test Signals.]
{\includegraphics[width=0.49\linewidth]{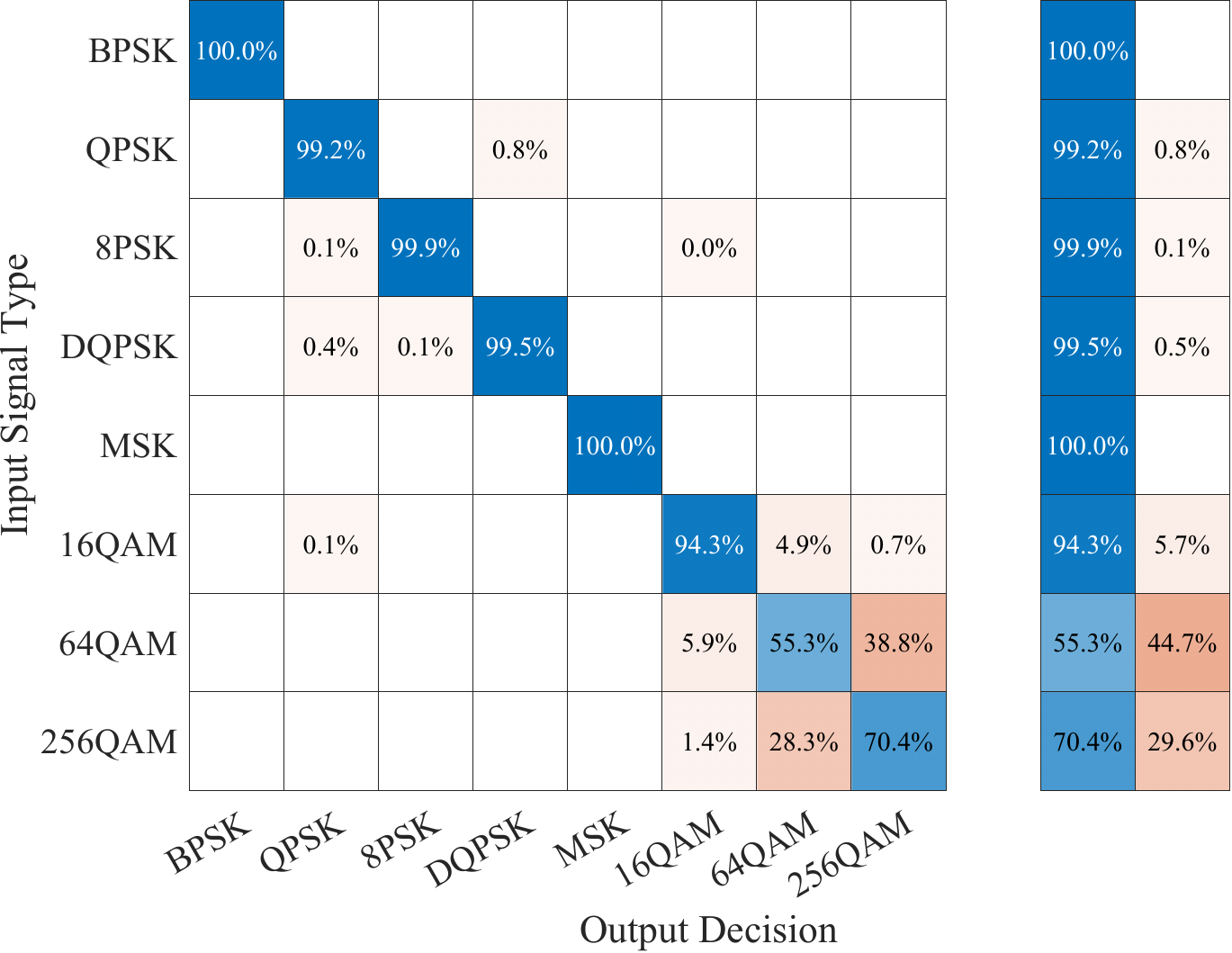}}
\subfigure[Confusion Matrix of the \texttt{CSPB.ML.2018} I/Q-BOI-UTP-trained Proposed CAP Classifying All \texttt{CSPB.ML.2022} Signals.]
{\includegraphics[width=0.49\linewidth]{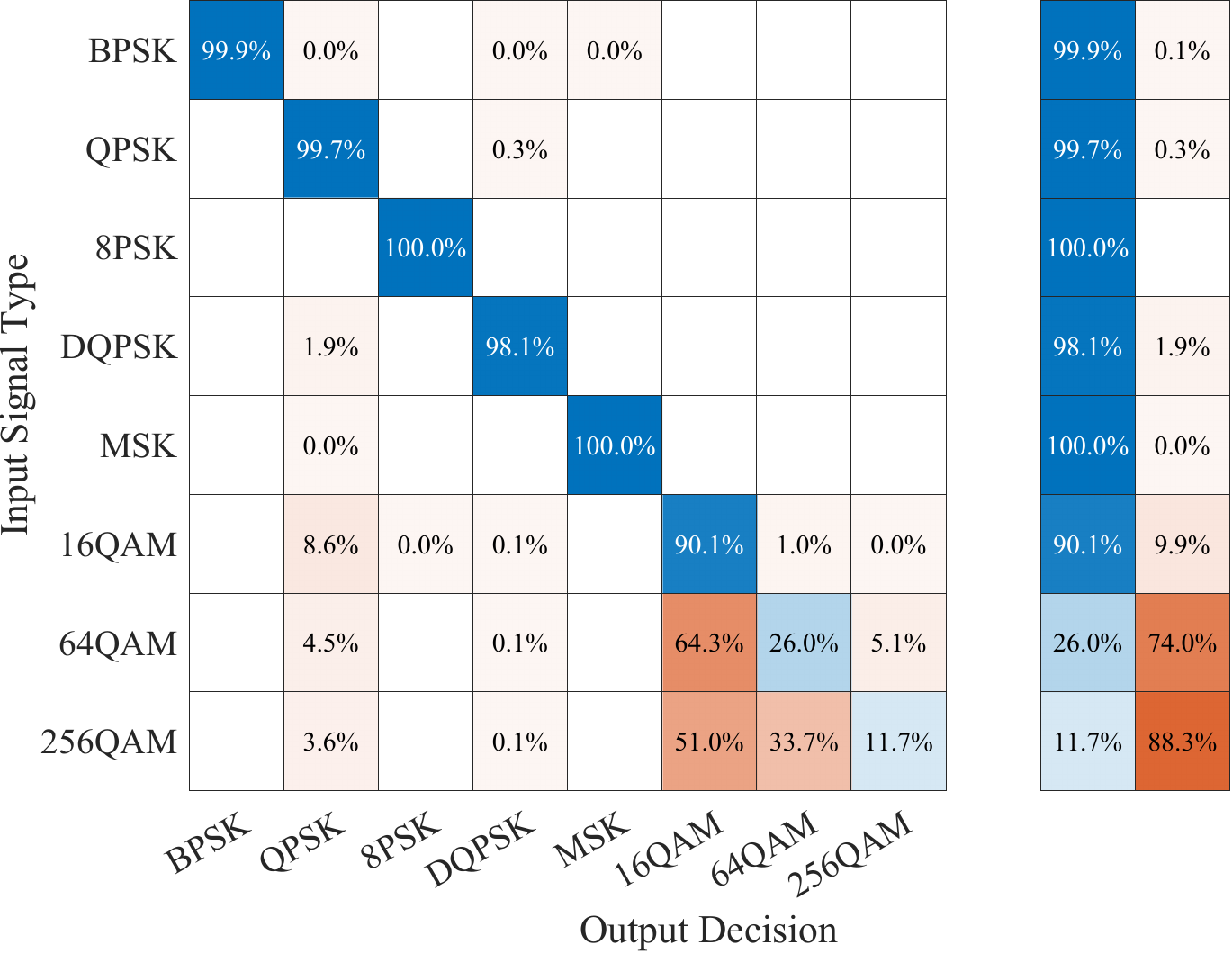}}
\caption[Confusion Matrices of the \texttt{CSPB.ML.2018} I/Q-BOI-UTP-trained Proposed CAP Classifying \texttt{CSPB.ML.2018} Test Signals and All \texttt{CSPB.ML.2022} Signals.]{\textbf{Confusion Matrices of the \texttt{CSPB.ML.2018} I/Q-BOI-UTP-trained Proposed CAP Classifying \texttt{CSPB.ML.2018} Test Signals and All \texttt{CSPB.ML.2022} Signals.}}\label{fig:CMs_2018_IQ_BOI_UTP_CAP}
\end{center}
\end{figure}

\begin{figure}
\begin{center}
\subfigure[Confusion Matrix of the \texttt{CSPB.ML.2022} I/Q-BOI-UTP-trained Proposed CAP Classifying \texttt{CSPB.ML.2022} Test Signals.]
{\includegraphics[width=0.49\linewidth]{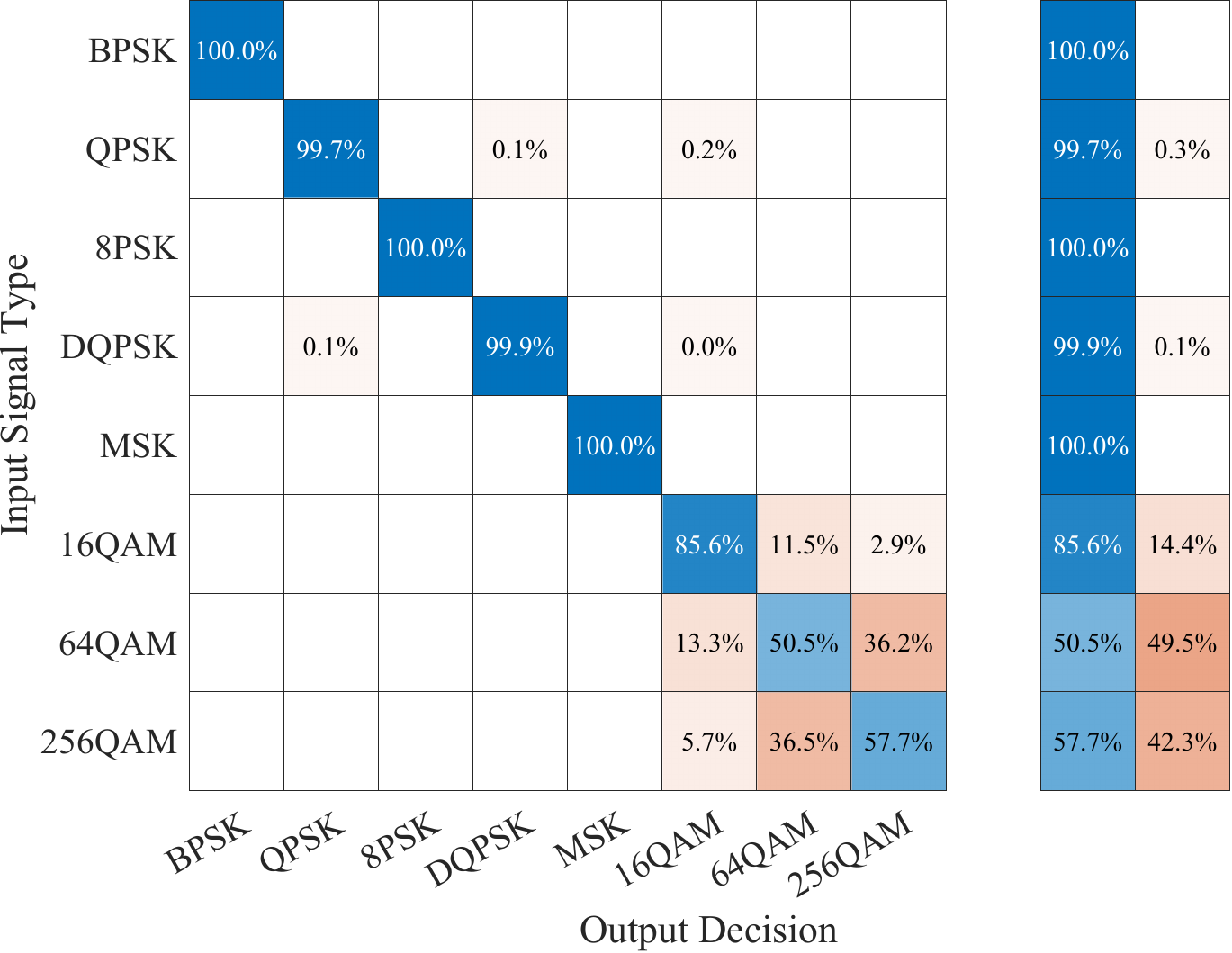}}
\subfigure[Confusion Matrix of the \texttt{CSPB.ML.2022} I/Q-BOI-UTP-trained Proposed CAP Classifying All \texttt{CSPB.ML.2018} Signals.]
{\includegraphics[width=0.49\linewidth]{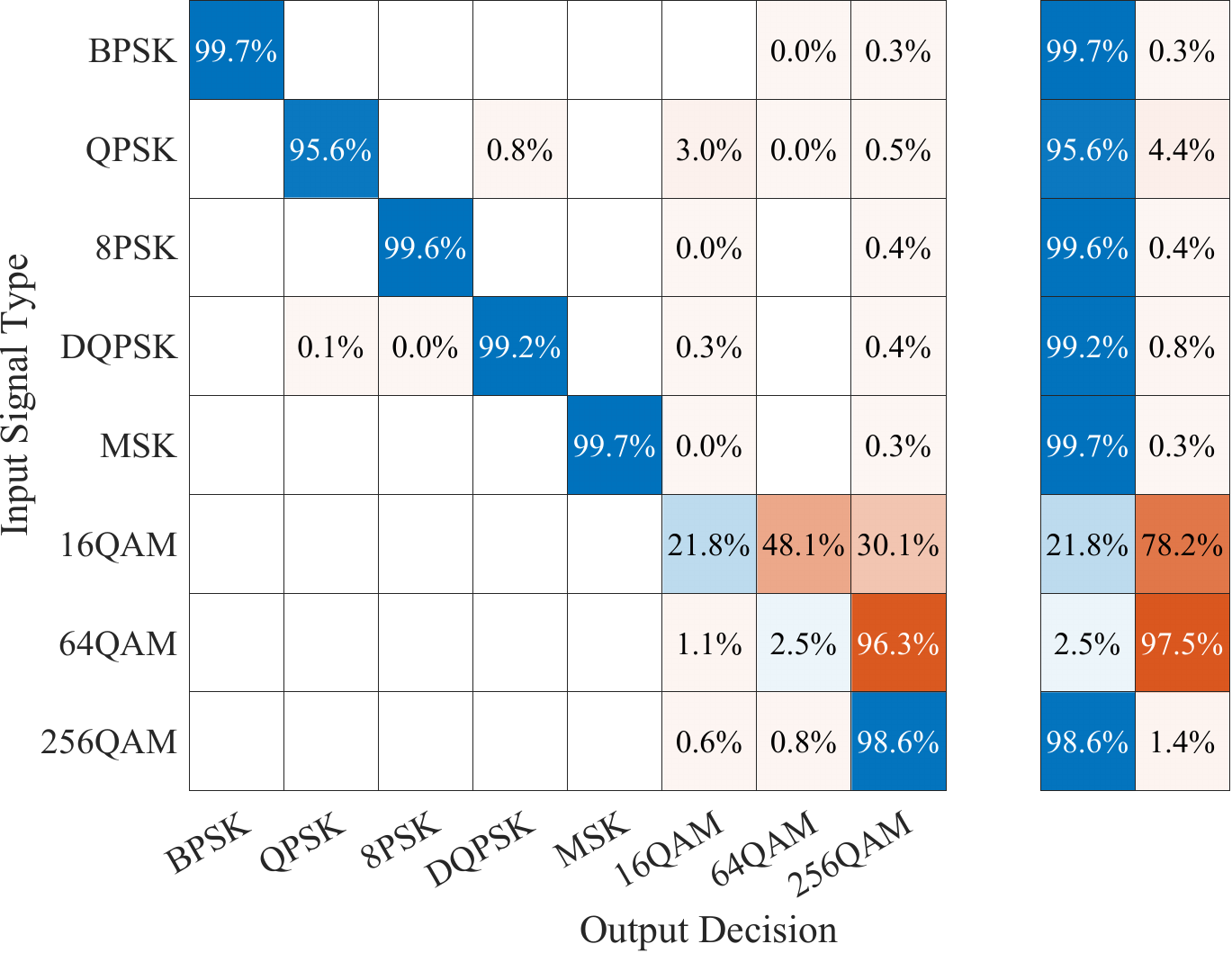}}
\caption[Confusion Matrices of the \texttt{CSPB.ML.2022} I/Q-BOI-UTP-trained Proposed CAP Classifying \texttt{CSPB.ML.2022} Test Signals and All \texttt{CSPB.ML.2018} Signals.]{\textbf{Confusion Matrices of the \texttt{CSPB.ML.2022} I/Q-BOI-UTP-trained Proposed CAP Classifying \texttt{CSPB.ML.2022} Test Signals and All \texttt{CSPB.ML.2018} Signals.}}\label{fig:CMs_2022_IQ_BOI_UTP_CAP}
\end{center}
\end{figure}

\begin{table}
\centering
\caption[Overall I/Q-BOI-UTP-trained CAP Classification Performance ($P_{CC}$).]{\textbf{Overall I/Q-BOI-UTP-trained CAP Classification Performance ($P_{CC}$).}}
{
\setlength{\tabcolsep}{6.5mm}
\begin{tabular}{ c c c }
\toprule
\textbf{Classification Model} & \begin{tabular}{@{}c@{}} \textbf{Results for Dataset} \\ \texttt{CSPB.ML.2018} \end{tabular} & \begin{tabular}{@{}c@{}} \textbf{Results for Dataset} \\ \texttt{CSPB.ML.2022} \end{tabular} \\
\midrule
\begin{tabular}{@{}c@{}} Image-Processing CAP \\ \texttt{CSPB.ML.2018} \\ I/Q-BOI-UTP-trained \end{tabular} & $67.7\%$ & $68.0\%$ \\
\hline
\begin{tabular}{@{}c@{}} Image-Processing CAP \\ \texttt{CSPB.ML.2022} \\ I/Q-BOI-UTP-trained \end{tabular} & $62.5\%$ & $72.1\%$ \\
\hline
\begin{tabular}{@{}c@{}} Proposed CAP \\ \texttt{CSPB.ML.2018} \\ I/Q-BOI-UTP-trained \end{tabular} & $89.8\%$ & $78.2\%$ \\
\hline
\begin{tabular}{@{}c@{}} Proposed CAP \\ \texttt{CSPB.ML.2022} \\ I/Q-BOI-UTP-trained \end{tabular} & $77.1\%$ & $86.7\%$ \\
\bottomrule
\end{tabular}
}\label{table:OverallPerformance2}
\end{table}

For additional insight into the classification performance of the proposed CAP with custom feature extracting layers, its corresponding confusion matrices are shown in Figures~\ref{fig:CMs_2018_IQ_BOI_UTP_CAP}--\ref{fig:CMs_2022_IQ_BOI_UTP_CAP}.  As can be observed from these figures, the classification performance for individual BPSK, QPSK, 8PSK, $\pi/4$-DQPSK, and MSK modulation schemes when the CAP is trained and tested on the \texttt{CSPB.ML.2018} dataset is within $1.5\%$ difference of the performance observed when the CAP is trained on the \texttt{CSPB.ML.2018} dataset and tested on the \texttt{CSPB.ML.2022} dataset, indicating excellent generalization performance for these five types of digital modulation schemes.  Similar confusion matrices are obtained for the cases when the CAP is trained and tested on the \texttt{CSPB.ML.2022} dataset and when the CAP is trained on the \texttt{CSPB.ML.2022} dataset and tested on the \texttt{CSPB.ML.2018} dataset as shown in Figure~\ref{fig:CMs_2022_IQ_BOI_UTP_CAP}.  Furthermore, the only modulation schemes for which the classification performance of the proposed CAP decreased when the testing dataset was changed from \texttt{CSPB.ML.2018} to \texttt{CSPB.ML.2022} are 16QAM, 64QAM, and 256QAM.

The reason the proposed CAP is unable to generalize its training well for the three QAM modulation types is due to the fact that these modulation types are only distinguishable from each other at higher orders when their signal power is known or accurately estimated.  Since the total (signal+noise) power was normalized rather than just the signal power, the relative differences between the power levels of the 16QAM, 64QAM, and 256QAM higher-order moments were removed.  Thus, the differences between the extracted features for 16QAM, 64QAM, and 256QAM were different enough between \texttt{CSPB.ML.2018} and \texttt{CSPB.ML.2022} that the proposed CAP could not obtain good generalization on these three modulation types.

In the following sections, additional custom NN function layers are investigated that process the I/Q signal data with signal power normalization, rather than total power normalization, in order to overcome the issue of poor generalization for QAM modulation types, without needing to resort to full estimates of the CCs.

\section{Second CAP with Custom Feature Extraction NN Layers for Modulation Classification}\label{sec:Details2}
\begin{figure}
\centering
\includegraphics[width=\linewidth]{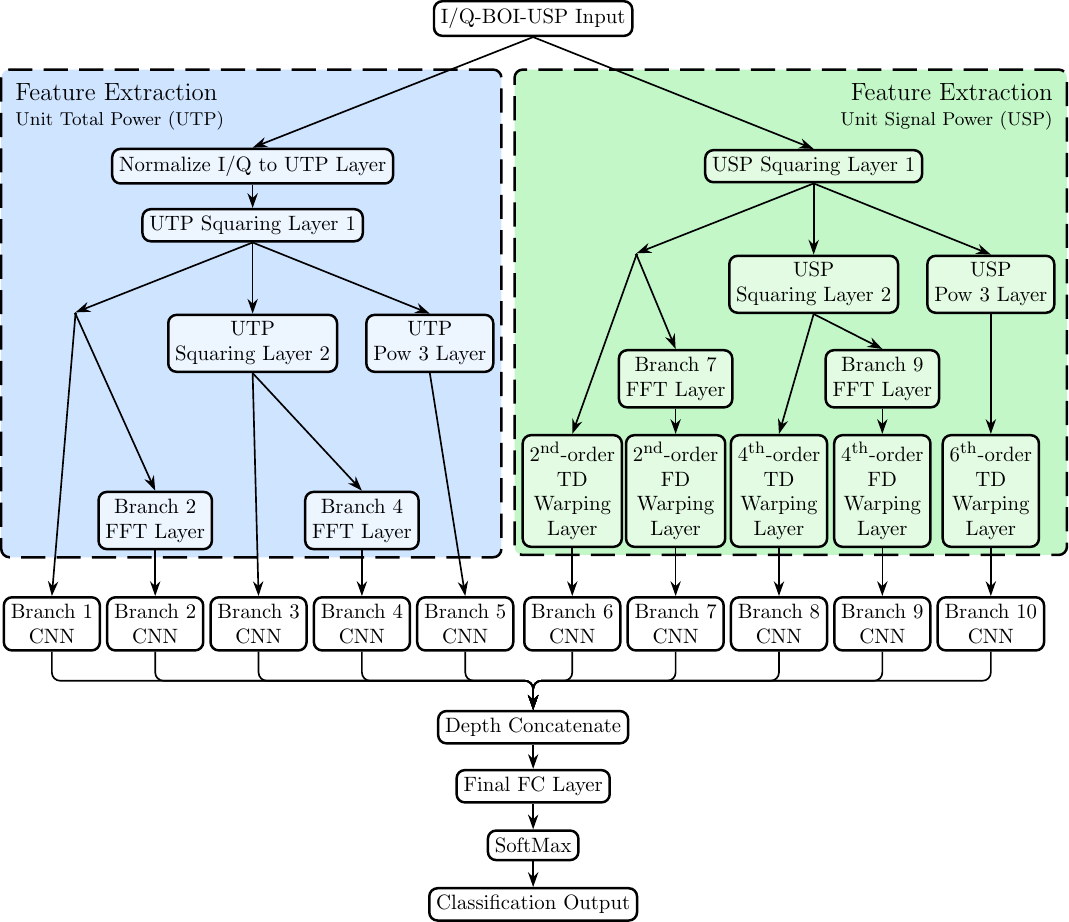}
\caption[I/Q-BOI-USP CAP Topology With Novel Custom NN Layers Performing Feature Extraction for Each Branch.]{\textbf{I/Q-BOI-USP CAP Topology With Novel Custom NN Layers Performing Feature Extraction for Each Branch.}}\label{fig:CAP_Topology3}
\end{figure}

As seen in Figure~\ref{fig:CAP_Topology3}, the input to the second custom ``feature-extraction'' layers is the I/Q signal data consisting of pairs of real numbers $(I,Q)$ that correspond to samples of the in-phase and quadrature components of the digitally modulated signal, which is assumed to have been pre-processed by a blind band-of-interest (BOI) detector~\cite{BOIdetector} to locate the signal bandwidth, filter out-of-band noise, center the signal's spectrum at zero frequency based on the CFO estimate provided by the BOI detector, and normalize the signal to unit signal power (USP).  The normalization to USP is possible because the blind BOI detector also provides an accurate estimate of the noise spectral density.

\subsection{Novel Feature Extraction Layers}
Two types of feature extraction layers are implemented:
\begin{itemize}
\item USP feature extraction layers, shown on the right side of the CAP topology diagram in Figure~\ref{fig:CAP_Topology3}, work directly on the input signal.
\item Unit total power (UTP) feature extraction layers, shown on the left side of the CAP topology diagram in Figure~\ref{fig:CAP_Topology3}, implement a normalization of the input signal to UTP prior to applying any other mathematical function to it.
\end{itemize}
This distinction is necessary because if the I/Q data were first normalized to UTP (that is the total power of the signal+noise is normalized to one), any custom-designed NN layers will be unable to distinguish the different types of QAM signals from each other because the signal power is an important characteristic that differentiates the QAM modulation types, because their probability distributions are highly similar.

For the UTP feature extraction layers, the UTP normalization ensures that large values of the signal power do not prevent the activation functions of the NN from converging and is accomplished for an input signal consisting of $N$~samples as follows:
\begin{align}
signal &= complex\left( I, Q \right), \nonumber \\
RMS &= \sqrt{ \frac{1}{N} \sum_{n=1}^{N}\left( {\left(|signal_{n}|\right)}^{2} \right) }, \\
signalRMS &= signal/RMS, \label{eq:UTP} \\
I_{Output} &= real\left( signalRMS \right), \nonumber \\
Q_{Output} &= imag\left( signalRMS \right). \nonumber
\end{align}

To consistently extract the second, fourth, and sixth-order signal features in both the time-domain and frequency-domain using the I/Q data, we only need to implement three custom NN layers for the needed nonlinear computations: a squaring layer, a raise-to-the-power-of-three (or Pow3) layer, and an FFT layer.

The squaring and power-of-three layers are implemented as follows:

--~Squaring Layer,
\begin{align}
I_{Output} &= \left( I \times I \right) - \left( Q \times Q \right), \\
Q_{Output} &= 2 \times I \times Q.
\end{align}

--~Pow3 Layer,
\begin{align}
I_{Output} &= \left( I \times I \times I \right) - \left( 3 \times I \times Q \times Q \right), \\
Q_{Output} &= \left( 3 \times I \times I \times Q \right) - \left( Q \times Q \times Q \right).
\end{align}

For the FFT layer, only the magnitude is needed at the output, with the zero frequency bin in the center, as this information is most likely to reveal CF patterns present in the signal.  Therefore, after taking the FFT, its absolute value is obtained to ensure the output magnitude is a real number, suitable for use by downstream trainable NN layers.

For the USP feature-extraction layers, the USP normalization performed in the pre-processing stage may result in large magnitude fluctuations of the I/Q signal data due to the noise variance, which will be further amplified by the subsequent squaring, Pow3, and FFT layers, resulting in potential issues with NN training.  Thus, we must implement additional custom layers ``warping'' the data using a warped scaling that is different at higher orders than at lower orders because the higher orders have higher variations in magnitude from the noise variance.  While the ``warping'' method employed on CC estimates by the CSP baseline model~\cite{Snoap_Sensors_2023} was an exponential function, these USP feature-extraction layers do not compute CC estimates.  So, employing the same exponential function as in~\cite{Snoap_Sensors_2023} would undo a squaring or Pow3 layer on a time-domain branch and it would undesirably increase the noise in an FFT magnitude on a frequency-domain branch.  Instead of directly using an exponential function in each USP branch, the exponential function was applied to the maximum magnitude to determine an appropriate ``warped'' scaling factor.  The way the warped scaling factors were obtained also differed between the temporal and the spectral domains, leading to the implementation of the following novel custom NN layers:

--~Sixth-Order Time-Domain Warping Layer (assume I/Q data is already at $6^{\mbox{\footnotesize th}}$-order),
\begin{align}
signal &= complex\left( I, Q \right), \nonumber \\
signalWarped &= \frac{signal}{max\left( | signal | \right)^{\left( 2/3 \right)}}, \label{eq:Warp6TD} \\
I_{Output} &= real\left( signalWarped \right), \nonumber \\
Q_{Output} &= imag\left( signalWarped \right). \nonumber
\end{align}

--~Fourth-Order Time-Domain Warping Layer (assume I/Q data is already at $4^{\mbox{\footnotesize th}}$-order),
\begin{align}
signal &= complex\left( I, Q \right), \nonumber \\
signalWarped &= \frac{signal}{max\left( | signal | \right)^{\left( 1/2 \right)}}, \label{eq:Warp4TD} \\
I_{Output} &= real\left( signalWarped \right), \nonumber \\
Q_{Output} &= imag\left( signalWarped \right). \nonumber
\end{align}

--~Fourth-Order Frequency-Domain Warping Layer (assume input is $4^{\mbox{\footnotesize th}}$-order frequency-domain magnitude),
\begin{align}
output &= \frac{input}{max\left( | input | \right)^{\left( 1/2 \right)}}. \label{eq:Warp4FD}
\end{align}

The second-order time-domain warping layer was implemented identically to the fourth-order time-domain warping layer.  Likewise, the second-order frequency-domain warping layer is identical to the fourth-order frequency-domain warping layer.  The implementation of these warping layers allows the relative signal powers to be retained for each input signal and at the same time prevents the downstream trainable NN layers from diverging to fatuous outcomes.

Each type of the two feature-extraction layers has five specific outputs:  second-order time-domain, second-order frequency-domain, fourth-order time-domain, fourth-order frequency-domain, and sixth-order time-domain, which align with the five domains used in the CSP baseline model that uses signal processing without any DL and is discussed in detail in~\cite{Snoap_Sensors_2023}.  Specifically, the second-order and fourth-order frequency domains are used to identify the basic CF pattern of the signal, which for typical digital modulation schemes is one of the following:  BPSK-like, QPSK-like, $\pi$/4-DQPSK-like, MPSK-like, and staggered QPSK (SQPSK)-like.  QPSK-like can also be referred to as QAM-like.  Illustrations of these CF patterns can be seen in Figure~\ref{fig:CFpatterns}.  After the basic modulation type has been estimated (when using the CSP baseline model), the second-, fourth-, and sixth-order time-domains are used to estimate the cyclic cumulants (CCs), which can then be used for modulation classification.  Thus, the five outputs of the feature extraction layers in Figure~\ref{fig:CAP_Topology3} correspond to a proxy of the CC features extracted in the CSP baseline model, where the UTP feature extraction layers can focus on classifying BPSK, QPSK, 8PSK, MSK, and $\pi$/4-DQPSK; and the USP feature extraction layers can focus on differentiating between 16QAM, 64QAM, and 256QAM.

\subsection{Classification Layers}
The CAP includes ten parallel CNN branches that implement the primary capsules, which will be trained to classify the eight common digital modulation schemes of interest: BPSK, QPSK, 8PSK, $\pi/4$-DQPSK, MSK, 16QAM, 64QAM, and 256QAM.  The primary capsules contain the majority of the network's learnable hyper-parameters, with each branch having as its input one of the ten desired features.  At the output of each branch, there is a fully connected layer with eight outputs, which is used to reduce the branch output size to the number of modulation classes in the training dataset; this is done to ensure that each branch potentially has the ability to distinguish between all modulation types in the training dataset.  This ability is important because, while not every branch will be able to fully distinguish between all modulation types, each branch will be able to identify which classes it can distinguish between and this determination can be made during NN training.  For example, due to the nature of the CF patterns, it is expected that the branches with second-order frequency-domain inputs will be able to distinguish between BPSK and MSK, but all other modulation types will not be distinguishable on this branch. Likewise, the branches with fourth-order frequency-domain inputs should be able to distinguish between 8PSK-like, $\pi/4$-DQPSK-like, and SQPSK-like patterns, but QAM-like and BPSK-like CF patterns may not be distinguishable to these branches.  Each branch, having opportunity to distinguish between all eight modulation types, then has its outputs concatenated together with the other branches followed by a final fully connected layer.  This last fully connected layer learns the appropriate weights to apply to the outputs of each CNN branch so that each branch's ability is combined together to obtain the minimum error on the training dataset.

The overall structure of the proposed CAP with custom feature-extracting NN layers is shown in Figure~\ref{fig:CAP_Topology3}, with the characteristics of the subsequent CNN branches for classifying digitally modulated signals outlined in Tables~\ref{table:CNN_Branch}~--~\ref{table:ConvAvgPoolLayer}.  The number of filters $A$ and filter size $\left[ B \times C \right]$ for each convolutional layer are defined in Table~\ref{table:CNN_Branch} as are the number of output activations (e.g., $X \times Y$) for each layer.

\section{Datasets Used for NN Training and Testing the Second CAP}\label{sec:Datasets4}
To train and test the performance of the proposed CAP (including its generalization ability) two publicly available datasets \cite{IEEE_DataPort} were used that both contain signals corresponding to the eight digital modulation schemes of interest (BPSK, QPSK, 8PSK, $\pi/4$-DQPSK, MSK, 16QAM, 64QAM, and 256QAM).  The two datasets are available as \texttt{CSPB.ML.2018} and \texttt{CSPB.ML.2022} \cite{IEEE_DataPort}, and their signal generation parameters are listed in Table~\ref{table:SigGenParms} in Chapter~\ref{ch:HQDatasets} of this dissertation.

The CFO ranges corresponding to signals in the two datasets are non-intersecting, which allows evaluation of the generalization ability of the trained CAP.  Specifically, if the CAP trained on a large portion of the \texttt{CSPB.ML.2018} dataset displays high classification accuracy on the remaining signals in the dataset, and its performance when classifying signals in the \texttt{CSPB.ML.2022} dataset is at similarly high levels, then the CAP is robust and has a high ability to generalize\footnote{Likewise, if its classification accuracy on signals in \texttt{CSPB.ML.2018} is high, while on signals in \texttt{CSPB.ML.2022} is low, then its generalization ability is low.}.

Prior to training the proposed CAP on these datasets, the signals are pre-processed as mentioned in Section~\ref{sec:Details2} using a blind BOI detector~\cite{BOIdetector} to locate the bandwidth for signals in the datasets, filter out-of-band noise, and center the I/Q data at zero frequency based on the CFO estimate provided by the BOI detector. Finally, the I/Q data is normalized to USP based on the signal power estimate provided by the BOI detector.  Our BOI detector estimated the CFO with accuracy such that the base-ten logarithm of the mean absolute error (MAE) was about $-3.01$ on \texttt{CSPB.ML.2018} and about $-3.10$ on \texttt{CSPB.ML.2022}, both of which correspond to an MAE of $\lessapprox 0.001$.  Similarly, the BOI detector estimated the SNR with an MAE of $0.49$~dB on \texttt{CSPB.ML.2018} and $0.35$~dB on \texttt{CSPB.ML.2022}.  Using a blind BOI detector to merely shift the signal to an estimated zero frequency does not necessarily enable an I/Q-trained CAP such as the one in~\cite{Latshaw_COMM2022} to generalize between the \texttt{CSPB.ML.2018} and \texttt{CSPB.ML.2022} datasets, as will be shown in the following section.  This I/Q-BOI-USP data was also used to retrain the image-processing CAP in~\cite{Latshaw_COMM2022} as an alternative to provide a point of comparison with the proposed CAP.

\begin{figure}
\begin{center}
\subfigure[Objective Function.]
{\includegraphics[width=\linewidth]{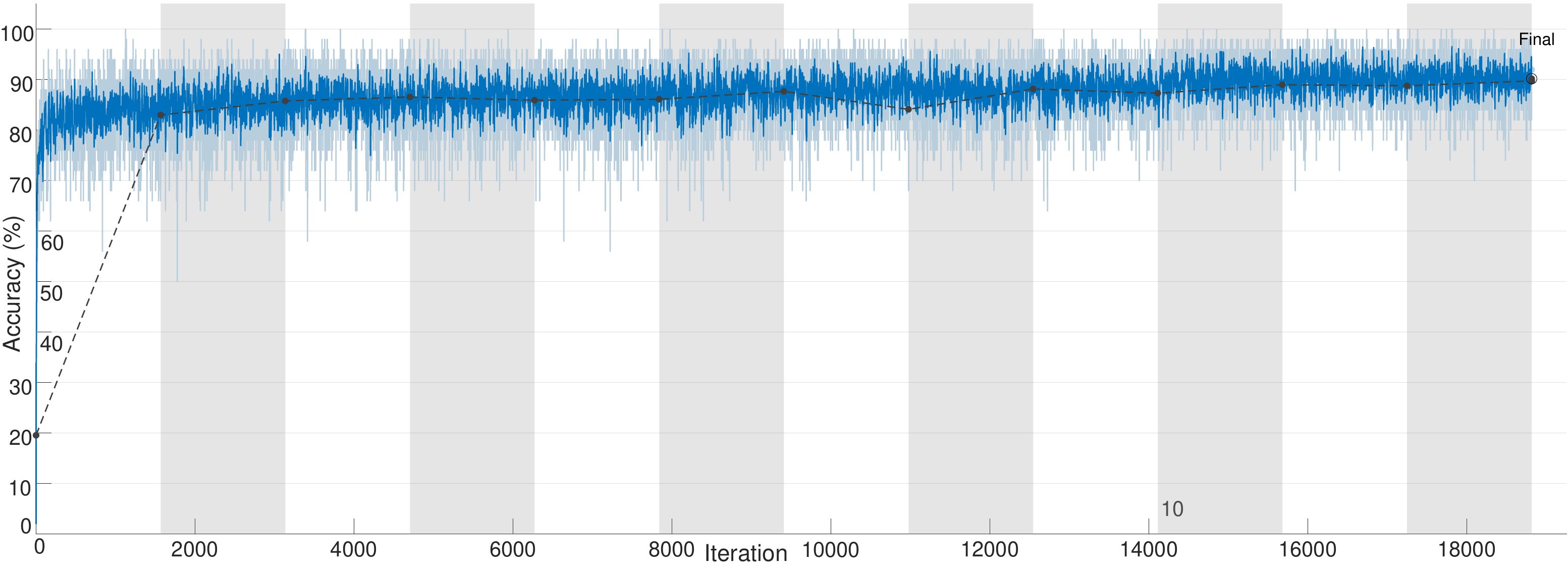}}
\subfigure[Loss Function.]
{\includegraphics[width=\linewidth]{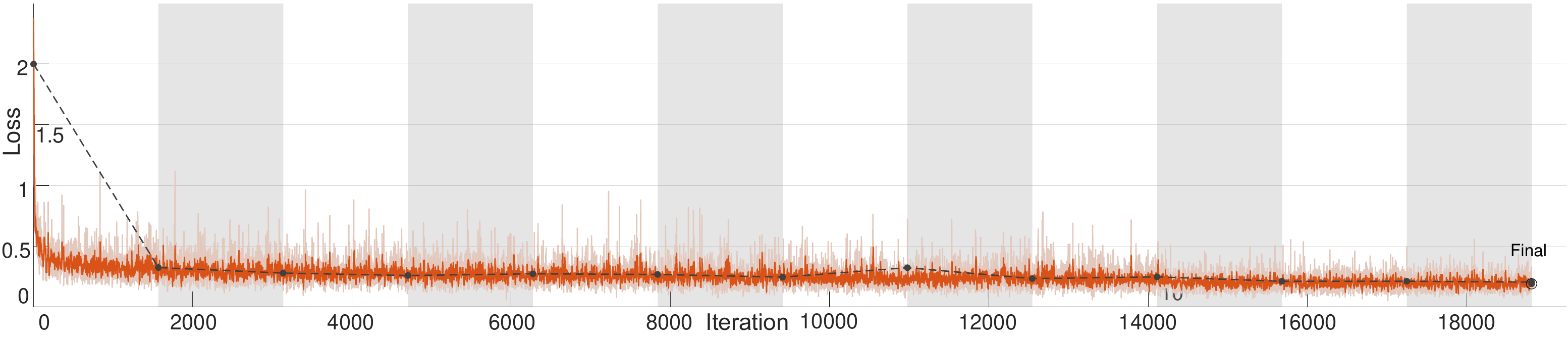}}
\caption[Objective and Loss Functions for the I/Q-BOI-USP-trained CAP Using the \texttt{CSPB.ML.2018} Dataset.]{\textbf{Objective and Loss Functions for the I/Q-BOI-USP-trained CAP Using the \texttt{CSPB.ML.2018} Dataset.}}\label{fig:obj_loss_2018_2}
\end{center}
\end{figure}

\begin{figure}
\begin{center}
\subfigure[Objective Function.]
{\includegraphics[width=\linewidth]{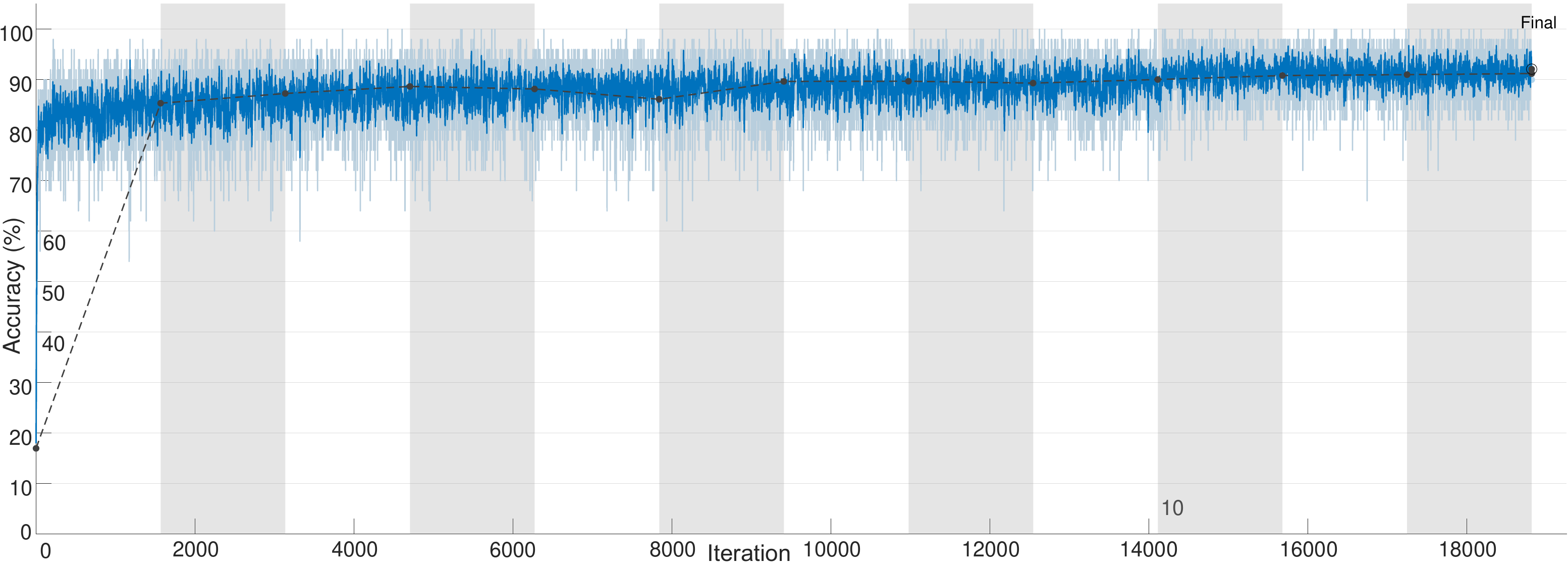}}
\subfigure[Loss Function.]
{\includegraphics[width=\linewidth]{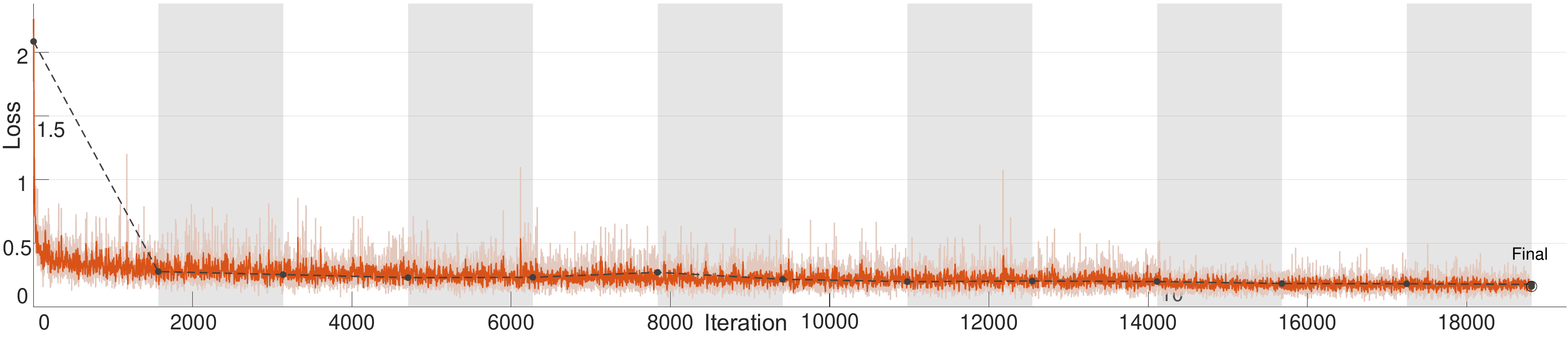}}
\caption[Objective and Loss Functions for the I/Q-BOI-USP-trained CAP Using the \texttt{CSPB.ML.2022} Dataset.]{\textbf{Objective and Loss Functions for the I/Q-BOI-USP-trained CAP Using the \texttt{CSPB.ML.2022} Dataset.}}\label{fig:obj_loss_2022_2}
\end{center}
\end{figure}

\section{Second CAP Training and Testing Results}\label{sec:NumericalResults2}
The proposed CAP with custom feature extraction layers along with the image-processing CAP in~\cite{Latshaw_COMM2022} have been implemented in MATLAB and trained on a high-performance computing cluster with 18~NVidia V100 graphical processing unit (GPU) nodes available, with each node having 128~GB of memory.  The training process is computationally intensive, but if the available computing resources are leveraged appropriately such that the entire training dataset is loaded into the available memory, training for the proposed CAP with custom feature extraction layers takes about $19$~hours, while for the image-processing CAP in~\cite{Latshaw_COMM2022}, it takes several hours.

For training the proposed CAP, a stochastic gradient descent with momentum (SGDM) optimizer~\cite{SGDM} was employed, using twelve epochs while shuffling the training data before each training epoch, a mini-batch size of $50$, an initial learning rate of $0.02$, a piecewise learning schedule involving a multiplicative learning rate drop factor of $0.1$ every nine epochs, an $L_2$ regularization factor of $0.0001$, a momentum of $0.9$, and a final batch normalization using the entire training data population statistics.  The objective and loss functions corresponding to the training of the proposed CAP are shown in Figures~\ref{fig:obj_loss_2018_2}--\ref{fig:obj_loss_2022_2}.

\begin{figure}
\begin{center}
\subfigure[Initial Test Results for CAPs Trained on \texttt{CSPB.ML.2018}.]
{\includegraphics[width=0.49\linewidth]{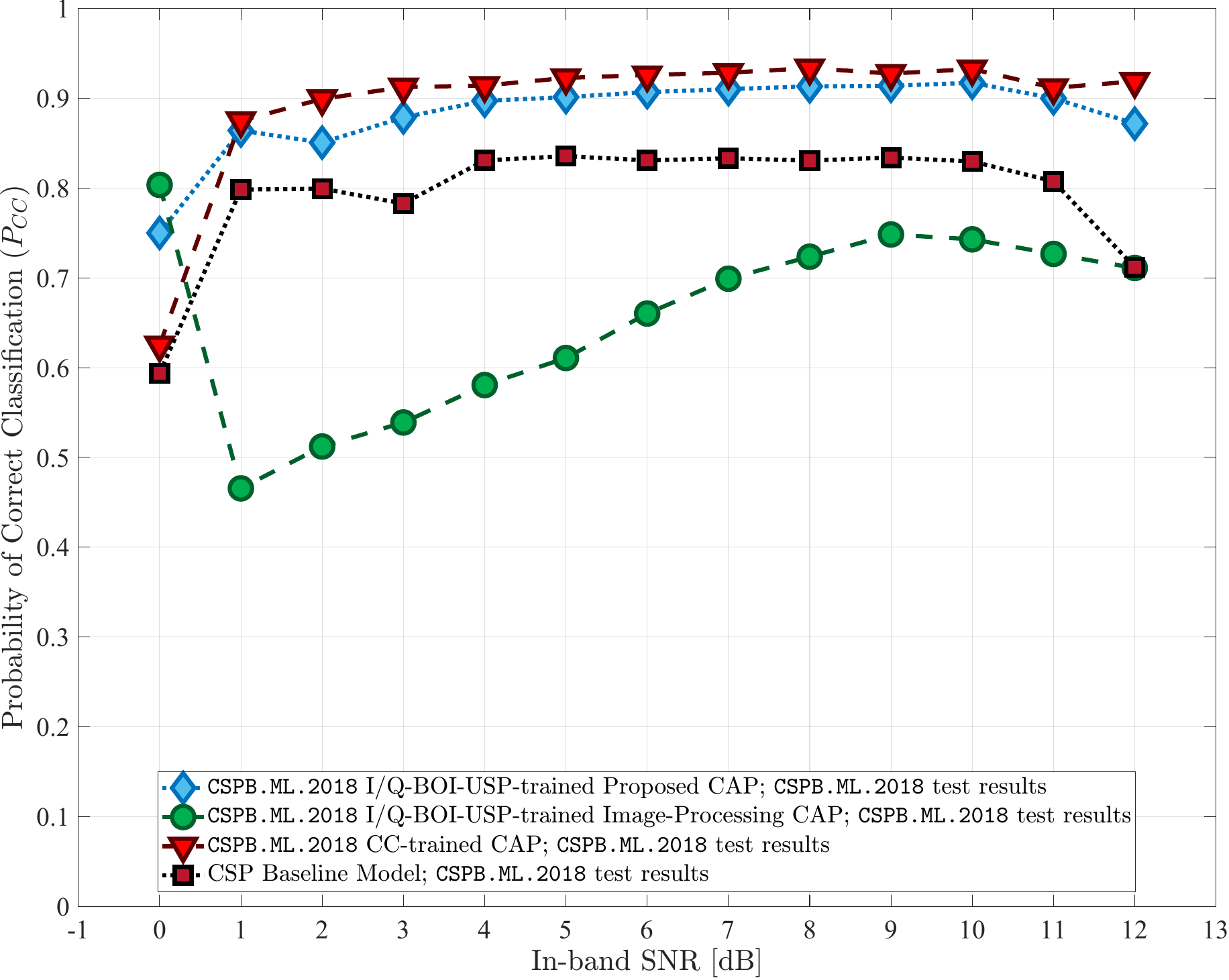}}
\subfigure[Generalization Test Results for CAPs Trained on \texttt{CSPB.ML.2018}.]
{\includegraphics[width=0.49\linewidth]{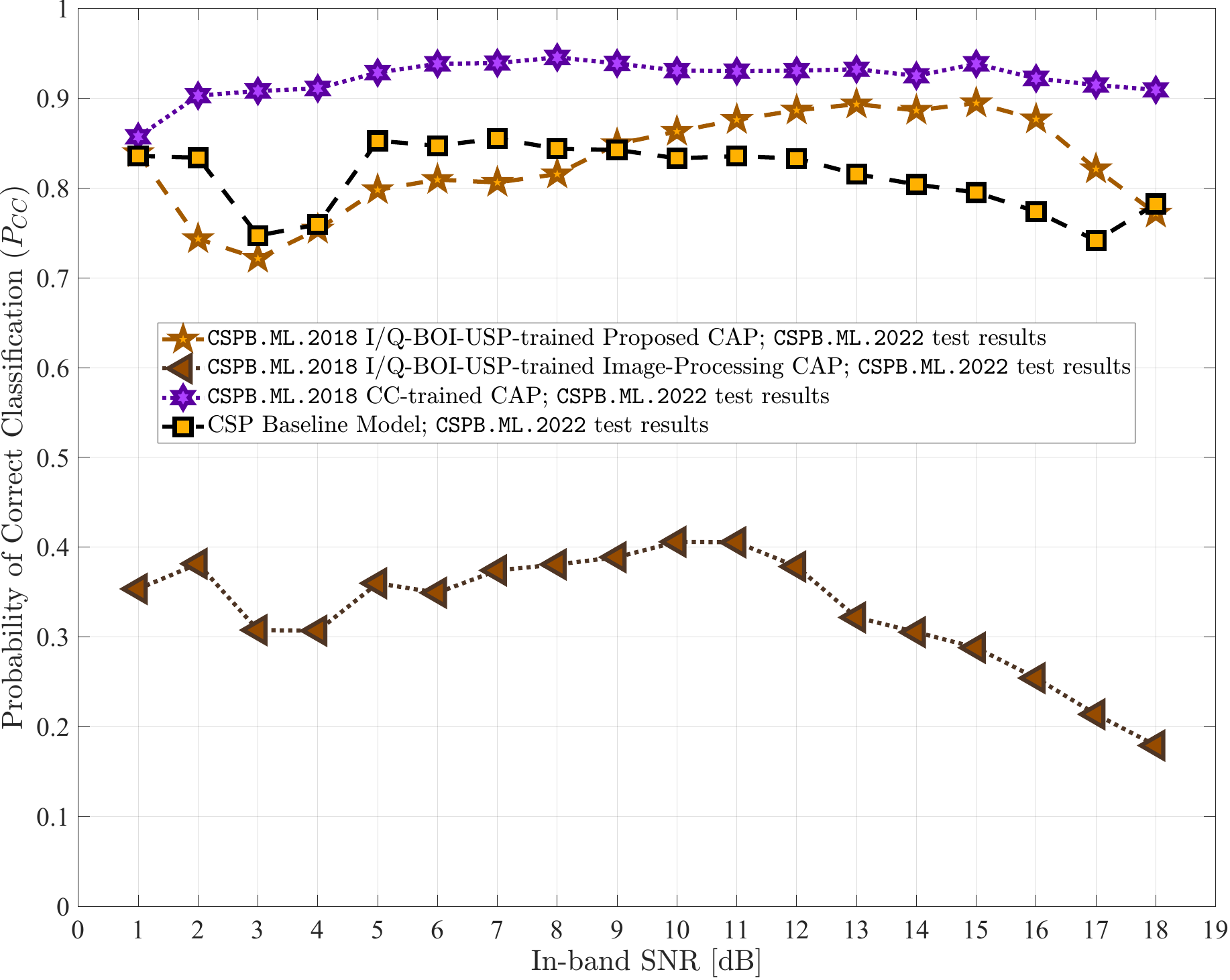}}
\caption[Initial and Generalization Test Results for CAPs Trained on \texttt{CSPB.ML.2018}.]{\textbf{Initial and Generalization Test Results for CAPs Trained on \texttt{CSPB.ML.2018}.}}\label{fig:SNR_CTest_CTrained_All_2}
\end{center}
\end{figure}

\begin{figure}
\begin{center}
\subfigure[Initial Test Results for CAPs Trained on \texttt{CSPB.ML.2022}.]
{\includegraphics[width=0.49\linewidth]{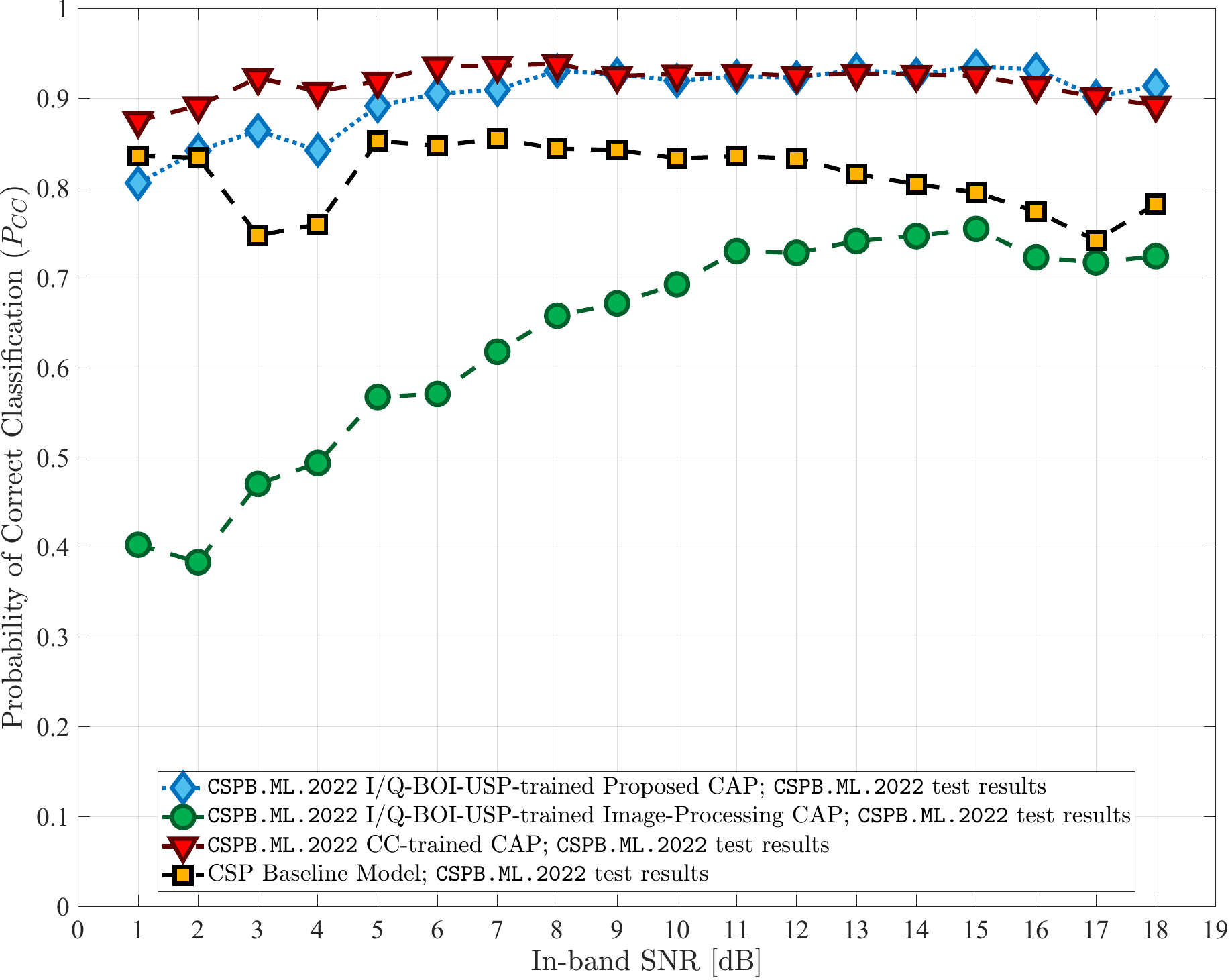}}
\subfigure[Generalization Test Results for CAPs Trained on \texttt{CSPB.ML.2022}.]
{\includegraphics[width=0.49\linewidth]{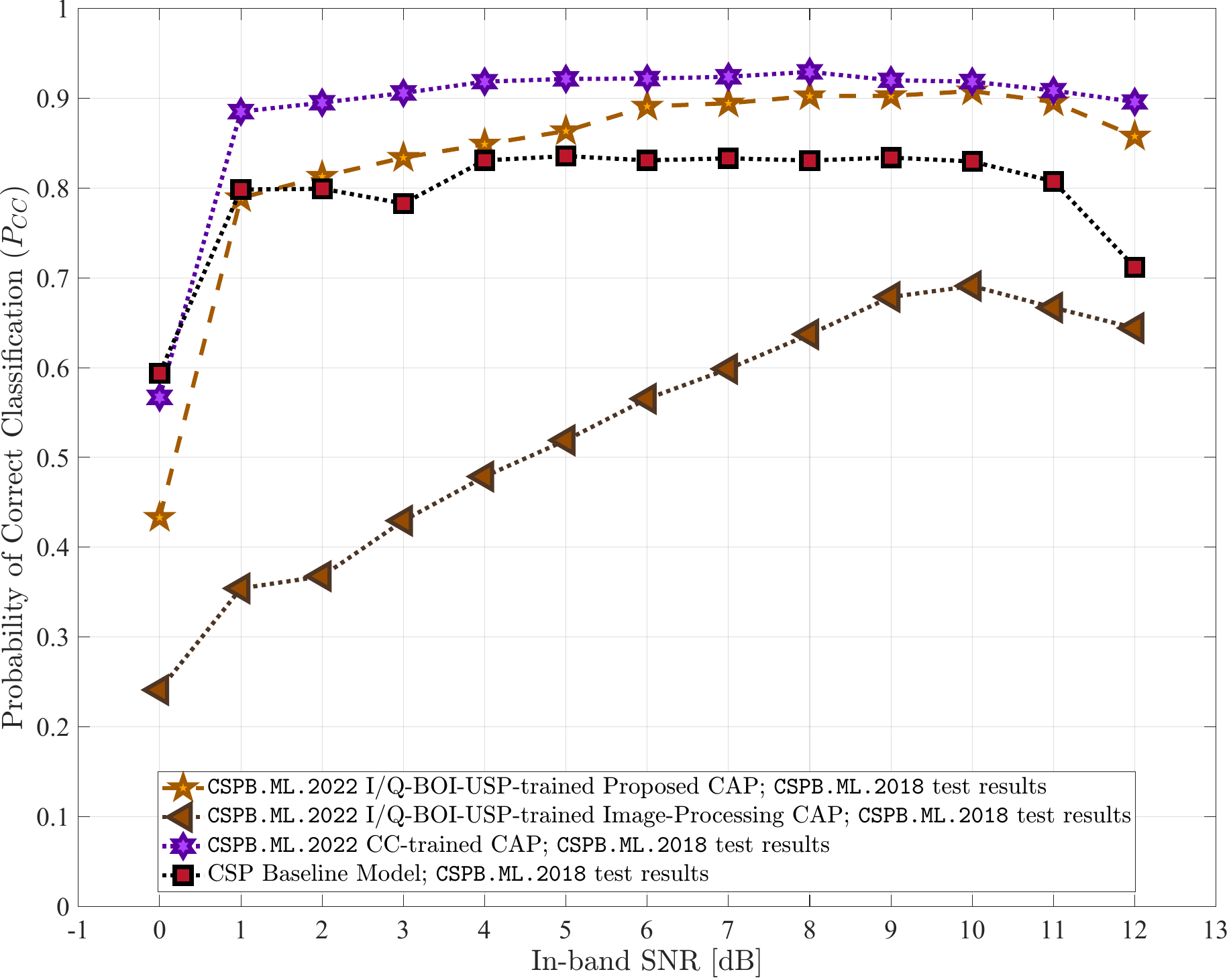}}
\caption[Initial and Generalization Test Results for CAPs Trained on \texttt{CSPB.ML.2022}.]{\textbf{Initial and Generalization Test Results for CAPs Trained on \texttt{CSPB.ML.2022}.}}\label{fig:SNR_GCTest_GCTrained_All_2}
\end{center}
\end{figure}

Each CAP was trained and tested two separate times:
\begin{itemize}
\item In the first training instance, dataset \texttt{CSPB.ML.2018} was used, splitting the available signals into $70\%$ for training, $5\%$ for validation, and $25\%$ for testing.  The probability of correct classification for the test results obtained using the $25\%$ test portion of signals in \texttt{CSPB.ML.2018} is shown in Figure~\ref{fig:SNR_CTest_CTrained_All_2}(a).

\hspace{0.25cm} The CAPs trained on \texttt{CSPB.ML.2018} are then tested on dataset \texttt{CSPB.ML.2022} to assess the generalization abilities of a trained CAP when classifying all signals available in \texttt{CSPB.ML.2022}.  The probability of correct classification for this test is shown in Figure~\ref{fig:SNR_CTest_CTrained_All_2}(b).
\item In the second training instance, the NN was reset and trained anew using signals in dataset \texttt{CSPB.ML.2022}, with a similar split of $70\%$ signals used for training, $5\%$ for validation, and $25\%$ for testing.  The probability of correct classification for the test results obtained using the $25\%$ test portion of signals in \texttt{CSPB.ML.2022} is shown in Figure~\ref{fig:SNR_GCTest_GCTrained_All_2}(a).

\hspace{0.25cm} The CAPs trained on \texttt{CSPB.ML.2022} are then tested on dataset \texttt{CSPB.ML.2018} to assess the generalization abilities of the trained NN when classifying all signals available in \texttt{CSPB.ML.2018}.  The probability of correct classification for this test is shown in Figure~\ref{fig:SNR_GCTest_GCTrained_All_2}(b).
\end{itemize}

The probability of correct classification corresponding to the CSP baseline model (which does not use any NN or training and makes decisions based on signal processing) and to the CC-trained CAP (which used the same training/validation/testing data) reported in~\cite{Snoap_Sensors_2023} has also been included on the plots shown in Figure~\ref{fig:SNR_CTest_CTrained_All_2}, from which the proposed CAP with custom feature extraction layers outperforms the image-processing CAP from~\cite{Latshaw_COMM2022} in all training and testing scenarios, displaying high classification performance and generalization abilities.  The proposed CAP outperforms the CSP baseline model and its performance is close to that of the CC-trained CAP, which is of significance because the proposed CAP does not require accurate CFO estimates, symbol rate estimates, or CC estimates prior to classification by the CAP, whereas the CC-trained CAP does.

The overall probability of correct classification ($P_{CC}$) for all experiments performed is summarized in Table~\ref{table:OverallPerformance3} and provides further confirmation that the proposed CAP containing the novel custom layers is able to perform feature extraction tailored to the modulated signals of interest more effectively than the image-processing CAP in~\cite{Latshaw_COMM2022}, which employs only the conventional NN layers used in the context of image processing.  The lowest classification performance seen by the proposed CAP came from training on \texttt{CSPB.ML.2018} and testing on \texttt{CSPB.ML.2022} with an overall correct classification probability $P_{CC} = 86.1\%$, while the highest classification performance observed for the image-processing CAP from~\cite{Latshaw_COMM2022} came from training on \texttt{CSPB.ML.2022} and testing on \texttt{CSPB.ML.2022}, resulting in a corresponding overall $P_{CC} = 69.9\%$.  Thus, even the lowest performance of the proposed CAP exceeds by $15\%$ the best performance of the image-processing CAP from~\cite{Latshaw_COMM2022}.

\begin{table}
\centering
\caption[Overall I/Q-BOI-USP-trained CAP Classification Performance ($P_{CC}$).]{\textbf{Overall I/Q-BOI-USP-trained CAP Classification Performance ($P_{CC}$).}}
{
\setlength{\tabcolsep}{6mm}
\begin{tabular}{ c c c }
\toprule
\textbf{Classification Model} & \begin{tabular}{@{}c@{}} \textbf{Results for Dataset} \\ \texttt{CSPB.ML.2018} \end{tabular} & \begin{tabular}{@{}c@{}} \textbf{Results for Dataset} \\ \texttt{CSPB.ML.2022} \end{tabular} \\
\midrule
CSP Baseline Model & $82.0\%$ & $82.0\%$ \\
\hline
\begin{tabular}{@{}c@{}} Image-Processing CAP \\ \texttt{CSPB.ML.2018} \\ I/Q-BOI-USP-trained \end{tabular} & $68.7\%$ & $34.5\%$ \\
\hline
\begin{tabular}{@{}c@{}} Image-Processing CAP \\ \texttt{CSPB.ML.2022} \\ I/Q-BOI-USP-trained \end{tabular} & $60.2\%$ & $69.9\%$ \\
\hline
\begin{tabular}{@{}c@{}} Proposed CAP \\ \texttt{CSPB.ML.2018} \\ I/Q-BOI-USP-trained \end{tabular} & $90.4\%$ & $86.1\%$ \\
\hline
\begin{tabular}{@{}c@{}} Proposed CAP \\ \texttt{CSPB.ML.2022} \\ I/Q-BOI-USP-trained\end{tabular} & $88.3\%$ & $92.1\%$ \\
\hline
\begin{tabular}{@{}c@{}} \texttt{CSPB.ML.2018} \\ CC-trained CAP \end{tabular} & $92.3\%$ & $93.1\%$ \\
\hline
\begin{tabular}{@{}c@{}} \texttt{CSPB.ML.2022} \\ CC-trained CAP \end{tabular} & $91.6\%$ & $92.5\%$ \\
\bottomrule
\end{tabular}
}\label{table:OverallPerformance3}
\end{table}

Additionally, the CSP baseline model obtained a classification performance of $P_{CC} = 82.0\%$ on both \texttt{CSPB.ML.2018} and \texttt{CSPB.ML.2022}.  So, even the lowest performance of the proposed CAP exceeds the performance of the CSP baseline model by $4\%$.  The CC-trained CAP obtained its best performance from training on \texttt{CSPB.ML.2018} and testing on \texttt{CSPB.ML.2022} with an overall correct classification probability $P_{CC} = 93.1\%$.  So, the I/Q-BOI-USP-trained proposed CAP is at most $7\%$ below in classification probability to the CC-trained CAP, however, it does not require CC estimates, high-quality CFO estimates, or symbol rate estimates; which are required for the CC-trained CAP.

\begin{figure}
\begin{center}
\subfigure[Confusion Matrix of the \texttt{CSPB.ML.2018} I/Q-BOI-USP-trained Proposed CAP Classifying \texttt{CSPB.ML.2018} Test Signals.]
{\includegraphics[width=0.49\linewidth]{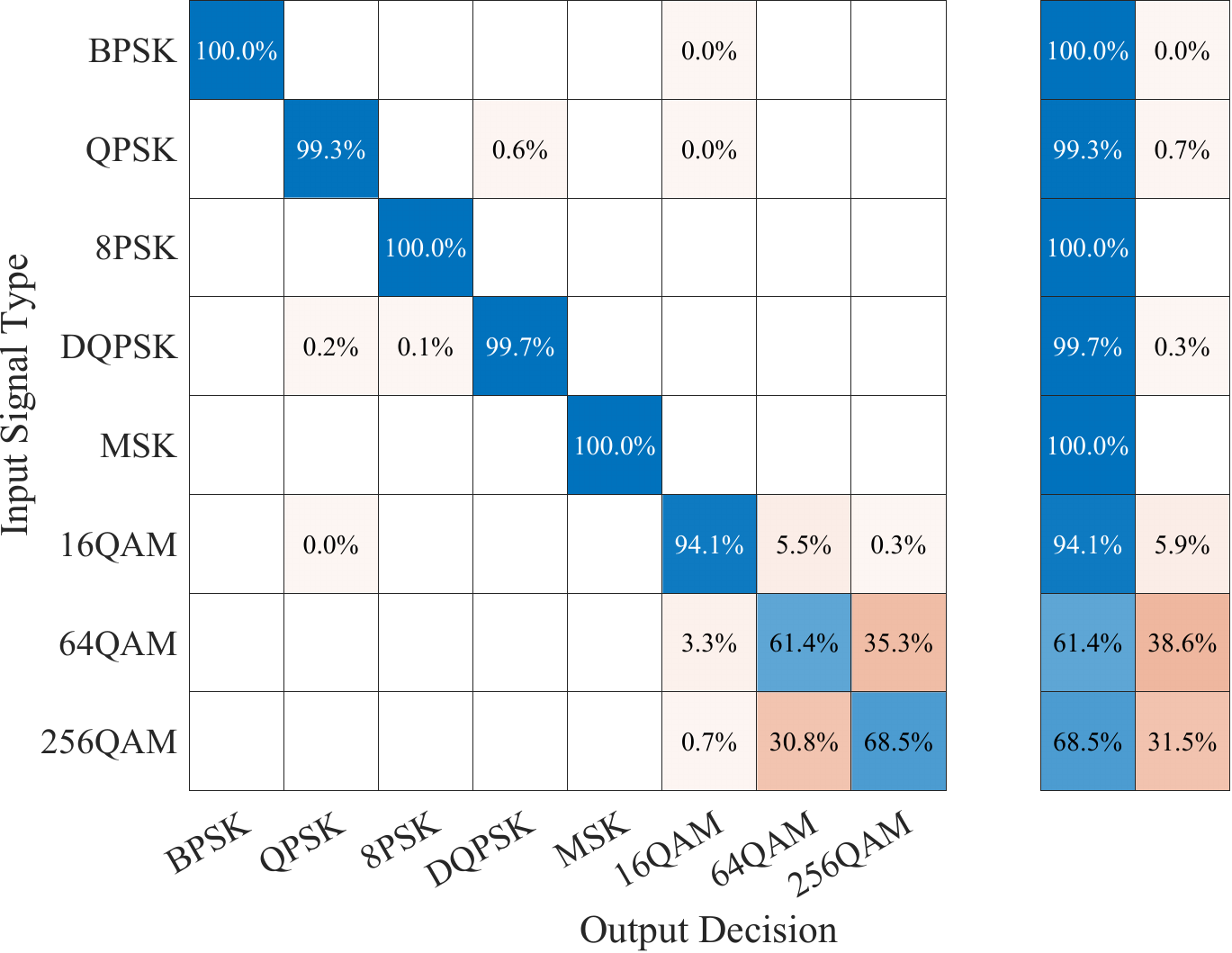}}
\subfigure[Confusion Matrix of the \texttt{CSPB.ML.2018} I/Q-BOI-USP-trained Proposed CAP Classifying All \texttt{CSPB.ML.2022} Signals.]
{\includegraphics[width=0.49\linewidth]{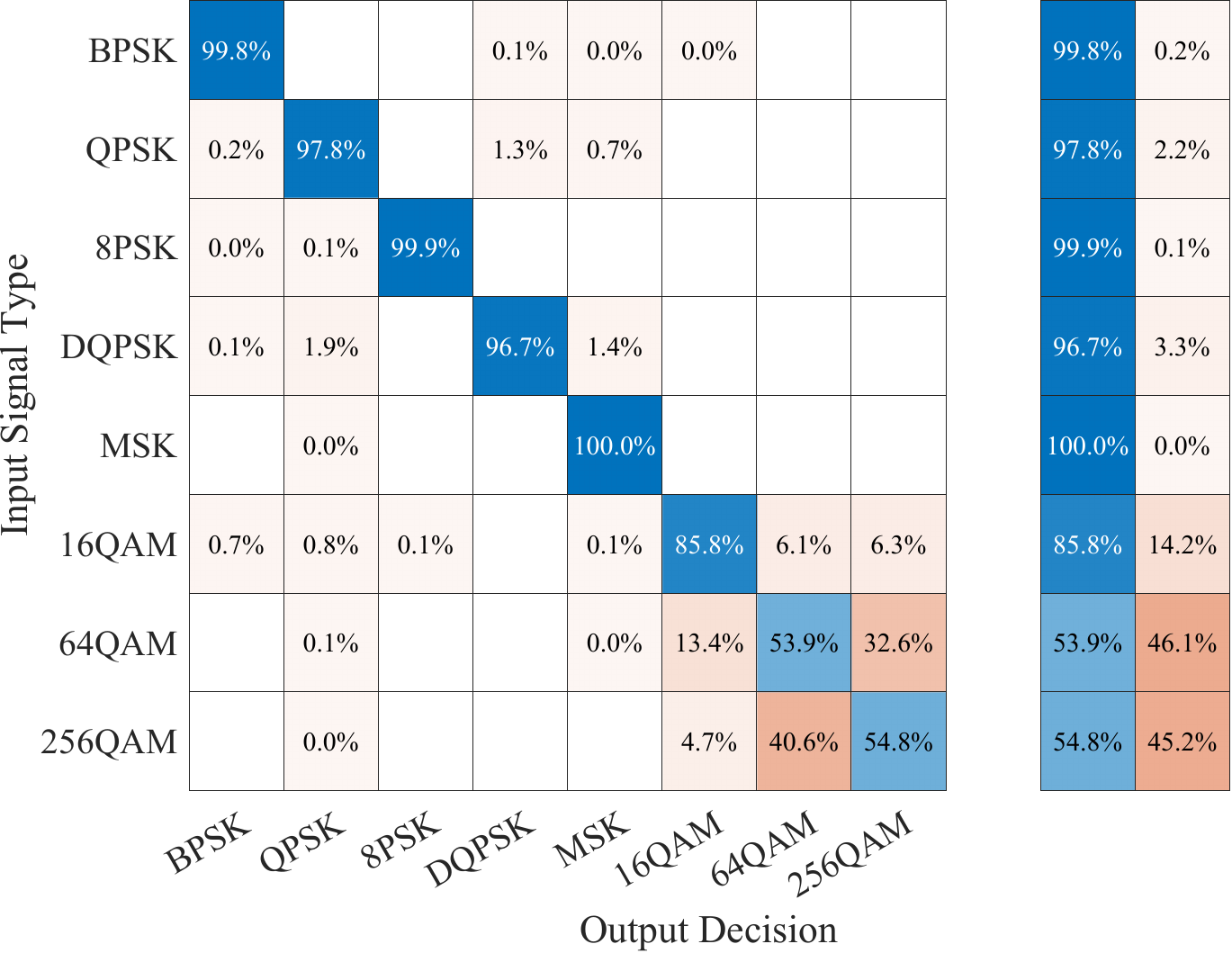}}
\caption[Confusion Matrices of the \texttt{CSPB.ML.2018} I/Q-BOI-USP-trained Proposed CAP Classifying \texttt{CSPB.ML.2018} Test Signals and All \texttt{CSPB.ML.2022} Signals.]{\textbf{Confusion Matrices of the \texttt{CSPB.ML.2018} I/Q-BOI-USP-trained Proposed CAP Classifying \texttt{CSPB.ML.2018} Test Signals and All \texttt{CSPB.ML.2022} Signals.}}\label{fig:CM_-_CTest_CTrained_-_IQ-Proposed-CAP}
\end{center}
\end{figure}

\begin{figure}
\begin{center}
\subfigure[Confusion Matrix of the \texttt{CSPB.ML.2022} I/Q-BOI-USP-trained Proposed CAP Classifying \texttt{CSPB.ML.2022} Test Signals.]
{\includegraphics[width=0.49\linewidth]{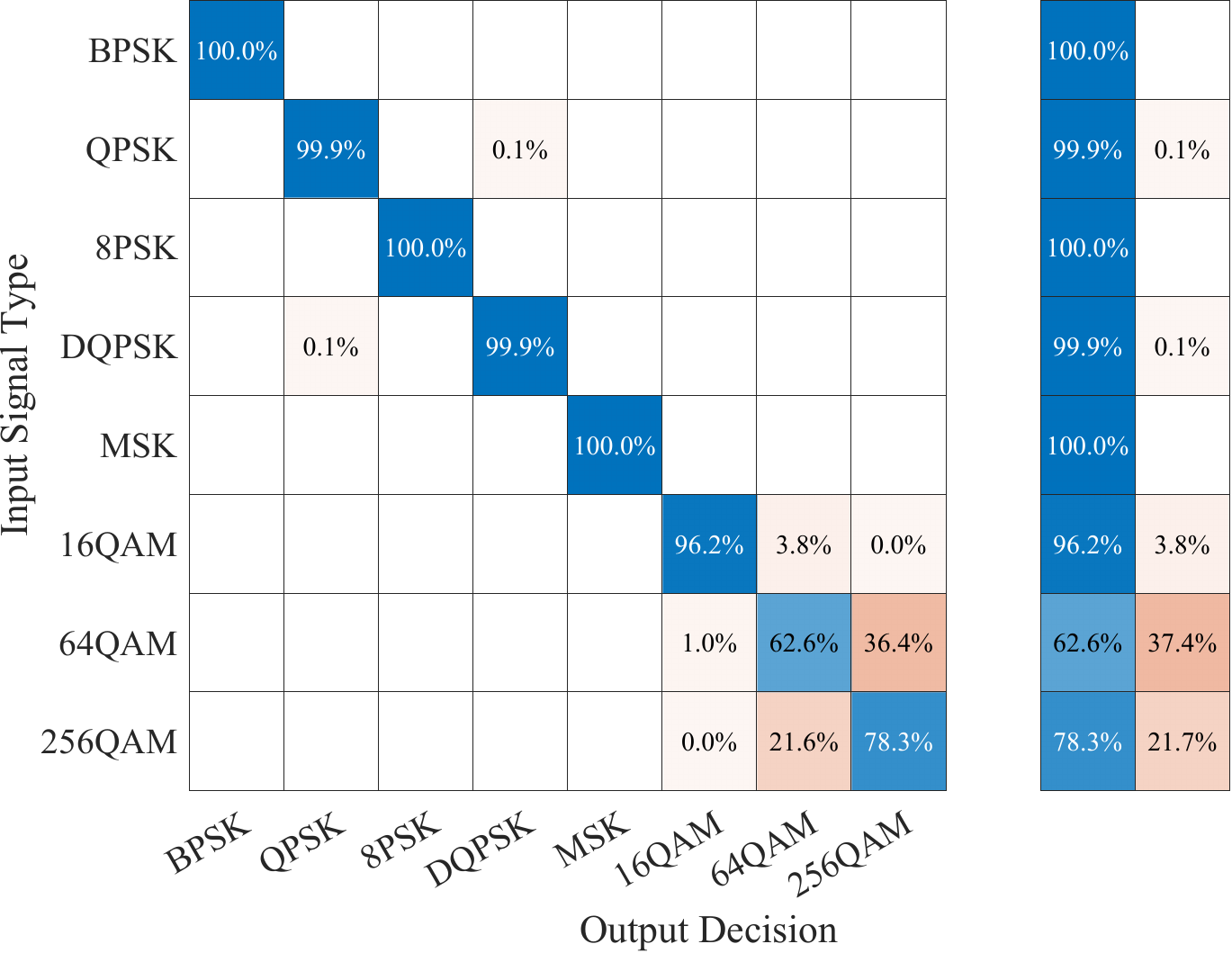}}
\subfigure[Confusion Matrix of the \texttt{CSPB.ML.2022} I/Q-BOI-USP-trained Proposed CAP Classifying All \texttt{CSPB.ML.2018} Signals.]
{\includegraphics[width=0.49\linewidth]{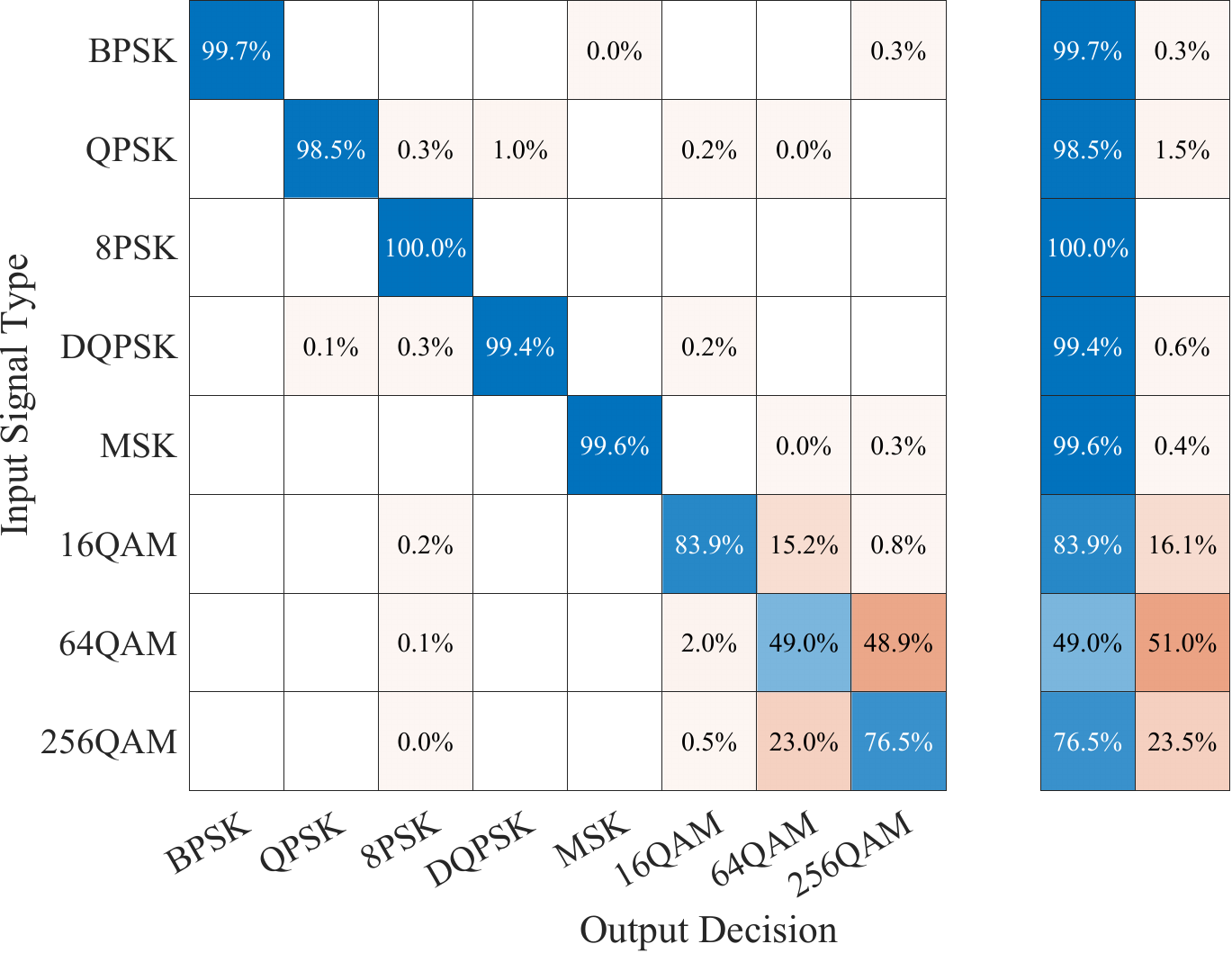}}
\caption[Confusion Matrices of the \texttt{CSPB.ML.2022} I/Q-BOI-USP-trained Proposed CAP Classifying \texttt{CSPB.ML.2022} Test Signals and All \texttt{CSPB.ML.2018} Signals.]{\textbf{Confusion Matrices of the \texttt{CSPB.ML.2022} I/Q-BOI-USP-trained Proposed CAP Classifying \texttt{CSPB.ML.2022} Test Signals and All \texttt{CSPB.ML.2018} Signals.}}\label{fig:CM_-_CTest_GCTrained_-_IQ-Proposed-CAP}
\end{center}
\end{figure}

For more insight into the generalization abilities of the proposed CAP with custom feature-extraction layers, confusion-matrix plots are shown in Figures~\ref{fig:CM_-_CTest_CTrained_-_IQ-Proposed-CAP}--\ref{fig:CM_-_CTest_GCTrained_-_IQ-Proposed-CAP} from where the following can be seen:
\begin{itemize}
\item For individual BPSK, QPSK, 8PSK, $\pi/4$-DQPSK, and MSK modulation schemes, when the CAP is trained on \texttt{CSPB.ML.2018} and tested on \texttt{CSPB.ML.2022} the classification performance is within $3.0\%$ difference of the performance obtained when the proposed CAP is trained and tested on \texttt{CSPB.ML.2018}.  When the CAP is trained on \texttt{CSPB.ML.2022} and tested on \texttt{CSPB.ML.2018}, the difference in performance for the same individual modulation schemes is within a $1.4\%$ difference, indicating excellent generalization performance for these five types of digital modulation schemes. 
\item A larger performance gap is observed for the QAM schemes, for which the difference in performance exceeds $13\%$ for specific cases ($13.7\%$ for 256QAM when the CAP is trained on \texttt{CSPB.ML.2018} and tested on \texttt{CSPB.ML.2022} versus the CAP trained and tested on \texttt{CSPB.ML.2018}; respectively, $13.6\%$ for 64QAM when the CAP is trained on \texttt{CSPB.ML.2022} and tested on \texttt{CSPB.ML.2018} versus the CAP trained and tested on \texttt{CSPB.ML.2018}).
\end{itemize}

\begin{figure}
\begin{center}
\subfigure[Confusion Matrix of the CSP Baseline Model on All \texttt{CSPB.ML.2018} Signals.]
{\includegraphics[width=0.49\linewidth]{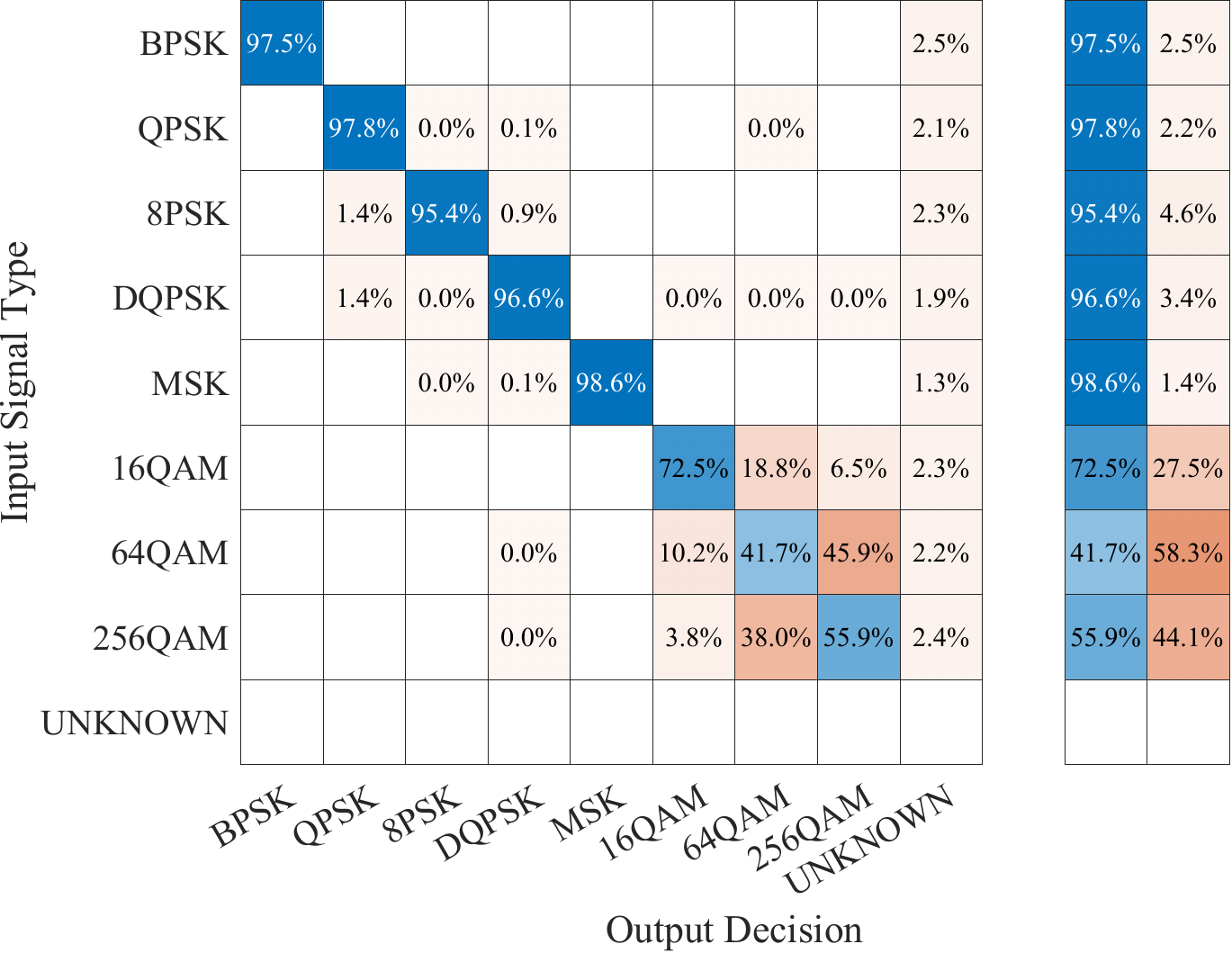}}
\subfigure[Confusion Matrix of the CSP Baseline Model on All \texttt{CSPB.ML.2022} Signals.]
{\includegraphics[width=0.49\linewidth]{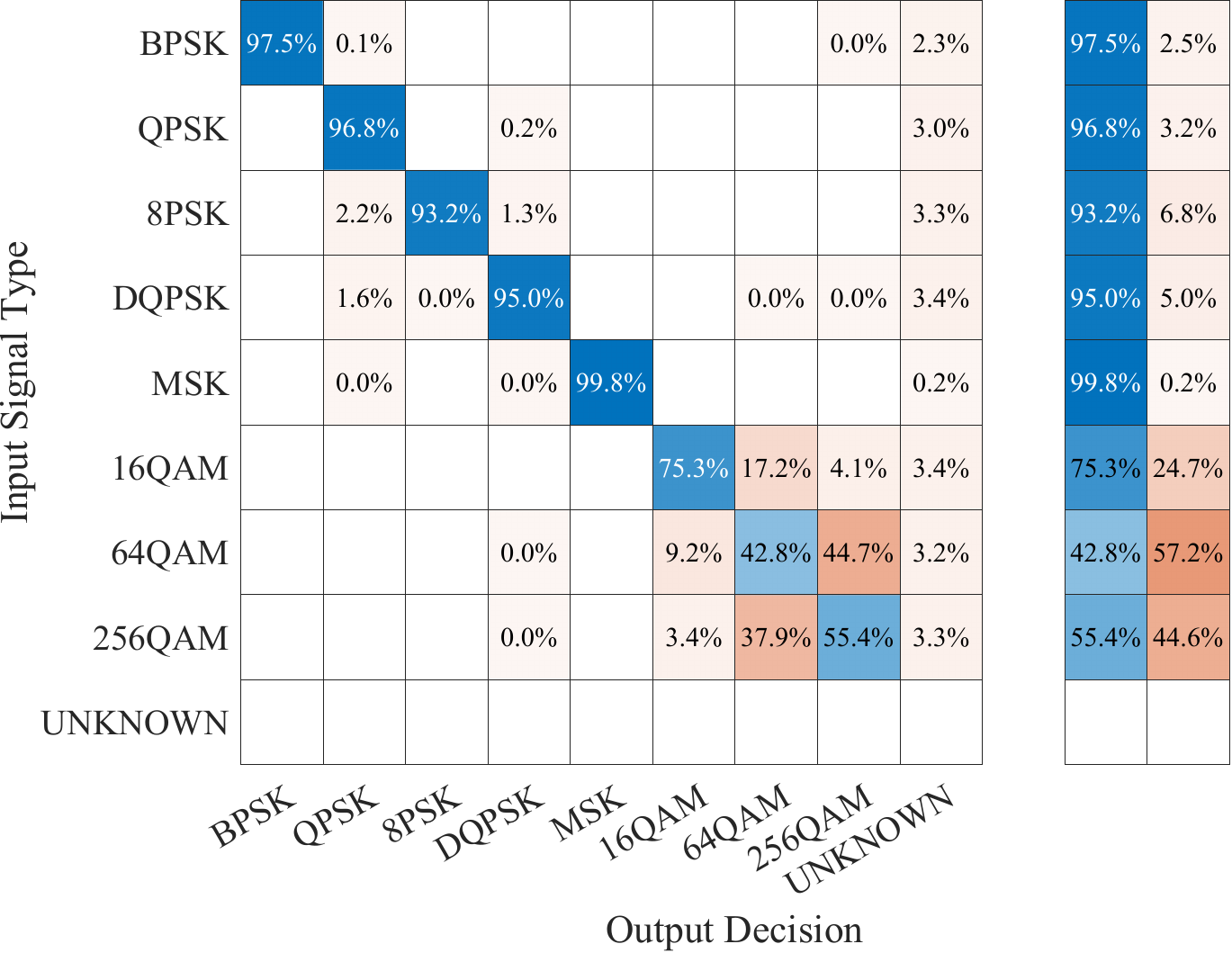}}
\caption[Confusion Matrices of the CSP Baseline Model on All \texttt{CSPB.ML.2018} and \texttt{CSPB.ML.2022} Signals.]{\textbf{Confusion Matrices of the CSP Baseline Model on All \texttt{CSPB.ML.2018} and \texttt{CSPB.ML.2022} Signals.}}\label{fig:CM_-_CSPB_MSSA_-_GC}
\end{center}
\end{figure}

The confusion matrices shown in Figures~\ref{fig:CM_-_CTest_CTrained_-_IQ-Proposed-CAP}--\ref{fig:CM_-_CTest_GCTrained_-_IQ-Proposed-CAP} compare favorably with the confusion matrix corresponding to the CSP baseline classification model that uses conventional CSP techniques for classifying the digitally modulated signals shown in Figure~\ref{fig:CM_-_CSPB_MSSA_-_GC}.  Similar results have been reported for the CSP baseline classification model on \texttt{CSPB.ML.2018} in~\cite{Snoap_Sensors_2023} (see Section 5.4 and Figure~6 in~\cite{Snoap_Sensors_2023} for full details), showing that the proposed DL-based classifier with custom feature extraction layers outperforms the conventional classifier and displays higher classification accuracy for all digital modulation schemes tested.  Figure~\ref{fig:CM_-_CSPB_MSSA_-_GC} also confirms that distinguishing between the three QAM schemes is challenging even for the conventional signal classifier that uses CSP techniques for signal classification.

\begin{figure}
\begin{center}
\subfigure[Confusion Matrix of the \texttt{CSPB.ML.2018} I/Q-BOI-USP-trained Conventional ``Image-Processing'' CAP Classifying \texttt{CSPB.ML.2018} Test Signals.]
{\includegraphics[width=0.49\linewidth]{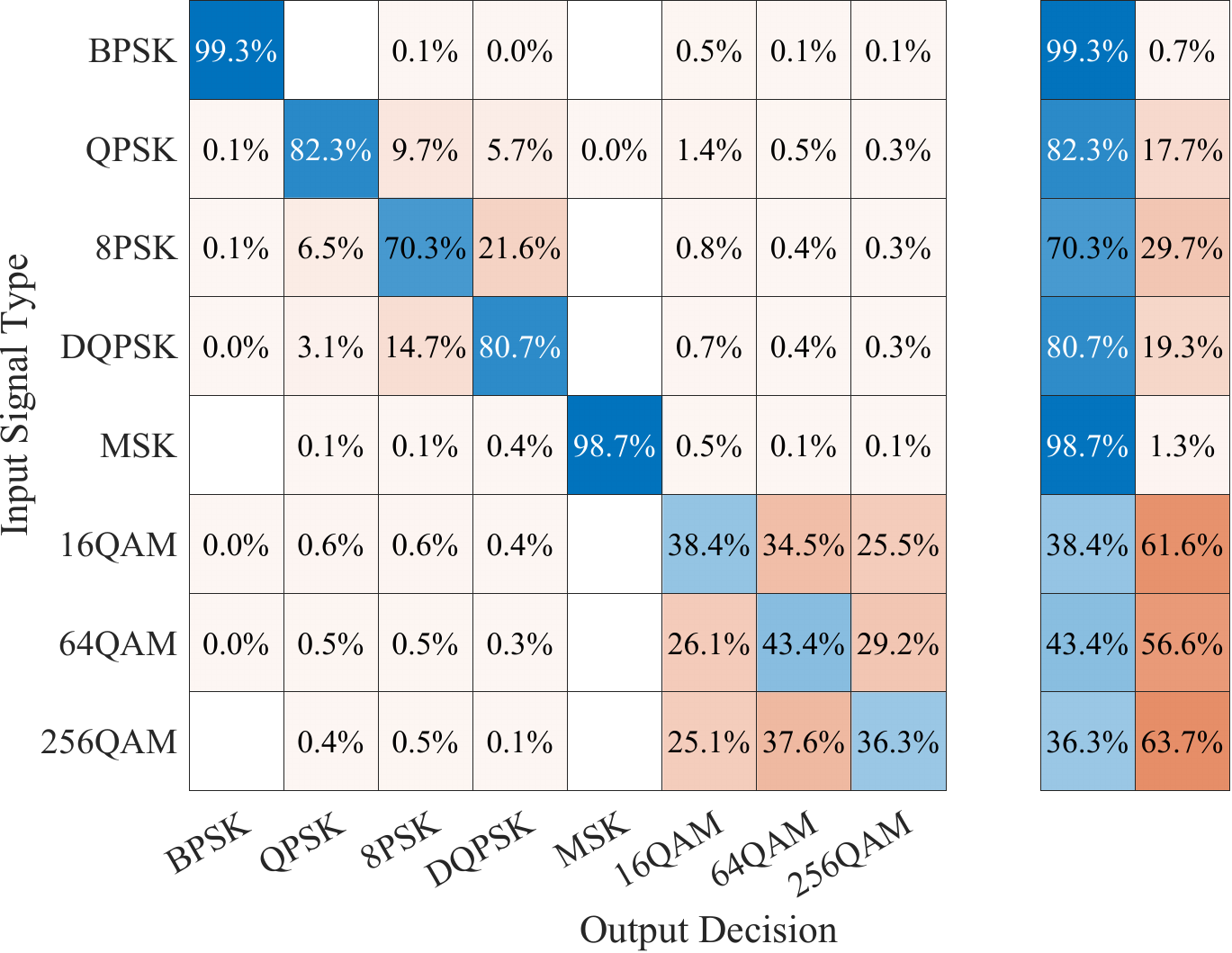}}
\subfigure[Confusion Matrix of the \texttt{CSPB.ML.2018} I/Q-BOI-USP-trained Conventional ``Image-Processing'' CAP Classifying All \texttt{CSPB.ML.2022} Signals.]
{\includegraphics[width=0.49\linewidth]{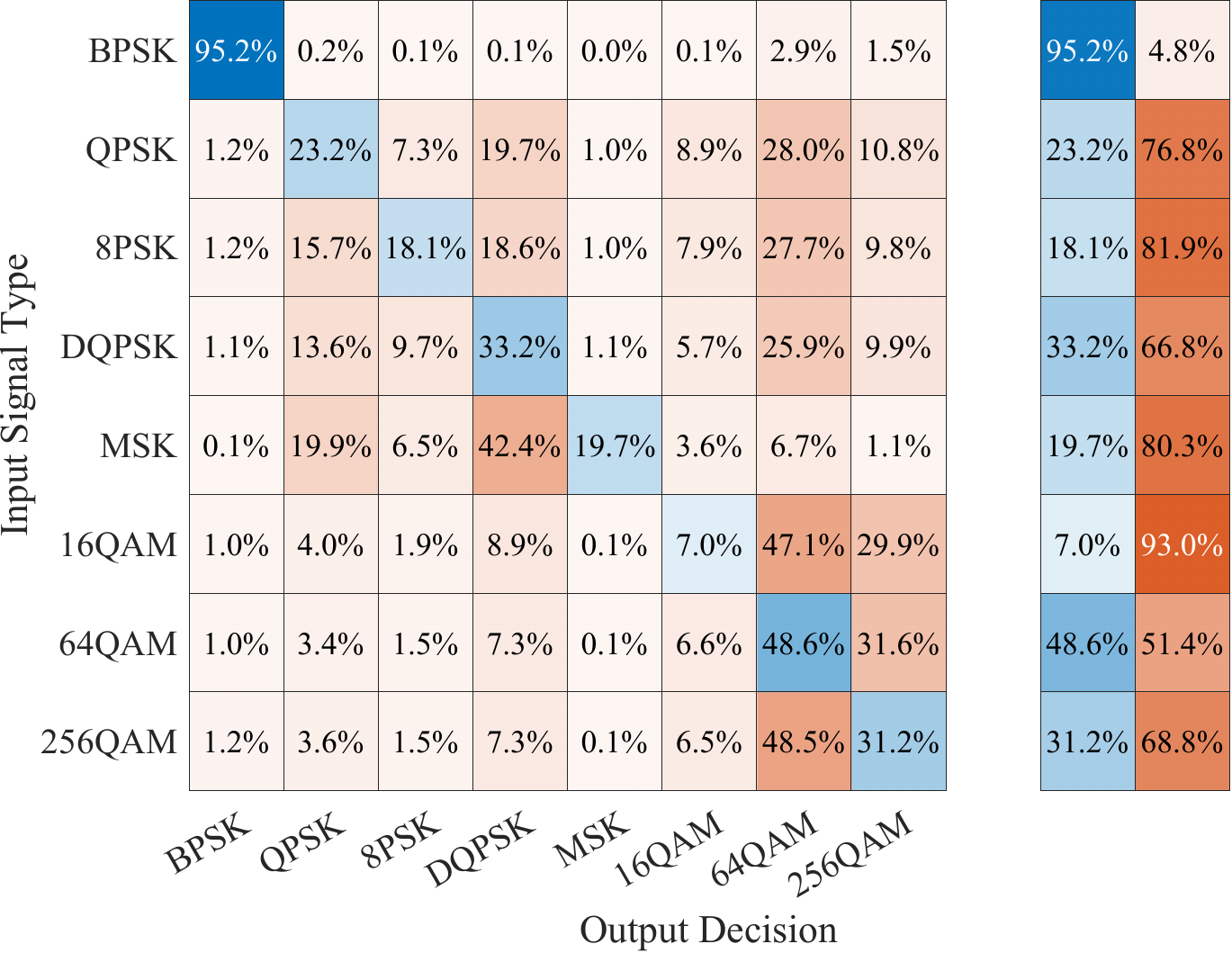}}
\caption[Confusion Matrices of the \texttt{CSPB.ML.2018} I/Q-BOI-USP-trained Conventional ``Image-Processing'' CAP Classifying \texttt{CSPB.ML.2018} Test Signals and All \texttt{CSPB.ML.2022} Signals.]{\textbf{Confusion Matrices of the \texttt{CSPB.ML.2018} I/Q-BOI-USP-trained Conventional ``Image-Processing'' CAP Classifying \texttt{CSPB.ML.2018} Test Signals and All \texttt{CSPB.ML.2022} Signals.}}\label{fig:CM_-_CTest_CTrained_-_IQ-Image-CAP}
\end{center}
\end{figure}

\begin{figure}
\begin{center}
\subfigure[Confusion Matrix of the \texttt{CSPB.ML.2022} I/Q-BOI-USP-trained Conventional ``Image-Processing'' CAP Classifying \texttt{CSPB.ML.2022} Test Signals.]
{\includegraphics[width=0.49\linewidth]{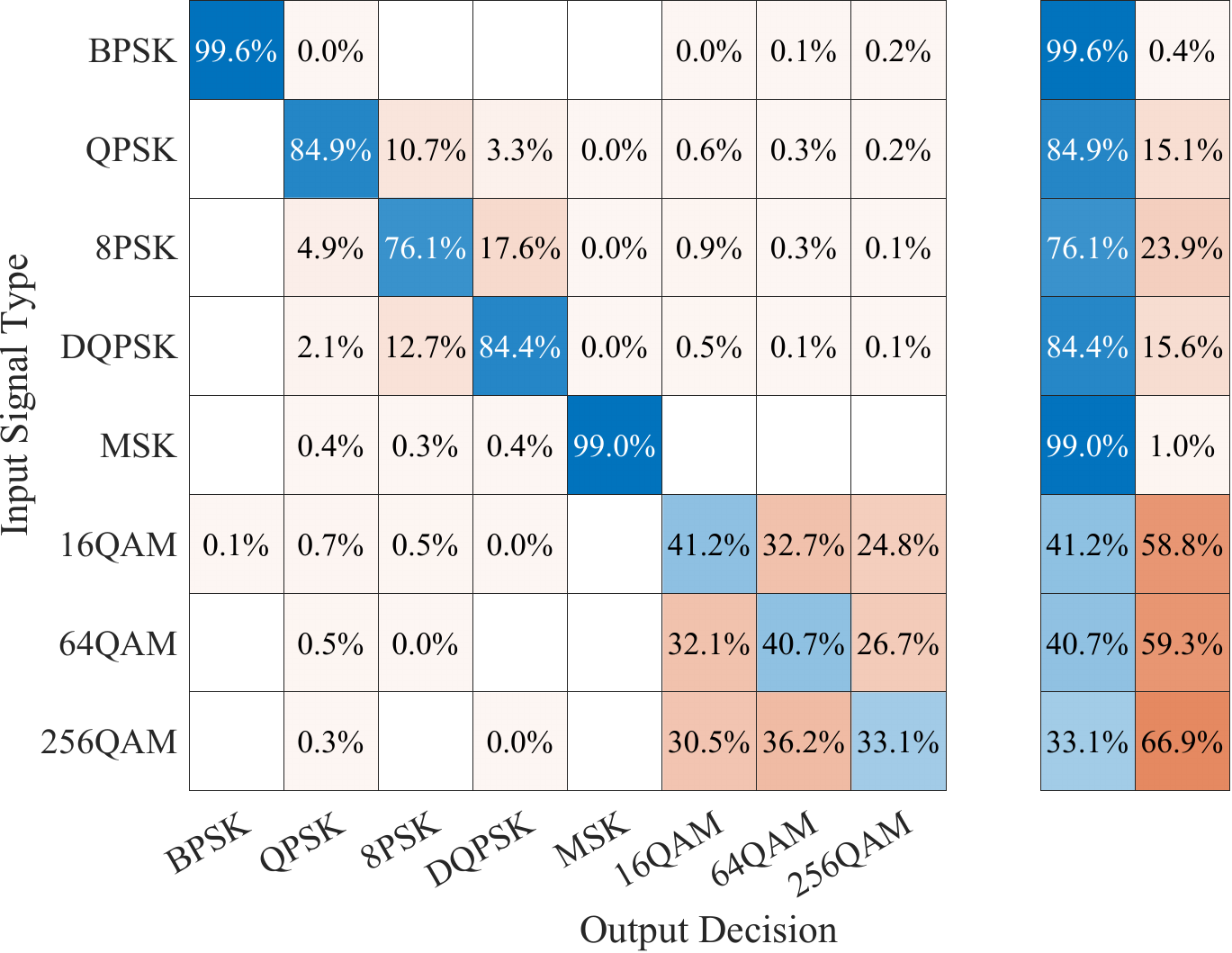}}
\subfigure[Confusion Matrix of the \texttt{CSPB.ML.2022} I/Q-BOI-USP-trained Conventional ``Image-Processing'' CAP Classifying All \texttt{CSPB.ML.2018} Signals.]
{\includegraphics[width=0.49\linewidth]{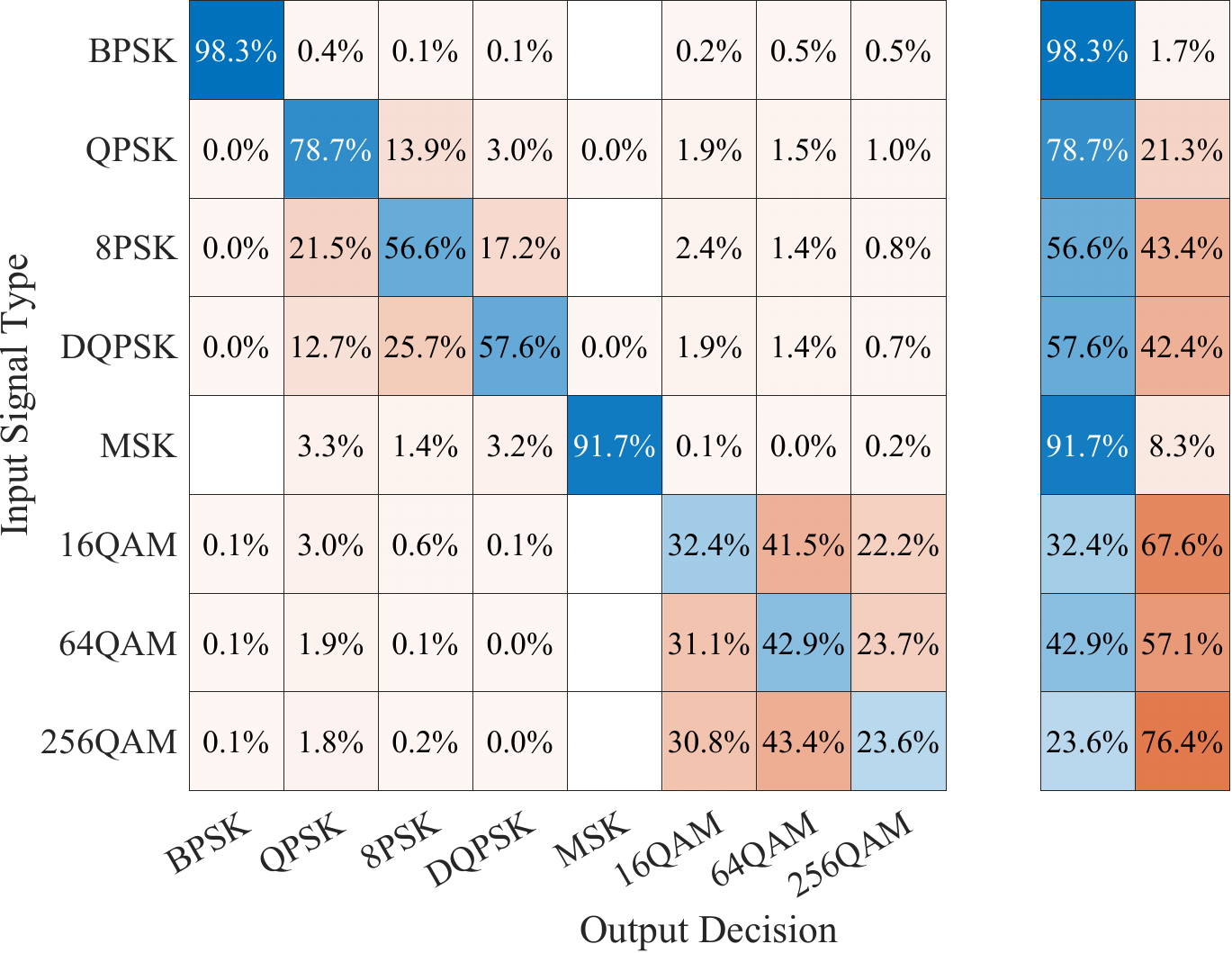}}
\caption[Confusion Matrices of the \texttt{CSPB.ML.2022} I/Q-BOI-USP-trained Conventional ``Image-Processing'' CAP Classifying \texttt{CSPB.ML.2022} Test Signals and All \texttt{CSPB.ML.2018} Signals.]{\textbf{Confusion Matrices of the \texttt{CSPB.ML.2022} I/Q-BOI-USP-trained Conventional ``Image-Processing'' CAP Classifying \texttt{CSPB.ML.2022} Test Signals and All \texttt{CSPB.ML.2018} Signals.}}\label{fig:CM_-_GCTest_CTrained_-_IQ-Image-CAP}
\end{center}
\end{figure}

To conclude the presentation of numerical testing results, the image-processing CAP~\cite{Latshaw_COMM2022} under-performs the CSP baseline model on each modulation type and displays very poor generalization performance, as can be seen from the confusion matrices shown in Figures~\ref{fig:CM_-_CTest_CTrained_-_IQ-Image-CAP} and \ref{fig:CM_-_GCTest_CTrained_-_IQ-Image-CAP}.  Similar confusion matrices to the ones shown in Figure~\ref{fig:CM_-_CTest_CTrained_-_IQ-Image-CAP} have been obtained for the image processing CAP when training was done using \texttt{CSPB.ML.2022}, and these are shown in Figure~\ref{fig:CM_-_GCTest_CTrained_-_IQ-Image-CAP}.

\section{Discussion}\label{sec:Discussion}
The questions probably arise in many readers minds of why CNNs are not able to generalize well, even for slight changes in the probability density functions for the underlying random variables involved in the generation and reception of modulation signals, and why don't these NNs synthesize their own squaring, raise-to-the-power-of-three, quartic, etc., operations if such operations are beneficial to both performance and generalization.  While these questions cannot definitively be answered in this chapter, speculatively, and by way of offering an interpretation of the intersection of modern ML for image recognition and conventional CSP for RF signal analysis, it appears that CNNs are good at, {\em and are designed for,} recognizing images embedded in scenes.  Multiple historical accounts of the birth of CNNs (``ConvNets''), such as Mitchell's  \cite{Mitchell_AI}, describe these computational engines as arising from simplified models of the human eye-brain system for image recognition, a system that is well-known for its remarkably accurate performance.  A key aspect of image recognizers is well-modeled by convolution-based elements such as edge and texture detectors (variants of matched filters).  However the RF modulation-recognition problem is not fundamentally an image-recognition problem, it is a sequence-recognition problem.  There is no subpart of the sequence that is of particular interest--the {\em BPSKness} of a BPSK signal is evenly distributed throughout the whole I/Q sequence.

\begin{figure}
\centering
\includegraphics[width=\linewidth]{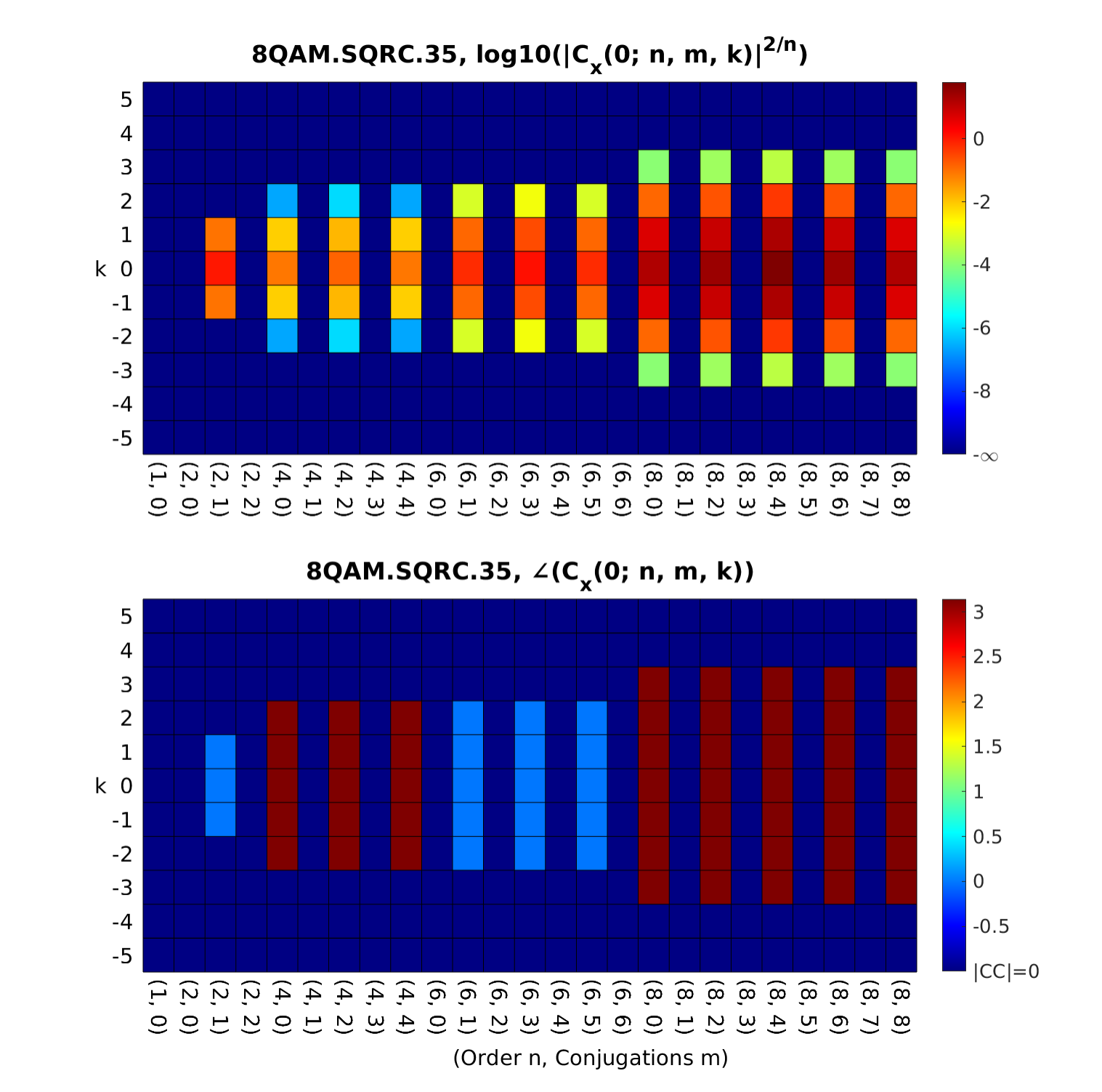}
\caption[Magnitudes and Phases of the CCs for Standard-Constellation 8QAM.]{\textbf{Magnitudes and Phases of the CCs for Standard-Constellation 8QAM.}  This CF Pattern Is QPSK-like.}\label{fig:8QAM_CC}
\end{figure}

\begin{figure}
\centering
\includegraphics[width=\linewidth]{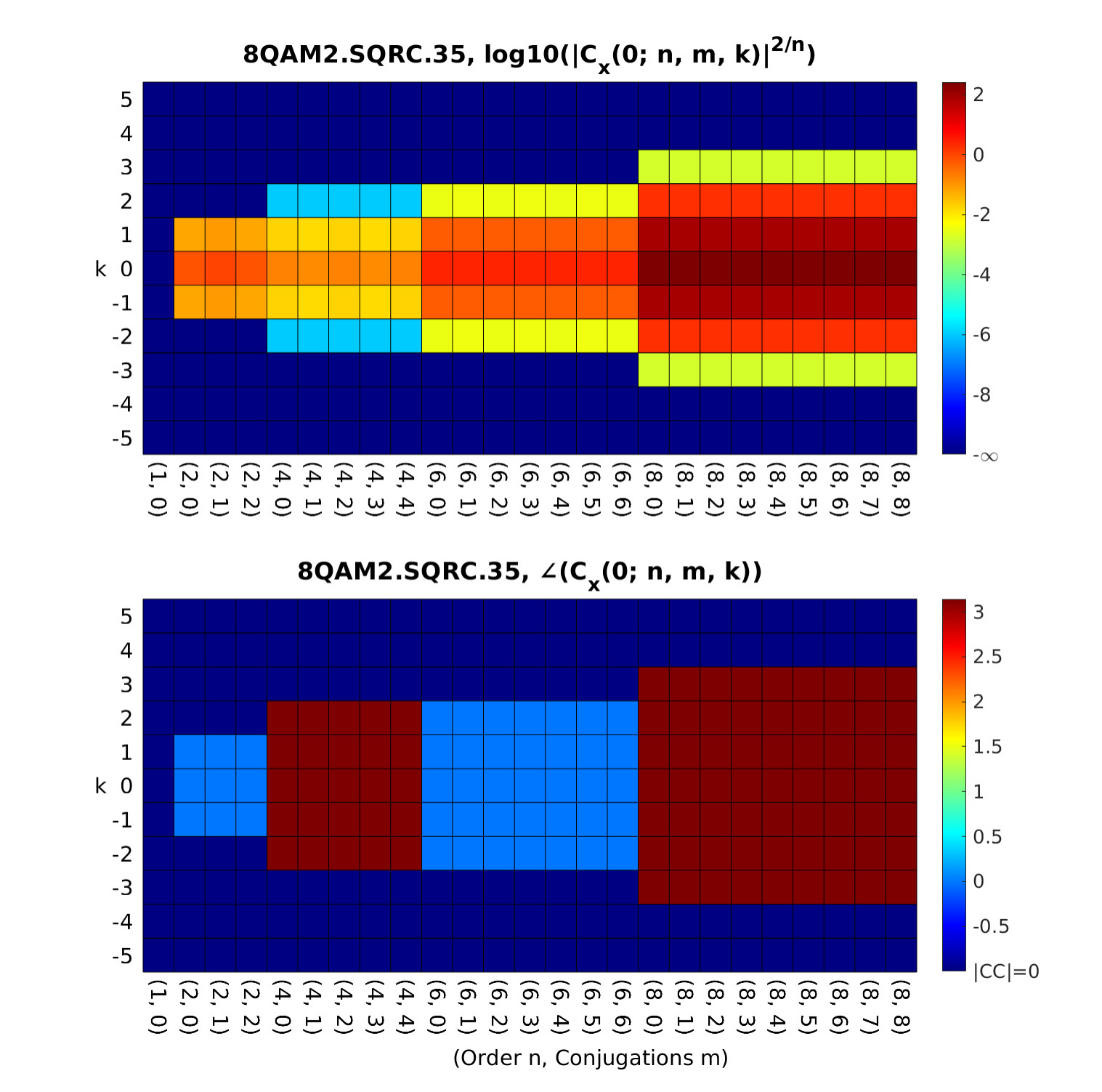}
\caption[Magnitudes and Phases of the CCs for Standard-Constellation 8APSK.]{\textbf{Magnitudes and Phases of the CCs for Standard-Constellation 8APSK.}  This CF Pattern Is BPSK-like.}\label{fig:8APSK_CC}
\end{figure}

\begin{figure}
\centering
\includegraphics[width=\linewidth]{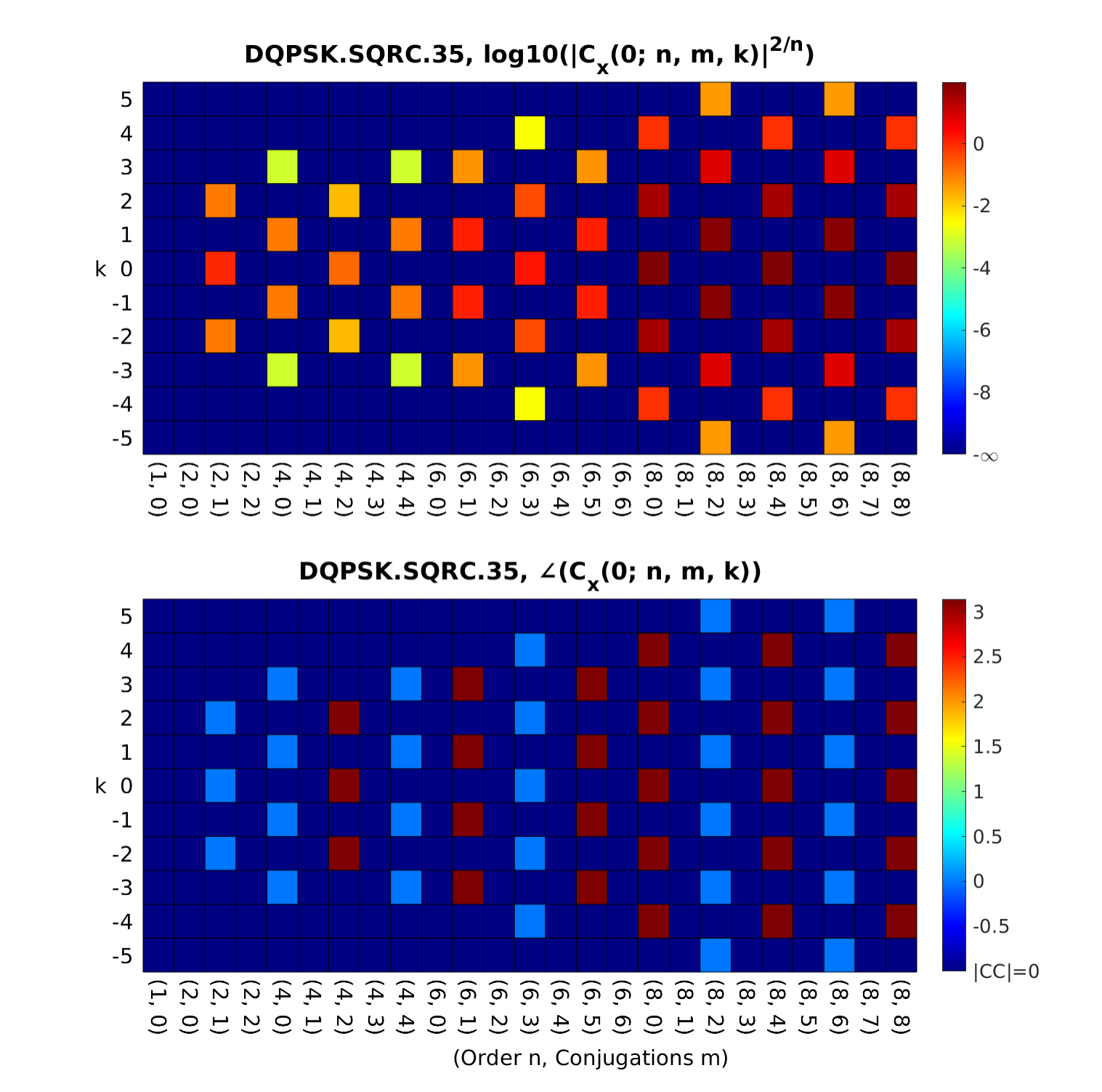}
\caption[Magnitudes and Phases of the CCs for $\pi/4$-DQPSK.]{\textbf{Magnitudes and Phases of the CCs for $\pi/4$-DQPSK.}  This CF Pattern Is DQPSK-like.}\label{fig:DQPSK_CC}
\end{figure}

For example, consider three distinct digital signals with eight points in their  constellations: 8QAM, 8APSK (8QAM2), and $\pi$/4-DQPSK.  These three modulation types have very different CCs, which is a consequence of the probability structure (cumulants) of the symbol random variable.  Here they have exactly the same pulse-shaping filters and all use IID symbol sequences.  A portion of their CC structure is visualized in Figures~\ref{fig:8QAM_CC}--\ref{fig:DQPSK_CC}.

\begin{figure}
\centering
\includegraphics[width=\linewidth]{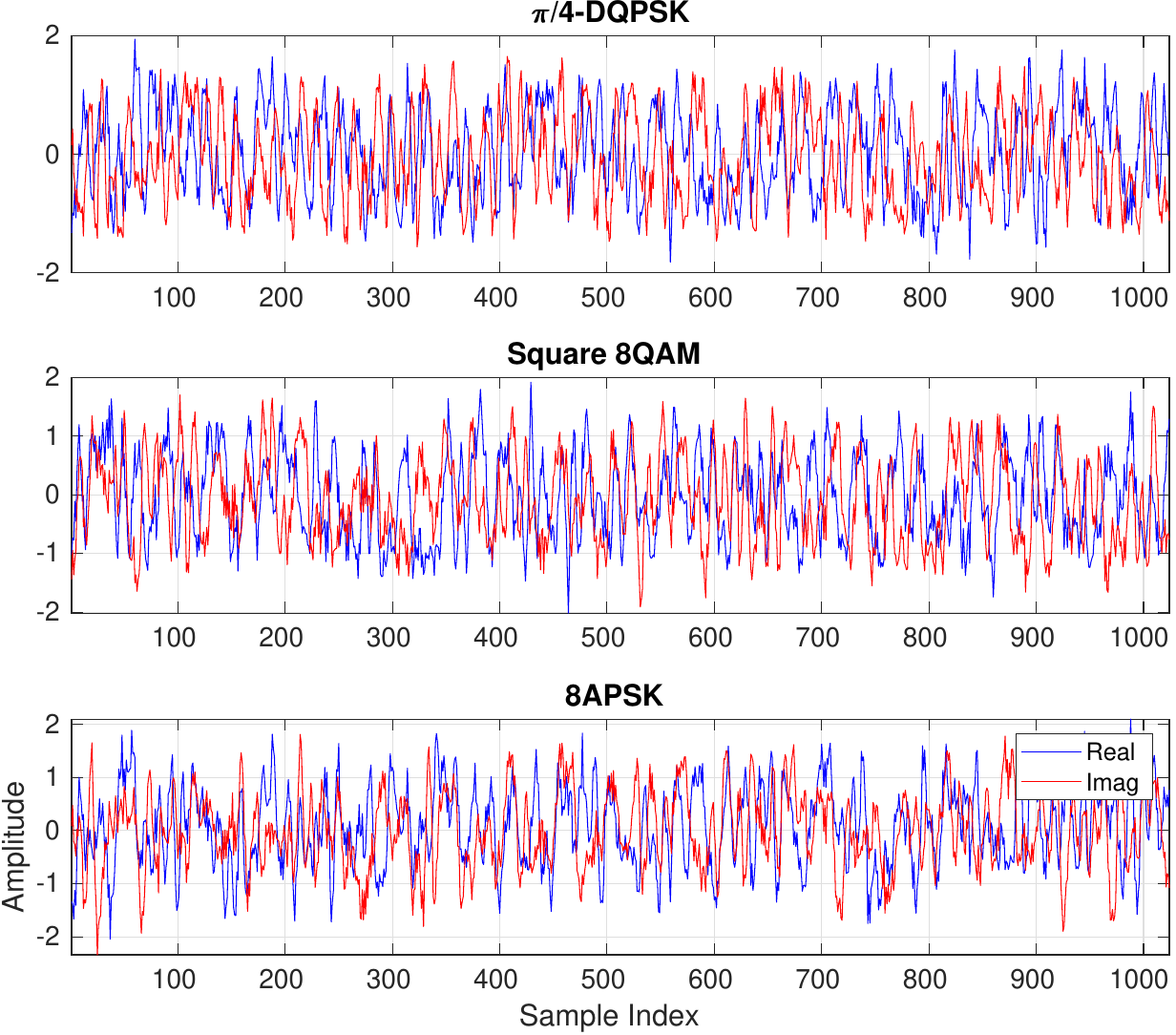}
\caption[In-Phase and Quadrature Components of the Three Eight-Constellation-Point Digital Signals With CCs Shown in Figures~\ref{fig:8QAM_CC}--\ref{fig:DQPSK_CC}.]{\textbf{In-Phase and Quadrature Components of the Three Eight-Constellation-Point Digital Signals With CCs Shown in Figures~\ref{fig:8QAM_CC}--\ref{fig:DQPSK_CC}.}}\label{fig:IQ_for_three_8pt_constellations}
\end{figure}

\begin{figure}
\centering
\includegraphics[width=\linewidth]{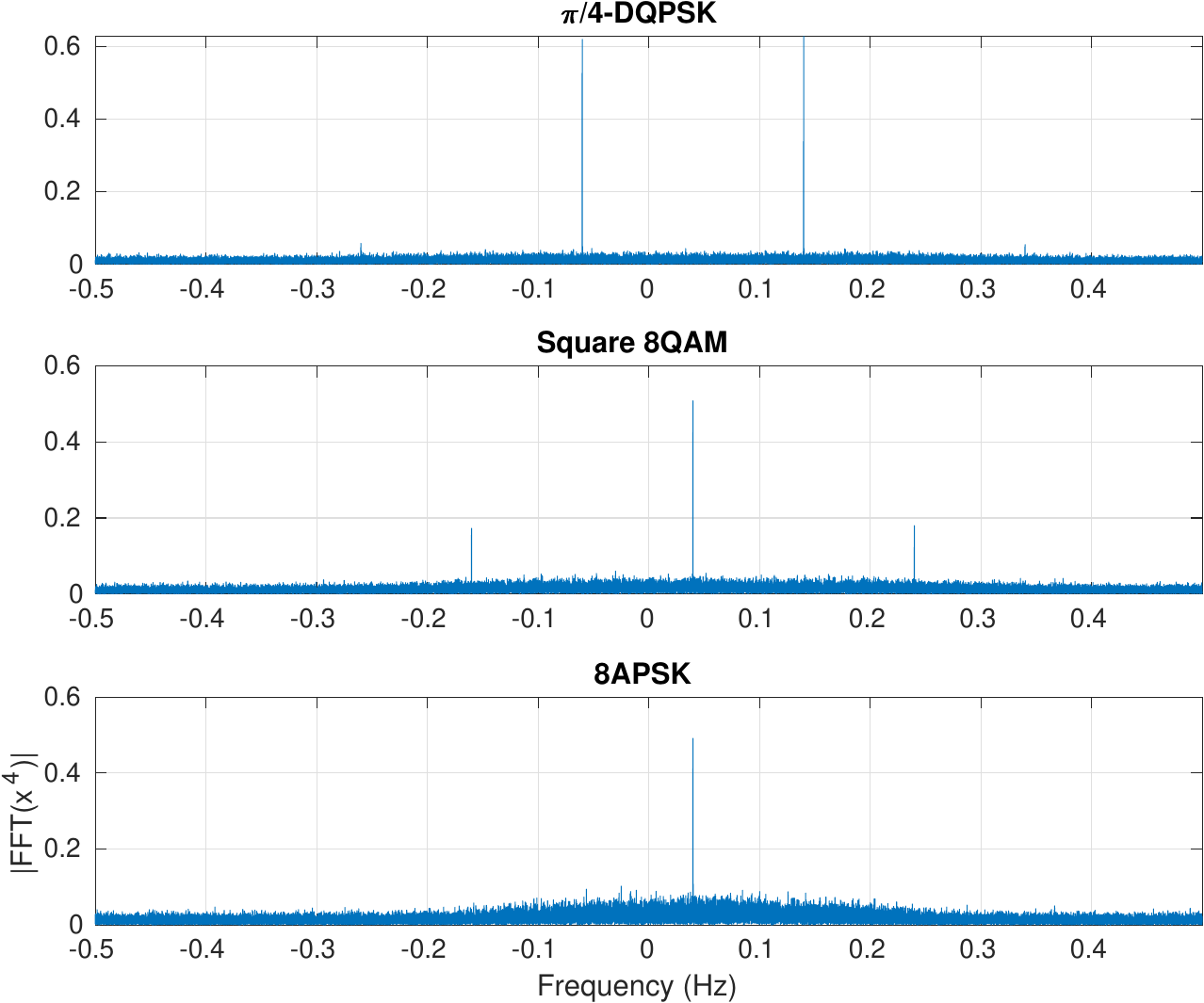}
\caption[Fourier Transforms of Nonlinear Functions of the Three Eight-Constellation-Point Digital Signals With CCs Shown in Figures~\ref{fig:8QAM_CC}--\ref{fig:DQPSK_CC}.]{\textbf{Fourier Transforms of Nonlinear Functions of the Three Eight-Constellation-Point Digital Signals With CCs Shown in Figures~\ref{fig:8QAM_CC}--\ref{fig:DQPSK_CC}.}}\label{fig:FFT_x4_for_three_8pt_constellations}
\end{figure}

The CF patterns in Figures \ref{fig:8QAM_CC}--\ref{fig:DQPSK_CC} are easy for humans to distinguish.  Contrast those figures with the plots of the I/Q data, which is the typical input to a CNN in the literature, as shown in Figure~\ref{fig:IQ_for_three_8pt_constellations}.  It is shown in \cite{Snoap_Sensors_2023}, and in Chapters~\ref{ch:CSPModClass}--\ref{ch:CCsNNsModClass}, that training CNNs with features such as those in Figures~\ref{fig:8QAM_CC}--\ref{fig:DQPSK_CC} leads to both excellent performance and generalization, but at substantial CSP cost.  In the present chapter, the desire is to obtain that same performance without the CSP steps, and with minimum network-layer complexity.  To do that, cyclic-moment {\em proxies} are used for the CC features.  Cyclic moments can be approximated by simply raising the signal to some even power, Fourier transforming, and finding the complex values of the peaks.  As an illustration, Figure~\ref{fig:FFT_x4_for_three_8pt_constellations} shows these Fourier transforms.  Thus a valid interpretation of the proposed novel layers is that they convert a sequence-classification problem based on Figure~\ref{fig:IQ_for_three_8pt_constellations}, which is difficult for CNNs, into an image-classification based on Figure~\ref{fig:FFT_x4_for_three_8pt_constellations}, which is known to be much easier for CNNs.

\section{Conclusions}\label{sec:Conclusion}
This chapter presents a novel DL-based classifier for digitally modulated signals that uses a capsule network (CAP) with custom feature-extraction layers.  The proposed CAP takes as input pre-processed I/Q data, on which blind band-of-interest (BOI) estimation has been applied to enable out-of-band noise filtering, CFO-estimate-based spectral centering, and USP normalization.  The proposed custom function layers are essentially homogeneous even-order nonlinear functions and perform feature generation and extraction on the pre-processed I/Q data more reliably and predictably than conventional CNN layers alone, resulting in very good classification and generalization performance for the eight digital modulation schemes commonly used in practical systems, which have been studied in this dissertation (BPSK, QPSK, 8PSK, $\pi/4$-DQPSK, MSK, 16QAM, 64QAM, and 256QAM).

To distinguish additional modulation schemes, future research may consider the inclusion of NN function layers that extract features proportional to non-conjugate CCs ($n = 2m$) and conjugate CCs, where the number of conjugations is greater than zero ($n \neq 2m$ and $m \neq 0$), as this may prove necessary to recognize modulation schemes such as FSK, CPM, or CPFSK, which can have little or no cyclostationarity for $n \neq 2m$.

\chapter{Conclusions}\label{sec:Con}
This dissertation presents several novel deep-learning (DL)-based classifiers for digitally modulated signals, all of which take advantage of capsule networks.  The first proposed classifier, presented in Chapter~\ref{ch:CCsNNsModClass}, takes blindly estimated cyclic cumulant (CC) features, described in Chapter~\ref{ch:CSPModClass}, as inputs, and it outperforms alternative classifiers for digitally modulated signals.  This first proposed classifier has very good generalization abilities, displaying high classification accuracy when tested with signals coming from datasets that have different characteristics from the signals in the training dataset.  Its generalization abilities are also superior to other DL-based classifiers which use in-phase and quadrature (I/Q) signal data for training and modulation classification.  However, this classifier requires highly accurate center frequency offset (CFO) and symbol rate estimates from the I/Q data in order for the CC estimates to be accurate enough to enable this high degree of generalization and the signal processing required to obtain these highly accurate estimates is computationally expensive.

The second proposed classifier, presented in Chapter~\ref{ch:NovelLayers}, takes blind band-of-interest (BOI) filtered, coarse CFO estimate spectral centered, and normalized to unit total power (UTP) I/Q data as inputs, and it improves upon the generalization abilities of DL-based classifiers which use I/Q data by employing novel neural network layers that raise the input data to $2^{\mbox{\footnotesize nd}}$, $4^{\mbox{\footnotesize th}}$, $6^{\mbox{\footnotesize th}}$, and $8^{\mbox{\footnotesize th}}$ orders in both temporal and spectral domains before the capsule network branches further extract information to perform modulation classification.  This classifier implements novel neural network layers for UTP I/Q data, described in Section~\ref{sec:Details}, to extract cyclic moment proxies in the neural network and enable the downstream capsules to train in a manner that more reliably generalizes to new unseen data.  However, due to the normalization to UTP, this method is unable to generalize what it has learned for the QAM modulations because the main distinction between the QAM modulations is their power level, and normalizing to unit total (signal plus noise) power removes this distinction.

The third proposed classifier, also presented in Chapter~\ref{ch:NovelLayers}, takes BOI filtered, coarse CFO estimate spectral centered, and normalized to unit signal power (USP) I/Q data as inputs, and it improves upon the generalization abilities of DL-based classifiers that use I/Q data by employing novel neural network layers that raise the input data to $2^{\mbox{\footnotesize nd}}$, $4^{\mbox{\footnotesize th}}$, and $6^{\mbox{\footnotesize th}}$ orders in both temporal and spectral domains before the capsule network branches further extract information to perform modulation classification.  This classifier implements novel neural network layers for both UTP and USP I/Q data, described in Section~\ref{sec:Details2}, to extract cyclic moment proxies in the neural network and enable the downstream capsules to train in a manner that more reliably generalizes to new unseen data.  This third proposed classifier outperforms the CSP baseline model, described in Chapter~\ref{ch:CSPModClass}, and does not require the computationally expensive estimation of the CFO, the symbol rate, or the CCs as the first proposed classifier does.

Several logical steps forward include further improving upon these capsule network digital modulation classifiers such that they can classify more modulation types and so that they can be deployed to devices with fewer resources, such as a Field Programmable Gate Array (FPGA) with limited random-access memory (RAM).  Several ideas that can be pursued are presented in the next section.

\section{Suggestions for Further Research}\label{sec:SFFR}
To distinguish additional modulation schemes, future research may consider the inclusion of neural network function layers that extract features proportional to non-conjugate CCs ($n = 2m$) and conjugate CCs, where the number of conjugations is greater than zero ($n \neq 2m$ and $m \neq 0$), as this may prove necessary to recognize modulation schemes such as FSK, CPM, or CPFSK, which can have little or no cyclostationarity for $n \neq 2m$.

To reduce the total number of learnable parameters needed in the I/Q-BOI-UTP-trained or I/Q-BOI-USP-trained CAPs and to reduce how much RAM is needed to perform classification, techniques such as the signature method~\cite{chevyrev2016primer, lyons2014rough, SignatureMethod_1, reizenstein2018iisignature}, the discrete signature method~\cite{adachi2022discrete, bellingeri2023discrete}, or the randomized signature method~\cite{compagnoni2023effectiveness_IEEE, compagnoni2023effectiveness} might be employed as a final layer prior to the start of each capsule branch, as this would reduce the total number of learnable parameters at the start of each branch from a size of $32,768$ to an arbitrary length from a truncated signature, perhaps $165$ to imitate the number of CCs, yet unique distinctions on the data of each branch could still be retained.

Employing the signature method after the I/Q-BOI-USP input has been raised to the even powers, Fourier transformed for the spectral branches, and warped for the USP branches, should maintain the feature sets that contain the cyclic moment proxies, yet reduce both the training time and the final RAM and computational footprint required by the neutral network.  This reduction in overall size for the neutral network could allow it to be implemented on resources like FPGAs, where processing power and memory are significantly more restricted than a modern personal computer with a dedicated GPU, let alone a high-performance computational cluster.


\bibliographystyle{IEEEtran}
\renewcommand\bibname{\uppercase{Bibliography}}
\cleardoublepage
\phantomsection
\addcontentsline{toc}{chapter}{\uppercase{Bibliography}}
\bibliography{./bib/refs}
%
%
%

\vitapage

\end{document}